\documentclass[10pt,epsfig]{article}

\usepackage{enumerate}
\usepackage{color}
\usepackage[table]{xcolor}
\usepackage{slashed}
\usepackage{cite}

\let\counterwithin\relax
\usepackage{chngcntr}
\usepackage{amssymb, amsmath,mathrsfs}
\usepackage{longtable}
\usepackage[figuresright]{rotating}
\usepackage{graphics}
\usepackage{graphicx}
\usepackage{subcaption}
\usepackage{epsf}
\usepackage{epsfig}
\usepackage{float}
\usepackage{makecell}
\usepackage{multirow}
\usepackage{pifont}
\usepackage{color}
\usepackage{xcolor}
\usepackage{simplewick}
\usepackage{adjustbox}
\usepackage[utf8]{inputenc}
\usepackage{amsmath}
\usepackage{amssymb}
\usepackage{arydshln}
\usepackage[normalem]{ulem}

\usepackage{feynmf}

\usepackage{mathtools}
\usepackage{amsmath}
\usepackage{diagbox}

\newcommand\undermat[2]{
	\makebox[0.5pt][l]{$\smash{\underbrace{\phantom{%
					\begin{matrix}#2\end{matrix}}}_{ \let\scriptstyle\textstyle\text{\large $#1$}}}$}#2}
\newcommand\overmat[2]{
	\makebox[-1pt][l]{$\smash{\overbrace{\phantom{%
					\begin{matrix}#2\end{matrix}}}^{ \let\scriptstyle\textstyle\text{\large $#1$}}}$}#2}    
\usepackage{tikz}
\usepackage[vcentermath]{youngtab}
\usepackage{slashed} 
\usepackage[font=small]{caption} 
\usepackage{times} 

\usepackage{bm}       
\usepackage{bbm} 

\usepackage{relsize}  

\usepackage{autobreak}  

\usepackage{makeidx} 
\usepackage{bbding} 
\usepackage{listings} 
\usepackage{ytableau}

\usepackage[colorlinks=true,
linkcolor=blue,
urlcolor=red,
citecolor=red]{hyperref}

\long\def\rpl#1!!#2!!{\textcolor{red}{#1} \textcolor{blue}{#2}}

\usepackage[top=1in, left=0.95in, bottom=1.1in, right=0.95in]{geometry}
\def\baselinestretch{1.27}
\usepackage[toc]{appendix}

\newcommand{\brkt}[1]{ \{#1\} }

\newcommand{\beq}{\begin {equation}}  
\newcommand{\eeq}{\end   {equation}} 
\newcommand{\bea}{\begin {eqnarray}} 
\newcommand{\eea}{\end   {eqnarray}}  
\newcommand{\baa}{\begin {array}   } 
\newcommand{\eaa}{\end   {array}   }     
\newcommand{\bit}{\begin {itemize} }
\newcommand{\eit}{\end   {itemize} }
\newcommand{\bet}{\begin {enumerate} }
\newcommand{\eet}{\end   {enumerate} }
\newcommand{\be }{\begin {equation}} 
\newcommand{\ee }{\end   {equation}}

\newcommand{\mc}[1]{\mathcal{#1}}
\newcommand{\mbf}[1]{\mathbf{#1}}

\newcommand{\ket}[1]{| #1 \rangle}
\newcommand{\bra}[1]{\langle #1 |}
\newcommand{\vev}[1]{ \left\langle {#1}  \right\rangle }
\newcommand{\inner}[2]{\langle #1 | #2 \rangle}

\newcommand{\eq}[1]{\begin{equation}\begin{split} #1 \end{split}\end{equation}}

\newcommand{\comment}[1]{}

\newcolumntype{M}[1]{>{\centering\arraybackslash}m{#1}}
\newcolumntype{N}{@{}m{0pt}@{}}

\newcommand{\hl}[1]{{\color{black} #1}}

\allowdisplaybreaks
\begin{document}

\begin{center}

{\Large \textbf  {The Bottom-Up EFT: Complete UV Resonances of the SMEFT Operators}}\\[10mm] 

Hao-Lin Li$^{a,c}$\footnote{haolin.li@uclouvain.be}, Yu-Han Ni$^{a, b}$\footnote{niyuhan@itp.ac.cn}, Ming-Lei Xiao$^{a,d,e}$\footnote{minglei.xiao@northwestern.edu}, Jiang-Hao Yu$^{a, b, f, g, h}$\footnote{jhyu@itp.ac.cn}\\[10mm]

\noindent 
$^a${\em \small CAS Key Laboratory of Theoretical Physics, Institute of Theoretical Physics, Chinese Academy of Sciences,    \\ Beijing 100190, P. R. China}  \\
$^b${\em \small School of Physical Sciences, University of Chinese Academy of Sciences,   Beijing 100049, P.R. China}   \\
$^c${\em \small Centre for Cosmology, Particle Physics and Phenomenology (CP3), Universite Catholique de Louvain,\\
Chem. du Cyclotron 2, 1348, Louvain-la-neuve, Belgium}\\
$^d${\em \small Department of Physics and Astronomy, Northwestern University, Evanston, Illinois 60208, USA}\\
$^e${\em \small High Energy Physics Division, Argonne National Laboratory, Lemont, Illinois 60439, USA}\\
$^f${\em \small Center for High Energy Physics, Peking University, Beijing 100871, China} \\
$^g${\em \small School of Fundamental Physics and Mathematical Sciences, Hangzhou Institute for Advanced Study, UCAS, \\ Hangzhou 310024, China} \\
$^h${\em \small International Centre for Theoretical Physics Asia-Pacific, Beijing/Hangzhou, China}\\[10mm]

\date{\today}   
          
\end{center}

\begin{abstract} 

The standard model effective field theory (SMEFT) provides systematic parameterization of all possible new physics above the electroweak scale. According to the amplitude-operator correspondence, an effective operator can be decomposed into a linear combination of several j-basis operators, which correspond to local amplitudes carrying certain spin and gauge quantum numbers in a particular scattering channel. Based on the Poincare and gauge symmetries of scattering amplitude, we construct the j-basis using the Casimir method for both the Lorentz and gauge sectors. The quantum numbers of the j-basis operators fix the quantum numbers of any intermediate state in the corresponding amplitudes, such as a UV resonance. This can be re-interpreted as the j-basis/UV correspondence, thus obtaining the j-bases in all partitions of fields for an operator amounts to finding {\it all} of its UV origins at tree level, constituting the central part of the bottom-up EFT framework. Applying the j-basis analysis to SMEFT, we obtain a complete list of possible tree-level UV origins of the effective operators at the dimension 5, 6, 7, and all the bosonic operators at dimension 8.

\end{abstract}

\newpage

\setcounter{tocdepth}{4}
\setcounter{secnumdepth}{4}

\tableofcontents

\setcounter{footnote}{0}

\def\baselinestretch{1.5}
\counterwithin{equation}{section}

\newpage

\section{Introduction: Top-down EFT vs Bottom-up EFT}
\label{sec:intro}

The standard model (SM) of particle physics has been established as a successful microscopic theory which explains almost all experimental results and precisely predicted a wide variety of phenomena at the sub-atomic scale. Despite its success, there are important questions that it does not answer, such as origin of the neutrino masses, the matter-antimatter asymmetry, and the nature of the dark matter, etc. There are different  ways to search for new physics beyond the standard model (BSM). During the past decades, various new physics models, such as new symmetries (supersymmetry), new interactions (gauge unification), or new spatial dimensions (extra dimension), were proposed to solve theoretical problems in the SM. The BSM searches at the large hadron collider (LHC) were mostly guided by new particle signatures predicted by these well-motivated models, but no signal has been found. Due to the complexity of the mass spectra and interactions in these new physics models, simplified model has been utilized to investigate a class of new physics models in which only the relevant new particles and interactions are retained and a universal but simple Lagrangian could characterize a key feature in various new physics models, named as {\it the new physics motivated simplified model}~\cite{LHCNewPhysicsWorkingGroup:2011mji, Alwall:2008ag}. Nowadays, searches for both new physics models and simplified models push the new physics scale up to TeV or several TeV, increasing the hierarchy between the electroweak scale and the new physics scale. Therefore, assuming new particles are too heavy to be directly produced at the LHC, it is possible to integrate out these heavy particles and obtain the effective Lagrangian at the electroweak scale up to certain order in the expansion of heavy masses. In this sense, the effective field theory (EFT) approach is quite suitable to capture new physics effects induced by these new particles based on the degrees of freedom at the electroweak scale. Recent years, the EFT approach attracts more and more attention and would take centre stage of the high energy theoretical and experimental programs over the coming decades.

The EFT framework provides a systematic parametrization of various kinds of ultraviolet (UV) physics using only the low energy field degrees of freedom with the proper power counting rules. Pioneered by Weinberg~\cite{Weinberg:1978kz}, starting from the field degrees of freedom at low energy scale, one writes down the most general Lagrangian, including {\it all} terms consistent with presumed symmetry principle. Applying to the electroweak scale physics, we should start from the fields and symmetries of the SM, and write down all possible operators order by order according to the canonical dimension power counting, which is the so-called standard model effective field theory (SMEFT). Since Weinberg wrote down the dimension 5 operator~\cite{Weinberg:1979sa}, lots of progress has been made on writing down complete and independent operators up to the mass dimension 9~\cite{Weinberg:1979sa, Buchmuller:1985jz, Grzadkowski:2010es, Lehman:2014jma, Liao:2016hru, Li:2020gnx, Murphy:2020rsh, Li:2020xlh, Liao:2020jmn}. Recently a general algorithm, implemented in a Mathematica package ABC4EFT~\cite{Li:2022tec}, has been proposed to construct the independent and complete SMEFT operator bases up to any mass dimension.

Given the complete set of the SMEFT operators, the centre task would be to determine the Wilson coefficients of these effective operators that parametrize the deviations from the SM using the existing experimental data in both the low energy precision experiments and the high energy colliders~\cite{Brivio:2017vri}. If the experimental result agrees with the SM prediction within statistical errors, it indicates no evidence of new physics and thus provides constraints on the relevant Wilson coefficients. For example, the LHC Run III could constrain these Wilson coefficients at the mass dimension 6 and above using the total cross section and tails of various distributions in various final states at the LHC. On the other hand, if the experimental data show significant deviation (anomaly) from the SM prediction, one can identify the relevant SMEFT operators causing such deviation.  Once the Wilson coefficients of the relevant operators are measured, the more important task would be to find the UV origin of such effective operators~\cite{Bechtle:2022tck}. 

%
To find the UV origin of an effective operator, a straightforward way is usually taken:  choose one well-motivated benchmark UV model, then perform a matching  procedure to know whether the operator could be generated. If a UV model could match to such an operator, then this UV model should be counted as one of the UV origins.
This procedure has been well-established in the path-integral formalism as the top-down approach~\cite{Georgi:1993mps, Skiba:2010xn}. A simple but typical example for this ``top-down'' approach is the three types of seesaw models~\cite{Yanagida:1979as, Gell-Mann:1979vob, Mohapatra:1979ia, Magg:1980ut, Schechter:1980gr, Foot:1988aq}. For example, in the type-I seesaw model, a heavy Majorana fermion is introduced, and integrating it out would generate the famous Weinberg operator at the dimension 5~\cite{Weinberg:1979sa}. From this matching procedure, one can conclude that one possible UV origin of the Weinberg operator would be a heavy Majorana fermion. However, one would not know whether there are other kinds of new particles generating the Weinberg operator. To find other kinds of UV origins, one has to explore other kinds of UV benchmark model and perform matching to verify whether the UV is feasible or not. Since the procedure is based on the top-down EFT, we call this method of finding UV origins {\it the top-down model approach}. This approach can only tell us whether the existing UV models could generate certain effective operator or not, and thus it is likely that the UV origin search is incomplete.

In order to obtain all possible UV origins, another more generic and unbiased approach has been proposed: trying to find all possible UV models generating the effective operator. In this case, starting from the possible topologies of an effective operator\footnote{A topology of the operator is represented by a diagram of which external lines are one-to-one corresponding to the fields in the operator, and internal lines represent the heavy UV resonances. }, one needs to perform an exhaustive search on various kinds of the UV Lagrangian based on the vertices in each topology, and pick up the UV models matching onto such operator. Here the explicit UV Lagrangian and the matching procedure are still needed, thus we call this method of finding all possible UV origin {\it the top-down exhaustive search approach}. Recent years, some efforts have been made along this direction towards finding the UV origins in the SMEFT framework: classification of the tree-level and loop-level generated operators~\cite{Arzt:1994gp, Einhorn:2013kja, Gargalionis:2020xvt, DasBakshi:2021xbl}, the complete tree-level dictionary for the dimension 6 operators~\cite{deBlas:2017xtg}, etc. Furthermore, there has also been some efforts towards the heavy resonance description in the chiral perturbation theory~\cite{Ecker:1988te, Ecker:1989yg} and the Higgs effective field theory~\cite{Pich:2016lew, Krause:2018cwe, Pich:2020xzo}. All of these works are trying to find all possible UV origins of the effective operators. Among these, let us lay out the procedures of the exhaustive search for the tree-level UV origins: 

\noindent\underline{\bf The procedure of the top-down exhaustive search:} 
\bet
\item Unfold an effective operator: draw all possible topologies of the UV Feynman diagrams with external lines appearing in the operator of interest, and internal lines representing heavy fields with assumed spins.

\item UV selection: scan over all possible gauge quantum numbers of the internal fields, select the ones allowed by gauge symmetries. In this way, one obtains various simplified models that potentially contribute to the operators. Here we call them {\it the EFT motivated simplified models} to distinguish from the new physics motivated simplified models.  


\item UV matching check: for each EFT motivated simplified model, one integrates out the heavy particles and check whether the resulting EFT indeed contains such operator or not.


\eet
In this approach, the allowed UV couplings and the heavy particles with various spins and gauge representations should be exhausted, which is a tedious and prone-to-error task. It is possible that the UV selection itself cannot judge some higher spin UV particles allowed or not. For example, for the dimension-5 Weinberg operator, the allowed gauge symmetry on the BSM Lagrangian tells us the gauge quantum numbers of the UV at the tree-level should be $SU(2)$ singlet or triplet, but it does not tell us the allowed spin quantum numbers of the heavy particles. For example, we do not know whether the spin-3/2 and spin-1 UV resonances are allowed or not. Therefore, the matching check is needed to eliminate the over-counted UV particles. For each UV model from the exhaustive search, performing the top-down matching is also quite tedious and converting the matching operators from an over-complete bases to the independent bases is also a painful process. On the other hand, finding all possible UV origin in a bottom-up approach would avoid such exhaustive search on the UV Lagrangian and the matching procedure.

In this work, we propose a bottom-up approach: without writing down the UV Lagrangian of the simplified model, we can still find all the possible tree-level UV origins (in other words, the UV resonances~\footnote{Hereafter, when we say {\it the UV resonance}, we really mean the tree-level UV particle, because only the tree-level amplitude with heavy immediate particle could behave as a resonance in high energy process. }) of the operators by simply examining the Lorentz and gauge quantum numbers of the local on-shell amplitude they generate. 
This is based on that the local amplitude at IR obtained by taking the low energy limit of the UV amplitude inherits the quantum number of the UV resonance that generates it. This can be seen from the fact that this limiting procedure keeps the structure of the numerator part of the heavy particle propagator unchanged, which determines the quantum numbers of the multi-particle states formed by particles connected on each side of the propagator.
%
The typical example is the traditional ``partial wave amplitude" that determines the angular momentum $J$ for the multi-particle state in a specific scattering channel,  then using the amplitude-operator correspondence one can translate these amplitudes to the corresponding operators which may be obtained by integrating out some heavy field of spin $J$ in the UV theory.
In order to investigate all kinds of UV origins that contribute to a particular operator, we should generalize the concept of partial wave amplitudes~\cite{Jacob:1959at} from the traditional $2\to 2$ or $2\to n$ scattering to $m\to n$ scattering and even multi-partite scattering\footnote{Multi-partite means that we divided the particles involving in a scattering into multiple sets, and each set has specific quantum numbers. See the definition around Eq.~\eqref{eq:non-ovl-ch} for detail. }. 
For each partition\footnote{ A partition represents a way of dividing the particles in the amplitudes (or equivalently the fields in the operator) into different sets. For example for a 6-pt amplitude, we can divided the particles into a 2-partite partition -- $\{123|456\}$ or a 3-partite partition -- $\{12|34|56\}$. Each parition also corresponds to a scattering diagram.}, we should construct a complete set of amplitudes with definite angular momenta $J$ and gauge quantum number ${\bf R}$ for each set of particles. Such amplitudes and their corresponding operators are called the j-basis \cite{Jiang:2020rwz,Li:2020zfq,Li:2022tec}, and their correspondence with UV resonances are thus named as the j-basis/UV correspondence. 

In this work, we describe two ways to construct the j-basis.  
First, we construct them in terms of the Clebsch-Gordan coefficients (CGCs) and the so-called spinor bridge rules, which basically glues together the ``subamplitudes'' (not in the usual on-shell sense) from all the vertices of the tree topology~\cite{Jiang:2020rwz,Li:2020zfq,Li:2022tec}. However it is not convenient in the analysis of amplitude basis at fixed mass dimensions, which is necessary for the corresponding effective operators organized by canonical power counting. 
Therefore in this work  we develop a systematic method using the Casimir operators of both the Poincar\'e and gauge groups to enumerate the partial waves at a given mass dimension, which relies on the Young tensor basis introduced in \cite{Li:2020gnx,Li:2020xlh,Li:2022tec}. 
Combining the partial wave basis from both Poincar\'e and gauge sectors, we get the j-basis of amplitudes/operators that are flavor-blind, which do not respect the spin-statistics for identical particles, because it is natural to distinguish all the external particles in the partial waves amplitude. The actual amplitudes that respect spin-statistics are organized as the f-basis in our language \cite{Li:2022tec}, and we have well-established routine to obtain the linear relations between the j-basis and f-basis, which is essential in determining the tree-level UV origins of the actual f-basis operators. In the following we lay out the procedure of obtaining the UV resonances for any kinds of the EFT operators. 

\noindent\underline{\bf The procedure of the bottom-up approach:} 
\bet
\item Unfold the effective operator: draw all the possible topologies of the UV tree amplitudes with external particles determined by the effective operator. Hence all the internal lines indicate possible heavy particle resonances.

\item J-basis/UV correspondence: by utilizing the Poincar\'e and gauge Casimir eigen-basis, we obtain the j-basis operators for a partition of a tree topology, and classify the UV resonance by the quantum numbers of the j-basis operators in each channel.


\item Dimension selection: eliminate the UV resonances from the j-basis operators with lower mass dimension than that required by dimensional analysis. It is usually necessary when non-renormalizable UV couplings are involved.

\eet
\hl{For comparison, we also summarize in table~\ref{tab:approaches} features of different approaches discussed above. The main difference between the top-down and bottom-up approach is that whether the explicit Lagrangian is needed in the procedure. The first three rows represent the steps or ingredients that are needed for each approach to realize the goal of the last row --``Complete UV". Here the terminology ``Complete UV" means the UV origins for a set of operators contributing to certain physical process at a given mass dimension. 
In the Lagrangian formalism, changing operator bases does not affect the physical amplitude for a certain physical process. Therefore it is more reasonable to consider 
the UV origins of the physical process involving a set of operators rather than considering the UV origins for a single operator, since the latter would be dependent on the choice of operator basis. 
Furthermore, the union of the UV origins of all the operators at a given dimension (the ``complete UV" for all operators) is basis independent, and can be obtained by our j-basis analysis for all types of operators at certain dimension, since it can be viewed as a combination of UV origins for all possible physical processes.
} 
 
\begin{table}[h]
\centering
\begin{tabular}{|c|c|c|c|}
\hline
\multirow{2}{*}{}   &   \multicolumn{2}{c|}{Top-down}    &   \multirow{2}{*}{Bottom-up} \\
\cline{2-3}
                            &   UV model    &   Exhaustive Search   &   \\
\hline
Scan over Topologies & \ding{56}    &   \ding{52}   &   \ding{52} \\
Explicit Lagrangian   &   \ding{52}       &   \ding{52}   &   \ding{56} \\
Matching    &       \ding{52}       &   \ding{52}   &   \ding{56} \\ \hdashline
Complete UV    &       \ding{56}       &   \ding{52}   &   \ding{52} \\
\hline
\end{tabular}
\caption{Comparison among the different approaches to obtain the UV resonance/operator relation.}
\label{tab:approaches}
\end{table}

In this work, we first illustrate that there are three different UV resonances contributing to the dimension-5 Weinberg operator that correspond to the Type-I, II and III seesaw models. What's more, we can also see that spin-3/2 resonance cannot give rise to the dimension-5 Weinberg operator\footnote{Here we choose a prescription for the UV couplings of the massive spin-3/2 (or other higher spin) particles so that some Lagrangian terms that do not contribute to on-shell amplitudes are converted to effective operators not involving the higher spin fields by equations of motions in the UV, and are hence not counted as contributing to the IR operator/UV resonance relation. We explain the details in section~\ref{sec:smeft}.}.
In the meantime, we list all possible UV resonances for the dimension 6 operators, which provides a proof and validation on the tree-level dictionary provided in Ref.~\cite{deBlas:2017xtg}. For the first time, we obtain the complete UV resonances for the dimension-7 and bosonic dimension-8 operators systematically and efficiently. We find that a new fermion fields transforming as the $SU(2)_L$ quadruplet, $SU(3)_C$ singlet and with the $U(1)_Y$ charge ${1}/{2}$ can contribute to the dimension-7 operator type $H^2L^2W_L$ and $H^3H^\dagger L^2$, and two new spin-$3/2$ fields transforming as singlet or doublet under the $SU(2)_L$ gauge group can contribute to the dimension-7 operator type $D^2H^2L^2$.  

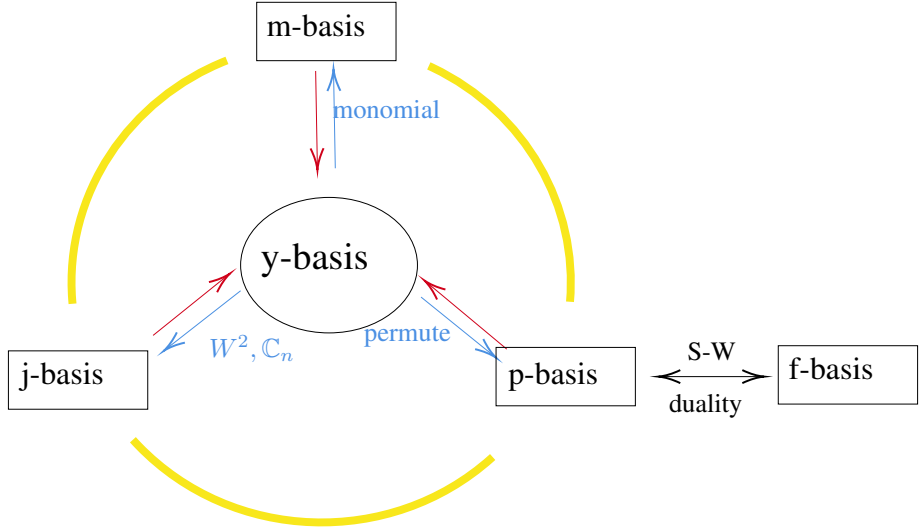
\begin{figure}
    \centering
\tikzset{every picture/.style={line width=0.3pt}} 

\begin{tikzpicture}[x=0.75pt,y=0.75pt,yscale=-1,xscale=1]

\draw   (205.9,150.02) .. controls (205.9,131.13) and (225.87,115.82) .. (250.49,115.82) .. controls (275.12,115.82) and (295.09,131.13) .. (295.09,150.02) .. controls (295.09,168.9) and (275.12,184.21) .. (250.49,184.21) .. controls (225.87,184.21) and (205.9,168.9) .. (205.9,150.02) -- cycle ;
\draw [color={rgb, 255:red, 74; green, 144; blue, 226 }  ,draw opacity=1 ]   (296.89,166.11) -- (333.17,195.48) ;
\draw [shift={(334.73,196.73)}, rotate = 218.98] [color={rgb, 255:red, 74; green, 144; blue, 226 }  ,draw opacity=1 ][line width=0.75]    (10.93,-3.29) .. controls (6.95,-1.4) and (3.31,-0.3) .. (0,0) .. controls (3.31,0.3) and (6.95,1.4) .. (10.93,3.29)   ;
\draw [color={rgb, 255:red, 74; green, 144; blue, 226 }  ,draw opacity=1 ]   (205.9,162.93) -- (168.74,192.03) ;
\draw [shift={(167.16,193.27)}, rotate = 321.94] [color={rgb, 255:red, 74; green, 144; blue, 226 }  ,draw opacity=1 ][line width=0.75]    (10.93,-3.29) .. controls (6.95,-1.4) and (3.31,-0.3) .. (0,0) .. controls (3.31,0.3) and (6.95,1.4) .. (10.93,3.29)   ;
\draw [color={rgb, 255:red, 74; green, 144; blue, 226 }  ,draw opacity=1 ]   (253.65,101.11) -- (252.78,53.13) ;
\draw [shift={(252.75,51.13)}, rotate = 88.97] [color={rgb, 255:red, 74; green, 144; blue, 226 }  ,draw opacity=1 ][line width=0.75]    (10.93,-3.29) .. controls (6.95,-1.4) and (3.31,-0.3) .. (0,0) .. controls (3.31,0.3) and (6.95,1.4) .. (10.93,3.29)   ;
\draw [color={rgb, 255:red, 208; green, 2; blue, 27 }  ,draw opacity=1 ]   (340.13,192.4) -- (300.23,159.1) ;
\draw [shift={(298.69,157.82)}, rotate = 39.85] [color={rgb, 255:red, 208; green, 2; blue, 27 }  ,draw opacity=1 ][line width=0.75]    (10.93,-3.29) .. controls (6.95,-1.4) and (3.31,-0.3) .. (0,0) .. controls (3.31,0.3) and (6.95,1.4) .. (10.93,3.29)   ;
\draw [color={rgb, 255:red, 208; green, 2; blue, 27 }  ,draw opacity=1 ]   (161.76,185.47) -- (198.94,155.52) ;
\draw [shift={(200.5,154.27)}, rotate = 141.15] [color={rgb, 255:red, 208; green, 2; blue, 27 }  ,draw opacity=1 ][line width=0.75]    (10.93,-3.29) .. controls (6.95,-1.4) and (3.31,-0.3) .. (0,0) .. controls (3.31,0.3) and (6.95,1.4) .. (10.93,3.29)   ;
\draw [color={rgb, 255:red, 208; green, 2; blue, 27 }  ,draw opacity=1 ]   (243.74,52) -- (244.6,99.98) ;
\draw [shift={(244.64,101.98)}, rotate = 268.97] [color={rgb, 255:red, 208; green, 2; blue, 27 }  ,draw opacity=1 ][line width=0.75]    (10.93,-3.29) .. controls (6.95,-1.4) and (3.31,-0.3) .. (0,0) .. controls (3.31,0.3) and (6.95,1.4) .. (10.93,3.29)   ;
\draw   (214.01,17.34) -- (284.28,17.34) -- (284.28,46.8) -- (214.01,46.8) -- cycle ;
\draw   (334.73,191.53) -- (404.99,191.53) -- (404.99,221) -- (334.73,221) -- cycle ;
\draw   (88.79,193.27) -- (159.06,193.27) -- (159.06,222.73) -- (88.79,222.73) -- cycle ;
\draw  [draw opacity=0][line width=3]  (121.13,169.5) .. controls (121.08,168.92) and (121.03,168.35) .. (120.98,167.77) .. controls (116.82,114.08) and (149.78,65.96) .. (199,46.67) -- (246.57,158.77) -- cycle ; \draw  [color={rgb, 255:red, 248; green, 231; blue, 28 }  ,draw opacity=1 ][line width=3]  (121.13,169.5) .. controls (121.08,168.92) and (121.03,168.35) .. (120.98,167.77) .. controls (116.82,114.08) and (149.78,65.96) .. (199,46.67) ;  
\draw  [draw opacity=0][line width=3]  (333.02,246.86) .. controls (332.77,247.08) and (332.53,247.3) .. (332.28,247.52) .. controls (281.37,293.02) and (201.73,290.17) .. (154.39,241.15) .. controls (153.44,240.16) and (152.5,239.16) .. (151.58,238.14) -- (246.57,158.77) -- cycle ; \draw  [color={rgb, 255:red, 248; green, 231; blue, 28 }  ,draw opacity=1 ][line width=3]  (333.02,246.86) .. controls (332.77,247.08) and (332.53,247.3) .. (332.28,247.52) .. controls (281.37,293.02) and (201.73,290.17) .. (154.39,241.15) .. controls (153.44,240.16) and (152.5,239.16) .. (151.58,238.14) ;  
\draw  [draw opacity=0][line width=3]  (300.82,49.44) .. controls (301.47,49.73) and (302.12,50.04) .. (302.77,50.35) .. controls (350.12,73.06) and (376.06,120.65) .. (372.02,168.49) -- (246.57,158.77) -- cycle ; \draw  [color={rgb, 255:red, 248; green, 231; blue, 28 }  ,draw opacity=1 ][line width=3]  (300.82,49.44) .. controls (301.47,49.73) and (302.12,50.04) .. (302.77,50.35) .. controls (350.12,73.06) and (376.06,120.65) .. (372.02,168.49) ;  
\draw    (418.71,206.27) -- (467.86,206.27) ;
\draw [shift={(469.86,206.27)}, rotate = 180] [color={rgb, 255:red, 0; green, 0; blue, 0 }  ][line width=0.75]    (10.93,-3.29) .. controls (6.95,-1.4) and (3.31,-0.3) .. (0,0) .. controls (3.31,0.3) and (6.95,1.4) .. (10.93,3.29)   ;
\draw [shift={(416.71,206.27)}, rotate = 0] [color={rgb, 255:red, 0; green, 0; blue, 0 }  ][line width=0.75]    (10.93,-3.29) .. controls (6.95,-1.4) and (3.31,-0.3) .. (0,0) .. controls (3.31,0.3) and (6.95,1.4) .. (10.93,3.29)   ;
\draw   (477.06,190.67) -- (547.33,190.67) -- (547.33,220.13) -- (477.06,220.13) -- cycle ;

\draw (214.95,137.88) node [anchor=north west][inner sep=0.75pt]   [align=left] {{\Large y-basis}};
\draw (217.99,22) node [anchor=north west][inner sep=0.75pt]   [align=left] {{\large m-basis}};
\draw (93.97,197.93) node [anchor=north west][inner sep=0.75pt]   [align=left] {{\large j-basis}};
\draw (339.76,197.93) node [anchor=north west][inner sep=0.75pt]   [align=left] {{\large p-basis}};
\draw (481.25,195.33) node [anchor=north west][inner sep=0.75pt]   [align=left] {{\large f-basis}};
\draw (251.23,65.68) node [anchor=north west][inner sep=0.75pt]   [align=left] {\textcolor[rgb]{0.29,0.56,0.89}{monomial}};
\draw (188.85,184.23) node [anchor=north west][inner sep=0.75pt]    {$\textcolor[rgb]{0.29,0.56,0.89}{W^{2},\mathbb{C}_n}$};
\draw (267.04,179.21) node [anchor=north west][inner sep=0.75pt]   [align=left] {\textcolor[rgb]{0.29,0.56,0.89}{permute}};
\draw (420.78,187.35) node [anchor=north west][inner sep=0.75pt]   [align=left] {\begin{minipage}[lt]{32.21pt}\setlength\topsep{0pt}
\begin{center}
S-W
\end{center}
duality
\end{minipage}};

\end{tikzpicture}
    
    \caption{Relations among the different bases~\cite{Li:2022tec}. The red arrow means that the basis can always be expressed as linear combination of the y-basis through a reduce procedure. The yellow circle indicates that the m-basis, the p-basis and the j-basis are related by linear transformations. The operators in the p-basis are tensors of the symmetric group $S_m$, and those in the f-basis are tensors of $SU(n_f)$ group, where $m$ is the number of repeated fields and $n_f$ is the flavor number of each repeated field. These two basis are connected by the Schur-Weyl duality.}
    \label{fig:relation}
\end{figure}

\hl{In our previous work, we have introduced different concepts for enumerating the operator basis~\cite{Li:2022tec}: the y-basis, the original complete and independent basis for Lorentz and gauge structures obtained by the young tableau method; the m-basis, the independent monomial basis with notations that are familiar to the phenomenology community; the p-basis, the re-organized basis in terms of irreducible representation of the permutation group for the flavor indices of 
the repeated fields\footnote{Repeated fields are those fields with the same Lorentz and gauge quantum numbers}; the f-basis, the operator basis that are allowed by the number of flavors.} In additional to these operator bases, we introduce the following central concepts for this work: 
\bit
\item[] \emph{Partial-Wave basis} (PW-basis): It is the basis of dimensionless amplitudes with definite total angular momentum that encode the dependence on all the scattering angles and energy ratios. Examples are the Legendre polynomials and Wigner-D matrices for $2\to 2$ scattering. We generalize it to arbitrary scattering processes. These particular basis amplitudes carry definite angular momentum quantum numbers and can be obtained by two methods: the spinor bridge rule of gluing several multi-particle states, and the Poincar\'e Casimir method.
\item[] \emph{Gauge PW-basis}: Similar to the PW-basis of the Lorentz sector, we can also write down a basis of gauge tensors in the amplitude with definite gauge quantum numbers associated to certain subsets of particles. This gauge PW-basis is obtained using the gauge Casimir method.
\item[] \emph{J-basis}: It is a basis of contact amplitudes (polynomials of spinor brackets) or the corresponding effective operators that consist of the PW-basis and the gauge PW-basis, labelled by the quantum numbers of the Poincar\'e and gauge groups. Note that they are defined in the flavor-blind sense, which means all the particles or fields are treated as distinguishable.
\eit
\hl{Finally, we use a schematic plot in figure.~\ref{fig:relation} to illustrate the relations between the y-basis, the m-basis, the p-basis, the f-basis and the j-basis. }

This paper is organized as follows. First we introduce partial wave basis and how to obtain it using either the spinor bridge rule or the Poincar\'e and gauge Casimir action in section 2. Then in section 3 applying the partial wave basis to the effective operator, we obtain the operator j-basis using the Casimir action on the y- and p- basis. Section 4 we propose the j-basis/UV correspondence and lay out the procedure to obtain complete list of UV resonances. In section 5, we list the complete tree-level UV resonances for the SMEFT operators at the dimension 5, 6, 7 and 8. We conclude in section 6. 
\hl{Finally, we provide here a guide for readers of different interest. For people who are interested in the technical details, it is advised to begin with section.~\ref{sec:partial} to get a pedagogical introduction to the concept and the algorithm to obtain the j-basis, which contains dedicated examples to help readers to understand each formula and how to implement them with the computer program. For people who are only interested in finding the corresponding UV for a specific operator, it is enough to start with section.~\ref{sec:dim-67} and with the help of the table of contents to locate the information they need.  }

\section{Scattering Amplitude: Generalized Partial Wave Analysis}
\label{sec:partial}

We start by introducing the generalization of partial wave expansion and the construction of the partial wave amplitude basis \cite{Jacob:1959at, Jiang:2020rwz,Shu:2021qlr}. 
Traditionally, partial waves are defined for two body scattering, and the basis amplitudes are functions of the scattering angles $(\theta,\phi)$ only. When the spins of the scattering particles are taken into account, the partial wave basis is given by the Wigner D-matrices $D^J_{\Delta\Delta'}(\theta,\phi) = d^J_{\Delta\Delta'}(\theta)e^{-i(\Delta+\Delta')\phi}$, where $\Delta$ and $\Delta'$ are the total helicities of the initial and final states, and $J$ characterize the total angular momentum of the process. For the purpose of analysing general tree topologies of the scattering amplitudes, we define the following expansion of a general $N$ point amplitude
\eq{\label{eq:pw_general}
    \mc{M}(\{\Psi\}_{N|m}) = \sum_{J_1,\dots,J_m} \sum_a \mc{M}^{J_1,\dots,J_m}_a(\{s\}_{N|m}) \overline{\mc{B}}^{J_1,\dots,J_m}_a(\{\Psi\}_{N|m}),
}
where $\{\Psi\}_{N|m}$ denotes $N$ particles divided into $m$ groups, each with total angular momentum $J_{i=1,\dots,m}$. $\overline{\mc{B}}_a$ are dimensionless partial wave basis with possible degeneracy labelled by $a$, and $\mc{M}^{J_1,\dots,J_m}_a$ are the partial wave coefficients which are functions of the Mandelstam variables $\{s\}_{N|m}$, a subset that commute with all the $J_i$ to be defined later. It reduces to the familiar two-partite partial wave expansion when $m=2$, such that
\eq{\label{eq:pw_2part}
    \overline{\mc{B}}^{J,J'}_a(\{\Psi\}_{N|2}) \equiv \overline{\mc{B}}^J_a \left(\{\Psi\}_{N_1} \to \{\Psi\}_{N_2} \right) \delta^{JJ'},
}
where the factor $\delta^{JJ'}$ is due to the angular momentum conservation. In particular, when $N_1=N_2=2$ we have $\overline{\mc{B}}^J \sim D^J$. In the following we are going to show the construction of the general partial wave amplitude basis.

\subsection{Partial Wave Amplitude Basis from Spinor Bridges}\label{sec:bridge}

In general, the multi-particle states used to define scattering amplitudes are the \emph{tensor basis} $\ket{\{\Psi\}_N} = \otimes_{i=1}^N \ket{\Psi_i}$, with no entanglement among the particles such that each particle $\Psi_i$ has independent momentum $p_i$ and helicity $h_i$. For the two-massless-particle states, in the Center of Mass (CoM) frame, the tensor state reads
\eq{
    \ket{p_1,h_1}\otimes\ket{p_2,h_2} \stackrel{\rm CoM}{\rightarrow} \ket{E,\theta,\varphi; h_1,h_2}.
}
Sun tensor basis state forms a tensor representation of the Poincar\'e group, which is usually reducible. Similar to how we do angular momentum superpositions for spins, we can define the irreducible representation of Poincar\'e group $\ket{P,J,J_z}$, while the \emph{Poincar\'e Clebsch-Gordan Coefficients} (CGC) are defined as the overlaps \cite{Jiang:2020rwz}
\eq{\label{eq:def_CGC}
    \inner{P,J,J_z}{\{\Psi\}_N} \equiv \mc{C}^{J,J_z}(p_i,h_i)\delta^4(P-\sum_i^N p_i)\,,
}
which is also a basis for the wave functions of the $N$-particle states.
For $N=2$, $\mc{C}^J$ are typically expressed as functions of the directions $\hat{p}_{1,2}$ in the Center of Mass (CoM) frame, given by the Wigner D-matrix $D^J(\theta,\varphi)$, 
\eq{ 
    \inner{J,J_z}{E,\theta,\varphi;h_1,h_2} = \sqrt{2J+1} D^J_{J_z,h_1-h_2}(\theta,\varphi) = \sqrt{2J+1} d^J_{J_z,h_1-h_2}(\theta)e^{-i(h_1-h_2)\varphi}.
}
However, it is not convenient for $N>2$ and the frame dependence is also disfavored in some situations. It was proposed in \cite{Jiang:2020rwz} that the CGC can be interpreted as a local amplitude between the $N$ particle state and an auxiliary particle with mass $\sqrt{P^2}\equiv \sqrt{s}$ and spin $J$, which can be naturally expressed in terms of spinor helicity variables \cite{Arkani-Hamed:2017jhn}. The spinor variables for both massless and massive particles are defined as
\bea
p_{i\alpha\dot{\alpha}}=p_{i\mu}\sigma^\mu_{\alpha\dot{\alpha}}=\lambda_{i\alpha}\tilde{\lambda}_{i\dot{\alpha}},\quad P_{\alpha\dot{\alpha}}=P_\mu\sigma^\mu_{\alpha\dot{\alpha}}=\chi_\alpha^I\tilde{\chi}_{\dot{\alpha}I},
\eea
where $\lambda_i$ may carry little group (LG) indices according to the mass $m_i$, but $\chi$ must be massive spinors. We refer to the Appendix~\ref{app:spinor} for more comprehensive discussion on the massless and massive spinor helicity notation. The CGC can now be expressed in terms of the spinor variables $\{\lambda_i,\tilde\lambda_i,\chi,\tilde\chi\}$. One could then write down Lorentz invariant contractions among the spinor variables that satisfy the LG representations of each particle. In particular,
\eq{\label{eq:LG}
    \text{Massless Particle: }\qquad & \mc{C}^{J,J_z}(\dots,e^{-i\varphi/2}\lambda_i,e^{i\varphi/2}\tilde\lambda_i,\dots) = e^{ih_i\varphi}\mc{C}^{J,J_z}(\dots,\lambda_i,\tilde\lambda_i,\dots), \\
    \text{Massive Particle: }\qquad & \mc{C}^{J,J_z}(\dots,\chi,\tilde\chi) \sim \chi^{\{I_1}_{\alpha_1}\cdots\chi^{I_{2J}\}}_{\alpha_{2J}} \quad,\qquad J_z = \sum_{i=1}^{2J} I_i,\ I_i = \pm\frac12 .
}
We see that $J_z$ is not represented by the LG components, so we may use $\mc{C}^J$ without superscript $J_z$ to denote the LG tensor, which will be used in the rest of the paper.
For 2-massless-particle states, the CGC is given by three particle amplitudes, which have the general formula based on the little-group scaling with the locality imposed \cite{Arkani-Hamed:2017jhn}:
\bea\label{eq:3-amp}
 \mathcal{C}^{J}(h_1,h_2) \sim 
 \frac{[12]^{J+h_1+h_2}}{s^{(3J+h_1+h_2)/2}}(\lambda_1^{J+h_2-h_1}\lambda_2^{J+h_1-h_2})^{\{\alpha_1,...,\alpha_{2J}\}}\chi_{\alpha_1}^{I_1}\cdots\chi_{\alpha_{2J}}^{I_{2J}},
\eea
where $[ij]\equiv\tilde\lambda_{i\dot\alpha}\tilde\lambda_j^{\dot\alpha}$ and $\vev{ij}\equiv\lambda_i^\alpha\lambda_{j\alpha}$ are understood, and the power of $s$ in the denominator is to make sure $\mc{C}^J$ is dimensionless. A more careful normalization could be defined to guarantee a unit norm under phase space integration \cite{Shu:2021qlr}. For $N>2$ particles, there is no unique solution to eq.\eqref{eq:LG}, and we need to use a subscript to label the degeneracy, for example, the $J=1$ amplitudes for three scalar particles can be written as:
\eq{\label{eq:ex_3cgc}
    & \mc{C}^{J=1}_1(0,0,0) \sim s^{-3/2}[12]\vev{1\chi}\vev{2\chi}, \\
    & \mc{C}^{J=1}_2(0,0,0) \sim s^{-3/2}[23]\vev{2\chi}\vev{3\chi}, \\
    & \mc{C}^{J=1}_3(0,0,0) \sim s^{-3/2}[31]\vev{3\chi}\vev{1\chi}.
}

So far we have discussed the Poincar\'e representations of the particles and the multi-particle states. We can do the same thing for the internal symmetry groups, like the state as a multiplet under a gauge group. For multi-particle states, there is also a \emph{tensor basis} where the particles are unentangled, $\otimes_i\ket{{\bf r}_i,a_i}$, and an irreducible state $\ket{{\bf R},b,A}$ of the irreducible representation $\bf R$, where $b$ labels the multiplicity of the ${\bf R}$ representation under the decomposition, while in the case where the multiplicity is one we usually omit this index. The gauge CGC should be a tensor with the set of indices $\{a_i\}$ of the representations participating in the direct product and an single index $A$ labeling the basis vectors in the subspace of the corresponding irreducible representation in the decomposition. Such a tensor should be invariant under the group action, i.e. an invariant tensor  $\mc{T}_b({\bf R};\{{\bf r}_i\})^A_{a_1,\dots,a_N}$ with appropriate normalization. Take the $SU(2)$ group as an example, suppose we have two particles under the fundamental representation, so that the tensor basis is $\{\ket{+,+},\ket{+,-},\ket{-,+},\ket{-,-}\}$. The two particles can form a singlet state with CGC $\mc{T}({\bf R = 1};{\bf 2,2})_{ij} = \epsilon_{ij}$, or a triplet state with CGC $\mc{T}({\bf R = 3};{\bf 2,2})^I_{ij} = \tau^I_{ij}$, such that
\eq{
    &\ket{{\bf 1}} = \frac{1}{\sqrt{2}}\sum_{i,j}\epsilon_{ij}\ket{i,j} = \frac{1}{\sqrt{2}}(\ket{+,-} - \ket{-,+}), \\
    &\ket{{\bf 3},I} = \frac{1}{\sqrt{2}}\sum_{i,j}\tau^I_{ij}\ket{i,j} = \left\{\begin{array}{ll}
    \ket{+,+}, & \tau^+ = \frac{1}{\sqrt{2}}(\sigma^1 + i\sigma^2)\epsilon \\
    \frac{1}{\sqrt{2}}(\ket{+,-} + \ket{-,+}), & \tau^0 = \sigma^3\epsilon \\
    \ket{-,-}, & \tau^- = \frac{1}{\sqrt{2}}(\sigma^1 - i\sigma^2)\epsilon
    \end{array}\right. ,
}
where we have omit the multiplicity index $k$, as the multiplicity is equal to one.
For more complicated cases, we would just need to construct the set of invariant tensors ${\cal T}_b({\bf R};\{{\bf r}_i\})$ to fully describe the possible representations of the multi-particle states. Together with the Poincar\'e group sector, we have the reduction of tensor representations for multi-particle states as
\eq{
    \ket{\Psi}_N \equiv \bigotimes_{i=1}^N\ket{p_i,h_i;\mathbf{r}_i,a_i} = \sum_{J,J_z} \sum_{\mathbf{R},b,A} \mc{C}^{J,J_z} \mc{T}_b(\mathbf{R})^A_{a_1,\dots,a_N} \ket{P=\textstyle\sum_i p_i;J,J_z;\mathbf{R},b,A} .
}

This decomposition turns out to be very helpful to define partial wave amplitude basis. Specifically, partial wave expansion is nothing but a change of basis for multi-particle states. 
\eq{
    {}_{N}\bra{\Psi} {\bf T} \ket{\Psi}_{N'} &= \sum_{J,J_z}\sum_{{\bf R},b,A}\left[\mc{C}^{J,J_z}\right]^* \mc{T}_b(\overline{\bf R})_{A}^{a_1,\dots,a_N} \sum_{J',J'_z}\sum_{{\bf R}',b',A'}\mc{C}'{}^{J',J'_z}\mc{T}'_{b'}({\bf R}')^{A'}_{a'_1,\dots,a'_{N'}} \\
    & \qquad \times \bra{P;J,J_z;\mathbf{R},A} {\bf T} \ket{P';J',J'_z;\mathbf{R}',A'}.
}
where we denoted the conjugate representation of ${\bf R}$ as $\overline{\bf R}$ and the indices are raised/lowered.
From the conservation laws of all the symmetry groups, the transfer matrix should be diagonal in the irreducible basis
\eq{
    \bra{P;J,J_z;\mathbf{R},A} {\bf T} \ket{P';J',J'_z;\mathbf{R}',A'} &= \mc{M}^{J,{\bf R}}(s)\delta^{JJ'}\delta^{J_z J'_z}\delta^{\bf RR'}\delta^{A}_{A'}\delta^{(4)}(P-P'). \\
}
Therefore we get
\eq{\label{eq:pwdefine}
    {}_{N}\bra{\Psi} {\bf T} \ket{\Psi}_{N'} &= \sum_{J,\mathbf{R}} \mc{M}^{J,\mathbf{R}}(s) \times \sum_{J_z} \left[\mc{C}^{J,J_z}\right]^*\mc{C}'{}^{J,J_z} \times \sum_A \mc{T}_b(\overline{\bf R})_{A}^{a_1,\dots,a_N}\mc{T}'_{b'}({\bf R})^{A}_{a'_1,\dots,a'_{N'}} \\
    &\equiv \sum_{J,\mathbf{R}} \mc{M}^{J,\mathbf{R}}(s) \overline{\mc{B}}^J \mc{T}_{bb'}(\mathbf{R})^{a_1,\dots,a_N}_{a'_1,\dots,a'_{N'}},
}
where we have defined the partial wave basis $\overline{\mc{B}}^J$ as shown in Eq.~\eqref{eq:pw_2part} and a gauge factor ${\cal T}_{bb'}(\mbf{R})$ as an invariant tensor indicating the particular intermediate irreducible representation ${\bf R}$ in the scattering channel $\Psi_{N_1} \to \Psi_{N_2}$. We will discuss the latter in detail in section~\ref{sec:gauge}.

We see from the above definition that the partial wave basis consists of the Poincar\'e CGC. Written in terms of the spinor helicity variables, the sum over $J_z$ becomes the contraction of little group indices, which we denote as a dot product, thus
\eq{\label{eq:2partite-pw}
    \overline{\mc{B}}^J(\Psi_{N_1} \to \Psi_{N_2}) = \mc{C}^J(\Psi_{N_1}) \cdot \mc{C}^J(\Psi_{N_2})^* = \mc{C}^J(\Psi_{N_1}) \cdot \mc{C}^J(\bar\Psi_{N_2}),
}
where the complex conjugate of the CGC is interpreted as the CGC of the CP conjugate states $\bar\Psi$, which have opposite helicities. 
In particular, one could verify \cite{Jiang:2020rwz} that using Eq.~\eqref{eq:3-amp} we have\footnote{The CoM values of the spinor variables are $\lambda = \sqrt{|p|}\begin{pmatrix}\cos(\hat\theta/2) \\ \sin(\hat\theta/2)e^{i\hat\phi} \end{pmatrix}$ and $\tilde\lambda = \lambda^*$ for momentum $p = (|p|,|p|\hat{n})$, with $\hat{n} = (\sin\hat\theta\cos\hat\phi,\sin\hat\theta\sin\hat\phi,\cos\hat\theta)$}
\eq{
    \overline{\mc{B}}^J(h_1,h_2 \to h_3,h_4) = \mathcal{C}^J(h_1,h_2) \cdot {\mathcal{C}^J(-h_3,-h_4}) \stackrel{\rm CoM}{\sim} D^J_{h_2-h_1,h_4-h_3}(\theta,\phi).
}
In the following, we are going to adopt the convention that all the momenta are incoming, thus the helicities should be defined accordingly, so that we do not need to flip signs as the above. Hence the partial wave in Eq.~\eqref{eq:2partite-pw} should be written as
\eq{
    \overline{\mc{B}}^J(\Psi_{N_1}|\bar{\Psi}_{N_2}) = \mc{C}^J(\Psi_{N_1}) \cdot \mc{C}^J(\bar\Psi_{N_2}).
}
There are various advantages, one of which is that we do not need to take extra CP conjugate when crossing to another scattering channel. The scattering $\mc{M}(\psi_1,\psi_2,\phi_3,\phi_4)$ can be understood as in $1,2\to3,4$ channel, denoted as $\mc{M}(\psi_1,\psi_2 | \phi_3,\phi_4) \equiv \mc{M}(\psi_1,\psi_2 \to \phi_3,\phi_4)$, or as in $1,3\to2,4$ channel, denoted as $\mc{M}(\psi_1,\phi_3 | \psi_2,\phi_4) \equiv \mc{M}(\psi_1,\phi_3 \to \bar\psi_2,\phi_4)$. They are actually the same function of the spinor variables up to analytic continuation, however they obviously have different partial wave amplitude basis. We list a few of the partial waves in each channel in table~\ref{tab:partialwave} and~\ref{tab:pw22-channel2} respectively.

\begin{table}[hb]
\centering
\begin{tabular}{c|c|c|c}
 \hline\hline
 spin $J$ & $\mc{C}^J(\psi_1,\psi_2)$ &  $\mathcal{C}^J(\phi_3,\phi_4)$ & partial-wave basis $\overline{\mathcal{B}}^{J}$ (not normalized) \\
 \hline
 $J=0$ & $\langle12\rangle$ & $1$ & $\langle12\rangle$\\
 \hline
 $J=1$ & \makecell{ $\langle1\chi\rangle \langle2\chi\rangle$ } & \makecell{ $-\langle\chi|p_3p_4|\chi\rangle$ } & \makecell{ $ [34] {\color{blue} \big(\langle 13\rangle  \langle 24\rangle + \langle 14\rangle  \langle 23\rangle \big) } $ }\\
 \hline
 $J=2$ & $\langle1\chi\rangle^2 \langle2\chi\rangle^2 [12]$ & $-\langle\chi|p_3p_4|\chi\rangle^2$ & \makecell{ $[12][34]^2 {\color{blue} \big( \langle13\rangle^2 \langle24\rangle^2 + 4\langle13\rangle\langle24\rangle \langle14\rangle\langle23\rangle + \langle23\rangle^2 \langle14\rangle^2 \big) }$}\\
 \hline
 $J=3$ & $\langle1\chi\rangle^3 \langle2\chi\rangle^3 [12]^2$ & $-\langle\chi|p_3p_4|\chi\rangle^3$ & \makecell{$[12]^2[34]^3 {\color{blue} \big(\langle13\rangle^3 \langle24\rangle^3 +9\langle13\rangle^2\langle24\rangle^2 \langle14\rangle\langle23\rangle } $ \\ 
 ${\color{blue} +9\langle13\rangle\langle24\rangle \langle14\rangle^2\langle23\rangle^2  +\langle23\rangle^3\langle14\rangle^3 \big) } $ }\\
 \hline\hline
\end{tabular}
\caption{Partial-wave basis of  process $\mc{M}(\psi_1,\psi_2|\phi_3,\phi_4)$. $\chi$ is an auxiliary massive spinor associated with the total momentum $P = p_1+p_2$. The factors in blue are the spinor bridge structure which has totally symmetric contractions, where the coefficients can be obtained from Eq.~\eqref{eq:bridge-comb}.}
\label{tab:partialwave}
\end{table}

\begin{table}[hbt]
	\centering
	\begin{tabular}{c|c|c|c}
	\hline\hline
		spin $J$ & $\mathcal{C}^J(\psi_1,\phi_3)$ &  $\mathcal{C}^J(\psi_2,\phi_4)$ & partial-wave basis $\overline{\mathcal{B}}^{J}$ (not normalized) \\
		\hline
		$J=1/2$ & $\langle1\chi\rangle$ & $\langle\chi2\rangle$ & $\langle12\rangle$\\
		\hline
		$J=3/2$ & $-\langle1\chi\rangle [\chi|p_3|\chi\rangle$ & $-\langle\chi2\rangle [\chi|p_4|\chi\rangle$ & $ [13][24] {\color{blue} \big( \langle12\rangle^2\langle34\rangle +2\langle12\rangle\langle14\rangle\langle32\rangle \big) } $ \\
		\hline
		$J=5/2$ & $\langle1\chi\rangle [\chi|p_3|\chi\rangle^2$ & $\langle\chi2\rangle [\chi|p_4|\chi\rangle^2$ & $ [13]^2[24]^2 {\color{blue} \big( \langle12\rangle^3\langle34\rangle^2 +6\langle12\rangle^2\langle34\rangle \langle14\rangle \langle32\rangle +3\langle12\rangle \langle14\rangle^2\langle32\rangle^2 \big) } $\\
		\hline
		$J=7/2$ & $-\langle1\chi\rangle [\chi|p_3|\chi\rangle^3$ & $-\langle\chi2\rangle [\chi|p_4|\chi\rangle^3$ & \makecell{$ [13]^3[24]^3 {\color{blue} \big( \langle12\rangle^4\langle34\rangle^3 +12\langle12\rangle^3\langle34\rangle^2 \langle14\rangle\langle32\rangle }$ \\
		$ {\color{blue} +18\langle12\rangle^2\vev{34} \langle14\rangle^2\langle32\rangle^2 + 4\vev{12}\vev{14}^3\vev{32}^3 \big) }$}\\
		\hline\hline
	\end{tabular}
	\caption{Partial-wave basis of  process $\mc{M}(\psi_1,\phi_3|\psi_2,\phi_4)$ is an auxiliary massive spinor associated with the total momentum $P=p_1+p_3$. The factors in blue are the spinor bridge structure which has totally symmetric contractions, where the coefficients can be obtained from Eq.~\eqref{eq:bridge-comb}.}
	\label{tab:pw22-channel2}
\end{table}

\begin{figure*}[htb]
\centering
\includegraphics[width=0.6\textwidth]{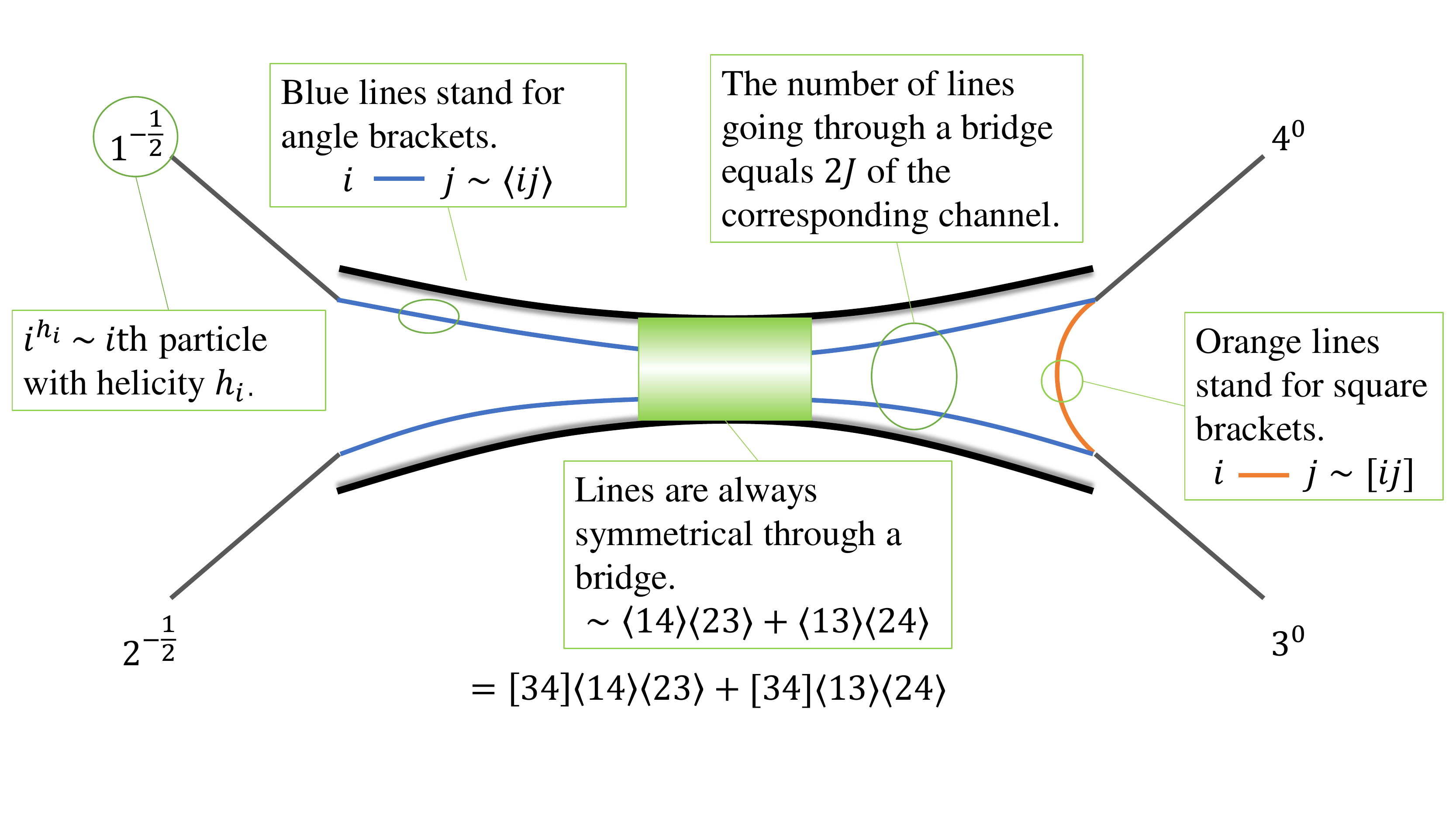}
\caption{Spinor bridge rules presented with a 2-partite example of a $J=1$ amplitude $[34]\langle14\rangle\langle23\rangle +[34]\langle13\rangle\langle24\rangle$. The bridge rule of partial wave expansion includes 3 main aspects. (1) All particles are labeled and categorized into their respective channels, and each categorization is called a partition, e.g. $\{12|34\}$ of this figure; (2) Each thread connecting two particles stands for a bracket, orange for square brackets and blue for angle brackets, while the number of orange and blue threads ending at a particle is constrained by its helicity; (3) A bridge is built up by threads connecting two channels, and all the threads through a bridge should be of the same type (orange/blue) and totally symmetrical. }\label{fig:bridge}
\end{figure*}

Such a construction of partial wave basis manifest a key feature of the structure, the dot product, which is a totally symmetric contractions between LG indices in the initial and final states. Note that
\eq{\label{eq:bridge_eval}
    \chi_{\alpha}^I\chi_{\beta I} =-\sqrt{s}\epsilon_{\alpha\beta},
}
the LG contractions can be converted to $2J$ spinor contractions between spinor variables of the external states in the initial and final states, and the total symmetry among the contractions is preserved as long as we have chosen the same type of spinors, either $\chi$ or $\tilde\chi$, in both CGC. 
We diagrammatically denote the spinor contractions in figure~\ref{fig:bridge} for an example of scattering $\mc{M}(\psi_1,\psi_2 | \phi_3,\phi_4)$, where the totally symmetric contraction is represented by a \emph{spinor bridge} across the scattering channel, through which threads are symmetrized. Therefore, the partial wave basis can be constructed by drawing such bridge diagrams which directly translate to spinor brackets. The partial wave basis of the scattering $\mc{M}(\psi_1,\psi_2 | \phi_3,\phi_4)$ for $J=0,1,2,3$ are listed in table~\ref{tab:partialwave}, which are obtained by spinor bridges, where the polynomials in the parentheses consist of totally symmetric spinor contractions, for instance
\eq{
    J=2:\qquad &\lambda_1^\alpha\lambda_1^{\beta}\lambda_2^\gamma\lambda_2^\delta \Big(\underbrace{\epsilon_{\alpha\alpha'}\epsilon_{\beta\beta'}\epsilon_{\gamma\gamma'}\epsilon_{\delta\delta'} + \text{perm}}_{\text{4! terms in }S_4} \Big) \lambda_3^{\alpha'}\lambda_3^{\beta'}\lambda_4^{\gamma'}\lambda_4^{\delta'} \\
    & = 4\big(\langle13\rangle^2 \langle24\rangle^2 + 4\langle13\rangle\langle24\rangle \langle14\rangle\langle23\rangle + \langle23\rangle^2 \langle14\rangle^2\big) .
}
The coefficients can be obtained by combinatorics. For example, if we want to connect a number of  (same type) spinor indices $n_1,n_2 \to n_3,n_4$, where $n_1+n_2=n_3+n_4$ is the total number of threads through the bridge, the coefficient of the structure $\vev{13}^a\vev{24}^{n_2-n_3+a}\vev{14}^{n_1-a}\vev{23}^{n_3-a}$ would be given by
\eq{\label{eq:bridge-comb}
    c_a \propto {n_1 \choose a}{n_2 \choose n_3-a}.
}
For more than $2\to2$ scattering, the combinatorics becomes more difficult, but it is still the most intuitive way to construct the general partial wave amplitude basis.

\begin{figure*}[hbt]
\centering
\includegraphics[width=\textwidth]{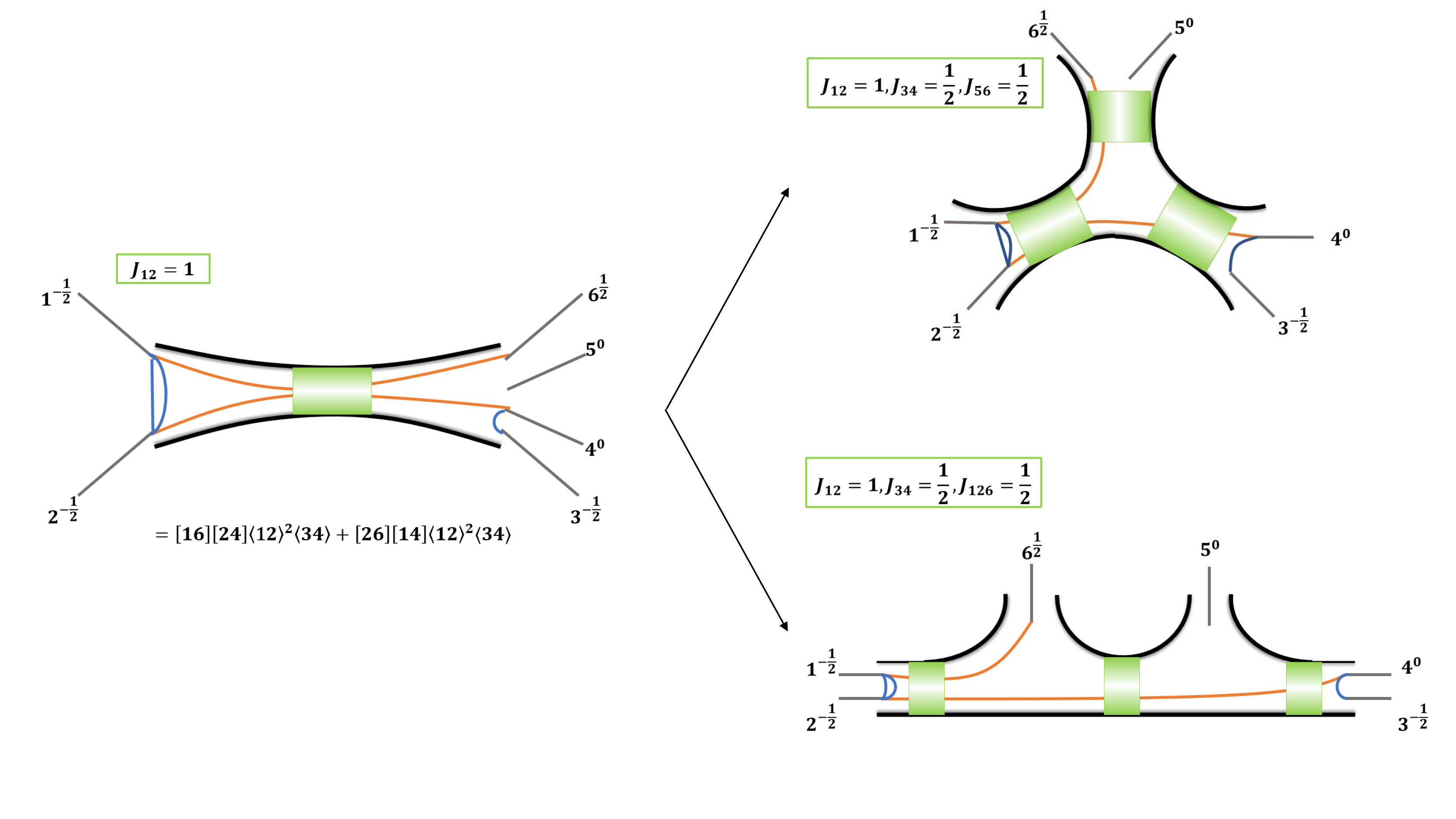}
\caption{ An example of 6-particle 2-partite and 3-partite partial wave basis constructed by the spinor bridge rule. We show that there can be different subdivisions of the partition $\{12|3456\}$. 
We see that the same amplitude satisfies the bridge conditions for both partitions, and hence is the  simultaneous partial wave: it has definite angular momenta for both the two-particle state $\{56\}$ and the three-particle state $\{345\}$. It is in general not true, because the angular momenta in the two channels do not commute, which will be specified in the next section. }\label{fig:2->3partite}
\end{figure*}

In the case when a multi-particle state involved in the amplitude has more than 2 particles, the Poincar\'e CGC is not unique as we showed in Eq.~\eqref{eq:ex_3cgc}, which induces degenerate partial waves for certain angular momentum $J$. 
We provide an example of a 2-to-4 scattering, $\psi^2\rightarrow\psi\phi^2\bar{\psi}$, in Table \ref{tab:12,34,56.}, where we list the 4-particle CGC $\mathcal{C}^J(\psi_3,\phi_4,\phi_5,\psi^\dagger_6)$ for $J=0,1$. It is actually hard to guarantee the independence and completeness of the CGC, which we shall deal with in the next section.
\begin{table}[ht]
\centering
\begin{tabular}{c|c|c|c}
\hline\hline
spin $J$ & $\mathcal{C}^J(\psi_1,\psi_2)$ & $\mathcal{C}^J_i(\psi_3,\phi_4,\phi_5,\psi^\dagger_6)$ & $\overline{\mathcal{B}}^J_i$ (not normalized)
 \\
\hline
$J=0$ & $\langle12\rangle$ & \makecell{ $[46]\langle34\rangle$ \\ $[56]\langle35\rangle$ }
& \makecell{ $ \langle 12\rangle [46]\langle 34\rangle $ \\ $ \langle 12\rangle  [56] \langle 35\rangle $ }  \\
\hline
$J=1$ & $\langle1\chi\rangle\langle2\chi\rangle$ & \makecell{ $[46]\langle3\chi\rangle\langle4\chi\rangle$ \\ $[56]\langle3\chi\rangle\langle5\chi\rangle$ \\ $[36]\langle3\chi\rangle^2$ }
& \makecell{$[46] {\color{blue} \big(\langle 13\rangle  \langle 24\rangle + \langle 14\rangle  \langle 23\rangle \big) }$ \\ 
$ [56] {\color{blue} \big( \langle 13\rangle  \langle 25\rangle + \langle 15\rangle  \langle 23\rangle \big) } $ \\ 
$[36] {\color{blue} \langle 13\rangle  \langle 23\rangle }$ } \\
\hline\hline
\end{tabular}
\caption{Partial-wave basis of process $\mc{M}(\psi_1\psi_2 | \psi_3\phi_4\phi_5\psi^{\dagger}_6)$. The factors in blue are the spinor bridge structure which has totally symmetric contractions. Note that for more than two particles, $\mc{C}^J$ have degeneracy labeled by a subscript, and so does the resulting partial waves. We are not providing a complete list of partial waves here, because there can be inequivalent $\mc{C}^J$ with more spinor brackets.}
\label{tab:12,3456}
\end{table}

\begin{figure}[htb]
\centering
\includegraphics[width=500pt]{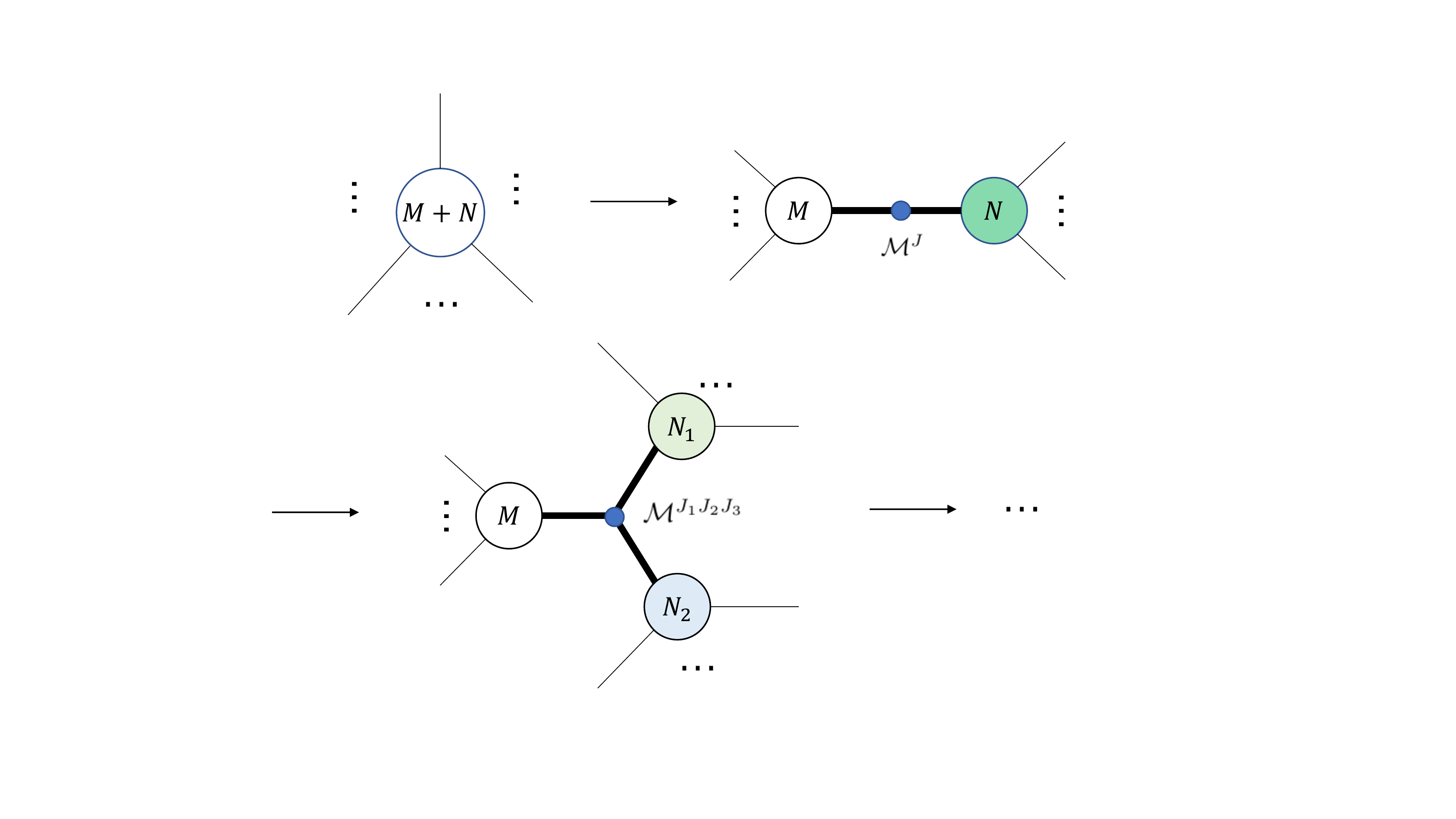}
\caption{Topology of a $M+N$-particle scattering amplitude splitting into 2 and 3 partitions and more.}
\label{fig:top}
\end{figure}

So far we have discussed the usual 2-partite partial wave amplitudes associated with a single scattering channel $\mc{I}\to\bar{\mc{I}}$. A special case is the $2 \to n$ scattering process for a back-to-back collision at a collider. From the collider physics, we note that a multi-particle final state can be divided into multi-partite via the recursive relation of the chain of the $1 \to 2 $ splitting. Motivating from such cascade partitions, we can further split the 2-partite scattering $\mc{I}\to\bar{\mc{I}}$ to multi-partite scattering. For illustration, Fig.~\ref{fig:top} shows how a $M+N$-particle amplitude is split into a 2-partite scattering $M \to N$, and further into a 3-partite scattering $\{M, N_1, N_2\}$, and so on.

We shall further introduce the $m$-partite amplitudes, by projecting the amplitude onto the eigenspaces of angular momentum for several scattering channels $N|m \equiv \mc{I}_{1,\dots,m} \subset \{1,\dots,N\}$. For any of the two channels to have definite angular momenta at the same time, the set of channels should be non-overlapping 
\eq{\label{eq:non-ovl-ch}
    & \mc{I}_i \cap \mc{I}_j = \emptyset \ \text{or}\ \mc{I}_i \cap \bar{\mc{I}}_j = \emptyset .
}
where $\bar{\mc{I}}$ is the compement of $\mc{I}$. An $N$ point scattering can have at most $N-3$ channels that satisfy Eq.~\eqref{eq:non-ovl-ch}, which we characterize by $m$ subsets of the external particles $\mc{P} = \{\mc{I}_i\}$ that we call a partition\footnote{A subset $\mc{I}$ and its complement $\bar{\mc{I}}$ are equivalent to indicate a scattering channel. One of them can be eliminated from $\mc{P}$, but for convenience of generating partitions, we usually keep a redundant one, making $|\mc{P}| \le N-2$ instead of $m \le N-3$. For example, ``two-partite'' is usually denoted by $\{\mc{I},\bar{\mc{I}}\}$, which has only a single channel.  }. Similar to the two-partite case, we define auxiliary massive states for the channels $\mc{I}_i$ with massive spinors $(\chi_i^I, \tilde\chi_i^I)$, and glue the sub-amplitudes from all the designated vertices via spinor bridges.
For example we can refine the partial waves in table~\ref{tab:12,3456} by adding channels in various ways, as shown in Fig.~\ref{fig:2->3partite}. 
In the partition $\{12|34|56\}$, we need to glue the three 2-particle CGC by a 3-point amplitude $\mc{B}^{J_{12},J_{34},J_{56}}$, which may have an independent basis $\mc{B}_i$. The partial wave expansion would be
\eq{\label{eq:3-partite}
    \mc{M} &= \sum_{\{J\},i}\mc{M}^{J_{12},J_{34},J_{56}}_i\overline{\mc{B}}^{J_{12},J_{34},J_{56}}_i \\
    &\quad \overline{\mc{B}}^{J_{12},J_{34},J_{56}}_i \sim \mc{B}_i^{J_{12},J_{34},J_{56}} \cdot \mathcal{C}^{J_{12}} \otimes \mathcal{C}^{J_{34}} \otimes \mathcal{C}^{J_{56}},
}
where $\otimes$ is a tensor product of different LG spaces, and the dot product means the symmetric contractions for each of them, which can also be carried out by Eq.~\eqref{eq:bridge_eval}. The amplitude in the Fig.~\ref{fig:2->3partite} corresponds to
\eq{
    \left.\begin{array}{l}
    \mc{B}^{1,\frac12,\frac12}(\chi_1,\chi_2,\chi_3) \sim [\chi_1\chi_3][\chi_1\chi_2] \\
    \mathcal{C}^1(1,2) \sim \vev{12}^2[1\chi_1][2\chi_1] \\
    \mathcal{C}^{\frac12}(3,4) \sim \vev{34}[4\chi_2] \\
    \mathcal{C}^{\frac12}(5,6) \sim [6\chi_3]
    \end{array}\right\} \Rightarrow \overline{\mc{B}}^{J_{12}=1,J_{34}=\frac12,J_{56}=\frac12} \sim {\color{blue} ([16][24]+[14][26])}\vev{12}^2\vev{34}.
}
The coefficient $\mc{M}^{\{J\}}$ in general may depend on any of the Mandelstam variables in the set 
\eq{
\{s\}_{\mc{P}} \equiv \{s_{\mc{I}}|\mc{I}\cap\mc{I}_i=\emptyset\text{ or }\mc{I}\cap\bar{\mc{I}}_i=\emptyset,\,\forall\,\mc{I}_i\in\mc{P}\}
}
These Mandelstam variables do not affect the angular momenta in the designated channels $\mc{I}_i\in\mc{P}$. In the above case we have $\{s\}_{\{12|34|56\}} = \{s_{12},s_{34},s_{56}\}$. 
The two-partite partial wave basis can be regarded as a special case due to the angular momentum conservation
\bea
\mathcal{B}^{J,J'} \cdot \mathcal{C}^{J} \otimes \mathcal{C}^{J'}
&=&  (\mathcal{C}^J_a \cdot \mathcal{C}^J_b) \delta^{JJ'}
\eea
Some partitions may not consist of disjoint sets of external particles, such as the second case in Fig.~\ref{fig:2->3partite} which we denote as $\{6|12|34|5\}$ {\color{yellow}}. The irreducible state in the middle channel should be regarded as a combination of tensor states from particle 6 and the 2-particle state $\{12\}$ with spinor variable $\chi_1$, which has CGC $\mc{C}^{J_{126}}(6,\chi_1)$. The same should be computed for the right half of the amplitude, and the final partial wave expansion is
\eq{
    \mc{M} = \sum_{\{J\},i} \mc{M}^{J_{12},J_{34},J_{126}}_{a,b}
    \mathcal{C}^{J_{12}}\cdot \mc{C}^{J_{126}}_a(6,\chi_1) \cdot \mc{C}_b^{J_{126}}(5,\chi_2) \cdot \mathcal{C}^{J_{34}}.
}
The amplitude in the figure~\ref{fig:2->3partite} corresponds to
\eq{
    \left.\begin{array}{l}
    \mathcal{C}^{J_{12}} \sim \vev{12}^2[1\chi_1][2\chi_1] \\
    \mc{C}^{\frac12}(6,\chi_1) \sim [\chi_16][\chi_1\chi_3] \\
    \mathcal{C}^{\frac12}(5,\chi_2) \sim [\chi_2\chi_3] \\
    \mathcal{C}^{\frac12} \sim \vev{34}[4\chi_2]
    \end{array}\right\} \Rightarrow \overline{\mc{B}}^{J_{12}=1,J_{34}=\frac12,J_{126}=\frac12} \sim {\color{blue} ([16][24]+[14][26])}\vev{12}^2\vev{34}.
}
We have chosen an amplitude that is simultaneously the eigenfunction of the three partitions as the simplest example of how the partial wave amplitude basis behaves when the partition refines. 
In general, for the partition $\{12|3456\}$, there is a degenerate eigenspace of angular momentum $J_{12}$ with a basis $\overline{\mc{B}}^{J_{12}}_i$. In a refined partition it splits into subspaces with various angular momenta in the additional channels $J_{34}$, $J_{56}$, with a basis that is generically a linear combination of the original basis such as
\eq{
    \overline{\mc{B}}^{J_{12},J_{34},J_{56}}_i = \sum_j\mc{K}^{J_{34},J_{56}}_{ij}\overline{\mc{B}}^{J_{12}}_j
}
Different refinements of the partition usually lead to different refined partial wave bases, which may not have common partial waves as in this example. 

\subsection{Poincar\'e Casimir and Partial Wave Basis}\label{sec:2.2}

We have just introduced the most intuitive constructions of the generalized partial wave amplitude basis by using the spinor bridges. However there are a few subtleties:
\begin{enumerate}
    \item Sub-amplitudes involving (auxiliary) massive states are involved, for which we do not yet have a complete basis from a general algorithm. A non-trivial example is the 3-massive-vector amplitudes, which is well known to have a basis of 7 independent local amplitudes (or form factors in traditional notation \cite{Hagiwara:1986vm}). It could appear in the construction of $J = (1,1,1)$ partial waves in the example Eq.~\eqref{eq:3-partite}. By naively writing down spinor brackets it is easy to get 8 amplitudes, among which there is a redundancy relation as pointed out in \cite{Durieux:2019eor,DeAngelis:2022qco,Dong:2022mcv}. To our knowledge, such redundancy is not easy to eliminate systematically, not mentioning amplitudes with more particles and higher spins may be involved.
    
    \item We have not taken care of the mass dimensions of the amplitudes yet, as the partial waves $\overline{\mc{B}}$ are dimensionless by definition. In the next section where we map the local amplitudes to the effective operators, it is important to know the minimum mass dimensions of the local amplitudes as the partial waves without normalization. The spinor bridge construction is not capable to provide a minimum-dimensional form of the partial waves. For instance the $\overline{\mc{B}}^{J=2}$ in table~\ref{tab:partialwave} could be written, after normalization, as
    \eq{
        \overline{\mc{B}}^{J=2} &= \frac{1}{{s_{34}}^{7/2}}[12][34]^2\Big( \langle13\rangle^2 \langle24\rangle^2 +4\langle13\rangle\langle24\rangle \langle14\rangle\langle23\rangle + \langle23\rangle^2 \langle14\rangle^2 \Big)\\ &=\frac{1}{{s_{34}}^{5/2}} \Big(6 \langle13\rangle \langle24\rangle^2 [24] [34] -6s_{34} \langle13\rangle \langle24\rangle [34]+s_{34}^2\vev{12}\Big)
    }
    which means a local amplitude with mass dimension as low as $d_{\mc{B}}=5$ can have $J=2$, not as directly indicated by the form of spinor bridges.
\end{enumerate}

Therefore it is also important to study an alternative way \cite{Jiang:2020rwz} to construct partial wave basis by use of the Poincar\'e Casimir operator ${\bf W}^2$, where ${\bf W}_\mu$ is the Pauli-Lubanski operator.
Notice that the irreducible representations of the Poincar\'e group used to define the CGC in Eq.~\eqref{eq:def_CGC} comes from the eigenvalues of the Poincar\'e Casimiar ${\bf P}^2$ and ${\bf W}^2$. The former gives the Mandelstam $s=P^2$ for multi-particle states, while the latter is $-s J(J+1)$, which defines the total angular momentum $J$ in a covariant way,
\eq{\label{eq:W2_eigen}
    {\bf W}^2|P,J,\sigma\rangle = -P^2 J(J+1)|P,J,\sigma\rangle.
}
Similar to how we define representation of quantum operator, such as the momentum ${\bf p}$, on the space of single-particle wave functions in the first course of quantum mechanics, we can define the representation of ${\bf W}^2$ acting on the space of wave functions of $N$-particle state $\ket{\Phi}$ as
\eq{\label{eq:W2eigeneq}
\langle \Phi |{\bf W}^2|\{\Psi_i\}_N\rangle \equiv W^2\langle \Phi |\{\Psi_i\}_N\rangle.
}
The specific form of $W^2$ is given in \cite{Jiang:2020rwz} as
\eq{\label{eq:W2casimir}
 W^2 &=\frac{1}{8} P^2(\mathrm{Tr}[M^2] +\mathrm{Tr}[\widetilde{M}^2]) -\frac14 \mathrm{Tr}[P^\intercal M P\widetilde{M}] \, .     
}
where $P=P_\mu \sigma^\mu_{\alpha\dot\alpha}$, $P^\intercal=P_\mu\bar\sigma^{\mu\dot\alpha\alpha}$, and $M$, $\widetilde{M}$ are the chiral components of the Lorentz generator $ M_{\mu\nu} \sigma^\mu_{\alpha\dot{\alpha}} \sigma^\nu_{\beta\dot{\beta}} =\epsilon_{\alpha\beta}\widetilde{M}_{\dot{\alpha}\dot{\beta}} +\tilde{\epsilon}_{\dot{\alpha}\dot{\beta}} M_{\alpha\beta} $, given as
\eq{\label{eq:LCasimir}
    M_{\alpha \beta}=&i \sum_{i=1}^N\left(\lambda_{i \alpha} \frac{\partial}{\partial \lambda_{i}^{\beta}}+\lambda_{i \beta} \frac{\partial}{\partial \lambda_{i}^{\alpha}}\right),\quad \widetilde{M}_{\dot{\alpha} \dot{\beta}}=i \sum_{i=1}^N\left(\tilde{\lambda}_{i \dot{\alpha}} \frac{\partial}{\partial \tilde{\lambda}_{i}^{\dot{\beta}}}+\tilde{\lambda}_{i \dot{\beta}} \frac{\partial}{\partial \tilde{\lambda}_{i}^{\dot{\alpha}}}\right),\\
    (M^2)_{\alpha}^{\beta} \equiv& M_\alpha^{\,\gamma}M_\gamma^{\,\beta} = -\sum_{i}\left( 3\lambda_{i\alpha}\frac{\partial}{\partial \lambda_{i\beta}} +\delta_\alpha^\beta \lambda_i^\gamma \frac{\partial}{\partial\lambda_i^\gamma} \right)\\ &-\sum_{i,j}\left( \lambda_{i\alpha}\lambda_{j\gamma} \frac{\partial}{\partial\lambda_{i\gamma}} \frac{\partial}{\partial\lambda_{j\beta}} -\langle ij\rangle \frac{\partial}{\partial\lambda_i^\alpha} \frac{\partial}{\partial\lambda_{j\beta}} -\lambda_{i\alpha}\lambda_j^\beta \frac{\partial}{\partial\lambda_{i\gamma}} \frac{\partial}{\partial\lambda_j^\gamma} +\lambda_i^\gamma \lambda_j^\beta \frac{\partial}{\partial\lambda_i^\alpha} \frac{\partial}{\partial\lambda_j^\gamma} \right),\\
    (\widetilde{M}^2)^{\dot{\alpha}}_{\dot{\beta}} \equiv& \widetilde{M}^{\dot\alpha}_{\,\dot\gamma}\widetilde{M}^{\dot\gamma}_{\,\dot\beta} = -\sum_{i}\left( 3\tilde{\lambda}^{i\dot{\alpha}}\frac{\partial}{\partial \tilde{\lambda}^{i\dot{\beta}}} +\delta^{\dot{\alpha}}_{\dot{\beta}} \tilde{\lambda}_{i\dot{\gamma}} \frac{\partial}{\partial\tilde{\lambda}_{i\dot{\gamma}}} \right)\\ &-\sum_{i,j}\left( \tilde{\lambda}_i^{\dot{\alpha}} \tilde{\lambda}_j^{\dot{\gamma}} \frac{\partial}{\partial\tilde{\lambda}_i^{\dot{\gamma}}} \frac{\partial}{\partial\tilde{\lambda}_j^{\dot{\beta}}} -[ij] \frac{\partial}{\partial\tilde{\lambda}_{i\dot{\alpha}}} \frac{\partial}{\partial\tilde{\lambda}_j^{\dot{\beta}}} -\tilde{\lambda}_i^{\dot{\alpha}} \tilde{\lambda}_{j\dot{\beta}} \frac{\partial}{\partial\tilde{\lambda}_i^{\dot{\gamma}}} \frac{\partial}{\partial\tilde{\lambda}_{j\dot{\gamma}}} +\tilde{\lambda}_{i\dot{\gamma}} \tilde{\lambda}_{j\dot{\beta}} \frac{\partial}{\partial\tilde{\lambda}_{i\dot{\alpha}}} \frac{\partial}{\partial\tilde{\lambda}_{j\dot{\gamma}}} \right).
}
while all the spinor derivatives are acting on the wave function.
$W^2$ should have the same eigenvalue as in Eq.~\eqref{eq:W2_eigen}, since acting on the wave functions of the Poincar\'e irreducible state $\ket{P,J,\sigma}$, namely the CGC, induce
\eq{\label{eq:W2eigeneq}
\langle P,J,\sigma|{\bf W}^2|\{\Psi_i\}_N\rangle \equiv W^2\mathcal{C}^J_N\delta^4(P-\sum_i^N p_i) =-P^2J(J+1)\mathcal{C}^J_N\delta^4(P-\sum_i^N p_i),
}
which can be directly checked.
It is then straightforward to define $W^2$ acting on an amplitude when the scattering channel $\mc{I} \to \bar{\mc{I}}'$ is specified
\eq{
    W^2_{\mc{I}} \mc{M} &\equiv \sum_J \mc{M}^J_{ab}(s_{\mathcal{I}}) \left[W^2\mc{C}^J_a(\mc{I})\right] \cdot \mc{C}^J_b(\mc{I}') \\
    &= \sum_J \mc{M}^J_{ab}(s_{\mathcal{I}}) \left[-s_{\mathcal{I}}J(J+1)\mc{C}^J_a(\mc{I})\right] \cdot \mc{C}^J_b(\mc{I}') \\
    &= -\sum_J J(J+1) \times s_{\mathcal{I}}\mc{M}^J_{ab}(s_{\mathcal{I}})\overline{\mc{B}}^J_{ab}(\mc{I} | \mc{I}'),
}
where $s_{\mc{I}} = (\sum_{i\in\mc{I}}p_i)^2$ is the Mandelstam variable in the scattering channel. Note that since ${\bf W}^2$ commutes with ${\bf P}^2$, $W_{\mc{I}}^2$ acting on $s_{\mc{I}}$ must be zero, hence it does not act on $\mc{M}^J(s_{\mc{I}})$.
The partial wave amplitude basis $\overline{\mc{B}}^J$ are indeed eigenfunctions of $W_{\mc{I}}^2$ as expected
\eq{
    W^2_{\mc{I}}\overline{\mc B}^J(\mc{I}\to \mc{I}') = -s_{\mc{I}}J(J+1)\overline{\mc B}^J(\mc{I}\to \mc{I}')
}

Let's look at how the $W^2$ helps us do partial wave analysis. 
First take the $2\to2$ scattering example we investigated $\mc{M}(\psi_1,\psi_2|\phi_3,\phi_4)$, with partial wave amplitudes listed in table~\ref{tab:partialwave}. We act $W^2$ on an arbitrary local amplitude $\mc{M} = \vev{12}s_{23}$:
\eq{
    W^2_{\{12\}} \mc{M} = \vev{12}s_{12}(s_{13}-s_{23})
}
There are two spinor contractions across the channel ($s_{13} = \vev{13}[31]$ counts as 2), which does not explicitly satisfy the bridge rule. One could use Schouten identity to convert it to the form that manifests as a $J=1$ partial wave, but we could also assume that the maximum $J$ component in $\mc{M}$ is $J=1$ due to the two contractions, and project it out by
\eq{
    \bar{P}_1\mc{M} \equiv (W^2_{\{12\}} + 2s_{12}) \mc{M}  = -\vev{12}s_{12}^2,
}
where we defined $\bar{P}_J = W^2 + J(J+1)s$ which precisely annihilate the $J$ partial wave component of an amplitude. The above result has explicitly $J=0$, which can be verified by it being annihilated by $W^2$. Note that
\eq{
    \bar{P}_{J'} \overline{\mc{B}}^J = [J'(J'+1) - J(J+1)]s\overline{\mc{B}}^J,
}
the original $J=0$ component is actually $-\frac{1}{2}\vev{12}s_{12}$.
Therefore we get the $J=0,1$ component of $\mc{M}$
\eq{
    \mc{M} = \underbrace{-\frac{1}{2}\vev{12}s_{12}}_{J=0} +  \underbrace{\vev{12}(s_{23}+\frac{1}{2}s_{12})}_{J=1} \,.
}
One could verify that the $J=1$ component is proportional to $\overline{\mc{B}}^{J=1}$. This partial wave expansion may be too trivial as one can manually convert the amplitude towards the partial wave amplitudes built by the spinor bridges, or even use the orthogonality of Wigner-d matrix in the traditional method.
The $3\to3$ scattering may provide a more sophisticated example. We take the simplest all-scalar amplitude $\mc{M}(\phi_1,\phi_2,\phi_3|\phi_4,\phi_5,\phi_6)$, and compute
\eq{
    \bar{P}_1 s_{14} = (W^2_{\{123\}} + 2s_{123})s_{14} = -(s_{12}+s_{13})(s_{45}+s_{46}),
}
which is a $J=0$ partial wave amplitude, verified by $W^2\bar{P}_1s_{14}=0$. 
The partial wave expansion is thus
\eq{
    s_{14} = \underbrace{-\frac{(s_{12}+s_{13})(s_{45}+s_{46})}{2s_{12}+2s_{13}+2s_{23}}}_{J=0} + \underbrace{\left[s_{14} + \frac{(s_{12}+s_{13})(s_{45}+s_{46})}{2s_{12}+2s_{13}+2s_{23}}\right]}_{J=1},
}
which does not seem to be easy to guess. For more general amplitudes, more than one projection may be needed to get the expansion, but still achievable by the $W^2$ action systematically, which we implement in the package ABC4EFT \cite{Li:2022tec}. This example has a feature that the partial wave components have a pole in the scattering channel, though the pole does not affect the angular momentum $J$ since the commutation relation $[W^2,s]=0$, and it cancels between the two components. It is not surprising because the normalized partial wave amplitudes by definition can have such structure. We have to keep in mind that partial wave amplitudes are pure kinematic functions that have nothing to do with dynamics like poles. We will come back to this issue in the next section.

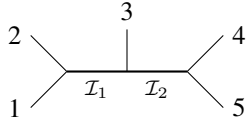
\begin{figure}
	\centering
	\begin{tikzpicture}[scale=0.8]
	\coordinate (v1) at (0,0) {};
	\coordinate (v2) at (1,0) {};
	\coordinate (v3) at (2,0) {};
	\draw (-0.6,-0.6) node[left]{1} -- (v1);
	\draw (-0.6,0.6) node[left]{2} -- (v1);
	\draw[thick] (v1) -- (v2);
	\draw (1,0.7) node[above]{3} -- (v2);
	\draw[thick] (v2) -- (v3);
	\draw (2.6,0.6) node[right]{4} -- (v3);
	\draw (2.6,-0.6) node[right]{5} -- (v3);
	\node at (0.5,-0.3) {\footnotesize $\mc{I}_1$};
	\node at (1.5,-0.3) {\footnotesize $\mc{I}_2$};
	\end{tikzpicture}
	\caption{A tree topology of a 5-particle amplitude. $\mc{I}_{1,2}$ denotes two scattering channels, which together defines a partition $\mc{P}=\{\mc{I}_1,\mc{I}_2\}$. The quantum numbers associated to them can be fixed simultaneously, which reflects the commutation relation $[W^2_{\mc{I}_1},W^2_{\mc{I}_2}]=0$.}\label{fig:multi-partite_topo}
\end{figure}
For multi-partite partial waves, we can act several $W_{\mc{I}_i}^2$ on the local amplitudes simultaneously, because Eq.~\eqref{eq:non-ovl-ch} guarantees that they commute
\eq{\label{eq:W2commute} [W^2_{\mc{I}_i},W^2_{\mc{I}_j}]=0.} 
A simple example is given in Figure~\ref{fig:multi-partite_topo}. Hence we can define the simultaneous eigenfunctions of them as the multi-partite partial wave amplitude basis
\eq{\label{eq:multi-partite}
    W_{\mc{I}_i}^2 \overline{\mc{B}}^{J_1,J_2,\dots} = -s_{\mc{I}_i}J_i(J_i+1)\overline{\mc{B}}^{J_1,J_2,\dots}, \quad \forall i.
}

\subsection{Gauge Eigen-basis and $SU(N)$ Casimir}\label{sec:2.3}

In above subsections, we have shown that for the Lorentz structure of amplitudes, the partial wave basis can be constructed via the spinor bridge method, while for the generalized partial wave basis two subtleties arise as described at the beginning of section.~\ref{sec:2.2}, thus the method of Casimir eigen-basis is introduced. For the scattering amplitude, there are several ways to present the gauge structure, such as the color basis~\cite{DelDuca:1999rs, DelDuca:1999iql}, the trace basis~\cite{Bern:1990ux, Berends:1987cv, Mangano:1988kk}, and the Young tensor basis~\cite{Li:2020gnx,Li:2020xlh,Li:2022tec}. To perform partial wave expansion on gauge structure, the gauge structure needs to be decomposed into the gauge eigen-basis. The multiplet basis, an orthogonal and normalized gauge eigen-basis, was proposed in Ref.~\cite{Keppeler:2012ih, Keppeler:2013yla}. However, it only applies to the amplitude with equal numbers of fundamental representation and anti-fundamental representation, and the form of the analytical expression could be very long. Thus to obtain the gauge eigen-basis for general scattering amplitude, we adopt the gauge Casimir method similar to the Lorentz structure. In this case, it does not have the subtleties related to the dimension between the amplitudes and the corresponding operators. Therefore we will consistently call the gauge basis $T({\bf R})$ in Eq.~\eqref{eq:pwdefine} for generalized partial wave and j-basis defined in the next section as gauge eigen-basis. We will start by illustrating the bridge method in the gauge sector and then introduce a similar Casimir operator technique for the gauge amplitude to obtain the generalized gauge eigen-basis for multi-partite channels. 

As a concrete example of the gauge sector of partial wave decomposition in Eq.~\eqref{eq:pwdefine}, we can have the partial wave expansion for a 2-to-2 gauge amplitude as:
\begin{eqnarray}
&&\langle (\mathbf{r}_1,I_1),(\mathbf{r}_2,I_2)| \mbf{T} |(\mathbf{r}_3,I_3),(\mathbf{r}_4,I_4)\rangle \nonumber\\
&=& 
\sum_{\mathbf{R},I,\mathbf{R}',I'} 
\langle (\mathbf{r}_1,I_1),(\mathbf{r}_2,I_2)|\mathbf{R},I\rangle \langle \mathbf{R},I| \mbf{T}|\mathbf{R}',I'\rangle\langle\mathbf{R}',I'|(\mathbf{r}_3,I_3),(\mathbf{r}_4,I_4)\rangle\nonumber\\
&=& \sum_{\mathbf{R}} c_{\mathbf{R}} \left(\sum_{I} \mathcal{T}(\mathbf{R};\mathbf{r}_1,\mathbf{r}_2)^{I}_{I_1,I_2} (\mathcal{T}(\mathbf{R};\mathbf{r}_3,\mathbf{r}_4)^{I}_{I_3,I_4})^*\right)\nonumber \\
&=& \sum_{\mathbf{R}} c_{\mathbf{R}} \left(\sum_{I} \mathcal{T}(\mathbf{R};\mathbf{r}_1,\mathbf{r}_2)^{I}_{I_1,I_2} \mathcal{T}(\overline{\mathbf{R}};\overline{\mathbf{r}}_3,\overline{\mathbf{r}}_4)^{I}_{I_3,I_4}\right)\nonumber \\
&=& \sum_{\mathbf{R}}c_{\mathbf{R}} \mc{T}(\mathbf{R})_{I_1I_2I_3I_4}, 
\end{eqnarray}
where from the second to the third line we have  $\langle \mathbf{R},I| \mbf{T}|\mathbf{R}',I'\rangle = c_{\mathbf{R}}\delta_{\mathbf{R}\mathbf{R}'}\delta_{II'}$ because of Wigner-Eckart theorem,\footnote{More specifically, this is because the $O$ transforms as a singlet under the gauge group, so $ \mbf{T}|\mathbf{R}\rangle$ is still a $\mathbf{R}$ representation.} and $c_{\mathbf{R}}$ is a constant that only depends on $\mathbf{R}$ characterizing the capability of $ \mbf{T}$ generating amplitudes in the $\mathbf{R}$-to-$\mathbf{R}$ scattering channel.
$\mathcal{T}(\mathbf{R};\mathbf{r}_i,\mathbf{r}_j)$'s are just CGCs for constructing representation $\mathbf{R}$ with $\mathbf{r}_1,\mathbf{r}_2$ or $\mathbf{r}_3,\mathbf{r}_4$, while $\mc{T}(\mathbf{R})$'s are gauge amplitude basis as counterparts of $\overline{\mc{B}}$'s in the Lorentz part of amplitudes, and the bridging can be viewed as contracting the two gauge indices of a pair of conjugated representations $\mbf{R}$ and $\overline{\mbf{R}}$. $\mc{T}(\mathbf{R})$'s can also be regarded as  CGCs that group the irreps $\mathbf{r}_1,\mathbf{r}_2,\overline{\mathbf{r}}_3,\overline{\mathbf{r}}_4$ to a singlet, and thus by definition  $\mc{T}(\mathbf{R})_{I_1I_2I_3I_4}$ is an invariant tensor such that under a general gauge group transformation, $\mc{T}(\mathbf{R})_{I_1I_2I_3I_4}$ does not change:
\begin{eqnarray}
\sum_{Z_i}(U_{\mathbf{r}_1})_{I_1Z_1}(U_{\mathbf{r}_2})_{I_2Z_2}(U_{\overline{\mathbf{r}}_3})_{I_3Z_3}(U_{\overline{\mathbf{r}}_4})_{I_4Z_4}\mc{T}(\mathbf{R})_{Z_1Z_2Z_3Z_4}=\mc{T}(\mathbf{R})_{I_1I_2I_3I_4},
\label{eq:invT}
\end{eqnarray}
where $U_{\mathbf{r}_i}$ are matrix representations for the transformation of the representation ${\mathbf{r}_i}$.
More specifically, We take the scattering between $\pi$ mesons in the chiral perturbation theory for a physical example:
\begin{eqnarray}
&&\langle \pi^{I_{1}},\pi^{I_2}| \mc{O} |\pi^{I_{3}},\pi^{I_{4}}\rangle \nonumber\\
&=&c_{\bf 1} \mc{T}({\bf 1;2,2})_{I_1I_2}(\mc{T}({\bf 1;2,2})_{I_3I_4})^*+c_{\bf 2} \mc{T}({\bf 2;2,2})^I_{I_1I_2}(\mc{T}({\bf 2;2,2})^I_{I_3I_4})^*\nonumber \\
&&+c_{\bf 3} \mc{T}({\bf 3;2,2})^I_{I_1I_2}(\mc{T}({\bf 3;2,2})^I_{I_3I_4})^*\nonumber \\
&=&c_{\bf 1}\mc{T}({\bf 1})_{I_1I_2I_3I_4}+c_{\bf 2}\mc{T}({\bf 2})_{I_1I_2I_3I_4}+c_{\bf 3}\mc{T}({\bf 3})_{I_1I_2I_3I_4}
\end{eqnarray}
where the CGCs $\mc{T}({\bf 1;2,2})_{IJ}$ and $\mc{T}({\bf 2;2,2})^I_{JK}$ can be expressed in nices forms with group invariant tensors:
\begin{eqnarray}
\mc{T}({\bf 1;2,2})_{IJ}=\delta^{IJ},\quad \mc{T}({\bf 2;2,2})^I_{JK}=\epsilon^{IJK}.
\end{eqnarray}
However, for the $\mc{T}({\bf 3;2,2})^I_{JK}$, one may only obtain a numerical form, and this feature is different from Lorentz amplitude, where the CGCs can be expressed as closed forms with spinor-helicity variables. 

Things become more complicated when we consider the gauge part of a general 2-partite $M\to N$ scattering as in Eq.~\eqref{eq:pwdefine}:

\begin{eqnarray}
\mc{M}&=&\langle \{\mathbf{r}_m,I_m\}| \mbf{T} |\{\mathbf{r}_n,I_n\}\rangle \nonumber\\
&=& 
\sum_{\substack{\mathbf{R},I,\mathbf{R}',I'\\b,b'}} 
\langle \{\mathbf{r}_m,I_m\}|\mathbf{R},I,b\rangle \langle \mathbf{R},I,b|\mbf{T}|\mathbf{R}',I',b'\rangle\langle\mathbf{R}',I',b'|\{\mathbf{r}_n,I_n\}\rangle\nonumber\\
&=& \sum_{\mathbf{R},b,b'} \mbf{T}({\mathbf{R}})_{bb'} \left(\sum_{I} \mathcal{T}_b(\mathbf{R},\{\mathbf{r}_m\})^{I}_{I_1I_2\dots I_M} (\mathcal{T}_{b'}(\mathbf{R},\{\mathbf{r}_n\})^{I}_{I_{M+1}\dots I_{M+N}})^*\right) \nonumber\\
&=& \sum_{\mathbf{R},b,b'} \mbf{T}({\mathbf{R}})_{bb'} \mc{T}_{bb'}(\mathbf{R})_{I_1I_2\dots I_{M+N}},
\end{eqnarray}
where from the second to the third line we have again used  $\mbf{T}({\mathbf{R}})_{bb'}=\langle \mathbf{R},I,b|\mbf{T}|\mathbf{R}',I',b\rangle\propto\delta_{\mathbf{R}\mathbf{R}'}\delta_{II'}$, while comparing to the 2-to-2 process, additional indices $b,b'$ label the degeneracy of the particular representation $\mathbf{R}$ constructed from the multiparticle states. 
$\mathcal{T}_b(\mathbf{R},\{\mathbf{r}_m\})$ and $\mc{T}_{b'}(\mathbf{R},\{\mathbf{r}_n\})$ are just generalized CGCs for constructing an $\mathbf{R}$ irreducible representation from  irreducible representations $\{\mathbf{r}_m\}$ and $\{\mathbf{r}_n\}$ respectively. 
As mentioned above, these CGCs are not easy to express in the form of invariant group tensors. Their numerical values can be in principle obtained by using the ordinary CGCs for combining two irreps to another irrep. Such 2-to-1 GCGs are well-known in the $SU(2)$ group and can be expressed in the Wigner $3-j$ symbols, while for a general $SU(N)$ group, finding the 2-to-1 GCGs needs dedicated algorithm introduced in textbooks.  This process might be extremely memory-cost as one needs to expand a very large tensor product constructed from the CGCs before contraction.
However, as we have illustrated in the previous works~\cite{Li:2020gnx,Li:2020xlh,Li:2022tec} and briefly summarized in section~\ref{sec:gauge}, a set of complete and independent monomial invariant tensors $\mc{T}^{\rm m}$'s for grouping any irreducible representation to singlets can be constructed using the modified Littlewood-Richardson Rules for Young Tableaux, 
and because all singlets form a basis of an invariant subspace, the corresponding invariant tensors also form a basis of invariant subspace. See appendix~\ref{app:E} for detailed discussions.  $\mc{T}_{ab}(\mathbf{R})$'s, similar to the $\mc{T}(\mathbf{R})$'s in Eq.~\eqref{eq:invT} as CGCs also grouping the same irreps into singlets is certainly an invariant tensor in the direct product space, and must be able to expressed as linear combinations of ${\mc{T}^{  m}}$'s. For example, for the $\pi\pi$ scattering, we have $\mc{T}^{  m}$ as:
\begin{eqnarray}
\mc{T}^{  m}_1={\delta_{I_1I_3}}\delta_{I_2I_4},\ \mc{T}^{  m}_2=\delta_{I_1I_2}\delta_{I_3I_4},\ \mc{T}^{  m}_3=\delta_{I_1I_4}\delta_{I_2I_3}.
\label{eq:Typipi}
\end{eqnarray}
In our previous work~\cite{Li:2020zfq}, we proposed a method for projecting out the ${\cal T}_{bb'}(\mathbf{R})$ by acting the  Young symmetrizer for the corresponding Young Tableau on the so-called Young tableau basis ${\cal T}^{\rm y}$'s expressed with the fundamental indices only, which is equivalent to the monomial basis in Eq.~\eqref{eq:Typipi}. 
For example, four $\pi$'s can be expressed with field with fundamental indices only:
\begin{eqnarray}
\pi^{I_k}\sim \lambda_{i_k}^{z}\epsilon_{z j_k}\pi^{I_k}=\pi^{i_kj_k},
\end{eqnarray}
which corresponds to the Young tableau $\young({{i_k}}{{j_k}})$.
One of the $\mc{T}^{\rm y}$'s becomes:
\begin{eqnarray}
\mc{T}^{\rm y}_{i_1j_1i_2j_2i_3j_3i_4j_4}=\epsilon^{i_1i_3}\epsilon^{j_1j_3}\epsilon^{i_2i_4}\epsilon^{j_2j_4}\sim \young({{i_1}}{{j_1}}{{i_2}}{{j_2}},{{i_3}}{{j_3}}{{i_4}}{{j_4}})
\end{eqnarray}
and projecting with the Young symmetrier of $\young({{i_1}}{{j_1}}{{i_2}}{{j_2}})$  gives you $T(2)$:
\begin{eqnarray}
{\cal Y}\left[{\young({{i_1}}{{j_1}}{{i_2}}{{j_2}})}\right]\circ \mc{T}^{\rm y}_{i_1j_1i_2j_2i_3j_3i_4j_4}=\frac{1}{24}\left(\epsilon^{i_1i_3}\epsilon^{j_1j_3}\epsilon^{i_2i_4}\epsilon^{j_2j_4}+\text{perm}(i_1,j_1,i_2,j_2)\right),\label{eq:ypermute}
\end{eqnarray}
after conctracting with $\lambda_{i_k}^{z}\epsilon_{z j_k}$ for each pion fields, one obtained the invariant tensors back to the adjoint indices as:
\begin{eqnarray}
\mc{T}(\mathbf{3})_{I_1I_2I_3I_4}=\frac{2}{3}\times(-3\delta_{I_1I_3}\delta_{I_2I_4}+2\delta_{I_1I_2}\delta_{I_3I_4}-3\delta_{I_1I_4}\delta_{I_2I_3}),
\end{eqnarray}
Analogously, one can obtain the $\mc{T}(\mbf{2})$ and $\mc{T}(\mbf{1})$:
\begin{eqnarray}
&&{\cal Y}\left[{\young({{i_1}}{{j_1}}{{j_2}},{{i_2}})}\right]\circ \mc{T}^{\rm y}_{i_1j_1i_2j_2i_3j_3i_4j_4}\sim \delta_{I_1I_3}\delta_{I_2I_4}-\delta_{I_1I_4}\delta_{I_3I_2}=\mc{T}(\mbf{2}),\\
&&{\cal Y}\left[{\young({{i_1}}{{j_1}},{{i_2}}{{j_2}})}\right]\circ T^{\rm y}_{i_1j_1i_2j_2i_3j_3i_4j_4}\sim \frac{4}{3}\delta_{I_1I_2}\delta_{I_3I_4}=\mc{T}(\mbf{1}),
\end{eqnarray}
This method circumvents the difficulties for constructing the intermediate CGCs like $\mc{T}({\bf 3;2,2})^I_{JK}$, however the computation complexity increases factorially with the number of fundamental indices, and it needs to check the independence of the resulting invariant tensor numerically, which also gets slow when the number of degeneracy of the corresponding intermediate irreducible representation is large. For example, it takes nearly one hour for an ordinary desktop computer to obtain all the $\mc{T}_{bb'}(\mathbf{R})$'s for the 2-to-2 scattering of gluons.

Due to the difficulties in the above methods, we propose a new method in this work to obtain $\mc{T}_{bb'}(\mathbf{R})$'s. Inspired by the Casimir method for the Lorentz amplitude,  the idea is to obtain a matrix representation of Casimir operators for each gauge group in the space of invariant tensors.
The only question remains here is to find the correct definition of Casimir operators on the invariant tensors space as counterparts of Eq.~\eqref{eq:LCasimir}. The $N-1$ independent Casimir operators for the $SU(N)$ group can be constructed by contracting $N-1$ generators with the structure constant. The general expressions for them can be found in appendix~\ref{app:casimir}. Here we only show the Casimirs for $SU(2)$ and $SU(3)$:
\begin{eqnarray}\label{eq:C2C3def}
\mathbbm{C}_2&=& \mathbbm{T}^a\mathbbm{T}^a,\ \text{for both $SU(2)$ and $SU(3)$,}\label{eq:CasimirC2}\\
\mathbbm{C}_3 &=& d^{abc}\mathbbm{T}^a\mathbbm{T}^b\mathbbm{T}^c,\ \text{for $SU(3)$ only},\label{eq:CasimirC3}
\end{eqnarray}
where $\mathbbm{T}$'s are generators of their corresponding group. It is well-known that Casimirs for irreducible representations are constant $C_i(\mathbf{R})$ times the identity matrix, and a table of  $C_i(\mathbf{R})$ for $SU(2)$ and $SU(3)$ and the general algorithm to calculate them 
can also be found in appendix~\ref{app:casimir}. While in our case, the generators $\mathbbm{T}$'s are usually for the direct product representations with the definition in the following formula:
\begin{eqnarray}
\mathbbm{T}^{A}_{\otimes \{{\bf r}_i\}} = \sum_{i=1}^N E_{\mathbf{r}_1}\times E_{\mathbf{r}_2}\times\dots\times T^A_{\mathbf{r}_i}\times\dots E_{\mathbf{r}_N},
\end{eqnarray}
where $T^A_{\mathbf{r}_i}$'s  and $E_{\mathbf{r}_i}$'s are generators and identity matrices for irreducible representation $r_i$ respectively. This formula can be easily obtain by expanding the following infinitesimal transformation for the direct product representation and truncate the expression to the first order:
\begin{eqnarray}
(E_{\mathbf{r}_1}+T_{\mathbf{r}_1}^Ad\alpha_A)\times (E_{\mathbf{r}_2}+T_{\mathbf{r}_2}^Ad\alpha_A)\dots \times (E_{\mathbf{r}_N}+T_{\mathbf{r}_N}^Ad\alpha_A).
\end{eqnarray}
Therefore the action of Casimir on a general tensor $\Theta_{I_1I_2\dots I_N}$ with indices $I_i$ transforming as irreducible representation $\mathbf{r}_i$ are defined by:
\begin{eqnarray}\label{eq:Trep}
\Theta'_{I_1I_2\dots I_N}= \mathbbm{T}^a_{\otimes \{\mathbf{r}_i\}}\circ \Theta_{I_1I_2\dots I_N}\equiv
\sum_{i=1}^N (T^a_{\mathbf{r}_i})_{I_{i}}^{Z}\Theta_{I_1\dots I_{i-1}Z I_{i+1} I_N},
\end{eqnarray}
where we have used the symbol $\circ$ to denote the action of the generator operator on a tensor.  For example, the generator $T^A_{ {\mathbf{3}^{\otimes 4}}}$ acting on the m-basis ${\cal T}^{\rm m}_1$ in Eq.~\eqref{eq:Typipi} reads:
\begin{eqnarray}
{\cal T}^{\rm m}{}'_{I_1I_2I_3I_4}&=&\mathbbm{T}^A_{ {\mathbf{3}^{\otimes 4}}}\circ \delta_{I_1I_3}\delta_{I_2I_4}\nonumber\\
&=&T^{A}_{I_1Z}\delta_{ZI_3}\delta_{I_2I_4}+T^{A}_{I_2Z}\delta_{I_1I_3}\delta_{ZI_4}+T^{A}_{I_3Z}\delta_{I_1Z}\delta_{I_2I_4}+T^{A}_{I_4Z}\delta_{I_1I_3}\delta_{I_2Z}\nonumber\\
&=& T^{A}_{I_1I_3}\delta_{I_2I_4}+T^{A}_{I_2I_4}\delta_{I_1I_3}+T^{A}_{I_3I_1}\delta_{I_2I_4}+T^{A}_{I_4I_2}\delta_{I_1I_3}
\end{eqnarray}
Following this definition, the action of Casimir on tensors is manifest.
Similar to the Lorentz amplitude, with the above definition of Casimir acting on arbitrary tensors, we can define the \textit{partial Casimir} that composed by the partial generator acting on a subset $\mathbbm{S}$ of indices as:
\begin{eqnarray}
 \underset{\mathbbm{S}}{\mathbbm{T}}^A\circ\Theta_{I_1I_2\dots I_N}=
\sum_{i\in \mathbbm{S}}^N (T^A_{r_i})_{I_{i}}^{Z}\Theta_{I_1\dots I_{i-1}Z I_{i+1} I_N}.\label{eq:partialT}
\end{eqnarray}
With the above definition of the action of generators, one can easily see that the complexity of computing the action of the Casimirs scales as polynomial with the number of indices, which is expected to be more efficient then the permutation method in Eq.~\eqref{eq:ypermute}.
By definition, $\mc{T}(\mathbf{R},\{{\bf r}_n\})^I_{I_1I_2\dots I_N}$ must be a  eigenvector of the partial Casimir operators $\mathbbm{C}_{i}$ acting on the indices $I_1I_2\dots I_N$:
\begin{eqnarray}
\mathbbm{C}_i\circ \mc{T}(\mathbf{R},\{\mathbf{r}_n\})^I_{I_1I_2\dots I_N}=C_i(\mathbf{R})\mc{T}(\mathbf{R},\{\mathbf{r}_n\})^I_{I_1I_2\dots I_N}.
\end{eqnarray}
Take the $\pi\pi$ scattering for an example, it is easy to varify that $\mc{T}({\bf 1;2,2})_{IJ}$ are eigenvector of $\mathbbm{C}_2$:
\begin{eqnarray}
&&\mathbbm{C}_2\circ\mc{T}({\bf 1;2,2})_{IJ}\nonumber\\
&=& (T^A)_{IL}\left((T^A)_{LK}\delta_{KJ}+(T^A)_{JK}\delta_{LK}\right)+(T^A)_{JL}\left((T^A)_{IK}\delta_{KL}+(T^A)_{LK}\delta_{IK}\right)\nonumber\\
&=& (T^AT^A)_{IJ}+(T^A(T^A)^{\rm T})_{JI}+(T^A(T^A)^{\rm T})_{JI}+(T^AT^A)_{JI}\nonumber \\
&=& 0\times \mc{T}({\bf 1;2,2})_{IJ},
\end{eqnarray}
where we have used $(T^A)_{IJ}=-i \epsilon^{AIJ}$.
In this way, the amplitude basis $\mc{T}_{bb'}(\mathbf{R})_{I_1I_2\dots I_{M+N}}$ is also a eigenvector for the partial Casimir acting on the first $M$ indices only: 
\begin{eqnarray}
\underset{\{\mathbf{r}_1,\dots, \mathbf{r}_M\}}{\mathbbm{C}_i}\circ\mc{T}_{bb'}(\mathbf{R})_{I_1I_2\dots I_{M+N}} = C_i(\mathbf{R})\mc{T}_{bb'}(\mathbf{R})_{I_1I_2\dots I_{M+N}}.
\end{eqnarray}
More generally, since  full Casimirs acting on all the indices and  partial Casimirs acting on only a part of indices commute with all the group elements, it is easy to verify that partial Casimirs $\underset{\mathbbm{S}}{\mathbbm{C}}$ acting on any part of indices of an invariant tensor is still an invariant tensor, by checking its transformation properties under an arbitrary $SU(N)$ transformation $\mathbbm{U}$:
\begin{eqnarray}
&&\mathbbm{U}\circ \underset{\mathbbm{S}}{\mathbbm{C}}\circ \Theta_{(i\in\mathbbm{S})(j\not\in \mathbbm{S})}\nonumber\\
&=&\mathbbm{U}\circ\sum_{(k)} C_{(i)(k)}\Theta_{(k)(j)}\nonumber \\
&=&\sum_{(k),(m),(l)}\underset{\mathbbm{S}}{U}{}_{(i)(l)}\underset{\slashed{\mathbbm{S}}}{U}{}_{(j)(m)}C_{(l)(k)}\Theta_{(k)(m)}\nonumber\\
&=&\sum_{(k),(m),(l)}C_{(i)(l)}\underset{\mathbbm{S}}{U}{}_{(l)(k)}\underset{\slashed{\mathbbm{S}}}{U}{}_{(j)(m)}\Theta_{(k)(m)}\nonumber\\
&=&\sum_{(l)}C_{(i)(l)}\Theta_{(l)(j)}=\underset{\mathbbm{S}}{\mathbbm{C}}\circ \Theta_{(i\in\mathbbm{S})(j\not\in \mathbbm{S})}
\end{eqnarray}
where we have collectively represented all the indices in $\mathbbm{S}$ as $(i)$ and those not in $\mathbbm{S}$ as $(j)$ in the first line. The representation matrices for Casimir operators on the set of indices in $\mathbbm{S}$ is given by $C_{(i)(k)}$, and two $U$'s are representation matrices of $SU(N)$ for collectively transforming the indices in and not in $\mathbbm{S}$.
From the third to the fourth line we use the fact that Casimir operators commute with all of the group elements, and from the fourth to the fifth line we use the properties that $\Theta$ is an invariant tensor. 
Therefore, we conclude that given a basis of invariant tensor $\{\mc{T}_j\}$, the partial Casimir has a matrix representation $C_{ji}$:
\begin{eqnarray}\label{eq:master-gauge}
\underset{\mathbbm{S}}{\mathbbm{C}}\circ \mc{T}_i=\sum_j \mc{T}_jC_{ji},
\end{eqnarray}
where $C_{ji}$ can be diagonalized with eigenvalues $C(\mathbf{R})$ and the corresponding eigenvectors. We emphasize here that Eq.~(\ref{eq:master-gauge}) is our master formula for the gauge sector to derive the gauge eigen-basis. Still, we take the $\pi\pi$ scattering as an example.
We have matrix $C_2$ as:
\begin{eqnarray}
\underset{\{1 2\}}{\mathbbm{C}_2}\circ\mc{T}^{ m}_i=\underset{\{1 2\}}{({C}^{\rm T}_2)_{ij}}{\cal T}^{ m}_j=
\begin{pmatrix}
4 & 0 & -2\\
-2 & 0 & -2\\
2 & 0 & 4
\end{pmatrix}
\begin{pmatrix}
{\cal T}^{ m}_1\\
{\cal T}^{ m}_2\\
{\cal T}^{ m}_3
\end{pmatrix},
\end{eqnarray}
and after diagonalization with the matrix ${\cal K}_{12}$, we get eigenvalues $6,2,0$ 
\begin{eqnarray}
{\cal K}^{jm}_{C_2,\{12\}}{({C}^{\rm T}_2)}({\cal K}^{jm}_{\{12\}})^{-1} = {\rm diag}\{6,2,0\},\ \text{with } {\cal K}^{jm}_{\{12\}}=\begin{pmatrix}
-3 & 2 & -3 \\
1 & 0 & -1 \\
0 & 1 & 0
\end{pmatrix},
\end{eqnarray}
and the corresponding coordinates of eigenvectors on the m-basis are:
\begin{eqnarray}
&&{\cal T}(\mathbf{3})_{I_1I_2I_3I_4}=-3{\cal T}^{m}_1+2{\cal T}^{ m}_2-3{\cal T}^{ m}_3=-3\delta_{I_1I_3}\delta_{I_2I_4}+2\delta_{I_1I_2}\delta_{I_3I_4}-3\delta_{I_1I_4}\delta_{I_2I_3}\nonumber \\
&&{\cal T}(\mathbf{2})_{I_1I_2I_3I_4}={\cal T}^{ m}_1-{\cal T}^{ m}_3=\delta_{I_1I_3}\delta_{I_2I_4}-\delta_{I_1I_4}\delta_{I_2I_3}\nonumber \\
&&{\cal T}(\mathbf{1})_{I_1I_2I_3I_4}={\cal T}^{ m}_2=\delta_{I_1I_2}\delta_{I_3I_4}.
\end{eqnarray}

\section{Effective Operator: J-basis from Partial Wave}
\label{sec:j-basis}

We have elaborated the construction of generalized partial wave basis as the irreducible representations $(J,\mathbf{R})$ of the Poincare and gauge groups. 
In this section, we are going to find the effective operators $\mc{O}^{J,\bf R}$ that correspond to these partial wave amplitudes, which we call the j-basis \cite{Jiang:2020rwz,Li:2020zfq,Shu:2021qlr,Li:2022tec}. 
In particular, these operators form eigen-basis of the symmetry algebra and exhibit non-interference property\footnote{The non-interference has several interpretations. Two operators of the same type but different quantum numbers, either $J$ or ${\bf R}$, contribute to amplitudes that do not interfere in the computation of cross section. Also, two operators with different quantum numbers cannot renormalize each other through particular loop diagram in the designated channel, namely the selection rule \cite{Jiang:2020rwz}. } due to the conservation laws
\eq{\label{eq:oper_sel}
    \mc{O}_{\mc{I}}^{J,{\bf R}}\ket{\Psi_\mc{I}}^{J',{\bf R'}} \sim \delta^{JJ'}\delta^{\bf RR'},
}
where $\mc{I}$ denotes the scattering channel in which we defined the partial waves, and $\Psi_\mc{I}$ is the scattering state in either the past or the future.
We will demonstrate a systematic way to construct j-basis by using Casimir operators and Young Tableau bases of both the Lorentz and gauge sector.

\subsection{Lorentz Eigen-Basis Construction}

\subsubsection{PW Amplitude Operator Correspondence}\label{subsubsec:ampop}

The correspondence between contact amplitudes and the effective operators are shown in
\cite{Ma:2019gtx,Li:2020gnx,Li:2020xlh}. 
The key idea is to define the leading vertex of an operator (with the lowest power of gauge couplings) as the corresponding amplitude. The mapping is one-to-one given the following assumption: operators containing kinematic terms of the fields or the covariant derivatives commutator (CDC) should be converted via equations of motion (EOM) or the identity $i[D,D]=F$ before taking the leading vertex. For example, the external state of the corresponding on-shell amplitude generated by the operator $\bar\psi\gamma^\mu\psi D^\nu F_{\mu\nu}$ does not involve the gauge boson coming from the field $F_{\mu\nu}$, but rather particles in the gauge current $J_\mu$ in the converted form $\bar\psi\gamma^\mu\psi J_{\mu}$. Then, amplitudes find their corresponding operators by the following translation
\eq{
    \begin{array}{l|c|c|c|c|c}
    \hline
    \text{Amplitude Blocks} & \lambda_i^n \tilde\lambda_i^n   &   \lambda_i^{n+1} \tilde\lambda_i^n & \lambda_i^n \tilde\lambda_i^{n+1} & \lambda_i^{n+2} \tilde\lambda_i^n & \lambda_i^n \tilde\lambda_i^{n+2} \\
    \hline
    \text{Operator Blocks} & D^n\phi_i &   D^n\psi_i &   D^n\psi^\dagger_i &   D^nF_{Li}    &   D^nF_{Ri} \\
    \hline
    \end{array}
}
Let's take an example of amplitude $\vev{12}[23][24]s_{14}$, the corresponding operator is found as
\eq{\label{eq:ex_amp-op}
    \vev{12}[23][24]s_{14} &= (\lambda_1^2 \tilde\lambda_1)^{\alpha\beta}_{\dot\alpha} (\lambda_2 \tilde\lambda_2^2)_{\alpha\dot\beta\dot\gamma} \tilde\lambda_3^{\dot\beta} (\lambda_4 \tilde\lambda_4^2)_\beta^{\dot\alpha\dot\gamma} \\
    &\stackrel{\text{amp/op}}{\longleftrightarrow} (D\psi_1)^{\alpha\beta}_{\dot\alpha}(D\psi^\dagger_2)_{\alpha\dot\beta\dot\gamma}(\psi^\dagger_3)^{\dot\beta}(D\psi^\dagger_4)_\beta^{\dot\alpha\dot\gamma} \\
    &\sim (D_\mu \psi_1)^\beta (D_\nu \psi^\dagger_2)_{\dot\beta} (\psi^\dagger_3)^{\dot\beta} (D_\rho \psi^\dagger_4)^{\dot\alpha} (\sigma^\rho\bar\sigma^\mu\sigma^\nu)_{\beta\dot\alpha} \\
    &= (D_\mu \psi_1 (g^{\rho\mu}\sigma^\nu -i\epsilon^{\mu\nu\rho\lambda}\sigma_\lambda) D_\rho \psi^\dagger_4) (D_\nu \psi^\dagger_2 \psi^\dagger_3) \,+\,\text{EOM}.
}
In the third line, the spinor indices are turned into usual Lorentz indices via the sigma matrices, with possibly multiple choices of parings, hence the $\sim$ sign. However, they are equivalent according to the above argument: the differences among the choices are counted in other types of amplitudes due to the EOM. One needs to be extra careful with the correspondence when the EOM is important. In the last line we convert the operator into the usual forms with Lorentz indices, but as is shown in the example, amplitude monomials do not correspond to operator monomials in general.

Therefore, any amplitude as a polynomial of spinor brackets, also known as contact amplitude denoted by $\mc{B}$ without the overline, corresponds to a local operator $\mc{O}$. On the other hand, the partial wave amplitudes $\overline{\mc B}^J$ normalized by negative powers of the Mandelstam variables are thus not contact amplitudes and do not directly correspond to operators. Since multiplying by the particular Mandelstam variables in the scattering channel do not alter the angular momentum of the scattering state, we directly obtain a family of operators corresponding to the partial wave amplitudes
\eq{
    \mc{B}^{J,d=N+2k} = s^k\overline{\mc B}^J \sim \mc{O}^{J,d}
}
where $N$ is the number of particles involved in the amplitude, and $d$ is the dimension of the corresponding operator on the right in the usual sense.

For example, the scattering $\psi^\dagger_1,\psi_2 \to \psi^\dagger_3,\psi_4$ has partial wave starting from $J=1$, which can be constructed as $\overline{\mc{B}}^{J=1}(\psi^\dagger_1,\psi_2 | \psi_3,\psi^\dagger_4) = s^{-2}[12]\vev{23}^2[34]$. The amplitude basis at dimension $d$ can be directly given
\eq{
    \mc{B}^{J=1,d=4+2k} \sim s^{k-2}[12]\vev{23}^2[34]
}
On the operator side, the correspondence gives
\eq{\label{eq:J=1_generald}
    \mc{O}^{J=1,d=4+2k} \sim (D^{k-2}_{\rho_1,\dots,\rho_{k-2}}\psi^\dagger_1 \bar\sigma_\mu D_\nu \psi_3) (D^{k-2,\rho_1,\dots,\rho_{k-2}}D^\mu\psi_2 \sigma^\nu \psi^\dagger_4).
}
It is easy to see that such operators start from $k=2$, i.e. dimension-8. However, there is also a dimension-6 operator with $J=1$ appearing as
\eq{
    \mc{O}^{J=1,d=6} \sim (\psi^\dagger_1 \bar\sigma^\mu \psi_2) (\psi_3 \sigma_\mu \psi^\dagger_4),
}
corresponding to the amplitude $[14]\vev{23} = s^{-1}[12]\vev{23}^2[34]$, where momentum conservation is used. One could verify that
\eq{
    (\psi^\dagger_1 \bar\sigma^\mu \psi_2)D^2(\psi_3 \sigma_\mu \psi^\dagger_4) \sim 2(\psi^\dagger_1 \bar\sigma_\mu D_\nu \psi_3) (D^\mu\psi_2 \sigma^\nu \psi^\dagger_4)
}
up to EOM (the factor of 2 comes from $\sigma^\mu\sigma_\mu=2\epsilon\epsilon$), where the right hand side is the $k=2$ case of Eq.~\eqref{eq:J=1_generald}.
This is a drawback of the bridge method -- it may miss some lowest dimensional j-basis operators. 
Similarly, for $J=2$ we have the j-basis amplitudes
\eq{
    \mc{B}^{J=2,d=4+2k} \sim s^{k-4}[12]^2\left( \vev{14}\vev{23}^3 + 3\vev{13}\vev{24}\vev{23}^2 \right)[34]^2
}
where the coefficients are given by the formula \eqref{eq:bridge-comb}. The corresponding operator is a very complicated polynomial of building blocks in terms of Lorentz indices. We only present the spinor index version here
\eq{
    \mc{O}^{J=2,d=4+2k} \sim 
    (D^\alpha_{\ \dot\alpha}\psi^\dagger_{\dot\beta})
    (D^{\beta\dot\beta}D^{\gamma\dot\alpha}\psi^\delta)D^{2k-8}
    \left[(D_{\gamma\dot\gamma}D_{\delta\dot\delta}\psi_{\beta})
    (D_\alpha^{\ \dot\gamma}\psi^{\dagger\,\dot\delta})
    +\text{perm}(\alpha,\beta,\gamma,\delta)\right]
}
Again, it starts from $k=4$ and dimension 12, which is not the lowest dimension that the $J=2$ operators appear. Note that there is a general principle that higher $J$ partial waves are ``unlocked'' at higher dimensions because more powers of momenta are needed to build up orbital angular momentum among the particles. 

\comment{\color{blue}
Also, for $J=3$ that first appears at dimension-10, 
\eq{
\mathcal{B}^{J=3,d=10+2k} =&s^k [14]\langle23\rangle (s^2+10st+15t^2),\\
\mathcal{O}^{J=3,d=10+2k} =&(\psi^\dagger_1 \bar\sigma^\mu \psi_2) D^{2k+4} (\psi_3 \sigma_\mu \psi^\dagger_4) +10(\psi^\dagger_1 \bar\sigma^\mu D^\nu \psi_2) D^{2k+2} (\psi_3 \sigma_\mu D_\nu \psi^\dagger_4) \\ &+15(\psi^\dagger_1 \bar\sigma^\mu D^\nu D^\rho \psi_2) D^{2k} (\psi_3 \sigma_\mu D_\nu D_\rho \psi^\dagger_4).
}
}

For more particles, the partial wave amplitudes have degeneracy such as in Eq.~\eqref{eq:ex_3cgc}. Consider the scattering $\psi_1,\psi_2 \to \phi_3,\phi_4,\phi_5$, the $J=1$ partial waves and the corresponding operators read
\eq{\begin{array}{ll}
    \mc{B}_1^{J=1,d=8} = \vev{13}\vev{24}[34] & \mc{O}_1^{J=1,d=8} = (\psi_1\sigma^{\mu\nu}\psi_2) D_\mu\phi_3 D_\nu\phi_4 \phi_5\\
    \mc{B}_2^{J=1,d=8} = \vev{14}\vev{25}[45] & \mc{O}_2^{J=1,d=8} = (\psi_1\sigma^{\mu\nu}\psi_2) \phi_3 D_\mu\phi_4 D_\nu\phi_5\\
    \mc{B}_3^{J=1,d=8} = \vev{15}\vev{23}[53] & \mc{O}_3^{J=1,d=8} = (\psi_1\sigma^{\mu\nu}\psi_2) D_\nu\phi_3 \phi_4 D_\mu\phi_5.
\end{array}}
Part of the degeneracy can also be organized as multiplication of Mandelstam variables that preserves the angular momentum. Note that in the bridge construction, the angular momentum is related to the number of threads across the channel. As long as a Mandelstam variable does not ``cross the channel'', it could be freely added to the partial wave functions and build higher dimensional j-basis. From the Casimir operator point of view, such Mandelstam variables satisfy
\eq{
    W^2_\mc{I} s_{\mc{I}'} = 0 \quad \text{if}\ \mc{I}' \subset \mc{I}\ \text{or} \ \mc{I}'\cap\mc{I}=\emptyset.
}
In the above example, all of the $s_{34},s_{35},s_{45}$ do not affect the total angular momentum, hence we can multiply a general degree-$k$ polynomial of the three variables to the partial wave. For instance the $J=0$ j-basis can be obtained as
\eq{
    & \mc{B}^{J=0,d=6+2k} = {\rm poly}^{(k)}(s_{34},s_{35},s_{45}) \vev{12} , \\
    & \mc{O}^{J=0,d=6+2k} = (\psi_1\psi_2)(D^{k_3}\phi_3)(D^{k_4}\phi_4)(D^{k_5}\phi_5),\ k_3+k_4+k_5=2k.
}
In \cite{Shu:2021qlr}, it is proved that the degeneracy of the lowest order partial waves must be finite. Therefore, if one focuses on j-basis with specific $J$ or those bounded by a maximum value, a complete solution up to arbitrary dimensions $d$ can be provided with $O(d^0)$ complexity.


\subsubsection{Poincare Casimir and Lorentz Eigen-Basis}

We have shown how partial wave amplitudes correspond to a family of j-basis operators in various dimensions. It is also interesting how to construct j-basis operators of different $J$ at a given dimension. The Casimir Poincar\'e operator $W^2$ defined in the previous section turns out to be most suitable for this task. Suppose the contact amplitudes of a given dimension $d$ form a linear space $V^d=[\mc{B}^d_i]$, we shall find the representation matrix of $W^2$ in this space such that $\mc{B}^J$ are nothing but its eigenvectors. In particular, if we find a matrix $\mc{W}$ that satisfies
\eq{\label{eq:W2action}
    W^2_{\mc{I}}\mc{B}_i = -s_{\mc{I}} \mc{W}_i{}^j\mc{B}_j,
}
where the superscript $d$ is omitted, we can diagonalize the matrix as
\eq{\label{eq:W2diagonalize}
    \mc{K}\cdot\mc{W}\cdot\mc{K}^{-1} = {\rm diag}\{J_i(J_i+1)\}
}
so that the j-basis amplitudes are given by
\eq{\label{eq:j-cnstr-diag}
    \mc{B}^{J_i} = \mc{K}_i^{\ j}\mc{B}_j.
}

\comment{
In the above, the effective operators are organized by the angular momentum, across the mass dimension of these operators. In this subsection, we investigate how the effective operators at the given mass dimension are re-organized by the angular momentum: the j-basis for a type of operator at given mass dimension.

According to the operator-amplitude correspondence, for a local amplitude $\mc{B}^{(d)}$ with mass dimension $d$ we have
\eq{
    W^2_{\mc{I}}\mc{B}^{(d)} = -s_{\mc{I}}\mc{B}^{'(d)} \equiv -s_{\mc{I}}\sum_{J} a_J J(J+1) \mc{B}^{J\,(d)}
}
Hence the action of ${\bf W}^2_{\mc{I}}$ is a map $V^d \to s_{\mc{I}} V^d$ where $V^d$ is the linear space of all dimension $d$ local amplitudes with particular external particles.
If we have an independent amplitude basis in $V^d$, denoted as $\{\mc{B}_i^{(d)}\}$, we can work out a representation matrix $\mc{W}$ such that
\eq{\label{eq:W2action}
    W^2_{\mc{I}}\mc{B}_i^{(d)} = -s_{\mc{I}} \mc{W}_i{}^j\mc{B}_j^{(d)},
}
thus the amplitude basis with definite angular momenta, denoted as j-basis, can be obtained by diagonalizing $\mc{W}$. 
}

In \cite{Li:2020gnx,Li:2020xlh}, a complete basis of local amplitudes and the corresponding operators are defined as the Young Tableau (Y-)basis. 
The name comes from the construction in terms of Young tableau of the $SU(N)$ group, where $N$ is the number of particles involved in the amplitude. Given the helicities of the particles $h_i$\footnote{The Young Tableau basis is only available for massless particles. Massive scalars and fermions can be included without essential modifications, but higher spin massive particles are not allowed. 
}, define the parameters 
\eq{
    n = \frac{k}{2} + \sum_{h_i<0}h_i \ ,\quad \tilde{n} = \frac{k}{2} + \sum_{h_i>0}h_i \ ,\quad
}
which are the number of angle brackets and square brackets in the amplitude, where $k$ is the number of momenta or the number of derivatives in the operator. These parameters determine a Young diagram
\newline
\bea
[\mathcal{M}]_{N,\tilde{n},n}= \arraycolsep=0pt\def\arraystretch{1}
 \rotatebox[]{90}{\text{$N-2$}} 
 \left\{
 \begin{array}{cccccc}
  \yng(1,1) &\ \ldots{}&\ \yng(1,1)& \overmat{n}{\yng(1,1)&\ \ldots{}\  &\yng(1,1)} \\
  \vdotswithin{}& & \vdotswithin{}&&&\\
  \undermat{\tilde{n}}{\yng(1,1)\ &\ldots{}&\ \yng(1,1)} &&&
 \end{array}
 \right. .\label{eq:yd}
\eea\newline\newline
A Young tableau is a particular way of filling labels $1\sim N$ into the Young diagram, while the number of each label is given by $\# i = \tilde{n} - 2h_i$ for the particular class of scattering state. The Young tableaux can be translated to amplitudes column by column via the following dictionary
\bea
\langle ij\rangle \sim \young(i,j),\quad 
[ij] \sim \mathcal{E}^{k_1...k_{N-2}ij} \tilde{\lambda}_i\tilde{\lambda}_j \sim \rotatebox[]{90}{\text{$N-2$}} \left\{ \begin{array}{c}
\ytableausetup{centertableaux, boxsize=2em} \begin{ytableau} k_1 \\ k_2 \end{ytableau}
 \\
\vdotswithin{}\\
\begin{ytableau}
\scriptstyle k_{N-2}
\end{ytableau}
\end{array}\right. ,
\eea
The y-basis of amplitudes is given by the Semi-Standard Young Tableau (SSYT): in each row the labels are non-decreasing from left to right; in each column the labels are increasing from top to bottom. Any amplitude/operators\footnote{Note that for operators this conversion is up to the ``EOM'' pieces, including the kinetic terms of all the fields and the covariant derivative commutators.} that does not satisfy this condition corresponds to non-SSYT, and can be converted to a linear combination of the y-basis after the following steps. 
\begin{description}
\item[1.] $p_i\ (i=1,2,3)$ is replaced by $-\sum_{j\neq i}p_j$ in the following cases.
\begin{description}
\item[(1)] $p_1$ is always replaced by $-\sum_{i\neq1}p_i$
\eq{
    \left[ i|p_1|j\right\rangle \sim \begin{array}{cc}
\multicolumn{2}{c}{\ytableausetup
{boxsize=1em}\begin{ytableau}
*(yellow) 2 & *(yellow) 1\\ 3 & j
\end{ytableau}}  \\
\vdotswithin{} & \\
\young(N) &
\end{array}
}
\item[(2)] $p_2$ is replaced for cases like $\langle 1|p_2|k]\ (k>2)$ or $[1|p_2|k\rangle\ (k>2)$
\eq{
    \left[1|p_2|k\right\rangle \sim \begin{array}{cc}
\multicolumn{2}{c}{\begin{ytableau}
*(yellow) 3 & *(yellow) 2 \\ 4 & k
\end{ytableau}}  \\
\vdotswithin{} & \\
\young(N) &
\end{array},\quad
\left[k|p_2|1\right\rangle \sim 
\begin{array}{cc}
\multicolumn{2}{c}{\begin{ytableau}
1 & 1 \\ *(yellow) 3 & *(yellow) 2
\end{ytableau}}  \\
\vdotswithin{} & \\
\young(N) &
\end{array};
}
\item[(3)] $p_3$ is replaced for cases like $\langle1|p_3|2]$ or $[1|p_3|2\rangle$
\eq{
    \left[1|p_3|2\right\rangle\sim \begin{array}{cc}
\multicolumn{2}{c}{\begin{ytableau}
2 & 2 \\ *(yellow) 4 & *(yellow) 3
\end{ytableau}}  \\
\vdotswithin{} & \\
\young(N) &
\end{array},\quad
\left[2|p_3|1\right\rangle\sim\begin{array}{cc}
\multicolumn{2}{c}{\begin{ytableau}
1 & 1 \\ *(yellow) 4 & *(yellow) 3
\end{ytableau}} \\
\vdotswithin{} & \\
\young(N) &
\end{array};
}
\item[(4)] since $\sum_{i,j\neq 1} s_{ij} = p_1^2=0$, $s_{23}$ is replaced by the other terms in the sum 
\eq{
    s_{23}\sim\begin{array}{cc}
\multicolumn{2}{c}{\begin{ytableau}
1 & 2 \\ *(yellow) 4 & *(yellow) 3
\end{ytableau}} \\
\vdotswithin{} & \\
\young(N) &
\end{array}.
}
\end{description}

\item[2.] Terms like $\langle il\rangle\langle jk\rangle$ and $[il][jk]\ (i<j<k<l)$ are replaced by Schouten identities as $ \langle il\rangle\langle jk\rangle \rightarrow -\langle ij\rangle \langle kl\rangle +\langle ik\rangle\langle jl\rangle$, because they represent nonstandard Young tableaux,
\eq{
    \langle il\rangle\langle jk\rangle\sim \begin{ytableau}
i & j \\ *(yellow) l & *(yellow) k
\end{ytableau},\quad
[jk][il]\sim \ytableausetup
{boxsize=1.5em}\begin{ytableau}
\none[\vdotswithin{}] & \none[\vdotswithin{}]\\
\scriptstyle k-1 & \scriptstyle k-1\\
*(yellow) \scriptstyle k+1 & *(yellow) k \\
\none[\vdotswithin{}] & \none[\vdotswithin{}]\\
*(yellow) l & *(yellow) \scriptstyle l-1\\
\scriptstyle l+1 & \scriptstyle l+1\\
\none[\vdotswithin{}] & \none[\vdotswithin{}]
\end{ytableau}; 
}
\end{description}
Now with the y-basis, we can try the Casimir method of constructing j-basis amplitudes/operators as in Eq.~\eqref{eq:j-cnstr-diag}. We shall refer back to the example of $\psi_1\psi_2\psi_3\phi_4\phi_5\bar{\psi}_6D$ in table \ref{tab:12,3456}. 
The first step is to determine a complete basis of the amplitude for the particles $h_i = \{-\frac12,-\frac12,-\frac12,0,0,\frac12\}$ with $k=1$ momentum/derivative.
The complete y-basis is given by the SSYT as
\bea
\mathcal{B}^y\equiv \begin{pmatrix}
\mathcal{B}^y_1 \\ \mathcal{B}^y_2 \\ \mathcal{B}^y_3 \\ \mathcal{B}^y_4 \\ \mathcal{B}^y_5
\end{pmatrix}= \begin{pmatrix}
[56]\langle 12\rangle \langle 35\rangle\\
-[46] \langle 12\rangle  \langle 34\rangle\\ 
[36] \langle 13\rangle  \langle 23\rangle \\
[56] \langle 13\rangle  \langle 25\rangle \\ 
-[46] \langle 13\rangle  \langle 24\rangle
\end{pmatrix}. 
\eea

Acting $W_{12}^2$ on the full basis $\mathcal{B}^y$ and reduce the results to get the coordinates, we get
\bea
W_{\{12\}}^2\mathcal{B}^y\equiv -s_{12}\mathcal{W}_{\{12\}}\mathcal{B}^y\ ,\qquad \mc{W}_{\{12\}} = \begin{pmatrix}
0&0&0&0&0\\0&0&0&0&0\\0&0&2&0&0\\-1&0&0&2&0\\0&-1&0&0&2
\end{pmatrix} .
\eea
Diagonalize the matrix to get $\mc{K}$ in Eq.~\eqref{eq:W2diagonalize}\footnote{The rows in $\mc{K}$ are the eigenvectors of $\mc{W}^\intercal$.}, 
\eq{
    \mc{K}_{\{12\}}^{\rm jy} = \begin{pmatrix}
    1&0&0&0&0 \\
    0&1&0&0&0 \\
    0&0&1&0&0 \\
    -1&0&0&2&0 \\
    0&-1&0&0&2
    \end{pmatrix} \ ,\quad \begin{split}\mc{K}_{\{12\}}^{\rm jy}\mc{W}_{\{12\}}(\mc{K}_{\{12\}}^{\rm jy})^{-1} &= {\rm diag}\{0,0,2,2,2\} \\
    &\Rightarrow\ J_i = \{0,0,1,1,1\}\end{split}.
\label{eq:Kjy12}}
where we have labeled $\mc{K}$ by the superscript ${\rm jy}$ as indicating the trasfer matrix from the y-basis to j-basis. Now we have
\eq{\label{eq:pw_12}
    \mc{B}^{\rm j}_{\{12\}} = \mc{K}_{\{12\}}^{\rm jy}\mc{B}^{\rm y}\ ,\qquad \Rightarrow\quad 
    \mc{B}^{J_{12}=0} = \begin{pmatrix}\mc{B}^y_1\\\mc{B}^y_2\end{pmatrix} ,\quad 
    \mc{B}^{J_{12}=1} = \begin{pmatrix}\mc{B}^y_3\\-\mc{B}^y_1+2\mc{B}^y_4\\-\mc{B}^y_2+2\mc{B}^y_5\end{pmatrix}
}
which can be verified to match with the result in table~\ref{tab:12,3456}.

We can further explore the multi-partite partial waves, for example the following partition
\begin{figure}[H]
	\centering
	\begin{tikzpicture}[scale=0.8]
	\coordinate (v1) at (0,0) {};
	\coordinate (v2) at (1,1.732) {};
	\coordinate (v3) at (2,0) {};
	\coordinate (v4) at (1,0.57735){};
	\draw (-1,1) node[left]{1} -- (v1);
	\draw (-1,-1) node[left]{2} -- (v1);
	\draw (0,2.732) node[above]{6} -- (v2);
	\draw (v2) -- (v4);
	\draw (2,2.732) node[above]{5} -- (v2);
	\draw (v3) -- (v4);
	\draw (v1) -- (v4);
	\draw (3,1) node[right]{4} -- (v3);
	\draw (3,-1) node[right]{3} -- (v3);
	\end{tikzpicture}.
\end{figure}
\noindent This topology contains three resonances, corresponding to channels $\{12\}$, $\{34\}$ and $\{56\}$ respectively. The multi-partite partial waves as defined in Eq.~\eqref{eq:multi-partite} are thus labelled by a tuple $\{J_{12},J_{34},J_{56}\}$, which are simultaneous eigenfunctions of $W^2$ in the three channels. It is possible due to the commutation relation Eq.~\eqref{eq:W2commute}. 
Adding an extra channel amounts to lift the degeneracy of the original eigenspaces. For instance, if we act $W^2_{\{34\}}$ on the 3-dim eigenspace $[J_{12}=1]$, we can get a restricted representation matrix
\eq{
    \mc{W}_{\{34\}}\big|_{J_{12}=1} &= \frac{1}{4}\begin{pmatrix}
    7 & 0 & 3 \\
    0 & 3 & 0 \\
    8 & 0 & 11
    \end{pmatrix} \\
    \mc{K}_{\{34\}}^{\rm jy}\big|_{J_{12}=1} &= \begin{pmatrix}
 -1 & 0 & -1 \\
 0 & 1 & 0 \\
 -2 & 0 & 1 
\end{pmatrix} \ ,\quad 
\begin{split}
  \mc{K}_{\{34\}}^{\rm jy}\mc{W}_{\{34\}}(\mc{K}_{\{34\}}^{\rm jy})^{-1}\big|_{J_{12}=1} &= {\rm diag}\{\frac{15}{4}, \frac{3}{4}, \frac{3}{4}\} \\[1em]
    &\Rightarrow\ J_{34,i} = \{\frac32,\frac12,\frac12\}.
\end{split}
}
In practice, instead of investigating the eigenspaces one by one, we deal with them altogether by using the trick of taking intersection of vector subspaces\footnote{The technique of taking intersections of vector subspaces are described in appendix~\ref{app:linearintersection}.}.
Therefore we need to first obtain the partial wave eigenspaces in each channel, specified by the rows of the $\mc{K}^{\rm jy}$ matrices, so that the multi-partite partial waves belong to the intersections among them. We have already got the eigenspaces for $\{12\}$ as in Eq.~\eqref{eq:pw_12}, which can be expressed as
\eq{
    [{J_{12}=0}] = \text{span}\{(\mc{K}_{\{12\}}^{\rm jy})_1,(\mc{K}_{\{12\}}^{\rm jy})_2\}\ ,\quad 
    [{J_{12}=1}] = \text{span}\{(\mc{K}_{\{12\}}^{\rm jy})_3,(\mc{K}_{\{12\}}^{\rm jy})_4,(\mc{K}_{\{12\}}^{\rm jy})_5\}.
}
\begin{table}[t]
\centering
\begin{tabular}{c|c}
\hline\hline
angular momenta & $\mathcal{B}^{J_{12},J_{34},J_{56}}$ \\
\hline
$J_{12}=0,J_{34}=J_{56}=\frac{1}{2}$ & \makecell{ $\mathcal{B}^y_1$ \\ $\mathcal{B}^y_2$ }\\
\hline
$J_{12}=1,J_{34}=J_{56}=\frac{1}{2}$ & \makecell{ $-\mathcal{B}^y_1 +2 \mathcal{B}^y_4 $ \\ $-\mathcal{B}^y_2 -2\mathcal{B}^y_3 +2\mathcal{B}^y_5$ }\\
\hline
$J_{12}=1,J_{34}=\frac{3}{2},J_{56}=\frac{1}{2}$ & \makecell{ $ \mathcal{B}^y_2-\mathcal{B}^y_3-2\mathcal{B}^y_5$ } \\
\hline\hline
\end{tabular}
\caption{Partial-wave basis of $\mc{M}(\psi_1,\psi_2|\psi_3,\phi_4|\phi_5,\psi^\dagger_6)$}
\label{tab:12,34,56.}
\end{table}
Now we aim at the $\brkt{34}$ channel, and following the same procedure above we get
\eq{
    \mc{K}_{\{34\}}^{\rm jy} = \begin{pmatrix}
 0 & -1 & 1 & 0 & 2 \\
 0 & 0 & -1 & 0 & 1 \\
 0 & 0 & 0 & 1 & 0 \\
 0 & 1 & 0 & 0 & 0 \\
 1 & 0 & 0 & 0 & 0
\end{pmatrix} \ ,\quad 
\begin{split}
  \mc{K}_{\{34\}}^{\rm jy}\mc{W}_{\{34\}}(\mc{K}_{\{34\}}^{\rm jy})^{-1} &= {\rm diag}\{\frac{15}{4}, \frac{3}{4}, \frac{3}{4}, \frac{3}{4}, \frac{3}{4}\} \\[1em]
    &\Rightarrow\ J_i = \{\frac32,\frac12,\frac12,\frac12,\frac12\}.
\end{split}
\label{eq:Kjy34}
}

which means the eigenspaces are
\eq{
    [{J_{34}=\frac32}] = (\mc{K}_{\{34\}}^{\rm jy})_1\ ,\quad 
    [{J_{34}=\frac12}] = \text{span}\{(\mc{K}_{\{34\}}^{\rm jy})_2,(\mc{K}_{\{34\}}^{\rm jy})_3,(\mc{K}_{\{34\}}^{\rm jy})_4,(\mc{K}_{\{34\}}^{\rm jy})_5\}.
}
The last channel $\{56\}$ is trivial, as one can verify that all the amplitudes are $J_{56}=\frac12$ partial waves.
Now we can take the intersections, for example the partial waves $\mc{B}^{J_{12}=1,J_{34}=\frac12}$ are given by the intersection
\eq{
    [J_{12}=1, J_{34}=\frac12] &\equiv [{J_{12}=1}] \cap [{J_{34}=\frac12}] = \begin{pmatrix}
    -1&0&0&2&0\\
    0&-1&-2&0&2
    \end{pmatrix} \\
    &\Rightarrow\ \mc{B}^{J_{12}=1,J_{34}=\frac12} = \begin{pmatrix}
    -\mc{B}^y_1+2\mc{B}^y_4 \\
    -\mc{B}^y_2-2\mc{B}^y_3+2\mc{B}^y_5
    \end{pmatrix}.
}
Hence we summarize the result in table~\ref{tab:12,34,56.}, and the full conversion matrix ${\cal K}_{\cal B}$ between the j-basis and the y-basis can be directly read out:
\begin{eqnarray}
{\cal B}^j=\begin{pmatrix}
{\cal B}^{0,1/2,1/2}_1\\
{\cal B}^{0,1/2,1/2}_2\\
{\cal B}^{1,1/2,1/2}_1\\
{\cal B}^{1,1/2,1/2}_2\\
{\cal B}^{1,3/2,1/2}_1\\
\end{pmatrix}=
\begin{pmatrix}
1 & 0 & 0 & 0 & 0\\
0 & 1 & 0 & 0 & 0\\
-1 & 0 & 0 & 2 & 0\\
0 & -1 & -2 & 2 & 0\\
0 & 1 & -1 & 0 & 2\\
\end{pmatrix}
\begin{pmatrix}
{\cal B}^{y}_1\\
{\cal B}^{y}_2\\
{\cal B}^{y}_3\\
{\cal B}^{y}_4\\
{\cal B}^{y}_5\\
\end{pmatrix}={\cal K}_{\cal B}\cdot {\cal B}^{y}.
\end{eqnarray}

Next we may look at a more non-trivial partition as the following
\begin{figure}[H]
	\centering
	\begin{tikzpicture}[scale=0.8]
	\coordinate (v1) at (0,0) {};
	\coordinate (v2) at (1,0) {};
	\coordinate (v3) at (2,0) {};
	\coordinate (v4) at (3,0) {};
	\draw (-1,1) node[left]{1} -- (v1);
	\draw (-1,-1) node[left]{2} -- (v1);
	\draw (v1) -- (v2);
	\draw (0.8,1.2) node[above]{6} -- (v2);
	\draw (v2) -- (v3);
	\draw (2.2,1.2) node[above]{5} -- (v3);
	\draw (v3) -- (v4);
	\draw (4,1) node[right]{4} -- (v4);
	\draw (4,-1) node[right]{3} -- (v4);
	\end{tikzpicture}
\end{figure}
\noindent which includes a new channel $\{1,2,6\}$.
However, we cannot find the matrix $\mc{W}_{126}$ defined in Eq.~\eqref{eq:W2action}. The reason is the following: suppose the dimension $d$ amplitude basis span a linear space $V^d$, the action of $W^2$ brings it to the space $V^{d+2}$, while the multiplication of $s_{126}$ also brings it to $V^{d+2}$. The images of the two mapping may not coincide, which is the necessary condition for the matrix $\mc{W}$ to exist. We have ignored this subtlety in the previous discussion, but in this case we have to deal with it. The consequence is, in short, that there will be less j-basis amplitudes than the y-basis, which live in the subspace $\tilde{V}_d$ that has the same image of the two maps
\eq{
    W^2_{\{126\}}\tilde{V}^d = s_{126}\tilde{V}^d.
}
We describe the detailed procedure of obtaining $\tilde{V}^d$ in appendix~\ref{app:jbasis} with explanation. Here we simply go over the computation for this particular example. First we find both images of the two maps, take the intersection and define a basis there
\eq{
    B_1\equiv W_{\{126\}}^2V^d \cap s_{126}V^d = \text{span}\,\mc{B}^{(B_1)}
}
We can find the projection map from both images to the intersection
\eq{
c. W_{\{126\}}^2\mathcal{B}^y =\mathcal{B}^{(B_1)},\quad c=\begin{pmatrix}
 0 & 0 & -\frac{4}{3} & -\frac{4}{3} & \frac{4}{3} \\
 0 & \frac{4}{3} & 0 & 0 & 0 \\
 \frac{4}{3} & 0 & 0 & 0 & 0
\end{pmatrix},\\
c'. s_{126} \mathcal{B}^y =\mathcal{B}^{(B_1)},\quad c'=\begin{pmatrix}
 0 & 0 & 1 & 1 & -1 \\
 0 & -1 & 0 & 0 & 0 \\
 -1 & 0 & 0 & 0 & 0 
\end{pmatrix}, 
}
The two projections $c$, $c'$ define two subspaces of $V^d$ that map to $B_1$ via $W^2$ and $s$, and luckily they are the same in this example, which is exactly the $\tilde{V}^d$ we are looking for. The representation matrix of $W^2$ restricted to $\tilde{V}^d$ is thus given by
\eq{
    c = \mc{W}.c' \ \Leftrightarrow\ W^2_{\{126\}}(c'.\mc{B}^y) = -s_{126}\mc{W}.(c'.\mc{B}^y).
}
Plug in $c$ and $c'$, we can solve for $\mc{W}$ as a constant $3\times3$ matrix $\mc{W} = -\frac{4}{3}\mathbf{1}_{3\times3}$, which means that all the three amplitudes defined by $c'$ are $J=\frac12$ partial waves
\eq{ 
\mc{K}_{\{126\}}^{\rm jy} = c' = \begin{pmatrix}
 0 & 0 & 1 & 1 & -1 \\
 0 & -1 & 0 & 0 & 0 \\
 -1 & 0 & 0 & 0 & 0 
\end{pmatrix} 
}
In general we need to diagonalize $\mc{W}$
\eq{
\tilde{\mc{K}}^{\rm jy}.\mc{W}.(\tilde{\mc{K}}^{\rm jy})^{-1} = {\rm diag}\{J_i(J_i+1)\}\ ,\quad \mc{B}^{\rm j} = \tilde{\mc{K}}^{\rm jy}.c'.\mc{B}^{\rm y}
}
therefore the full transfer matrix from y-basis to j-basis is 
\eq{\mc{K}^{\rm jy} = \tilde{\mc{K}}^{\rm jy}.c'}

Finally we take the intersections among the eigenspaces $[J_{12}]$, $[J_{34}]$ and $[J_{126}]$ and get the result in table~\ref{tab:12,126,34,345}.
\comment{
Then we shall obtain the eigenvectors of each combination of $\{J_{12},J_{34},J_{126}\}$ by calculating the linear intersection of the corresponding j-bases. For example, Eq.~(\ref{eq:Kjy12}) shows that the $[J_{12}=1]$ is spanned by $\begin{pmatrix}
0&0&1&0&0\\-1&0&0&2&0\\0&-1&0&0&2
\end{pmatrix}.\mathcal{B}^y$, $[J_{34}=\frac{1}{2}]$ by $\begin{pmatrix}
0&0&-1&0&1\\0&0&0&1&0\\0&1&0&0&0\\1&0&0&0&0
\end{pmatrix}.\mathcal{B}^y$, and $[J_{126}=\frac{1}{2}]$ by $\begin{pmatrix}
 0 & 0 & 1 & 1 & -1 \\
 0 & -1 & 0 & 0 & 0 \\
 -1 & 0 & 0 & 0 & 0 
\end{pmatrix}.\mathcal{B}^y$. The linear intersection of $[J_{12}=1], [J_{34}=\frac{1}{2}], [J_{126}=\frac{1}{2}]$ is $[\begin{pmatrix}-1&1&2&2&-2\end{pmatrix}. \mathcal{B}^y]$. The final result is in table~\ref{tab:12,126,34,345}.
}
\begin{table}[htbp]
\centering
\begin{tabular}{c|c}
\hline\hline
angular momenta & $\mathcal{B}^{J_{12},J_{126},J_{45}}$ \\
\hline
\multirow{2}{*}{$J_{12}=0,J_{126}=J_{34}=\frac{1}{2}$} & $\mathcal{B}^y_1$\\
& $\mathcal{B}^y_2$ \\
\hline
$J_{12}=1,J_{126}=J_{34}=\frac{1}{2}$ & $ -\mathcal{B}^y_1 +\mathcal{B}^y_2 +2\mathcal{B}^y_3 +2\mathcal{B}^y_4 -2\mathcal{B}^y_5$ \\
\hline\hline
\end{tabular}
\caption{Partial-wave basis of partition $\mc{M}(\psi_1,\psi_2|\psi^\dagger_6|\psi_3,\phi_4)$.}
\label{tab:12,126,34,345}
\end{table}

\subsection{Gauge j-basis from Gauge Casimir} \label{sec:gauge}

The correspondence of the gauge structures between operators and amplitudes are simple, the invariant tensors of group factors in the amplitudes exactly correspond to the invariant tensors that are used to contract the fields in operators to form gauge singlets. As we have already discussed in section~\ref{sec:2.3}, 
the gauge structure of an operator type can also be decomposed into a set of gauge eigen-bases of a gauge partial Casimir operators $\underset{\{\mathbbm{S}\}}{\mathbbm{C}}$ for two partitions. The representation of the gauge partial Casimir for a type of operator can be found on a basis of monomial invariant tensors $\{\mc{T}_j\}$ that can be derived using the generalized Littlewood-Richardson (L-R) rule invented in Ref.~\cite{Li:2020gnx}, which yields our master formular for finding the gauge eigen-basis in Eq.~\eqref{eq:master-gauge}:
\begin{eqnarray}
\underset{\{\mathbbm{S}\}}{\mathbbm{C}}\circ \mc{T}_i=\sum_j \mc{T}_jC_{ji},\nonumber
\end{eqnarray}
where we have omit the superscript $m$ to indicate that $T_j$ are m-basis, and we will adopt this convention in the following discussion in this section.
We have seen that diagonalizing $C_{ji}^T$ for particular channel with the similar transformation ${\cal K}^{jm}$ gives the eigenvalues $C(\mathbf{R})$ and the corresponding gauge eigenbasis $\mc{T}(\mathbf{R})$ of which coordinates on the original basis ${\cal T}_i$ are encoded in ${\cal K}^{jm}$:
\begin{eqnarray}
{\cal K}^{jm}. C^{T}. ({\cal K}^{jm})^{-1} = {\rm diag}\{C(\mathbf{R}_i)\}.
\end{eqnarray}
In this section, we shall extend the discussion and generalize the idea to multi-partition scenarios and the gauge group which involve finding common eigenvectors of more than one Casimir.

Let us first review the algorithm for constructing the independent and complete gauge y-basis --- the generalized Littlewood-Richardson (L-R) rule. As a counterpart of the ordinary L-R rule which applies to Young diagrams to find the multiplicities of irreducible representations in the decomposition of the direct product representation, our generalized L-R rule applies to the Young tableaux instead, and constructs a series of singlet Young tableaux of the gauge group $SU(M)$, of which each column can be identified as an $\epsilon$ tensor, and the y-basis are just the product of these  $\epsilon$ tensors. The algorithm requires that  fields in the operator as gauge representations must be written in  forms of tensor representation with fundamental gauge indices only, so that they can be map to Young tableaux.
The fields of adjoint and anti-fundamental representations are of most interest in physical models, we can write them in terms of tensors with $M$ and $M-1$ fundamental indices respectively using the following conversions: 
\bea\label{eq:adjconversion}
\mathcal{F}^A&\rightarrow& \mathcal{F}_{a_1...a_{M-1}i}\equiv \mathcal{F}^A (T^A)^{a_M}_i \epsilon_{a_1...a_M} \sim \ytableausetup{boxsize=2.5em} \begin{ytableau}
a_1 & i \\ \none[\vdots] \\ a_{M-1}
\end{ytableau};\\
\mathcal{F}^a&\rightarrow& \mathcal{F}_{a_1...a_{M-1}}\equiv \mathcal{F}^a \epsilon_{a_1...a_{M-1}a} \sim \begin{ytableau}
a_1 \\ \none[\vdots] \\ a_{M-1}
\end{ytableau}.
\eea
To illustrate the alogrithm and to show that our framework for finding the gauge eigen-basis introduced in section.~\ref{sec:2.3} also applies to the gauge group with more than one Casimir operator, we focus the example of ${G_L}^4$ with $SU(3)$ representation $\{\mathbf{8},\mathbf{8},\mathbf{8},\mathbf{8}\}$ for the partition $\{12|34\}$.
As an concrete example for Eq.~(\ref{eq:adjconversion}), the gluon field can be expressed in terms of a tensor of three fundamental indices with the relation to the ordinary gluon field in the following equation,
\begin{eqnarray}
\ytableausetup{boxsize=1.2em} 
G_{abc} =G^A {\lambda^A}^x_c \epsilon_{abx} \sim \begin{ytableau}
a & b \\ c
\end{ytableau}.
\end{eqnarray}
Then one can construct the singlet Young tableaux with four copies of this Young tableau  with different subscripts of gauge indices using the generalized L-R rule, which yields eight complete and independent y-basis group factors:
\bea
\begin{ytableau}
a_1 & b_1 \\ c_1
\end{ytableau} \mathop{\longrightarrow}\limits^{\begin{ytableau}
a_2 & b_2 \\ c_2
\end{ytableau}} \begin{ytableau}
a_1 & b_1 & a_2 & b_2 \\ c_1 & c_2
\end{ytableau} \mathop{\longrightarrow}\limits^{\begin{ytableau}
a_3 & b_3 \\ c_3
\end{ytableau}} \begin{ytableau}
a_1 & b_1 & a_2 & b_2 \\ c_1 & c_2 & a_3 \\ b_3 & c_3
\end{ytableau} \mathop{\longrightarrow}\limits^{\begin{ytableau}
a_4 & b_4 \\ c_4
\end{ytableau}} \begin{ytableau}
a_1 & b_1 & a_2 & b_2 \\ c_1 & c_2 & a_3 & a_4 \\ b_3 & c_3 & b_4 & c_4
\end{ytableau} =\epsilon^{a_1c_1b_3} \epsilon^{a_2a_3b_4} \epsilon^{b_1c_2c_3} \epsilon^{b_2a_4c_4}\nonumber \\
\sim -8 i f^{{A_1}{A_2}B} d^{{A_3}{A_4}B}+8 i d^{{A_1}{A_2}B} f^{{A_3}{A_4}B}+8 d^{{A_1}{A_2}B} d^{{A_3}{A_4}B}\nonumber\\ +16 \delta^{{A_1}{A_3}} \delta^{{A_2}{A_4}}+\frac{16}{3} \delta^{{A_1}{A_2}} \delta^{{A_3}{A_4}}+8 f^{{A_1}{A_2}B} f^{{A_3}{A_4}B},\\
\begin{ytableau}
a_1 & b_1 \\ c_1
\end{ytableau} \mathop{\longrightarrow}\limits^{\begin{ytableau}
a_2 & b_2 \\ c_2
\end{ytableau}} \begin{ytableau}
a_1 & b_1 & a_2 \\ c_1 & b_2 & c_2
\end{ytableau} \mathop{\longrightarrow}\limits^{\begin{ytableau}
a_3 & b_3 \\ c_3
\end{ytableau}} \begin{ytableau}
a_1 & b_1 & a_2 & a_3 \\ c_1 & b_2 & c_2 \\b_3 & c_3
\end{ytableau} \mathop{\longrightarrow}\limits^{\begin{ytableau}
a_4 & b_4 \\ c_4
\end{ytableau}} \begin{ytableau}
a_1 & b_1 & a_2 & a_3 \\ c_1 & b_2 & c_2 & a_4 \\b_3 & c_3 & b_4 & c_4
\end{ytableau} =\epsilon^{a_1c_1b_3} \epsilon^{a_2c_2b_4} \epsilon^{a_3a_4c_4} \epsilon^{b_1b_2c_3}\nonumber \\
\sim -16 i f^{{A_1}{A_3}B} d^{{A_2}{A_4}B}-16 i d^{{A_1}{A_3}B} f^{{A_2}{A_4}B}-16 d^{{A_1}{A_3}B} d^{{A_2}{A_4}B}\nonumber\\ +\frac{64}{3} \delta^{{A_1}{A_3}} \delta^{{A_2}{A_4}}+16 f^{{A_1}{A_3}B} f^{{A_2}{A_4}B},\\
\begin{ytableau}
a_1 & b_1 \\ c_1
\end{ytableau} \mathop{\longrightarrow}\limits^{\begin{ytableau}
a_2 & b_2 \\ c_2
\end{ytableau}} \begin{ytableau}
a_1 & b_1 & a_2 & b_2 \\ c_1 \\ c_2
\end{ytableau} \mathop{\longrightarrow}\limits^{\begin{ytableau}
a_3 & b_3 \\ c_3
\end{ytableau}} \begin{ytableau}
a_1 & b_1 & a_2 & b_2 \\ c_1 & a_3 & b_3 \\ c_2 & c_3
\end{ytableau} \mathop{\longrightarrow}\limits^{\begin{ytableau}
a_4 & b_4 \\ c_4
\end{ytableau}} \begin{ytableau}
a_1 & b_1 & a_2 & b_2 \\ c_1 & a_3 & b_3 & a_4 \\ c_2 & c_3 & b_4 & c_4
\end{ytableau} =\epsilon^{a_1c_1c_2} \epsilon^{a_2b_3b_4} \epsilon^{b_1a_3c_3} \epsilon^{b_2a_4c_4}\nonumber \\
\sim 16 i f^{{A_1}{A_3}B} d^{{A_2}{A_4}B}+16 i d^{{A_1}{A_3}B} f^{{A_2}{A_4}B}-16 d^{{A_1}{A_3}B} d^{{A_2}{A_4}B}\nonumber\\ +\frac{64}{3} \delta^{{A_1}{A_3}} \delta^{{A_2}{A_4}}+16 f^{{A_1}{A_3}B} f^{{A_2}{A_4}B},\\
\begin{ytableau}
a_1 & b_1 \\ c_1
\end{ytableau} \mathop{\longrightarrow}\limits^{\begin{ytableau}
a_2 & b_2 \\ c_2
\end{ytableau}} \begin{ytableau}
a_1 & b_1 \\ c_1 & a_2 \\ b_2 & c_2
\end{ytableau} \mathop{\longrightarrow}\limits^{\begin{ytableau}
a_3 & b_3 \\ c_3
\end{ytableau}} \begin{ytableau}
a_1 & b_1 & a_3 & b_3 \\ c_1 & a_2 & c_3 \\ b_2 & c_2
\end{ytableau} \mathop{\longrightarrow}\limits^{\begin{ytableau}
a_4 & b_4 \\ c_4
\end{ytableau}} \begin{ytableau}
a_1 & b_1 & a_3 & b_3 \\ c_1 & a_2 & c_3 & a_4 \\ b_2 & c_2 & b_4 & c_4
\end{ytableau} =\epsilon^{a_1c_1b_2} \epsilon^{a_3c_3b_4} \epsilon^{b_1a_2c_2} \epsilon^{b_3a_4c_4}\nonumber \\ 
\sim 64 \delta^{{A_1}{A_2}} \delta^{{A_3}{A_4}},\\
\begin{ytableau}
a_1 & b_1 \\ c_1
\end{ytableau} \mathop{\longrightarrow}\limits^{\begin{ytableau}
a_2 & b_2 \\ c_2
\end{ytableau}} \begin{ytableau}
a_1 & b_1 & a_2 \\ c_1 & b_2 \\ c_2
\end{ytableau} \mathop{\longrightarrow}\limits^{\begin{ytableau}
a_3 & b_3 \\ c_3
\end{ytableau}} \begin{ytableau}
a_1 & b_1 & a_2 & a_3 \\ c_1 & b_2 & b_3 \\ c_2 & c_3
\end{ytableau} \mathop{\longrightarrow}\limits^{\begin{ytableau}
a_4 & b_4 \\ c_4
\end{ytableau}} \begin{ytableau}
a_1 & b_1 & a_2 & a_3 \\ c_1 & b_2 & b_3 & a_4 \\ c_2 & c_3 & b_4 & c_4
\end{ytableau} =\epsilon^{a_1c_1c_2} \epsilon^{a_2b_3b_4} \epsilon^{a_3a_4c_4} \epsilon^{b_1b_2c_3}\nonumber \\
\sim 32 d^{{A_1}{A_2}B} d^{{A_3}{A_4}B}+\frac{16}{3} \delta^{{A_1}{A_2}} \delta^{{A_3}{A_4}},\\
\begin{ytableau}
a_1 & b_1 \\ c_1
\end{ytableau} \mathop{\longrightarrow}\limits^{\begin{ytableau}
a_2 & b_2 \\ c_2
\end{ytableau}} \begin{ytableau}
a_1 & b_1 & a_2 \\ c_1 & b_2 \\ c_2
\end{ytableau} \mathop{\longrightarrow}\limits^{\begin{ytableau}
a_3 & b_3 \\ c_3
\end{ytableau}} \begin{ytableau}
a_1 & b_1 & a_2 & a_3 \\ c_1 & b_2 & c_3 \\ c_2 & b_3
\end{ytableau} \mathop{\longrightarrow}\limits^{\begin{ytableau}
a_4 & b_4 \\ c_4
\end{ytableau}} \begin{ytableau}
a_1 & b_1 & a_2 & a_3 \\ c_1 & b_2 & c_3 & a_4 \\ c_2 & b_3 & b_4 & c_4
\end{ytableau} =\epsilon^{a_1c_1c_2} \epsilon^{a_2c_3b_4} \epsilon^{a_3a_4c_4} \epsilon^{b_1b_2b_3}\nonumber\\
\sim 16 i d^{{A_1}{A_2}B} f^{{A_3}{A_4}B}+16 d^{{A_1}{A_2}B} d^{{A_3}{A_4}B}-\frac{16}{3} \delta^{{A_1}{A_2}} \delta^{{A_3}{A_4}},\\
\begin{ytableau}
a_1 & b_1 \\ c_1
\end{ytableau} \mathop{\longrightarrow}\limits^{\begin{ytableau}
a_2 & b_2 \\ c_2
\end{ytableau}} \begin{ytableau}
a_1 & b_1 & a_2 \\ c_1 & c_2 \\ b_2
\end{ytableau} \mathop{\longrightarrow}\limits^{\begin{ytableau}
a_3 & b_3 \\ c_3
\end{ytableau}} \begin{ytableau}
a_1 & b_1 & a_2 & a_3 \\ c_1 & c_2 & b_3 \\ b_2 & c_3
\end{ytableau} \mathop{\longrightarrow}\limits^{\begin{ytableau}
a_4 & b_4 \\ c_4
\end{ytableau}} \begin{ytableau}
a_1 & b_1 & a_2 & a_3 \\ c_1 & c_2 & b_3 & a_4 \\ b_2 & c_3 & b_4 & c_4
\end{ytableau} =\epsilon^{a_1c_1b_2} \epsilon^{a_2b_3b_4} \epsilon^{a_3a_4c_4} \epsilon^{b_1c_2c_3}\nonumber\\
\sim 16 i f^{{A_1}{A_2}B} d^{{A_3}{A_4}B}+16 d^{{A_1}{A_2}B} d^{{A_3}{A_4}B}+\frac{32}{3} \delta^{{A_1}{A_2}} \delta^{{A_3}{A_4}},\\
\begin{ytableau}
a_1 & b_1 \\ c_1
\end{ytableau} \mathop{\longrightarrow}\limits^{\begin{ytableau}
a_2 & b_2 \\ c_2
\end{ytableau}} \begin{ytableau}
a_1 & b_1 & a_2 \\ c_1 & c_2 \\ b_2
\end{ytableau} \mathop{\longrightarrow}\limits^{\begin{ytableau}
a_3 & b_3 \\ c_3
\end{ytableau}} \begin{ytableau}
a_1 & b_1 & a_2 & a_3 \\ c_1 & c_2 & c_3 \\ b_2 & b_3
\end{ytableau} \mathop{\longrightarrow}\limits^{\begin{ytableau}
a_4 & b_4 \\ c_4
\end{ytableau}} \begin{ytableau}
a_1 & b_1 & a_2 & a_3 \\ c_1 & c_2 & c_3 & a_4 \\ b_2 & b_3 & b_4 & c_4
\end{ytableau} =\epsilon^{a_1c_1b_2} \epsilon^{a_2c_3b_4} \epsilon^{a_3a_4c_4} \epsilon^{b_1c_2b_3} \nonumber\\
\sim 8 i f^{{A_1}{A_2}B} d^{{A_3}{A_4}B}+8 i d^{{A_1}{A_2}B} f^{{A_3}{A_4}B}+8 d^{{A_1}{A_2}B} d^{{A_3}{A_4}B}-\frac{32}{3} \delta^{{A_1}{A_2}} \delta^{{A_3}{A_4}}-8 f^{{A_1}{A_2}B} f^{{A_3}{A_4}B}.
\eea
As one can see, the y-basis group factors when converted to the invariant tensors with adjoint indices may become polynomials, while a complete and independent monomial basis, named the gauge m-basis, can always be chosen from these polynomials, and an efficient algorithm to find the gauge m-basis has been proposed in Ref.~\cite{Li:2022tec}.  Here we select a series of independent monomials to form the m-basis and index them in the following equation, 
\bea
\mathcal{T} \equiv\begin{pmatrix}
\mathcal{T}_1\\ \mathcal{T}_2\\ \mathcal{T}_3\\ \mathcal{T}_4\\ \mathcal{T}_5\\ \mathcal{T}_6\\ \mathcal{T}_7\\ \mathcal{T}_8
\end{pmatrix}
=\begin{pmatrix}
d^{A_1A_2B} d^{A_3A_4B}\\
d^{A_1A_2B} f^{A_3A_4B}\\
f^{A_1A_2B} f^{A_3A_4B}\\
\delta^{A_1A_2} \delta^{A_3A_4}\\
d^{A_3A_4B} f^{A_1A_2B}\\
\delta^{A_1A_3} \delta^{A_2A_4}\\
d^{A_1A_3B} d^{A_2A_4B}\\
d^{A_1A_3B} f^{A_2A_4B}
\end{pmatrix},\label{eq:basis1}
\eea
where again the subscripts of the gauge indices denote the label of particles. For instance, $A_1$ denotes adjoint $SU(3)$ indices for particle $1$. 
To find out the matrix representation of the partial Casimir operator $\mathop{\mathbb{C}_2}\limits_{\{12\}}$ on such a basis, we need to act it on each basis vector in $\mathcal{T}$ and then convert the result back into the form of linear combination of these basis vectors. For example, $\mathop{\mathbb{C}_2}\limits_{\{12\}}$ acting on ${\cal T}_1$ reads:
\bea
\mathop{\mathbb{C}_2}\limits_{\{12\}}\circ \mathcal{T}_1 &=& \mathop{T^{D}}\limits_{\{12\}}\circ (\mathop{T^D}\limits_{\{12\}}\circ \mathcal{T}_1)\nonumber\\
&=&\mathop{T^D}\limits_{\{12\}} (-if^{DA_1E}d^{EA_2B}d^{A_3A_4B} -if^{DA_2E}d^{A_1EB}d^{A_3A_4B})\nonumber\\ &=& -f^{DA_1F}f^{DFE}d^{EA_2B}d^{A_3A_4B} -f^{DA_2F}f^{DA_1E}d^{EFB}d^{A_3A_4B}\nonumber\\&{}& -f^{DA_1F}f^{DA_2E}d^{FEB}d^{A_3A_4B} -f^{DA_2F}f^{DFE}d^{A_1EB}d^{A_3A_4B}\\ &=&6 d^{A_1A_2E} d^{A_3A_4E}-2 d^{A_3A_4E} d^{EFG} f^{A_1FH} f^{A_2GH}=3{\cal T}_1.\nonumber
\eea
The action on the rest of the basis vectors can be derived in a similar way and we obtain: $\mathop{\mathbb{C}_2}\limits_{\{12\}} \circ \mathcal{T}$  as
\bea
\mathop{\mathbb{C}_2}\limits_{\{12\}} \circ \mathcal{T} =
\begin{pmatrix}
6 d^{A_1A_2E} d^{A_3A_4E}-2 d^{A_3A_4E} d^{EFG} f^{A_1FH} f^{A_2GH}\\
6 d^{A_1A_2E} f^{A_3A_4E}-2 d^{EFG} f^{A_1EH} f^{A_2FH} f^{A_3A_4G}\\
6 f^{A_1A_2E} f^{A_3A_4E}-2 f^{A_1EF} f^{A_2EG} f^{A_3A_4H} f^{FGH}\\
0\\
6 d^{A_3A_4E} f^{A_1A_2E}-2 d^{A_3A_4E} f^{A_1FG} f^{A_2FH} f^{EGH}\\
6 \delta^{A_1A_3} \delta^{A_2A_4}-2 f^{A_1A_3E} f^{A_2A_4E}\\
6 d^{A_1A_3E} d^{A_2A_4E}-2 d^{A_3EF} d^{A_4EG} f^{A_1FH} f^{A_2GH}\\
6 d^{A_1A_3E} f^{A_2A_4E}+2 d^{A_3EF} f^{A_1EG} f^{A_2GH} f^{A_4FH}
\end{pmatrix},
\eea
whose coordinate matrix on basis Eq.~(\ref{eq:basis1}) is
\bea
(\mathop{C_2}\limits_{\{12\}})_{ji}=
\left(
\begin{array}{cccccccc}
 3 & 0 & 0 & 0 & 0 & 0 & 0 & 0 \\
 0 & 3 & 0 & 0 & 0 & 0 & 0 & 0 \\
 0 & 0 & 3 & 0 & 0 & 0 & 0 & 0 \\
 0 & 0 & 0 & 0 & 0 & 0 & 0 & 0 \\
 0 & 0 & 0 & 0 & 3 & 0 & 0 & 0 \\
 -2 & 0 & -2 & -\frac{4}{3} & 0 & \frac{22}{3} & 2 & 0 \\
 \frac{5}{6} & 0 & -\frac{7}{6} & -\frac{13}{9} & 0 & \frac{4}{9} & \frac{20}{3} & 0 \\
 0 & \frac{3}{2} & 0 & 0 & -\frac{3}{2} & 0 & 0 & 6 \\
\end{array}
\right).
\eea
By diagnoalizing $(\mathop{C_2}\limits_{\{12\}})_{ji}^{\mathtt{T}}$, we obtain the corresponding eigenvalues,  eigenvectors and the conversion matrix ${\cal K}_{C_2,\{12\}}^{jm}$:
\begin{eqnarray}
&&{\cal K}_{C_2,\{12\}}^{jm}.(\mathop{C_2}\limits_{\{12\}})^{\mathtt{T}},({\cal K}_{C_2,\{12\}}^{jm})^{-1}={\rm diag}\{8,6,6,3,3,3,3,0\},\\
&&{\cal T}^j=\begin{pmatrix}
{\cal T}^{j}_{1,C_2=8}\\
{\cal T}^{j}_{1,C_2=6}\\
{\cal T}^{j}_{2,C_2=6}\\
{\cal T}^{j}_{1,C_2=3}\\
{\cal T}^{j}_{2,C_2=3}\\
{\cal T}^{j}_{3,C_2=3}\\
{\cal T}^{j}_{4,C_2=3}\\
{\cal T}^{j}_{1,C_2=0}
\end{pmatrix}
 = \begin{pmatrix}
{-\frac{1}{10}} & 0 & {-\frac{1}{2}} & {-\frac{7}{24}} & 0 & 0 & 1 & 0 \\
0 & \frac{1}{2} & 0 & 0 & -\frac{1}{2} & 0 & 0 & 1 \\
\frac{1}{2} & 0 & -\frac{1}{6} & -\frac{1}{6} & 0 & -\frac{1}{3} & 1 & 0 \\
1 & 0 & 0 & 0 & 0 & 0 & 0 & 0 \\
0 & 1 & 0 & 0 & 0 & 0 & 0 & 0 \\
0 & 0 & 1 & 0 & 0 & 0 & 0 & 0 \\
0 & 0 & 0 & 0 & 1 & 0 & 0 & 0 \\
0 & 0 & 0 & 1 & 0 & 0 & 0 & 0 
\end{pmatrix}
\begin{pmatrix}
{\cal T}_{1}\\
{\cal T}_{2}\\
{\cal T}_{3}\\
{\cal T}_{4}\\
{\cal T}_{5}\\
{\cal T}_{6}\\
{\cal T}_{7}\\
{\cal T}_{8}
\end{pmatrix}=
{\cal K}_{C_2,\{12\}}^{jm}\cdot {\cal T}, \label{eq:k12jm}
\end{eqnarray}
which partitions the invariant tensors into four invariant subspaces:
\begin{eqnarray}
&&{\cal T}^{j}_{C_2=8}=\mathcal{T}_7-\frac{1}{10} \mathcal{T}_1+\frac{2}{3} \mathcal{T}_6-\frac{7}{24} \mathcal{T}_4-\frac{1}{2} \mathcal{T}_3
,\quad \nonumber \\
&&{\cal T}^{j}_{C_2=6}=\begin{pmatrix}
-\frac{1}{2} \mathcal{T}_5+\mathcal{T}_8+\frac{1}{2} \mathcal{T}_2\\
\mathcal{T}_7+\frac{1}{2} \mathcal{T}_1-\frac{1}{3} \mathcal{T}_6-\frac{1}{6} \mathcal{T}_4-\frac{1}{6} \mathcal{T}_3
\end{pmatrix},\quad {\cal T}^{j}_{C_2=3}=\begin{pmatrix}
\mathcal{T}_1\\
\mathcal{T}_2\\
\mathcal{T}_3\\
\mathcal{T}_5
\end{pmatrix},\quad {\cal T}^{j}_{C_2=0}=\mathcal{T}_4,\nonumber
\end{eqnarray}
or equivalently, we can represent each eigenvectors in terms of coordinates on the m-basis:
\begin{eqnarray}
&&[(\mathop{C_2}\limits_{\{12\}})=8]={\rm span}\{({\cal K}_{C_2,\{12\}}^{jm})_1\},\quad [(\mathop{C_2}\limits_{\{12\}})=6]={\rm span}\{({\cal K}_{C_2,\{12\}}^{jm})_{2},({\cal K}_{C_2,\{12\}}^{jm})_{3}\},\nonumber\\ 
&&[(\mathop{C_2}\limits_{\{12\}})=3]={\rm span}\{({\cal K}_{C_2,\{12\}}^{jm})_{4},({\cal K}_{C_2,\{12\}}^{jm})_{5},({\cal K}_{C_2,\{12\}}^{jm})_{6},({\cal K}_{C_2,\{12\}}^{jm})_{7}\},\quad [(\mathop{C_2}\limits_{\{12\}})=0]={\rm span}\{({\cal K}_{C_2,\{12\}}^{jm})_{8}\},\nonumber
\end{eqnarray}
where again the subscripts of ${\cal K}_{C_2,\{12\}}^{jm}$ represents the row number, which indicates the coordinates of the corresponding eigenvectors.
\begin{table}[ht]
\centering
\begin{tabular}{c|c}
\makecell{$(C_2(\mbf{R}), \mathbf{R})$} & Eigenstates \\
\hline
\makecell{($8$, $\mbf{27}$)} & \makecell{ $\mathcal{T}_7-\frac{1}{10} \mathcal{T}_1+\frac{2}{3} \mathcal{T}_6-\frac{7}{24} \mathcal{T}_4-\frac{1}{2} \mathcal{T}_3$ } \\
\hline
\makecell{($6$,  $\overline{\mbf{10}}$ or $\mbf{10}$)} & \makecell{$-\frac{1}{2} \mathcal{T}_5+\mathcal{T}_8+\frac{1}{2} \mathcal{T}_2$ \\
$\mathcal{T}_7+\frac{1}{2} \mathcal{T}_1-\frac{1}{3} \mathcal{T}_6-\frac{1}{6} \mathcal{T}_4-\frac{1}{6} \mathcal{T}_3$} \\
\hline
\makecell{($3$, $\mbf{8}$)} & \makecell{$\mathcal{T}_5$, $\mathcal{T}_3$, $\mathcal{T}_2$, $\mathcal{T}_1$}\\
\hline
\makecell{($0$, $\mbf{1}$)} & $\mathcal{T}_4$
\end{tabular}
\caption{Representations of $SU(3)$ gluons $\{8,8,8,8\}$ at channel $\{12\}$.}
\label{tab:12e1}
\end{table}
As one can see in Table~\ref{tab:12e1}, both the representation $\mbf{\overline{10}}$ and $\mbf{10}$ have eigenvalue $C_2=6$, therefore we need the values of a second Casimir $\mathbbm{C}_3$ to resolve such a degeneracy.
The action of $\mathop{\mathbb{C}_3}\limits_{\{12\}}$ is similar to that of $\mathop{\mathbb{C}_2}\limits_{\{12\}}$, for example the action of $\mathop{\mathbb{C}_3}\limits_{\{12\}}$ on ${\cal T}_1$ gives
\bea
\mathop{\mathbb{C}_3}\limits_{\{12\}}\circ \mathcal{T}_1 &=& d^{DEF} \mathop{T^D}\limits_{\{12\}} \circ \mathop{T^E}\limits_{\{12\}} \circ \mathop{T^F}\limits_{\{12\}}  \mathcal{T}_1\nonumber\\
&=& d^{DEF} \mathop{T^D}\limits_{\{12\}} \circ \mathop{T^E}\limits_{\{12\}} (-if^{FA_1G}d^{GA_2C}d^{A_3A_4C} -if^{FA_2G}d^{A_1GC}d^{A_3A_4C})\nonumber\\
&=& d^{DEF} \mathop{T^D}\limits_{\{12\}} (-f^{EA_1H}f^{FHG}d^{GA_2C}d^{A_3A_4C} -f^{EA_2H}f^{FA_1G}d^{GHC}d^{A_3A_4C}\nonumber \\ &{}& -f^{EA_1H}f^{FA_2G}d^{HGC}d^{A_3A_4C} -f^{EA_2H}f^{FHG}d^{A_1GC}d^{A_3A_4C})\\
&=& 3i d^{DGF} d^{A_3A_4C} d^{CHE} f^{A_1HF} f^{DA_2B} f^{GBE}
+3i d^{DGF} d^{A_3A_4C} d^{CHE} f^{A_1BF} f^{DA_2H} f^{GBE}\nonumber\\
&{}& +i d^{DEF} d^{A_2CH} d^{A_3A_4C} f^{A_1GE} f^{DGB} f^{BHF}
+i d^{DEF} d^{A_1CH} d^{A_3A_4C} f^{BDG} f^{A_2GE} f^{BHF}.\nonumber
\eea
If we focus on only resolving the degeneracy between $\mbf{\overline{10}}$ and $\mbf{10}$, then it is enough to find the representation matrix for $\mathop{\mathbb{C}_3}\limits_{\{12\}}$ in the subspace of eigenvectors of  $[(\mathop{C_2}\limits_{\{12\}})=6]$.  Acting $\mathop{\mathbb{C}_3}\limits_{\{12\}}$ on these  two vectors gives
\bea
\mathop{\mathbb{C}_3}\limits_{\{12\}}
\circ\begin{pmatrix}
{\cal T}^{j}_{1,C_2=6}\\
{\cal T}^{j}_{2,C_2=6}
\end{pmatrix}&=&
\mathop{\mathbb{C}_3}\limits_{\{12\}}
\circ\begin{pmatrix}
-\frac{1}{2} \mathcal{T}_5+\mathcal{T}_8+\frac{1}{2} \mathcal{T}_2\\
\mathcal{T}_7+\frac{1}{2} \mathcal{T}_1-\frac{1}{3} \mathcal{T}_6-\frac{1}{6} \mathcal{T}_4-\frac{1}{6} \mathcal{T}_3
\end{pmatrix}\nonumber \\ &=&\begin{pmatrix}
-\frac{9i}{2}\mathcal{T}_1+ \frac{3 i}{2}\mathcal{T}_3+ \frac{3 i}{2}\mathcal{T}_4 +3 i\mathcal{T}_6 -9 i\mathcal{T}_7\\
\frac{9 i}{2}\mathcal{T}_2 -\frac{9 i}{2}\mathcal{T}_5 +9 i\mathcal{T}_8
\end{pmatrix}\nonumber \\
&=&\begin{pmatrix}
0 & -9i\\ 9i & 0
\end{pmatrix}\begin{pmatrix}
{\cal T}^{j}_{1,C_2=6}\\
{\cal T}^{j}_{2,C_2=6}
\end{pmatrix}={\mathop{C_3^{\mathtt{T}}}\limits_{\{12\}}}.T^j_{C_2=6},
\eea
The representation  matrix ${\mathop{C_3^{\mathtt{T}}}\limits_{\{12\}}}$ can be diagonalized by ${\cal K}^{jm}_{C_3,\{12\}}$ to reorganize the basis vector in $[(\mathop{C_2}\limits_{\{12\}})=6]$ into eigenvectors of $\mathop{\mathbb{C}_3}\limits_{\{12\}}$, which yields
\bea
&&{\cal K}^{jm}_{C_3,\{12\}}.{\mathop{C_3^{\mathtt{T}}}\limits_{\{12\}}}.({\cal K}^{jm}_{C_3,\{12\}})^{-1} = {\rm diag}\{9,-9\},\quad {\cal K}^{jm}_{C_3,\{12\}}=\begin{pmatrix}
i& 1\\
-i & 1
\end{pmatrix}.\nonumber \\
&& {\cal T}^{j}_{C_3=9,C_2=6}=-\frac{1}{2} i \mathcal{T}_5+i \mathcal{T}_8+\frac{1}{2} i \mathcal{T}_2+\mathcal{T}_7+\frac{1}{2} \mathcal{T}_1-\frac{1}{3} \mathcal{T}_6-\frac{1}{6} \mathcal{T}_4-\frac{1}{6} \mathcal{T}_3,  \\
&&  {\cal T}^{j}_{C_3=-9,C_2=6}= \frac{i}{2}  \mathcal{T}_5-i \mathcal{T}_8-\frac{i}{2}  \mathcal{T}_2+\mathcal{T}_7+\frac{1}{2} \mathcal{T}_1-\frac{1}{3} \mathcal{T}_6-\frac{1}{6} \mathcal{T}_4-\frac{1}{6} \mathcal{T}_3.
\eea
In this way we find the common eigenvectors for $\mathop{\mathbb{C}_2}\limits_{\{12\}}$ and $\mathop{\mathbb{C}_3}\limits_{\{12\}}$, thus resolving the degeneracy between $\overline{\mbf{10}}$ and $\mbf{10}$. The comprehensive results for the gauge basis of the partition $\{12|34\}$ with definite quantum numbers for each part are given in table.~\ref{tab:gggg}, and the full conversion matrix ${\cal K}_{G}^{jm}$ can be read out as:
\begin{eqnarray}
{\cal K}_{G}^{jm}
=\begin{pmatrix}
{-\frac{1}{10}} & 0 & {-\frac{1}{2}} & {-\frac{7}{24}} & 0 & 0 & 1 & 0 \\
\frac{1}{2} & \frac{i}{2} & -\frac{1}{6} & -\frac{1}{6} & -\frac{i}{2} & -\frac{1}{3} & 1 & i \\
\frac{1}{2} & -\frac{i}{2} & -\frac{1}{6} & -\frac{1}{6} & \frac{i}{2} & -\frac{1}{3} & 1 & -i \\
1 & 0 & 0 & 0 & 0 & 0 & 0 & 0 \\
0 & 1 & 0 & 0 & 0 & 0 & 0 & 0 \\
0 & 0 & 1 & 0 & 0 & 0 & 0 & 0 \\
0 & 0 & 0 & 0 & 1 & 0 & 0 & 0 \\
0 & 0 & 0 & 1 & 0 & 0 & 0 & 0 
\end{pmatrix},
\end{eqnarray}
which only differ from Eq.~\eqref{eq:k12jm} by the second and the third rows as expected.
\begin{table}[ht]
\centering
\begin{tabular}{c|c}
$\mathop{C_2}\limits_{\{12\}}, \mathop{C_3}\limits_{\{12\}}$, $\mathbf{R}_{\{12\}}$ & basis \\
\hline
$8,0,\mbf{27}$ & \makecell{ $\mathcal{T}_7-\frac{1}{10} \mathcal{T}_1+\frac{2}{3} \mathcal{T}_6-\frac{7}{24} \mathcal{T}_4-\frac{1}{2} \mathcal{T}_3$ }\\
\hline
$6,9,\mbf{10}$ & \makecell{$-\frac{i}{2}  \mathcal{T}_5 +i \mathcal{T}_8 +\frac{1}{2} i \mathcal{T}_2 +\mathcal{T}_7 +\frac{1}{2} \mathcal{T}_1-\frac{1}{3} \mathcal{T}_6-\frac{1}{6} \mathcal{T}_4-\frac{1}{6} \mathcal{T}_3$}\\
\hline
$6,-9,\overline{\mbf{10}}$ & \makecell{$\frac{i}{2}  \mathcal{T}_5-i \mathcal{T}_8-\frac{i}{2}  \mathcal{T}_2+\mathcal{T}_7+\frac{1}{2} \mathcal{T}_1-\frac{1}{3} \mathcal{T}_6-\frac{1}{6} \mathcal{T}_4-\frac{1}{6} \mathcal{T}_3$}\\
\hline
$3,0,\mbf{8}$ & \makecell{$\mathcal{T}_5$, $\mathcal{T}_3$, $\mathcal{T}_2$, $\mathcal{T}_1$}\\
\hline
$0,0,\mbf{1}$ & $\mathcal{T}_4$
\end{tabular}
\caption{ Gauge partial-wave basis of $SU(3)$ particles $\{8,8,8,8\}$ of partition $\{12|34\}$ }
\label{tab:gggg}
\end{table}

Up to now we have finished our discussion on finding the gauge eigen-basis for two-partite cases. It is straightforward to generalize such a method into multi-partite. The methodology is essentially the same as the above example, as long as a set of Casimir operators no matter of different parts or of different type commute with each other, one can always construct common eigenvectors specified by a set of eigenvalues $\{\underset{\mathbbm{S}_k}{C_i}\}$.
To illustrate this point, we shall take another example of $SU(3)$ for finding the gauge amplitude basis for  
$\{\mbf{3},\mbf{3},\bar{\mbf{3}},\bar{\mbf{3}},\mbf{8},\mbf{8}\}$ with partition $\{12|34|56\}$. It has $13 $ independent monomial invariant tensors
\bea
\mathcal{T}\equiv\begin{pmatrix}
\mathcal{T}_1\\ \mathcal{T}_2\\ \mathcal{T}_3\\ \mathcal{T}_4\\ \mathcal{T}_5\\ \mathcal{T}_6\\ \mathcal{T}_7\\ \mathcal{T}_8 \\ \mathcal{T}_9\\ \mathcal{T}_{10}\\ \mathcal{T}_{11}\\ \mathcal{T}_{12}\\ \mathcal{T}_{13}
\end{pmatrix}
=\begin{pmatrix}
d^{A_5A_6C} \delta^{a_1}_{b_3} \tau^{C,a_2}_{b_4}\\
d^{A_5A_6C} \delta^{a_1}_{b_4} \tau^{C,a_2}_{b_3}\\
d^{A_5A_6C} \delta^{a_2}_{b_4} \tau^{C,a_1}_{b_3}\\
\delta^{a_1}_{b_3} f^{A_5,A_6,C} \tau^{C,a_2}_{b_4}\\
\delta^{a_1}_{b_4} f^{A_5,A_6,C} \tau^{C,a_2}_{b_3}\\
\delta^{a_2}_{b_4} f^{A_5,A_6,C} \tau^{C,a_1}_{b_3}\\
\tau^{A_5,a_1}_{b_3} \tau^{A_6,a_2}_{b_4}\\
\delta^{a_1}_{b_3} \delta^{a_2}_{b_4} \delta^{A_5A_6}\\
\delta^{a_1}_{b_4} \delta^{a_2}_{b_3} \delta^{A_5A_6}\\
\tau^{A_5,a_2}_{b_3} \tau^{A_6,a_1}_{b_4}\\
\tau^{A_5,a_2}_{b_4} \tau^{A_6,a_1}_{b_3}\\
\tau^{A_5,a_1}_{b_4} \tau^{A_6,a_2}_{b_3}\\
\delta^{a_2}_{b_3} f^{A_5A_6C} \tau^{C,a_1}_{b_4}
\end{pmatrix}.\label{eq:gbasis}
\eea

The action of $\mathop{\mathbb{C}_2}\limits_{\{12\}}$ on basis $\mathcal{T}$ gives:
\bea
\mathop{\mathbb{C}_2}_{\{12\}}\circ \mathcal{T} =
\left(
\begin{array}{ccccccccccccc}
 \frac{10}{3} & -1 & 1 & 0 & 0 & 0 & \frac{1}{2} & \frac{1}{3} & -\frac{1}{3} & -\frac{1}{2} & \frac{1}{2} & -\frac{1}{2} & 0 \\
 0 & \frac{7}{3} & 1 & 0 & 0 & 0 & 0 & 0 & 0 & 0 & 0 & 0 & 0 \\
 0 & 1 & \frac{7}{3} & 0 & 0 & 0 & 0 & 0 & 0 & 0 & 0 & 0 & 0 \\
 0 & 0 & 0 & \frac{7}{3} & 0 & 0 & 0 & 0 & 0 & 0 & 0 & 0 & 1 \\
 0 & 0 & 0 & 0 & \frac{7}{3} & 1 & 0 & 0 & 0 & 0 & 0 & 0 & 0 \\
 0 & 0 & 0 & 0 & 1 & \frac{7}{3} & 0 & 0 & 0 & 0 & 0 & 0 & 0 \\
 0 & 0 & 0 & 0 & 0 & 0 & \frac{7}{3} & 0 & 0 & 1 & 0 & 0 & 0 \\
 0 & 0 & 0 & 0 & 0 & 0 & 0 & \frac{7}{3} & 1 & 0 & 0 & 0 & 0 \\
 0 & 0 & 0 & 0 & 0 & 0 & 0 & 1 & \frac{7}{3} & 0 & 0 & 0 & 0 \\
 0 & 0 & 0 & 0 & 0 & 0 & 1 & 0 & 0 & \frac{7}{3} & 0 & 0 & 0 \\
 0 & 0 & 0 & 0 & 0 & 0 & 0 & 0 & 0 & 0 & \frac{7}{3} & 1 & 0 \\
 0 & 0 & 0 & 0 & 0 & 0 & 0 & 0 & 0 & 0 & 1 & \frac{7}{3} & 0 \\
 0 & 0 & 0 & 1 & 0 & 0 & 0 & 0 & 0 & 0 & 0 & 0 & \frac{7}{3} \\
\end{array}
\right) \begin{pmatrix}
\mathcal{T}_1\\ \mathcal{T}_2\\ \mathcal{T}_3\\ \mathcal{T}_4\\ \mathcal{T}_5\\ \mathcal{T}_6\\ \mathcal{T}_7\\ \mathcal{T}_8 \\ \mathcal{T}_9\\ \mathcal{T}_{10}\\ \mathcal{T}_{11}\\ \mathcal{T}_{12}\\ \mathcal{T}_{13}
\end{pmatrix}.
\eea
Similar representation matrices can be obtained for channels $\{34\}$ and $\{56\}$.
By diagonalizing these matrices, we obtain the eigenvalues and eigenvectors for partial Casimir operators $\mathop{\mathbb{C}_2}$ for channel $\{12\}$, $\{34\}$ and $\{56\}$, which are listed in table \ref{tab:12}. We can perform the same analysis for Casimir operator $\mathop{\mathbb{C}_3}$, and finally obtain the total six sets of eigenvectors and eigenvalues.

\begin{table}[htb]
\centering
\begin{tabular}{c|c}
\makecell{$(\mathop{\mathbb{C}_2}\limits_{\{12\}}(\mbf{R}),\ \mbf{R})$} & Eigenstates \\
\hline
($10/3$, $\bar{\mbf{6}}$) & \makecell{ $\mathcal{T}_4+\mathcal{T}_{13}$,\quad $-2\mathcal{T}_1+2\mathcal{T}_2-\mathcal{T}_7 -\frac{2}{3}\mathcal{T}_8+\mathcal{T}_{12}$, \\$2\mathcal{T}_1 -2\mathcal{T}_2 +\mathcal{T}_7+\frac{2}{3}\mathcal{T}_8 +\mathcal{T}_{11}$,\\ $\mathcal{T}_7+\mathcal{T}_{10}$,\quad $\mathcal{T}_8+\mathcal{T}_9$,\quad $\mathcal{T}_5+\mathcal{T}_6$,\quad $\mathcal{T}_2+\mathcal{T}_3$ }\\
\hline
($4/3$, $\bar{\mbf{3}}$) & \makecell{$-\mathcal{T}_4+\mathcal{T}_{13}$,\quad $-\mathcal{T}_{11}+\mathcal{T}_{12}$,\quad $-\mathcal{T}_7+\mathcal{T}_{10}$,\\ $-\mathcal{T}_8+\mathcal{T}_9$,\quad $-\mathcal{T}_5+\mathcal{T}_6$,\quad $-\mathcal{T}_2+\mathcal{T}_3$}\\
\hline\hline
\makecell{($\mathop{\mathbb{C}_2}\limits_{\{34\}}(\mbf{R})$, $\mbf{R}$)} & Eigenstates \\
\hline
($10/3$, ${\mbf{6}}$) & \makecell{ $\mathcal{T}_6+\mathcal{T}_{13}$,\quad $\mathcal{T}_7+\mathcal{T}_{12}$,\quad \\ $2\mathcal{T}_1+2\mathcal{T}_3+\mathcal{T}_7 +\frac{2}{3}\mathcal{T}_8+\mathcal{T}_{11}$,\\
$-2\mathcal{T}_1-2\mathcal{T}_3-\mathcal{T}_7 -\frac{2}{3}\mathcal{T}_8+\mathcal{T}_{10}$, \\ $\mathcal{T}_8+\mathcal{T}_9$,\quad $\mathcal{T}_4+\mathcal{T}_5$,\quad $\mathcal{T}_1+\mathcal{T}_2$ }\\
\hline
($4/3$, ${\mbf{3}}$) & \makecell{$-\mathcal{T}_6+\mathcal{T}_{13}$,\quad $-\mathcal{T}_7+\mathcal{T}_{12}$,\quad $-\mathcal{T}_{10}+\mathcal{T}_{11}$,\\ $-\mathcal{T}_8+\mathcal{T}_9$,\quad $-\mathcal{T}_4+\mathcal{T}_5$,\quad $-\mathcal{T}_1+\mathcal{T}_2$}\\
\hline\hline
\makecell{($\mathop{\mathbb{C}_2}\limits_{\{56\}}(\mbf{R})$, $\mbf{R}$)} & Eigenstates \\
\hline
$(8,\ \mbf{27})$ & $-\frac{2}{3}\mathcal{T}_1 -\frac{2}{3}\mathcal{T}_3 +\frac{2}{3}\mathcal{T}_7 -\frac{7}{18}\mathcal{T}_8 -\frac{1}{6}\mathcal{T}_9+\mathcal{T}_{10} +\frac{2}{3}\mathcal{T}_{11}+\mathcal{T}_{12}$ \\
\hline
($6$, $\mbf{10}$ or $\bar{\mbf{10}}$) & \makecell{$-\mathcal{T}_5-\frac{3i}{2}\mathcal{T}_7 +\frac{3i}{2}\mathcal{T}_{11}+\mathcal{T}_{13}$,\\ $-\frac{2i}{3}\mathcal{T}_4+\frac{2i}{3}\mathcal{T}_6 -\mathcal{T}_{10}+\mathcal{T}_{12}$} \\
\hline
($3,\ \mbf{8}$) & \makecell{ $-\mathcal{T}_7-\frac{2}{3}\mathcal{T}_8 +\frac{2}{3}\mathcal{T}_9,+\mathcal{T}_{10} -\mathcal{T}_{11}+\mathcal{T}_{12}$,\\ $\mathcal{T}_{13}$,\quad $\mathcal{T}_6$,\quad $\mathcal{T}_5$,\quad $\mathcal{T}_4$,\quad $\mathcal{T}_3$,\quad $\mathcal{T}_2$,\quad $\mathcal{T}_1$} \\
\hline
$(0,\ \mbf{1})$ & $\mathcal{T}_9$,\quad $\mathcal{T}_8$
\end{tabular}
\caption{Representations of $SU(3)$ particles $\{3,3,\bar{3},\bar{3},8,8\}$ at channel $\{12\}$, $\{34\}$ and $\{56\}$ respectively.}
\label{tab:12}
\end{table}

The next step is to find the common eigenvectors for these six of partial Casimir operators. In contrast to the above method of diagonalizing the representation matrices in the subspaces recursively, we implement a more systematic method discussed in appendix~\ref{app:linearintersection} to find these common eigenvectors --- calculating the basis of the linear intersections space of different eigenvalue subspaces with basis of each subspace. This method can provide the basis vectors for the intersection of an arbitrary number of subspaces, and the final result is shown in table \ref{tab:g12,34,56}.
\begin{table}[H]
\centering
\begin{tabular}{c|c}
Representation $\mathbf{R}_{\mathcal{I}}$ & basis \\
\hline
$\mathbf{R}_{12}=\bar{6},\mathbf{R}_{34}=\mbf{6},\mathbf{R}_{56}=27$ & \makecell{ $-\frac{2}{3}\mathcal{T}_1-\frac{2}{3}\mathcal{T}_3 +\frac{2}{3}\mathcal{T}_7-\frac{7}{18}\mathcal{T}_8 -\frac{1}{6}\mathcal{T}_9+\mathcal{T}_{10} +\frac{2}{3}\mathcal{T}_{11}+\mathcal{T}_{12}$}\\
\hline
$\mathbf{R}_{12}=\bar{6},\mathbf{R}_{34}=3,\mathbf{R}_{56}=\bar{10}$ & \makecell{ $-\mathcal{T}_4+\mathcal{T}_5+\mathcal{T}_6 +\frac{3i}{2}\mathcal{T}_7 +\frac{3i}{2}\mathcal{T}_{10} -\frac{3i}{2}\mathcal{T}_{11} -\frac{3i}{2}\mathcal{T}_{12}-\mathcal{T}_{13}$ }\\
\hline
$\mathbf{R}_{12}=\bar{3},\mathbf{R}_{34}=6,\mathbf{R}_{56}=10$ & \makecell{ $-\mathcal{T}_4-\mathcal{T}_5+\mathcal{T}_6 -\frac{3i}{2}\mathcal{T}_7 +\frac{3i}{2}\mathcal{T}_{10} +\frac{3i}{2}\mathcal{T}_{11} -\frac{3i}{2}\mathcal{T}_{12}+\mathcal{T}_{13}$ }\\
\hline
$\mathbf{R}_{12}=\bar{6},\mathbf{R}_{34}=6,\mathbf{R}_{56}=1$ & \makecell{ $\mathcal{T}_8+\mathcal{T}_9$ }\\
\hline
$\mathbf{R}_{12}=\bar{3},\mathbf{R}_{34}=3, \mathbf{R}_{56}=1$ & \makecell{ $\mathcal{T}_8-\mathcal{T}_9$ }\\
\hline
$\mathbf{R}_{12}=\bar{6},\mathbf{R}_{34}=6,\mathbf{R}_{56}=8$ & \makecell{ $\mathcal{T}_4+\mathcal{T}_5+\mathcal{T}_6 +\mathcal{T}_{13}$, $-4\mathcal{T}_1-4\mathcal{T}_3-\mathcal{T}_7 -\frac{2}{3}\mathcal{T}_8+\frac{2}{3}\mathcal{T}_9 +\mathcal{T}_{10}-\mathcal{T}_{11}+\mathcal{T}_{12}$ }\\
\hline
$\mathbf{R}_{12}=\bar{6},\mathbf{R}_{34}=3,\mathbf{R}_{56}=8$ & \makecell{ $-\mathcal{T}_4+\mathcal{T}_5+\mathcal{T}_6 -\mathcal{T}_{13}$, $4\mathcal{T}_1-4\mathcal{T}_2+\mathcal{T}_7 +\frac{2}{3}\mathcal{T}_8 -\frac{2}{3}\mathcal{T}_9-\mathcal{T}_{10} +\mathcal{T}_{11}-\mathcal{T}_{12}$ }\\
\hline
$\mathbf{R}_{12}=\bar{3},\mathbf{R}_{34}=6,\mathbf{R}_{56}=8$ & \makecell{ $-\mathcal{T}_4-\mathcal{T}_5+\mathcal{T}_6 +\mathcal{T}_{13}$, $4\mathcal{T}_2-4\mathcal{T}_3-\mathcal{T}_7 -\frac{2}{3}\mathcal{T}_8 +\frac{2}{3}\mathcal{T}_9+\mathcal{T}_{10} -\mathcal{T}_{11}+\mathcal{T}_{12}$ }\\
\hline
$\mathbf{R}_{12}=\bar{3},\mathbf{R}_{34}=3,\mathbf{R}_{56}=8$ & \makecell{ $\mathcal{T}_4-\mathcal{T}_5+\mathcal{T}_6 -\mathcal{T}_{13}$, $\mathcal{T}_7+\frac{2}{3}\mathcal{T}_8 -\frac{2}{3}\mathcal{T}_9-\mathcal{T}_{10} +\mathcal{T}_{11}-\mathcal{T}_{12}$ }
\end{tabular}
\caption{Gauge partial-wave basis of $SU(3)$ particles $\{3,3,\bar{3},\bar{3},8,8\}$ of partition $\{12|34|56\}$ }
\label{tab:g12,34,56}
\end{table}

\subsection{The j-basis Operator and p-basis Conversion}

So far we have obtained the Lorentz j-basis and the gauge j-basis, and they can be respectively expressed as linear combinations of the corresponding Lorentz y-basis and gauge m-basis (y-basis is equivalent with m-basis for the following example) with the conversion matrices ${\cal K}^{jy}_{\cal B}$ and ${\cal K}^{jm}_{G}$, which can be for example read from  table.~\ref{tab:12,34,56.} and table.~\ref{tab:g12,34,56}. Combining the Lorentz and gauge structures and contracting the corresponding gauge indices, one can obtain the operator y-basis and j-basis defined as follows:
\begin{eqnarray}
    \mc{O}^{y}_{(l,n)} &\equiv& {\cal T}^m_l{\cal B}^{y}_n\\
    \mc{O}^j_{(k,m)}&\equiv& {\cal K}^{jm}_{G,kl}{\cal T}^m_l{\cal K}^{jy}_{{\cal B},qn}{\cal B}^{y}_n
    =\mc{K}^{jy}_{(k,q),(l,n)}\mc{O}^{y}_{(l,n)}.
\end{eqnarray}
In the following we will collectively label the tuple $(k,m)$ and $(l,n)$ as single indices, such that $K^{jy}$, the conversion matrix between operator j-basis and y-basis appears as a matrix from an outer product of the two matrices ${\cal K}^{jy}_{G}$ and ${\cal  K}^{jy}_{{\cal B}}$.

Notice that both the j-basis and the y-basis operators are the so-called \emph{flavor-blind operators}, which assumes that all the fields and particles are distinguishable. For repeated fields within an operator, they are supposed to be distinguished by (spurious) \emph{flavor indices}, even if there is no flavor degrees of freedom. 
Partial waves are flavor blind because we are specifying the particles by labels $1,2,\dots$. For physical amplitudes where identical particles are present, permutation symmetries are imposed by quantum statistics, and partial waves alone are usually not valid physical amplitudes. For fields or particles with only one flavor generation, such as gauge bosons and the Higgs, the constraints from the amplitude directly translates into constraints on operators, i.e. the permutation symmetry among the flavor indices for these fields in the operator should be totally symmetric if they are bosons, or anti-symmetric if they are fermions. 
However, for the fields with flavor numbers larger than one, the constraints from the amplitude point of view translates to the constraints in Lagraingian terms when contracting all the flavor indices in the Wilson coefficients and those in operators, which can be further viewed as flavor relations among the Wilson coefficients or equivalently among the flavor tensor components of the operators. 
As discussed in Ref.~\cite{Li:2020gnx,Li:2020xlh}, it is better to organize the flavor-blind operators into p-basis where they have definite permutation symmetries among the flavor indices of repeated fields to identify their independent components. Since p-basis operators as also flavor blind operators, there must be a full-rank conversion matrix $\mc{K}_{py}$ relating them to the y-basis operators, such that: $\mc{O}^p= \mc{K}^{py}\cdot \mc{O}^y$.
In this sense, it is necessary to find the relation between the j-basis and p-basis operators, which naturally introduces the conversion matrix between j-basis and p-basis as $\mc{K}^{jp}$, such that the change between j-basis and p-basis reads:
\begin{equation}
    \mc{O}^j=\mc{K}^{jy}. \mc{O}^{y}=\mc{K}^{jy}.(\mc{K}^{py})^{-1}. \mc{O}^{p} \equiv \mc{K}^{jp}. \mc{O}^{p}.
\end{equation}

To illustrate the conversion between j-basis and p-basis operators, we take the example operator type ${d_C}^2{Q^\dagger}^4$ --- with flavor and gauge indices explicitly written out as ${d_C}_p^a {d_C}_r^b {Q^\dagger}_s^{ci} {Q^\dagger}_t^{dj} {Q^\dagger}_u^{ek} {Q^\dagger}_v^{fl}$, where $p,r,s,t,u,v$ denote flavor indices of fields, $a,b,c,d,e,f$ denote their $SU(3)$ gauge indices, and $i,j,k,l$ denote $SU(2)$ gauge indices of four quark doublets.  The $SU(2)$ and the $SU(3)$ gauge y-bases can be obtained by the generalized L-R rule in the previous subsection:
\bea
T_{SU(2)}=\begin{pmatrix}
\epsilon_{ik}\epsilon_{jl}\\
\epsilon_{ij}\epsilon_{kl}
\end{pmatrix},\qquad
T_{SU(3)}=\begin{pmatrix}
\epsilon_{ace}\epsilon_{bdf}\\
\epsilon_{abe}\epsilon_{cdf}\\
\epsilon_{abf}\epsilon_{cde}\\
\epsilon_{ade}\epsilon_{bcf}\\
\epsilon_{adf}\epsilon_{bce}
\end{pmatrix},
\eea
which yield the y-basis operator of type ${d_C}^2{Q^\dagger}^4$ as direct product of the gauge and Lorentz y-bases:
\bea
\mathcal{O}^y=\mathcal{M}\otimes T_{SU(2)}\otimes T_{SU(3)}=\begin{pmatrix}
\epsilon_{ace} \epsilon_{bdf} \epsilon_{ik} \epsilon_{jl} ({d_C}_p^a {d_C}_r^b) ({Q^\dagger}_s^{ci} {Q^\dagger}_t^{dj}) ({Q^\dagger}_u^{ek} {Q^\dagger}_v^{fl}) \\
\epsilon_{ace}\epsilon_{bdf} \epsilon_{ik}\epsilon_{jl} ({d_C}_p^a {d_C}_r^b) ({Q^\dagger}_s^{ci} {Q^\dagger}_u^{ek}) ({Q^\dagger}_t^{dj} {Q^\dagger}_v^{fl})\\
\epsilon_{ace}\epsilon_{bdf}\epsilon_{il}\epsilon_{jk} ({d_C}_p^a {d_C}_r^b) ({Q^\dagger}_s^{ci} {Q^\dagger}_t^{dj}) ({Q^\dagger}_u^{ek} {Q^\dagger}_v^{fl})\\
\epsilon_{ace}\epsilon_{bdf}\epsilon_{il}\epsilon_{jk} ({d_C}_p^a {d_C}_r^b) ({Q^\dagger}_s^{ci} {Q^\dagger}_u^{ek}) ({Q^\dagger}_t^{dj} {Q^\dagger}_v^{fl})\\
\epsilon_{abe}\epsilon_{cdf}\epsilon_{ik}\epsilon_{jl} ({d_C}_p^a {d_C}_r^b) ({Q^\dagger}_s^{ci} {Q^\dagger}_t^{dj}) ({Q^\dagger}_u^{ek} {Q^\dagger}_v^{fl})\\
\epsilon_{abe}\epsilon_{cdf}\epsilon_{ik}\epsilon_{jl} ({d_C}_p^a {d_C}_r^b) ({Q^\dagger}_s^{ci} {Q^\dagger}_u^{ek}) ({Q^\dagger}_t^{dj} {Q^\dagger}_v^{fl})\\
\epsilon_{abe}\epsilon_{cdf}\epsilon_{il}\epsilon_{jk} ({d_C}_p^a {d_C}_r^b) ({Q^\dagger}_s^{ci} {Q^\dagger}_t^{dj}) ({Q^\dagger}_u^{ek} {Q^\dagger}_v^{fl})\\
\epsilon_{abe}\epsilon_{cdf}\epsilon_{il}\epsilon_{jk} ({d_C}_p^a {d_C}_r^b) ({Q^\dagger}_s^{ci} {Q^\dagger}_u^{ek}) ({Q^\dagger}_t^{dj} {Q^\dagger}_v^{fl})\\
\epsilon_{abf}\epsilon_{cde}\epsilon_{ik}\epsilon_{jl} ({d_C}_p^a {d_C}_r^b) ({Q^\dagger}_s^{ci} {Q^\dagger}_t^{dj}) ({Q^\dagger}_u^{ek} {Q^\dagger}_v^{fl})\\
\epsilon_{abf}\epsilon_{cde}\epsilon_{ik}\epsilon_{jl} ({d_C}_p^a {d_C}_r^b) ({Q^\dagger}_s^{ci} {Q^\dagger}_u^{ek}) ({Q^\dagger}_t^{dj} {Q^\dagger}_v^{fl})\\
\epsilon_{abf}\epsilon_{cde}\epsilon_{il}\epsilon_{jk} ({d_C}_p^a {d_C}_r^b) ({Q^\dagger}_s^{ci} {Q^\dagger}_t^{dj}) ({Q^\dagger}_u^{ek} {Q^\dagger}_v^{fl})\\
\epsilon_{abf}\epsilon_{cde}\epsilon_{il}\epsilon_{jk} ({d_C}_p^a {d_C}_r^b) ({Q^\dagger}_s^{ci} {Q^\dagger}_u^{ek}) ({Q^\dagger}_t^{dj} {Q^\dagger}_v^{fl})\\
\epsilon_{ade}\epsilon_{bcf}\epsilon_{ik}\epsilon_{jl} ({d_C}_p^a {d_C}_r^b) ({Q^\dagger}_s^{ci} {Q^\dagger}_t^{dj}) ({Q^\dagger}_u^{ek} {Q^\dagger}_v^{fl})\\
\epsilon_{ade}\epsilon_{bcf}\epsilon_{ik}\epsilon_{jl} ({d_C}_p^a {d_C}_r^b) ({Q^\dagger}_s^{ci} {Q^\dagger}_u^{ek}) ({Q^\dagger}_t^{dj} {Q^\dagger}_v^{fl})\\
\epsilon_{ade}\epsilon_{bcf}\epsilon_{il}\epsilon_{jk} ({d_C}_p^a {d_C}_r^b) ({Q^\dagger}_s^{ci} {Q^\dagger}_t^{dj}) ({Q^\dagger}_u^{ek} {Q^\dagger}_v^{fl})\\
\epsilon_{ade}\epsilon_{bcf}\epsilon_{il}\epsilon_{jk} ({d_C}_p^a {d_C}_r^b) ({Q^\dagger}_s^{ci} {Q^\dagger}_u^{ek}) ({Q^\dagger}_t^{dj} {Q^\dagger}_v^{fl})\\
\epsilon_{adf}\epsilon_{bce}\epsilon_{ik}\epsilon_{jl} ({d_C}_p^a {d_C}_r^b) ({Q^\dagger}_s^{ci} {Q^\dagger}_t^{dj}) ({Q^\dagger}_u^{ek} {Q^\dagger}_v^{fl})\\
\epsilon_{adf}\epsilon_{bce}\epsilon_{ik}\epsilon_{jl} ({d_C}_p^a {d_C}_r^b) ({Q^\dagger}_s^{ci} {Q^\dagger}_u^{ek}) ({Q^\dagger}_t^{dj} {Q^\dagger}_v^{fl})\\
\epsilon_{adf}\epsilon_{bce}\epsilon_{il}\epsilon_{jk} ({d_C}_p^a {d_C}_r^b) ({Q^\dagger}_s^{ci} {Q^\dagger}_t^{dj}) ({Q^\dagger}_u^{ek} {Q^\dagger}_v^{fl})\\
\epsilon_{adf}\epsilon_{bce}\epsilon_{il}\epsilon_{jk} ({d_C}_p^a {d_C}_r^b) ({Q^\dagger}_s^{ci} {Q^\dagger}_u^{ek}) ({Q^\dagger}_t^{dj} {Q^\dagger}_v^{fl})
\end{pmatrix},
\eea
where $p,r,s,t,u,v$ are flavor indices. 

The p-basis operator of this type is obtained by projecting our the y-basis operators with the group algebra projectors for the corresponding irreducible representations of the symmetry group. Here we index the p-basis operators and label the permutation symmetries of the repeated fields of ${d_C}$ and $Q^\dagger$ in terms of Young diagram in the subscripts:
\bea
\mathcal{O}^{p}_1 =\mathcal{O}^{p}_{\tiny\yng(2),\tiny\yng(4)} = \frac{1}{48}\mathcal{Y}[\young(pr)] \mathcal{Y}[\young(stuv)] \mathcal{O}^y_1,\\
\mathcal{O}^{p}_{\{2,...,7\}}=\mathcal{O}^{p}_{\tiny\yng(2),\tiny\yng(2,2)} =\frac{1}{24}\begin{pmatrix}
\mathcal{Y}[\young(pr)] \mathcal{Y}[\young(st,uv)] \mathcal{O}^y_1\\
\mathcal{Y}[\young(pr)] (t\ u)\mathcal{Y}[\young(st,uv)] \mathcal{O}^y_1\\
\mathcal{Y}[\young(pr)] \mathcal{Y}[\young(st,uv)] \mathcal{O}^y_3\\
\mathcal{Y}[\young(pr)] (t\ u)\mathcal{Y}[\young(st,uv)] \mathcal{O}^y_3\\
\mathcal{Y}[\young(pr)] \mathcal{Y}[\young(st,uv)] \mathcal{O}^y_{13}\\
\mathcal{Y}[\young(pr)] (t\ u)\mathcal{Y}[\young(st,uv)] \mathcal{O}^y_{13}
\end{pmatrix},\\
\mathcal{O}^{p}_{8} =\mathcal{O}^{p}_{\tiny\yng(2),\tiny\yng(1,1,1,1)} =\frac{1}{48}\mathcal{Y}[\young(pr)] \mathcal{Y}[\young(s,t,u,v)] \epsilon_{ace} \epsilon_{bdf} \epsilon_{ik} \epsilon_{jl} ({d_C}_p^a {d_C}_r^b) ({Q^\dagger}_s^{ci} {Q^\dagger}_t^{dj}) ({Q^\dagger}_u^{ek} {Q^\dagger}_v^{fl}),\\
\mathcal{O}^{p}_{\{9,...,14\}} =\mathcal{O}^{p}_{\tiny\yng(1,1),\tiny\yng(3,1)} =\frac{1}{16}\begin{pmatrix}
\mathcal{Y}[\young(p,r)] \mathcal{Y}[\young(stu,v)] \mathcal{O}^y_1\\
\mathcal{Y}[\young(p,r)] (u\ v)\mathcal{Y}[\young(stu,v)] \mathcal{O}^y_1\\
\mathcal{Y}[\young(p,r)] (t\ u\ v)\mathcal{Y}[\young(stu,v)] \mathcal{O}^y_1\\
\mathcal{Y}[\young(p,r)] \mathcal{Y}[\young(stu,v)] \mathcal{O}^y_3\\
\mathcal{Y}[\young(p,r)] (u\ v)\mathcal{Y}[\young(stu,v)] \mathcal{O}^y_3\\
\mathcal{Y}[\young(p,r)] (t\ u\ v)\mathcal{Y}[\young(stu,v)] \mathcal{O}^y_3
\end{pmatrix},\\
\mathcal{O}^{p}_{\{15,...,20\}} =\mathcal{O}^{p}_{\tiny\yng(1,1),\tiny\yng(2,1,1)} =\frac{1}{16}\begin{pmatrix}
\mathcal{Y}[\young(p,r)] \mathcal{Y}[\young(st,u,v)] \mathcal{O}^y_1\\
\mathcal{Y}[\young(p,r)] (t\ u)\mathcal{Y}[\young(st,u,v)] \mathcal{O}^y_1\\
\mathcal{Y}[\young(p,r)] (t\ v\ u)\mathcal{Y}[\young(st,u,v)] \mathcal{O}^y_1\\
\mathcal{Y}[\young(p,r)] \mathcal{Y}[\young(st,u,v)] \mathcal{O}^y_3\\
\mathcal{Y}[\young(p,r)] (t\ u)\mathcal{Y}[\young(st,u,v)] \mathcal{O}^y_3\\
\mathcal{Y}[\young(p,r)] (t\ v\ u)\mathcal{Y}[\young(st,u,v)] \mathcal{O}^y_3
\end{pmatrix}, \label{eq:perm211}
\eea
where $\cal Y$ is the Young symmetrizer, and flavor indices in the parenthesis represent the permutation of flavor indices in the cycle notation.
Among the five classes of p-basis operators,  the one with totally anti-symmetrizing of the flavor indices of $Q^\dagger$, $\mathcal{O}^{p}_{\tiny\yng(2),\tiny\yng(1,1,1,1)}$, vanishes if the flavor generation of the quark doublet is less than four. 

Since j-basis and p-basis  are all flavor-blind operators, as representations of the symmetric group, one can derive the convertion matrix between the two using $\mc{K}^{jp}=\mc{K}^{jy}(\mc{K}^{py})^{-1}$,
\bea\label{eq:Kjpexample}
\tiny
\mc{K}^{jp}=\left(
\begin{array}{cccccccccccccccccccc}
 -12 & 3 & 18 & -6 & -12 & -6 & -12 & 6 & 0 & 0 & 0 & 0 & 0 & 0 & 0 & 0 & 0 & 0 & 0 & 0 \\
 6 & 3 & 0 & 0 & 0 & 0 & 0 & 0 & 0 & 0 & 0 & 0 & 0 & 0 & 0 & 0 & 0 & 0 & 0 & 0 \\
 6 & 1 & -4 & -6 & 0 & 4 & 8 & 4 & 0 & 0 & 0 & 0 & 0 & 0 & 0 & 0 & 0 & 0 & 0 & 0 \\
 0 & 1 & 2 & 0 & 0 & -2 & -4 & -2 & 0 & 0 & 0 & 0 & 0 & 0 & 0 & 0 & 0 & 0 & 0 & 0 \\
 0 & 0 & 0 & 0 & 0 & 0 & 0 & 0 & -14 & 6 & -8 & -2 & -6 & -8 & 12 & -8 & -8 & -12 & -8 & -8 \\
 0 & 0 & 0 & 0 & 0 & 0 & 0 & 0 & 2 & 2 & 4 & 2 & 2 & 4 & 0 & 4 & 4 & 0 & 4 & 4 \\
 0 & 0 & 0 & 0 & 0 & 0 & 0 & 0 & 6 & -2 & 4 & -2 & 6 & 4 & 8 & 4 & 4 & -8 & -12 & -12 \\
 0 & 0 & 0 & 0 & 0 & 0 & 0 & 0 & -2 & 2 & 0 & 2 & -2 & 0 & -4 & 0 & 0 & 4 & 0 & 0 \\
 -6 & -1 & 4 & -4 & -8 & 6 & 0 & -4 & 0 & 0 & 0 & 0 & 0 & 0 & 0 & 0 & 0 & 0 & 0 & 0 \\
 0 & -1 & -2 & 2 & 4 & 0 & 0 & 2 & 0 & 0 & 0 & 0 & 0 & 0 & 0 & 0 & 0 & 0 & 0 & 0 \\
 -4 & 5 & -2 & -4 & 0 & -4 & 0 & 2 & 0 & 0 & 0 & 0 & 0 & 0 & 0 & 0 & 0 & 0 & 0 & 0 \\
 2 & -3 & 0 & 2 & 0 & 2 & 0 & 0 & 0 & 0 & 0 & 0 & 0 & 0 & 0 & 0 & 0 & 0 & 0 & 0 \\
 0 & 0 & 0 & 0 & 0 & 0 & 0 & 0 & 7 & 3 & 10 & 1 & -3 & -2 & 4 & -6 & -6 & 4 & 6 & 6 \\
 0 & 0 & 0 & 0 & 0 & 0 & 0 & 0 & -1 & 1 & 0 & -1 & 1 & 0 & -2 & 0 & 0 & -2 & 0 & 0 \\
 0 & 0 & 0 & 0 & 0 & 0 & 0 & 0 & -3 & -1 & -4 & 1 & 3 & 4 & -2 & -4 & -4 & 6 & 4 & 4 \\
 0 & 0 & 0 & 0 & 0 & 0 & 0 & 0 & 1 & 1 & 2 & -1 & -1 & -2 & 0 & 2 & 2 & 0 & -2 & -2 \\
 0 & 0 & 0 & 0 & 0 & 0 & 0 & 0 & 4 & 4 & 0 & -8 & -8 & 0 & 8 & 8 & -8 & -16 & -16 & 16 \\
 0 & 0 & 0 & 0 & 0 & 0 & 0 & 0 & 0 & 0 & 0 & 0 & 0 & 0 & -4 & -4 & 4 & 8 & 8 & -8 \\
 0 & 0 & 0 & 0 & 0 & 0 & 0 & 0 & -8 & -8 & 0 & 0 & 0 & 0 & -4 & -4 & 4 & 0 & 0 & 0 \\
 0 & 0 & 0 & 0 & 0 & 0 & 0 & 0 & 4 & 4 & 0 & 0 & 0 & 0 & 0 & 0 & 0 & 0 & 0 & 0 \\
\end{array}
\right).
\eea

In the mean time, group theory tells us that when viewing the flavor-blind operators as flavor tensors, the operators reside in the same irreducible representation of the symmetric group will span the same tensor space~\cite{Fonseca:2019yya,Li:2020gnx,Li:2020xlh,Li:2022tec}.  
Thus in terms of flavor-specified operators, which span independent flavor $SU(n_f)$ tensor spaces, the p-basis is reduced to what we called the f-basis~\cite{Li:2022tec}:
\bea
\mathcal{O}^f_{1}&=&\frac{1}{48}\mathcal{Y}[\young(pr)] \mathcal{Y}[\young(stuv)] \mathcal{O}^y_1, \\
\mathcal{O}^f_{2}&=&\frac{1}{24}\mathcal{Y}[\young(pr)] \mathcal{Y}[\young(st,uv)] \mathcal{O}^y_1, \\
\mathcal{O}^f_{3}&=&\frac{1}{24}\mathcal{Y}[\young(pr)] \mathcal{Y}[\young(st,uv)] \mathcal{O}^y_3, \\
\mathcal{O}^f_{4}&=&\frac{1}{24}\mathcal{Y}[\young(pr)] \mathcal{Y}[\young(st,uv)] \mathcal{O}^y_{13}, \\
\mathcal{O}^f_5&=&\frac{1}{16}\mathcal{Y}[\young(p,r)] \mathcal{Y}[\young(stu,v)] \mathcal{O}^y_1, \\
\mathcal{O}^f_{6}&=&\frac{1}{16}\mathcal{Y}[\young(p,r)] \mathcal{Y}[\young(stu,v)] \mathcal{O}^y_3, \\
\mathcal{O}^f_{7}&=&\frac{1}{16}\mathcal{Y}[\young(p,r)] \mathcal{Y}[\young(st,u,v)] \mathcal{O}^y_1, \\
\mathcal{O}^f_{8}&=&\frac{1}{16}\mathcal{Y}[\young(p,r)] \mathcal{Y}[\young(st,u,v)] \mathcal{O}^y_3.
\eea
The extra vectors within the same irreducible representation of the symmetric group in the p-basis can be obtained by permuting the flavor indices of f-basis  operators. For example, the multiplicity of ${\tiny\yng(1,1),\yng(2,1,1)}$ irreducible representations for permuting $d_C$ and $Q^\dagger$ is two, so the p-basis for this particular irreducible representation $\mathcal{O}^{p}_{\tiny\yng(1,1),\yng(2,1,1)}$ in Eq.~(\ref{eq:perm211}) reduces to
\bea
\mathcal{O}^{f}_{\tiny\yng(1,1),\yng(2,1,1)} 
=\begin{pmatrix}
\mathcal{O}^{p}_{\tiny\yng(1,1),\yng(2,1,1),1}\\
\mathcal{O}^{p}_{\tiny\yng(1,1),\yng(2,1,1),4}
\end{pmatrix}
=\frac{1}{16}\begin{pmatrix}
\mathcal{Y}[\young(p,r)] \mathcal{Y}[\young(st,u,v)] \mathcal{O}^y_1\\
\mathcal{Y}[\young(p,r)] \mathcal{Y}[\young(st,u,v)] \mathcal{O}^y_3
\end{pmatrix}. 
\eea
The rest vectors in $\mathcal{O}^{p}_{\tiny\yng(1,1),\yng(2,1,1)}$ can be expressed in terms of  the basis vector in $\mathcal{O}^{f}_{\tiny\yng(1,1),\yng(2,1,1)}$ with permutations of  flavor indices:
\bea
\mathcal{O}^{p}_{\tiny\yng(1,1),\yng(2,1,1),2} =(t\ u)\mathcal{O}^{f}_{\tiny\yng(1,1),\yng(2,1,1),1},\quad
\mathcal{O}^{p}_{\tiny\yng(1,1),\yng(2,1,1),3} =(t\ v\ u)\mathcal{O}^{f}_{\tiny\yng(1,1),\yng(2,1,1),1},\nonumber\\
\mathcal{O}^{p}_{\tiny\yng(1,1),\yng(2,1,1),5} =(t\ u)\mathcal{O}^{f}_{\tiny\yng(1,1),\yng(2,1,1),2},\quad
\mathcal{O}^{p}_{\tiny\yng(1,1),\yng(2,1,1),6} =(t\ v\ u)\mathcal{O}^{f}_{\tiny\yng(1,1),\yng(2,1,1),2}. 
\eea

In the above discussions, we assume that the number of flavors of fermions are all three. The p-basis operator is particularly useful for selecting possible operators if we have additional restrictions on the number of flavors that each fermion field can take. For example, if we take a further step to restrict ${Q^\dagger}_s$ and ${Q^\dagger}_t$ to be the first generation, and ${Q^\dagger}_u$ and ${Q^\dagger}_v$ can be generation 2 or 3, then the f-basis can only take the following independent components
\bea
\mathcal{O}^f_{\tiny\yng(2),\tiny\yng(2)} =\frac{1}{4}\begin{pmatrix}
\mathcal{Y}[\young(pr)] \mathcal{Y}[\young(uv)] \mathcal{O}^y_1\\
\mathcal{Y}[\young(pr)] \mathcal{Y}[\young(uv)] \mathcal{O}^y_2\\
\mathcal{Y}[\young(pr)] \mathcal{Y}[\young(uv)] \mathcal{O}^y_3\\
\mathcal{Y}[\young(pr)] \mathcal{Y}[\young(uv)] \mathcal{O}^y_4
\end{pmatrix},\\
\mathcal{O}^f_{\tiny\yng(1,1),\tiny\yng(2)} =\frac{1}{4}\begin{pmatrix}
\mathcal{Y}[\young(p,r)] \mathcal{Y}[\young(uv)]  \mathcal{O}^y_5\\
\mathcal{Y}[\young(p,r)] \mathcal{Y}[\young(uv)] \mathcal{O}^y_6
\end{pmatrix},\\
\mathcal{O}^f_{\tiny\yng(1,1),\tiny\yng(1,1)} =\frac{1}{4}\begin{pmatrix}
\mathcal{Y}[\young(p,r)] \mathcal{Y}[\young(u,v)] \mathcal{O}^y_1\\
\mathcal{Y}[\young(p,r)] \mathcal{Y}[\young(u,v)] \mathcal{O}^y_2\\
\mathcal{Y}[\young(p,r)] \mathcal{Y}[\young(u,v)] \mathcal{O}^y_3\\
\mathcal{Y}[\young(p,r)] \mathcal{Y}[\young(u,v)] \mathcal{O}^y_4
\end{pmatrix},
\eea
where we omit the flavor indices $s,t$ as they are set to 1, $u$ and $v$ can be chosen from $2$ and $3$, while $p$ and $r$, with no restriction, can still be chosen from $1$ to $3$. 
Since the dimension of f-basis is usually less then j-basis and p-basis, it makes no sense to introduce the conversion matrix between j-basis and f-basis $K^{jf}$. However, with the aforementioned relation between j-basis and f-basis operators, it is still a good practice to express j-basis operators in terms of f-basis operator with possible permutation of the flavor indices, after all the f-basis is our final non-redundant flavor-specified basis. For example, we can expresse the first row of j-basis operators with quantum numbers of partitions as $J=\{0,1,1\},\ \mathbf{R}_{SU(2)}=\{1,3,3\},\ \mathbf{R}_{SU(3)}=\{6,6,6\}$ in Eq.~\eqref{eq:Kjpexample} in terms of our f-basis operators in the following equation:   
\bea
\mathcal{O}^j_1 &=&-12\mathcal{O}^f_{\tiny\yng(2),\yng(4)} +\left(3 +18(t\ u)\right)\circ\mathcal{O}^f_{\tiny\yng(2),\yng(2,2),1}\nonumber\\ &{}&+\left(-6 -12(t\ u)\right)\circ\mathcal{O}^f_{\tiny\yng(2),\yng(2,2),2} +\left(-6 -12(t\ u)\right)\circ\mathcal{O}^f_{\tiny\yng(2),\yng(2,2),3}. 
\eea
Especially when only one generation of each fermion flavor is allowed, all the p-basis operators with Young diagram with more than two rows vanish, which results in $\mathcal{O}^j_1=-12\mathcal{O}^f_{\tiny\yng(2),\yng(4)}$. 
The rest of $\mathcal{O}^j$ spanned on flavor-specified operators $\mathcal{O}^f$ can be presented in a similar way.

\section{UV-IR Correspondence: UV Resonances from j-basis Operators}
\label{sec:bottom}

The previous section discusses the construction of j-basis operators as eigen-basis of the Poincar\'e and gauge Casimirs, and its relation with the real f-basis operators.
In this section we focus on the most straightforward implication of the j-basis operator: each j-basis operator has certain spin and gauge quantum numbers, and this can be interpreted as the UV resonance mediating the scattering amplitude in the designated channel with the same quantum numbers. 
For a type of operators, we obtain the j-bases for all channels amounts to find all the possible UV resonances that generate such operators in the infrared, which is laid out as the \emph{j-basis/UV correspondence}. 
Taking the dimension 5 Weinberg operator as an example, we build the correspondence between the j-basis and seesaw models. 
The j-basis/UV correspondence is also applied to the dimension 7 operators relevant to neutrino masses, for which we obtain the complete UV resonances.

\subsection{The Weinberg Operator: Why only 3 types of Seesaw?}

The existence of neutrino masses provides the first evidence of new physics beyond the standard model. Without introducing any new particle, it was Weinberg who wrote down the first higher dimension operator in the SMEFT: $LLHH$, the so-called Weinberg operator~\cite{Weinberg:1979sa}. It is the leading operator in the SMEFT and arises at dimension-five, suppressed by one power of an inverse mass scale. We know it violates lepton number by two units and generates neutrino masses, thus neutrino masses and oscillations can be explained in terms of the celebrated Weinberg operator.

Nearly at the same time, the first seesaw mechanism~\cite{Yanagida:1979as, Gell-Mann:1979vob} was proposed to generate neutrino masses naturally in the ultraviolet model building way. Since then, three types of seesaw models have been proposed~\cite{Yanagida:1979as, Gell-Mann:1979vob, Mohapatra:1979ia, Magg:1980ut, Schechter:1980gr, Foot:1988aq}, in which each type of seesaw model introduces the following resonances respectively: 
\bea
\textrm{Type-I seesaw}:  & N: (1/2, 1, 1, 0), \\
\textrm{Type-II seesaw}: & \Delta: (0, 1, 3, -1), \\
\textrm{Type-III seesaw}:& \Psi: (1/2, 1, 3, 0),
\eea
where the quantum numbers in the bracket are the spin, $SU(3)_c$ and $SU(2)_L$ representations and $U(1)$ charge. 
Integrating out resonance in the above seesaw model at tree-level, the leading non-renormalizable operator obtained at the electroweak scale is the dimension 5 Weinberg operator. Starting from a seesaw model, integrating out heavy resonance, and matching to the effective operators is the typical way of doing the top-down approach.

Let us ask the following questions. Firstly, can we find other tree-level seesaw mechanism than the above three-type canonical seesaw models? Based on the argument from gauge symmetry, a singlet scalar with hypercharge $-1$ should be allowed, but why it cannot give rise to the Weinberg operator? The type-3/2 seesaw model was proposed in literature, but we would like to ask whether a spin-3/2 or spin-1 UV resonance gives rise to the Weinberg operator? Even more general, can we find out all possible UV resonances contributing the Weinberg operator at the tree-level? All of these questions should be answered when we focus on the bottom-up approach to analyze the Weinberg operator. 

We first lay out how to find the UV resonances for the $LLHH$ operator in the mixed bottom-up and top-down approach:
\bet
\item draw all possible topologies that can generate $LLHH$ operators, which results in one topology and two partitions. 

\item For each vertices involving new physics field in the partition, one can scan over all possible spin (spin-$0,1/2,1,3/2,\cdots$) and gauge quantum numbers of the new physics field such that a corresponding interaction Lagrangian can be constructed obeying gauge and Lorentz symmetries.
For example, when we write $LHF$ and $LLS$, $HHS$ and $LLV$, $HHV$, in which $F, S, V$ are heavy new particles carrying $SU(2)$ gauge quantum number 1 or 3, there is no way to know whether the $LLV$ and  $HHV$ interactions are able to generate $LLHH$ operator. In other words, we do not know whether the spin-3/2 and spin-1 UV resonances are allowed or not.

\item 
To know whether the new heavy particle is allowed or not, one needs to concretely write down the $LLV$ and  $HHV$ interactions and then perform a top-down matching to know whether it would contribute to the Weinberg operator.

\eet
As is mentioned in the introduction, this approach is essentially a mixture of the bottom-up and top-down approach. It depends on the assumption of the BSM-SM interactions and the spin information, and there is no angular momentum selection rule to guarantee that one does not miss or over-count the UV resonances for a certain operator.

On the other hand, the bottom-up approach is taken to determine the complete UV resonances from j-basis construction of the $LLHH$ operator, based on Poincare and gauge symmetries of the effective operator. The first step is the same: draw all possible topologies of the $LLHH$ operators and find out that there are two kinds of independent partitions. Starting from the second step, we would like to utilize the j-basis operators to find the UV resonances.
In the following, we propose that the j-basis operator corresponds to the scattering amplitude mediated by the heavy particles with the same spin and gauge quantum number, and thus the UV resonance of the effective operator is obtained. Finding all the j-basis operators using the Casimir action and express them in terms of f-basis operators would determine all possible UV resonances of the effective operators. 
In the following we will discuss the example of the Weinberg operator to find the correspondence between the seesaw models and the j-basis operators.




Let us start from the y- and p- basis of the  operator type $LLHH$. 
Utilizing the Young Tensor method, we write down the p-basis of the Weinberg operator as
\bea
\mathcal{O}^p_{LLHH} =\begin{pmatrix}
\mathcal{O}^p_{LLHH,1} \\ {\color{gray} \mathcal{O}^p_{LLHH,2}}
\end{pmatrix} =\begin{pmatrix}
\frac{1}{4}\mathcal{Y}[\young(pr)] \mathcal{Y}[\yng(2)_{H}] \epsilon^{ik}\epsilon^{jl}H_kH_l (L_{pi}L_{rj}) \\
{\color{gray} \frac{1}{4}\mathcal{Y}[\young(p,r)] \mathcal{Y}[\yng(1,1)_{H}] \epsilon^{ik}\epsilon^{jl}H_kH_l (L_{pi}L_{rj})}
\end{pmatrix}, 
\eea
where we assume both lepton and Higgs boson can carry flavor indices and can be formally treated as different objects under the permutation group. 
To obtain the j-basis, we need first to identify the scattering channels or equivalently define the partition for this operator. 
The two scattering channels are $\{L_{1,i},L_{2,j}| {H}_{3}^k,{H}_{4}^l\}$, and $\{L_{1,i},H_{3,k}| {L}_{2}^j,{H}_{4}^l\}$\footnote{the j-basis information for the partition $\{L_{1,i},H_{4,k}| L_{2,j},H_{3,l}\}$ is the same as that obtained from $\{L_{1,i},H_{3,k}| L_{2,j},H_{4,l}\}$.}.
Let us take one partition $\{L_{1,i},H_{3,k}| L_{2,j},H_{4,l}\}$, equivalently $\{13|24\}$, as an example for the j-basis analysis. The Lorentz j-basis is obtained from
\begin{eqnarray}
\mathcal{B}^y_{LLHH} = \langle12\rangle ,\quad
W_{\{13\}}^2\mathcal{B}^y = 
	-\frac{3}{4} \mathcal{B}^y, \qquad
\Rightarrow \mathcal{B}^{J=1/2}=\mathcal{B}^y= \langle 12\rangle, 
\end{eqnarray}
while the gauge basis and the corresponding matrix representation for the Casimir operator $\underset{\{1 3\}}{\mathbbm{C}_{2}}$ read
\begin{eqnarray}
\mathcal{T}^m_{LLHH}=\begin{pmatrix}
	\epsilon^{ik}\epsilon^{jl} \\ \epsilon^{ij}\epsilon^{kl}
\end{pmatrix},\quad 
\underset{\{1 3\}}{\mathbbm{C}_{2}}\circ\mathcal{T}^m=(\underset{\{1 3\}}{C_2})^{\rm T}.\mathcal{T}^m=\begin{pmatrix}
	0 & 0 \\ -1 & 2 
\end{pmatrix}
\begin{pmatrix}
	\epsilon^{ik}\epsilon^{jl} \\ \epsilon^{ij}\epsilon^{kl}
\end{pmatrix}. 
\end{eqnarray}
The representation matrix $(\underset{\{1 3\}}{C_2})^{\rm T}$ can be diagonalized with ${\mathcal{K}}^{jm}_{G}$ and yields: 
\begin{eqnarray}
{\mathcal{K}}^{jm}_{G}.(\underset{\{1 3\}}{C_2})^{\rm T}({\mathcal{K}}^{jm}_{G})^{-1} = {\rm diag}\{0,6\}\ \text{with }
{\mathcal{K}}^{jm}_{G} = \begin{pmatrix}
	1 & 0 \\ 1 & -2 
\end{pmatrix}
\\
\Rightarrow \mathcal{T}^{j}= {\mathcal{K}}^{jm}_{G} \mathcal{T}^m =  \begin{cases}
	\epsilon^{ik}\epsilon^{jl} & \mathbf{R}=1\\
	\epsilon^{ik}\epsilon^{jl}-2\epsilon^{ij}\epsilon^{kl} & \mathbf{R}=3
\end{cases}
\label{eq:LLHHjbasis}
\end{eqnarray}
The j-basis operators are obtained as $\mc{O}^j = {\mathcal{K}}^{jm}_{G}\mathcal{T}^m \mathcal{B}^y $. Repeat the above analysis for the partition $\{L_{1,i},L_{2,j}| H_{3,k},H_{4,l}\}$, we obtain the complete j-basis operators for the operator type $LLHH$ as shown in table~\ref{tab:llhh}. 
From the Table, by identifying the j-basis quantum numbers as those of intermediate UV resonances, we discover that there are three types of seesaw. And more than that, from the last column we note that there is one additional UV: scalar charged singlet resonance. 
On the other hand, in the SMEFT, there is only one Higgs doublet, thus the p-basis $\mathcal{O}^p_{LLHH,2}$ with totally anti-symmetric $H$ fields should vanish in the SMEFT. 
Thus in the SMEFT, scalar charged singlet UV resonance is not allowed~\footnote{In the two Higgs doublet model, the type-IV seesaw in which a singlet scalar resonance is assumed can be allowed.}.
In summary, there are only three types of the tree-level seesaw models that can generate dim-5 Weinberg operator.

\begin{table}[htb]
\centering
\begin{tabular}{c|c|c|c}
\hline
Topology & j-basis & Quantum numbers $\{J,\mathbf{R},Y\}$ & Model \\
\hline
\multirow{4}{*}{ \begin{fmffile}{4x}
\begin{fmfgraph*}(30,18)
\fmfpen{thin} \fmfleft{e1,e2} \fmfv{}{g1} \fmfv{}{g2} \fmfright{e4,e3}
\fmf{plain}{e1,g1} \fmf{dashes}{g1,e2} \fmf{plain}{g1,g2} \fmf{dashes}{e4,g2} \fmf{plain}{g2,e3}
\fmflabel{$L$}{e1} \fmflabel{$H$}{e2} \fmflabel{$L$}{e3} \fmflabel{$H$}{e4}
\end{fmfgraph*}
\end{fmffile} \ } &  \multirowcell{2}{$\mathcal{O}^{J=1/2,\mathbf{R}=1}_{\{13\}} =\mathcal{O}^p_{1}+\mathcal{O}^p_{2}$} & \multirowcell{2}{$\{\frac{1}{2},1,0\}$} & \multirowcell{2}{Type I} \\
&& & \\
\cline{2-4}
&\multirowcell{2}{$\mathcal{O}^{J=1/2,\mathbf{R}=3}_{\{13\}} =-\mathcal{O}^p_{1}+3\mathcal{O}^p_{2}$} & \multirowcell{2}{$\{\frac{1}{2},3,0\}$} & \multirowcell{2}{Type III} \\
& && \\
\hline
\multirow{4}{*}{ \begin{fmffile}{4y}
\begin{fmfgraph*}(30,18)
\fmfpen{thin} \fmfleft{e1,e2} \fmfv{}{g1} \fmfv{}{g2} \fmfright{e4,e3}
\fmf{plain}{e1,g1} \fmf{plain}{g1,e2} \fmf{dashes}{g1,g2} \fmf{dashes}{e4,g2} \fmf{dashes}{g2,e3}
\fmflabel{$L$}{e1} \fmflabel{$L$}{e2} \fmflabel{$H$}{e3} \fmflabel{$H$}{e4}
\end{fmfgraph*}
\end{fmffile} \ } & \multirowcell{2}{$\mathcal{O}^{J=0,\mathbf{R}=3}_{\{12\}} =-2\mathcal{O}^p_{1}$} & \multirowcell{2}{$\{0,3,-1\}$} & \multirowcell{2}{Type II} \\
&& & \\
\cline{2-4}
& \multirowcell{2}{$\mathcal{O}^{J=0,\mathbf{R}=1}_{\{12\}} =2\mathcal{O}^p_{2}$} & \multirowcell{2}{$\{0,1,-1\}$} & \multirowcell{2}{N/A} \\
&& & \\
\hline
\end{tabular}
\caption{The complete j-basis for the Weinberg operator $LLHH$ in two different partitions. Here three UV resonances correspond to the three types of the seesaw models. The last one vanishes because of the repeated field in the operator p-basis. }
\label{tab:llhh}
\end{table}

\begin{table}[htb]
\centering
\begin{tabular}{c|c|c}
\hline
Topology & j-basis & Quantum numbers $\{J,\mathbf{R},Y\}$ \\
\hline
\multirow{4}{*}{ \begin{fmffile}{4x}
\begin{fmfgraph*}(30,18)
\fmfpen{thin} \fmfleft{e1,e2} \fmfv{}{g1} \fmfv{}{g2} \fmfright{e4,e3}
\fmf{plain}{e1,g1} \fmf{dashes}{g1,e2} \fmf{plain}{g1,g2} \fmf{dashes}{e4,g2} \fmf{plain}{g2,e3}
\fmflabel{$L$}{e1} \fmflabel{$H$}{e2} \fmflabel{$L$}{e3} \fmflabel{$H$}{e4}
\end{fmfgraph*}
\end{fmffile} } & $\mathcal{O}_{\{13\},1} =3\mathcal{O}^p_1+6\mathcal{O}^p_2 -9\mathcal{O}^p_3-2\mathcal{O}^p_4$ & $\{\frac{3}{2},3,0\}$ \\
\cline{2-3}
& $\mathcal{O}_{\{13\},2} =3\mathcal{O}^p_2 -\mathcal{O}^p_4$ & $\{\frac{1}{2},3,0\}$ \\
\cline{2-3}
& $\mathcal{O}_{\{13\},3} =-3\mathcal{O}^p_1+2\mathcal{O}^p_2 -3\mathcal{O}^p_3+2\mathcal{O}^p_4$ & $\{\frac{3}{2},1,0\}$ \\
\cline{2-3}
& $\mathcal{O}_{\{13\},4} =\mathcal{O}^p_2 +\mathcal{O}^p_4$ & $\{\frac{1}{2},1,0\}$ \\
\hline
\multirow{4}{*}{ \begin{fmffile}{4y}
\begin{fmfgraph*}(30,18)
\fmfpen{thin} \fmfleft{e1,e2} \fmfv{}{g1} \fmfv{}{g2} \fmfright{e4,e3}
\fmf{plain}{e1,g1} \fmf{plain}{g1,e2} \fmf{dashes}{g1,g2} \fmf{dashes}{e4,g2} \fmf{dashes}{g2,e3}
\fmflabel{$L$}{e1} \fmflabel{$L$}{e2} \fmflabel{$H$}{e3} \fmflabel{$H$}{e4}
\end{fmfgraph*}
\end{fmffile} } & $\mathcal{O}_{\{12\},1} =2\mathcal{O}^p_1-4\mathcal{O}^p_4$ & $\{1,3,-1\}$ \\
\cline{2-3}
& $\mathcal{O}_{\{12\}}=-2\mathcal{O}^p_1$ & $\{0,3,-1\}$ \\
\cline{2-3}
& $\mathcal{O}_{\{12\}} =4\mathcal{O}^p_2-2\mathcal{O}^p_3$ & $\{1,1,-1\}$ \\
\cline{2-3}
& $\mathcal{O}_{\{12\}} =2\mathcal{O}^p_3$ & $\{0,1,-1\}$ \ N/A \\
\hline
\end{tabular}
\caption{The complete j-basis for operator $LLHHD^2$ in two different topologies. Among the 7 UV resonances for this operator, three of them are the same as the seesaw model resonances, and the other four carry higher spin quantum numbers. Still the last one vanishes because of the repeated field in the operator p-basis. }
\label{tab:llhhdd}
\end{table}

From the above j-basis and UV resonance correspondence, we note that it is not possible to obtain the Weinberg operator from the tree-level UV resonances with spin-$3/2$ or spin-1.
For the partial waves with four particles involved, such as $\{LL | HH\}$ or $\{LH | LH\}$, to obtain the immediate resonance with higher angular momentum, the operator needs to carry derivatives. 
Thus let us consider the operator $LLHHD^2$ and perform the j-basis analysis. 
Our algorithm yields the p-basis operators of the type $LLHHD^2$ as follows:
\bea
\mathcal{O}^p_{LLHHD^2} =\begin{pmatrix}
\mathcal{O}^p_1\\ \mathcal{O}^p_2\\ {\color{gray}\mathcal{O}^p_3}\\ {\color{gray}\mathcal{O}^p_4}
\end{pmatrix} =\begin{pmatrix}
\frac{1}{4}\mathcal{Y}[\young(pr)] \mathcal{Y}[\yng(2)_{H}] \epsilon^{ik}\epsilon^{jl}(D_\mu H_k)(D^\mu H_l) (L_{pi}L_{rj}) \\
\frac{1}{4}\mathcal{Y}[\young(pr)] \mathcal{Y}[\yng(2)_{H}] i\epsilon^{ik}\epsilon^{jl} (D^\mu H_k)(D^\nu H_l)(L_{pi}\sigma_{\mu\nu}L_{rj})\\
\color{gray}\frac{1}{4}\mathcal{Y}[\young(p,r)] \mathcal{Y}[\yng(1,1)_{H}] \epsilon^{ik}\epsilon^{jl}(D_\mu H_k)(D^\mu H_l) (L_{pi}L_{rj}) \\
\color{gray}\frac{1}{4}\mathcal{Y}[\young(p,r)] \mathcal{Y}[\yng(1,1)_{H}] i\epsilon^{ik}\epsilon^{jl} (D^\mu H_k)(D^\nu H_l)(L_{pi}\sigma_{\mu\nu}L_{rj})
\end{pmatrix}.
\eea
Considering that Young symmetrizer in $\mathcal{O}^p_3$ and $\mathcal{O}^p_4$  force Higgs fields to be anti-symmetric and that Higgs field only have one generation, both of the operators vanish, and that is why we put them gray.
This type of operators could give rise to the neutrino masses via the radiative processes at one-loop order.
We still take the same channel $\{L_{1,i},H_{3,k}| L_{2,j},H_{4,l}\}$ as an example for the j-basis analysis. 
The Lorentz y-basis and j-basis and their relation are given as follows
\begin{eqnarray}
\mathcal{B}^y_{\psi^2\phi^2D^2} =\begin{pmatrix}
s_{34} \langle12\rangle\\ [34]\langle13\rangle\langle24\rangle
\end{pmatrix},\quad
W_{\{13\}}^2\mathcal{B}^y =s_{13}\begin{pmatrix}
	-\frac{15}{4} & 2 \\ 0 & -\frac{3}{4}
\end{pmatrix}\mathcal{B}^y, \quad {\mathcal{K}}^{jy}_{\cal B} = \begin{pmatrix}
	3 & 2 \\ 0 & 1
\end{pmatrix}\\
\Rightarrow \mathcal{B}^j= {\mathcal{K}}^{jy}_{\cal B} \mathcal{B}^y =  \begin{cases}
	3 s_{34} \langle 12\rangle +2 [34] \langle 13\rangle  \langle 24\rangle & J=\frac{3}{2}\\
\langle 13\rangle\langle 24\rangle & J=\frac{1}{2}
\end{cases}.
\end{eqnarray}
The gauge j-basis is the same as the one in Eq.~\eqref{eq:LLHHjbasis}, thus
the j-basis operator is obtained as $\mc{O}^j = {\mathcal{K}}^{jy}_{\cal B}\mathcal{B}^y_{\psi^2\phi^2D^2} {\mathcal{K}}^{jm}_{G}  \mathcal{T}^m_{LLHH}$.
Repeat the above analysis for the $L_{1,i},L_{2,j} \rightarrow H_{3,k},H_{4,l}$, we obtain the complete j-basis operators for the operator type $LLHHD^2$, as is shown in table~\ref{tab:llhhdd}. 
The above j-basis analysis tells us that the same three resonances as the seesaw models exist, but there are additional higher-spin resonances with the spin-$3/2$ or spin-$1$ quantum number which could give rise to the $LLHHD^2$ operator through tree level processes. Note that although the $SU(2)$ singlet scalar is still not allowed, the $SU(2)$ singlet vector boson could induce the $LLHHD^2$ operator.

\subsection{J-basis and UV Resonance Correspondence}\label{sec:corandtopo}

From the above example, we find that the j-basis operators are strong indicators of UV resonance structure. The reason is simple: by definition or Eq.~\eqref{eq:oper_sel}, the j-basis has definite quantum numbers -- angular momentum, isospin, etc. -- in a given scattering channel. Hence if a UV resonance contributes in this scattering channel, the quantum numbers of the state is determined by the resonance, which further determines the operator that is generated in the infrared to be the corresponding j-basis operator. 
Thus there is a correspondence between the j-basis and the scattering amplitude mediated by the heavy resonance in the same channel
\footnote{Of course, this is only a possible source of effective operators. The heavy particles could run in loops, indicating multi-particle intermediate states on the unitarity cut, which result in superposition of j-basis operators. But these are considered subdominant when the tree-level generation is available. }. 
%
In other words, among the different possible bases of operators, the j-basis is a preferred basis when we integrate out heavy particles, and should be the prioritized target of phenomenological searches.
%

This correspondence provides a bottom-up approach to the UV origins of the EFT. Traditionally, one needs to write down a UV Lagrangian or simplified model with heavy particles, and integrate them out to match with operators in the EFT. In such a way, even the EFT motivated simplified models have to be computed one by one, so it is possible to miss some UV resonances and furthermore, it is quite tedious to perform the matching procedure. From the bottom-up viewpoint, we want to construct all the possible UV origins of a given effective operator at the tree-level. Now we can perform a j-basis analysis and immediately list all the possible UV resonances that could contribute to the operator. The analysis consists of three steps:
\begin{enumerate}
    \item Draw all the topologies of UV tree amplitudes that reduce to the local amplitude basis of the same type of the operator. All the internal lines are therefore heavy particles.
    
    \item Obtain the j-basis operators for the given tree topology, as is described in section~\ref{sec:j-basis}. 
    
    \item Note that it is not a sufficient condition for a j-basis operator to be generated from the resonance with the same quantum numbers. 
    It is due to the restriction from the minimal mass dimensions of the implied UV couplings, which rules out some of the low-dimensional j-basis operators.
    This process could be summarized as \emph{dimension selection} (Dim selection).
    
\end{enumerate}

\begin{figure}[h]
\centering
\includegraphics[width=0.8\textwidth]{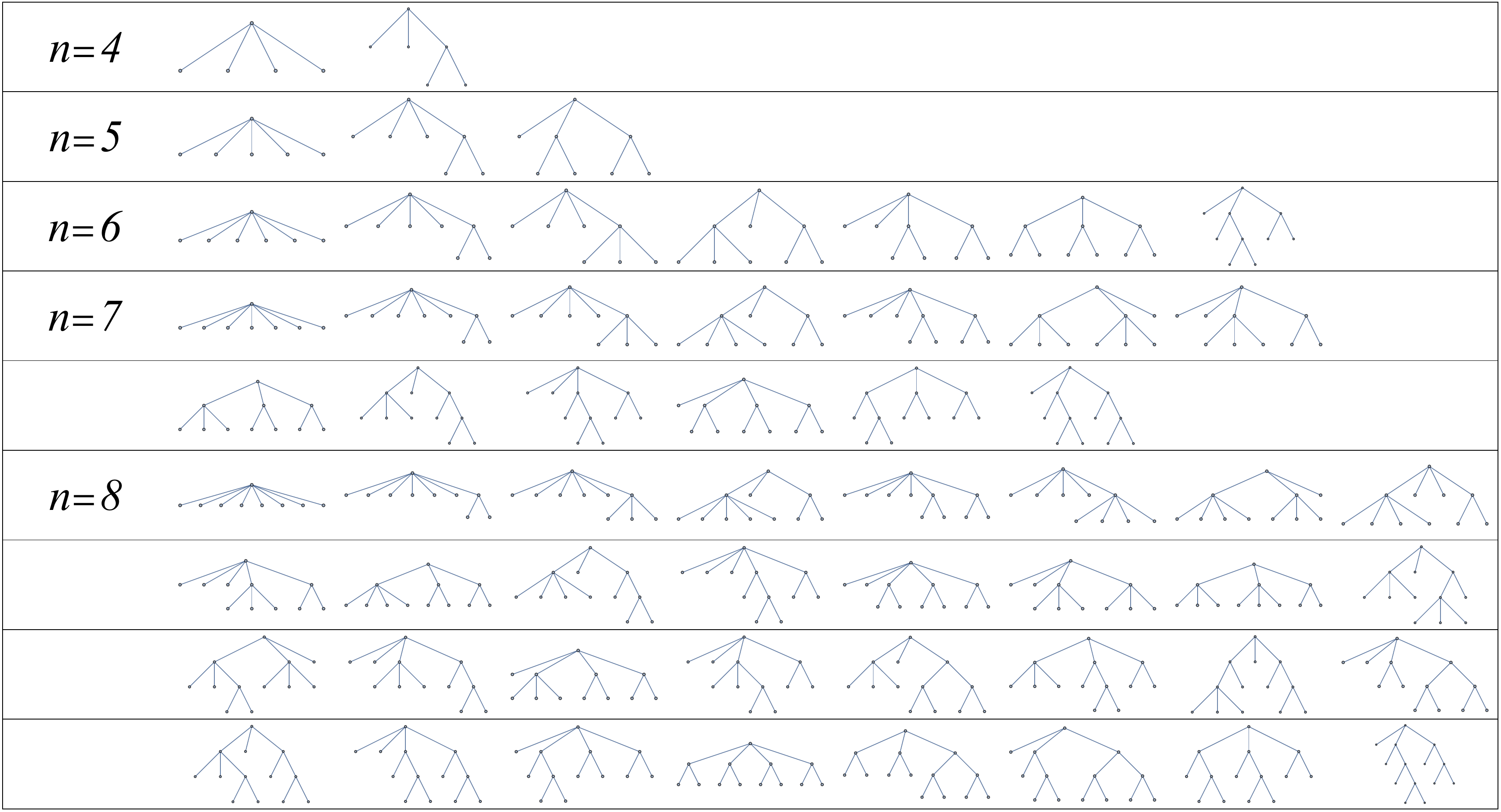}
\caption{The tree level topologies up to 8 external particles obtaind by the \texttt{qgraf-3.5.0}~\cite{Nogueira:1991ex} package.}\label{fig:treetopo}
\end{figure}

\underline{Topology generation}: As j-basis is defined for certain partition of external particles, it actually corresponds to a particular tree diagram, where internal lines represent resonances with quantum numbers given along with the j-basis operator. We summarize the tree topologies organized by the number of external particles, which starts from 4 for the only type $F^3$.
Given number of external particles (no need to know the type of particle), it is possible to find all possible topologies. 
Here we use the package \texttt{qgraf-3.5.0}~\cite{Nogueira:1991ex} to generate all possible topologies. To fullfill this purpose, we construct a qgraf model of a single scalar field $\phi$ with contact interaction involving $3,4,\dots,n$ number of $\phi$, then the independent topologies with $n$ external particles can be generated with process of $n$ in-coming $\phi$'s and $0$ out-going $\phi$. Here we list all the possible topologies generated by qgraf up to 8 external particles in figure.~\ref{fig:treetopo}.



\begin{figure}[h]
    \centering
    \includegraphics[scale=0.7]{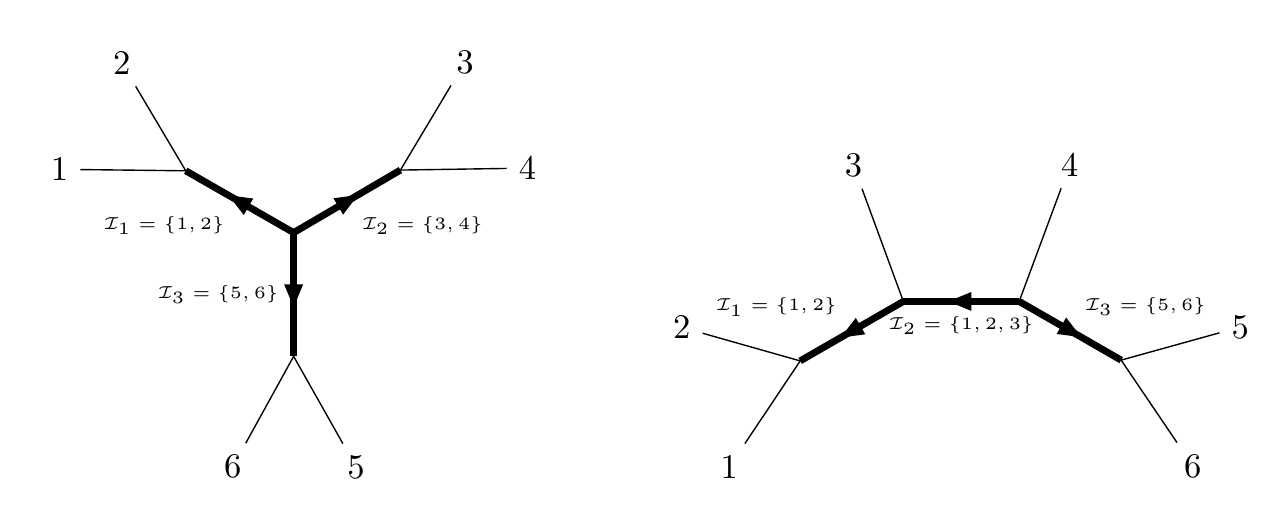}
    \caption{Two distinct 3-partitions of 6-point amplitudes specified by three subsets $\{\mc{I}_1,\mc{I}_2,\mc{I}_3\}$. The arrows suggest a clarifying way to associate each subset with an internal line that potentially represents a resonance.}\label{fig:part}
\end{figure}

Further, when the topology is fixed, one can assign different particles for each external line to form different diagrams which are what we called partitions. Thus a partition contains not only the information of the topology of the diagram but also the detailed information of which group of external particles shares the same interaction vertex in a tree level diagram. As we have already shown in the previous work, a partition corresponds to a set of subsets, where subsets satisfy the following two conditions: 1) each subset and its complement contains at least two particles; 2) for any two subsets, either one contains another or they do not intersect. Such conditions guarantee that the physical observables of different subsets commute with each other. Therefore each subset in the partition corresponds to an internal line of heavy fields. 
To illustrate the correspondence between diagrams and partitions, we show in figure.~\ref{fig:part} two possible partitions for different partitions of 6-point amplitudes. The left diagram corresponds to the partition $\{12|34|56\}$, and the right diagram corresponds to  $\{12|123|56\}$. Once fixing the partition, we have a well-defined channel and are able to discuss the j-basis of such a partition.

\underline{Dim selection}: We know that integrating out a resonance usually results in a tower of operators at different dimensions due to the expansion of the massive propagator
\eq{\label{eq:prop_expand}
    - \frac{1}{p^2-M^2} \sim \frac{1}{M^2}\mc{O}^{(d_{\rm min})} + \frac{1}{M^4}\mc{O}^{(d_{\rm min}+2)} + \frac{1}{M^6}\mc{O}^{(d_{\rm min}+4)} + \dots,
}
all of which should belong to the j-basis of the same quantum numbers at their dimensions, c.f. section~\ref{subsubsec:ampop}. However, $d_{\rm min}$ may not be the minimum dimension that such j-basis first appears. An example is the four-fermion operator with angular momentum $J=1$ in the channel $\{12|34\}$:
\eq{
    \mc{O}_{\psi^4}^{(J=1)} \sim (\psi_1 \sigma^{\mu\nu} \psi_2)D^{2k}(\psi_3 \sigma_{\mu\nu} \psi_4).
}
It obviously appears first at dimension 6 when $k=0$. However, integrating out a heavy vector particle mediating this scattering results in $d_{\rm min} = 8$. Therefore the dimension 6 $J=1$ operator could not be generated by a tree-level resonance. There are several ways to understand this, but the easiest one is to realize that the couplings required in the UV tree-level diagram are high dimensional by themselves. In this case, the UV coupling involved is among two same helicity fermions and a massive vector, described by a new physics term
\eq{
    g(\psi_1\sigma^{\mu\nu}\psi_2)V_{\mu\nu} \equiv g\mc{O}^{\rm NP} \subset \mc{L}^{\rm NP}
}
where $V_{\mu\nu} = \partial_\mu V_\nu - \partial_\nu V_\mu$ for the vector field $V_\mu$. The operator also corresponds to a local amplitude in terms of the massive spinors introduced in Ref.\cite{Arkani-Hamed:2017jhn}
\eq{\label{eq:uv_cpl_example}
    \mc{B}(\psi_1,\psi_2,V_3) = \vev{1{\bf3}}\vev{2{\bf3}} \quad \Leftrightarrow\quad \mc{O}^{\rm NP}.
}
The mass dimension of the coupling $[g]$ can be found by reading the mass dimension of $\mc{B}$, without referring to the explicit form of the operator $\mc{O}^{\rm NP}$
\eq{
    [g] = 4 - [\mc{O}] = 4 - N - [\mc{B}],
}
with $N$ the number of particles. In this case $[\mc{B}] = 2$ as the number of spinor brackets, hence we have $[g]=-1$. Therefore, the resulting effective operator in the IR has a Wilson coefficient proportional to $g^2/M^2$, indicating a minimum dimension $d_{\rm min}=8$, forbidding the generation of the dimension 6 operator $(\psi_1 \sigma^{\mu\nu} \psi_2)(\psi_3 \sigma_{\mu\nu} \psi_4)$ for $J=1$.

There are several problems to solve in this procedure. First of all, how do we get the form of $\mc{B}$? While the three-particle amplitudes are thoroughly studied in Ref.\cite{Arkani-Hamed:2017jhn}, and the four-point amplitudes also studied in Ref.\cite{Durieux:2020gip}, we need a more general treatment for arbitrary number of massless and massive particles. However, we do not need an independent basis of amplitudes as the above literature aim for. It is much more straightforward to only solve for a minimum mass dimension of $\mc{B}$ by solving the linear constraints on the non-negative powers of the spinor brackets. The constraints come from the little group representations of the particles, the helicities $h_i$ for massless particles and the spins $s_i$ for massive particles, which requires $\tilde{n}_i - n_i = 2h_i$ and $\tilde{n}_i + n_i = 2s_i$ where $(\tilde{n}_i,n_i)$ are the numbers of $|i]$ and $\ket{i}$ respectively. The example of Eq.~\eqref{eq:uv_cpl_example} involves the following constraints for the general form of amplitude $\vev{12}^{b_{12}}\vev{1{\bf 3}}^{a_1}\vev{2{\bf 3}}^{a_2}[1{\bf 3}]^{s_1}[2{\bf 3}]^{s_2}$\footnote{$b_{12}$ could be negative, indicating $-b_{12}$ power of $[12]$. The same convention applies for all the spinor brackets between massless particles.}
\eq{
    & \tilde{n}_1 - n_1 = s_1-a_1-b_{12} = 2h_1 = -1, \\
    & \tilde{n}_2 - n_2 = s_2-a_2-b_{12} = 2h_2 = -1, \\
    & \tilde{n}_3 + n_3 = s_1+s_2+a_1+a_2 = 2s_3 = 2, \qquad s_i,a_i \ge 0.
}
One could easily get 4 solutions\footnote{In Ref.\cite{Arkani-Hamed:2017jhn}, general formula is given for three-particle amplitudes under the chiral basis, which is $\mc{B}_1$ in Eq.~\eqref{eq:sol1}. However the minimum dimension does not necessarily appear under chiral basis. For example if the fermion helicities are positive, the amplitude under chiral basis would be higher dimensional, while anti-chiral basis provides minimum dimensional amplitudes. Moreover, some minimal dimensional amplitudes appear under non-chiral basis, which are not given directly by the general formula.}
\eq{\label{eq:sol1}
    &\mc{B}_1 = \vev{1{\bf 3}}\vev{2{\bf 3}}, \\
    &\mc{B}_2 = \vev{12}\vev{1{\bf 3}}[1{\bf 3}], \\
    &\mc{B}_3 = \vev{12}\vev{2{\bf 3}}[2{\bf 3}], \\
    &\mc{B}_4 = \vev{12}^2[1{\bf 3}][2{\bf 3}].
}
Again, these amplitudes may not be independent, which does not matter, as we only need to read off the minimum dimension of them. To do that, there is a small subtlety to tackle in advance, which was pointed out in \cite{Durieux:2020gip} as well. There are different spinor bases specified by the tuple $(\tilde{n}_i,n_i)$ for massive particles in an amplitude, among which $(0,2s_i)$ is called the chiral basis and $(2s_i,0)$ the anti-chiral basis, which are equivalent by the Dirac equation $p_i\ket{\bf i} = m_i |{\bf i}]$. But they are preferred in different cases in regard of minimum mass dimension. In particular, when we have massive particles with $s\ge 1$, the non-chiral basis with $\tilde{n}_i\neq0$ and $n_i\neq0$, or the \emph{non-minimal} helicity category as denoted in \cite{Durieux:2020gip}, should be normalized by a factor of $(1/m_i)^x$ to match with the operator, such as
\eq{
\begin{array}{l|rl}
    s=1 & V_\mu & \simeq \epsilon_\mu = \dfrac{[{\bf i}|\sigma_\mu\ket{\bf i}}{\sqrt{2}m_V^{x=1}}, \\
    \hline
    s=2 & h_{\mu\nu} & \simeq \epsilon_\mu \epsilon_\nu = \dfrac{[{\bf i}|\sigma_\mu\ket{\bf i}[{\bf i}|\sigma_\nu\ket{\bf i}}{2m_h^{x=2}}, \\
    & (\partial_\mu h_{\nu\rho})\sigma^{\mu\nu}_{\alpha\beta} & \simeq
    \dfrac{{}_\alpha\ket{{\bf i}}[{\bf i}|\sigma_\rho\ket{\bf i}\bra{\bf i}_\beta}{\sqrt{2}m_h^{x=1}}.
\end{array}}
In principle, the power $x$ depends on the normalization of the fields by their definitions, and conventionally we have $x = \min(\tilde{n}_i,n_i)$. Therefore the solution in Eq.~\eqref{eq:sol1} should be rewritten as
\eq{
\begin{array}{lll}
    \mc{B}    &   [\mc{B}_i] &    \mc{O}^{\rm NP}\\
    \hline
    \vev{1{\bf 3}}\vev{2{\bf 3}} & 2 &  (\psi_1\sigma^{\mu\nu}\psi_2)V_{\mu\nu}\\
    \dfrac{1}{m_3}\vev{12}\vev{1{\bf 3}}[1{\bf 3}] & 2 & (D^\mu\psi_1\psi_2)V_\mu\\
    \dfrac{1}{m_3}\vev{12}\vev{2{\bf 3}}[2{\bf 3}] & 2 & (\psi_1D^\mu\psi_2)V_\mu\\
    \vev{12}^2[1{\bf 3}][2{\bf 3}] & 4 & (D^\mu\psi_1 D^\nu\psi_2)V_{\mu\nu}
\end{array}
}
where the mass dimensions and the corresponding operators are also listed. The constraints can be worked out for general external states and the minimum mass dimension $[\mc{B}]_{\rm min}$ can be similarly obtained.

Finally, the minimum dimension of the generated IR operators is $d_{\rm min} = 4-[C]_{\rm max}$, where the Wilson coefficient $C$ must be proportional to $(\prod_{i=1}^{N_r+1} g_i)/(\prod_{r=1}^{N_r} M_r^2)$, where $N_{r}$ is the number of resonances in the topology.
However, the mass dimension $[C]$ depends on one more factor, which is the possible mass insertions at fermion lines
\eq{
    [C] = \sum_{i=1}^{N_r+1} [g_i] - 2N_r + N_{\rm insert}.
}
While $N_{\rm insert}$ is bounded by the number of fermion propagators, it is also constraint by the helicities of external fermions at the ends of each fermion line, which determines the parity of the number of helicity flips along the line coming from either mass insertions or the vertices. Finally we have
\eq{
    d_{\rm min} = 4 + 2N_r - \sum_{i=1}^{N_r+1} [g_i]_{\rm max} - N_{\rm insert}.
} 







\subsection{Higher Dim Operators: Genuine dim-7 Seesaw}

After identifying the topology and building the j-basis and UV resonance correspondence, let us study more complicated examples on UV tree-level resonances. We still use the seesaw related operators as example, in this case, the dimension 7 seesaw operator $L^2H^3H^\dagger$, involving in more than four particles. This operator contains six particles and 19 partitions are identified, so the UV structure is quite rich. Let us ask whether it is possible to have the tree-level UV resonances which can only induce the dim-7 $L^2H^3H^\dagger$ but not the dim-5 $LLHH$. In another way, we would like to find the genuine dim-7 seesaw models.

Utilizing the Young tensor method, we obtain the p-basis of the $L^2H^3H^\dagger$ operator as follows
\bea
\mathcal{O}^p_{L^2H^3H^{\dagger}} \equiv\begin{pmatrix}
\mathcal{O}^p_1\\ \color{gray} \mathcal{O}^p_2\\ \color{gray} \mathcal{O}^p_3\\ \color{gray} \mathcal{O}^p_4\\ \color{gray} \mathcal{O}^p_5  
\end{pmatrix} =\begin{pmatrix}
\frac{1}{12}\mathcal{Y}[\young(pr)] \mathcal{Y}[\yng(3)_H] \epsilon^{ik}\epsilon^{jm} H_kH_lH_m {H^\dagger}^l (L_{pi}L_{rj}) \\
{\color{gray}-\frac{1}{6}\mathcal{Y}[\young(pr)] \mathcal{Y}[\yng(2,1)_H] \epsilon^{jm}\epsilon^{kl} H_kH_lH_m {H^\dagger}^i (L_{pi}L_{rj})}\\
{\color{gray}-\frac{1}{6}\mathcal{Y}[\young(pr)] (2\ 3)_H\mathcal{Y}[\yng(2,1)_H] \epsilon^{jm}\epsilon^{kl} H_kH_lH_m {H^\dagger}^i (L_{pi}L_{rj})}\\
{\color{gray}-\frac{1}{6}\mathcal{Y}[\young(p,r)] \mathcal{Y}[\yng(2,1)_H] \epsilon^{jm}\epsilon^{kl} H_kH_lH_m {H^\dagger}^i (L_{pi}L_{rj})}\\
{\color{gray}-\frac{1}{6}\mathcal{Y}[\young(pr)] (2\ 3)_H\mathcal{Y}[\yng(2,1)_H] \epsilon^{jm}\epsilon^{kl} H_kH_lH_m {H^\dagger}^i (L_{pi}L_{rj})}
\end{pmatrix}.
\eea
From the above, we note that the operator $L^2H^3H^\dagger$ contains 19 kinds of paritions. Applying the j-basis analysis, we obtain the j-basis operator for each topology. In each topology, the j-basis gives rise to several UV resonances as shown in table.~\ref{tab:llhhhh}. Taking the repeated field into account, there are 51 kinds of the UV theories with different interactions between new particles and SM ones, much more than the UV particles listed in Ref.~\cite{Bonnet:2009ej}, where only 34 kinds of the UV particles are listed.

Among all kinds of UV resonances, compared to the dimension-5 UV particles, we find new SU(2) quantum number: the SU(2) quadruplet, and new U(1) quantum numbers: $1/2$, $3/2$ (and $1$ for spin-$1/2$, $0$ for spin-$0$ resonance). If we want to find the UV resonance which can only generate the dimension-7 operators instead of the dimension-5 one, we need to find the resonances in which all quantum numbers cannot be in table~\ref{tab:llhh}. Fortunately we find that there is one topology in which all the UV resonances carry different quantum numbers than the ones for the three types of seesaw, which is marked as the blue color in the table~\ref{tab:llhhhh}, as the genuine dimension-7 seesaw model~\cite{Babu:2009aq}.  Among the three kinds of resonances, in the SMEFT, only one Higgs doublet indicates that only the first kind of the UV resonances survive. Thus the seesaw mechanism including both particles with the quantum numbers $({1/2,3,-1})$ and $({0,4,-3/2})$ provides the genuine dim-7 seesaw in the SMEFT. From the table, in the two Higgs doublet model, due to the flavor symmetry in the Higgs fields, there are two additional genuine dim-7 seesaw models.

\begin{longtable}[H]{c|c|c}
\hline
Topology & j-basis & Quantum numbers $\{J,\mathbf{R},Y\}$ \\
\hline
\multirow{5}{*}{ \begin{fmffile}{61x}
\begin{fmfgraph*}(30,30)
\fmfpen{thin} \fmfsurroundn{e}{6} \fmfvn{}{g}{4} 
\fmf{plain}{e1,g1} \fmf{plain}{g1,e2} \fmf{dashes}{g1,g4} \fmf{dashes}{e3,g2} \fmf{dashes}{g2,e4} \fmf{dashes}{g2,g4} \fmf{dashes}{e5,g3} \fmf{dashes}{g3,g4} \fmf{dashes}{g3,e6}
\fmflabel{$L$}{e1} \fmflabel{$L$}{e2} \fmflabel{$H$}{e3} \fmflabel{$H$}{e4} \fmflabel{$H$}{e5} \fmflabel{$H^\dagger$}{e6} 
\end{fmfgraph*}
\end{fmffile} } \ \ & $\mathcal{O}_{\{12|34|56\},1} =2\mathcal{O}^p_1-4\mathcal{O}^p_2$ & $\{0,3,-1\},\{0,3,1\},\{0,3,0\}$ \\
\cline{2-3}
& $\mathcal{O}_{\{12|34|56\},2} =2\mathcal{O}^p_2+4\mathcal{O}^p_3$ & $\{0,3,-1\},\{0,1,1\},\{0,3,0\}$ \\
\cline{2-3}
& $\mathcal{O}_{\{12|34|56\},3} =12\mathcal{O}^p_4$ & $\{0,1,-1\},\{0,3,1\},\{0,3,0\}$ \\
\cline{2-3}
& $\mathcal{O}_{\{12|34|56\},4} =4\mathcal{O}^p_1+4\mathcal{O}^p_2$ & $\{0,3,-1\},\{0,3,1\},\{0,1,0\}$ \\
\cline{2-3}
& $\mathcal{O}_{\{12|34|56\},5} =2\mathcal{O}^p_4+4\mathcal{O}^p_5$ & $\{0,1,-1\},\{0,1,1\},\{0,1,0\}$ \\
\hline
\multirow{5}{*}{ \begin{fmffile}{61y}
\begin{fmfgraph*}(30,30)
\fmfpen{thin} \fmfsurroundn{e}{6} \fmfvn{}{g}{4} 
\fmf{plain}{e1,g1} \fmf{dashes}{g1,e2} \fmf{plain}{g1,g4} \fmf{plain}{e3,g2} \fmf{dashes}{g2,e4} \fmf{plain}{g2,g4} \fmf{dashes}{e5,g3} \fmf{dashes}{g3,g4} \fmf{dashes}{g3,e6}
\fmflabel{$L$}{e1} \fmflabel{$H$}{e2} \fmflabel{$L$}{e3} \fmflabel{$H$}{e4} \fmflabel{$H$}{e5} \fmflabel{$H^\dagger$}{e6} 
\end{fmfgraph*}
\end{fmffile} } \ \ & $\mathcal{O}_{\{13|24|56\},1} =-\mathcal{O}^p_2-2 \mathcal{O}^p_3+3 \mathcal{O}^p_4$ & $\{\frac{1}{2},3,0\},\{\frac{1}{2},3,0\},\{0,3,0\}$ \\
\cline{2-3}
& $\mathcal{O}_{\{13|24|56\},2}= -\mathcal{O}^p_1+3 \mathcal{O}^p_2+2 \mathcal{O}^p_3+3 \mathcal{O}^p_4$ & $\{\frac{1}{2},3,0\},\{\frac{1}{2},1,0\},\{0,3,0\}$ \\
\end{longtable}

Finally let us mention that for operators involving in six particles and more, the above j-basis analysis should be extended, as is mentioned in the appendix.~\ref{app:jbasis}. This caveat appears when we consider the j-basis for the dim-9 seesaw operator $L^2H^3H^\dagger D^2$.
The difficulty appears in the Lorentz sector. Furthermore, the gauge structure of type $L^2H^3H^\dagger D^2$ is the same as $L^2H^3H^\dagger$, so we will only focus on the Lorentz structure of such operator. The Lorentz y-basis of $L^2H^3H^\dagger D^2$ is written down as
\bea
\mathcal{B}^y_{L^2H^3H^\dagger D^2}\equiv \begin{pmatrix}
\mathcal{B}^y_1 \\ \mathcal{B}^y_2 \\ \mathcal{B}^y_3 \\ \mathcal{B}^y_4 \\ \mathcal{B}^y_5 \\ \mathcal{B}^y_6 \\ \mathcal{B}^y_7 \\ \mathcal{B}^y_8 \\ \mathcal{B}^y_9 \\ \mathcal{B}^y_{10} \\ 
\mathcal{B}^y_{11} \\ \mathcal{B}^y_{12} \\ 
\end{pmatrix} =\begin{pmatrix}
[56] \langle12\rangle \langle56\rangle \\ -[46] \langle12\rangle \langle46\rangle \\ [45] \langle12\rangle \langle45\rangle \\ [36] \langle12\rangle \langle36\rangle \\ -[35] \langle12\rangle \langle35\rangle \\ [34] \langle12\rangle \langle34\rangle \\ [56] \langle15\rangle \langle26\rangle \\ -[46] \langle14\rangle \langle26\rangle \\ [45] \langle14\rangle \langle25\rangle \\ [36] \langle13\rangle \langle26\rangle \\ -[35] \langle13\rangle \langle25\rangle \\ [34] \langle13\rangle \langle24\rangle
\end{pmatrix}. 
\eea
Similar to the above topology analysis, there are 19 topologies in total. Let us only focus on the topology with the caveat, as shown in table~\ref{tab:llhhhh2}. The Lorentz j-basis at channel $\{134\}$ is
\bea
\mathcal{B}^{J=1/2}_{\{134\}} =\begin{pmatrix}
\frac{4}{3}\mathcal{B}^y_{1} \\ \frac{4}{3}\mathcal{B}^y_{6} \\ \frac{4}{3}\mathcal{B}^y_{7} \\ \frac{4}{3}\mathcal{B}^y_{8} \\ \frac{4}{3}\mathcal{B}^y_{9} \\ \frac{4}{3}\mathcal{B}^y_{10} \\ \frac{4}{3}\mathcal{B}^y_{11} \\ \frac{4}{3}\mathcal{B}^y_{12} \\ 
\end{pmatrix}. 
\eea
After performing the j-basis analysis in the appendix, we obtain the Lorentz j-basis in table~\ref{tab:llhhhh2}. 

\begin{longtable}[H]{c|c|c}
\hline
Topology & Lorentz j-basis & Spin $J$ \\
\hline\multirow{8}{*}{\begin{fmffile}{622lll}
\begin{fmfgraph*}(50,30)
\fmfpen{thin} \fmfsurroundn{e}{10} \fmfvn{}{g}{4} 
\fmf{dashes}{e7,g1} \fmf{plain}{g1,e5} \fmf{plain}{g1,g2} \fmf{dashes}{e4,g2} \fmf{phantom}{g2,e8} \fmf{phantom}{g3,e9} \fmf{plain}{g2,g3} \fmf{dashes}{e3,g3} \fmf{plain}{g3,g4} \fmf{dashes}{e2,g4} \fmf{plain}{g4,e10}
\end{fmfgraph*}
\end{fmffile}}  & $\mathcal{B}_{\{13|135|24\},{{1}}}={2 \mathcal{B}^y_{5} -\frac{4}{3}\mathcal{B}^y_{11}  -\frac{2}{3}\mathcal{B}^y_{3} -\frac{2}{3}\mathcal{B}^y_{7} +  \frac{2}{3}\mathcal{B}^y_{9}}$ & $\left\{\frac{3}{2},\frac{1}{2},\frac{1}{2}\right\}$ \\
\cline{2-3} & $\mathcal{B}_{\{13|135|24\},{{2}}}= {-\frac{4}{3}\mathcal{B}^y_{10}  -\frac{4}{3}\mathcal{B}^y_{7}  +  \frac{4}{3}\mathcal{B}^y_{2} +  \frac{8}{3}\mathcal{B}^y_{8}}$ & $\left\{\frac{1}{2},\frac{1}{2},\frac{3}{2}\right\}$ \\
\cline{2-3} & \multirow{6}{*}{$\mathcal{B}_{\{13|135|24\},{{{3}}, {{4}}, {{5}}, {{6}}, {{7}}, {{8}}}}=\begin{pmatrix} -\frac{4}{3}\mathcal{B}^y_{12} \\
 -\frac{4}{3}\mathcal{B}^y_{11}\\
 -\frac{4}{3}\mathcal{B}^y_{10}\\
 -\frac{4}{3}\mathcal{B}^y_{9} +  \frac{4}{3}\mathcal{B}^y_{3}\\
 -\frac{4}{3}\mathcal{B}^y_{8} + \frac{4}{3}\mathcal{B}^y_{2} \\
 -\frac{4}{3}\mathcal{B}^y_{7} \end{pmatrix}$} & \multirow{6}{*}{$\left\{\frac{1}{2},\frac{1}{2},\frac{1}{2}\right\}$} \\
&&\\
&&\\
&&\\
&&\\
&&\\
\hline\multirow{8}{*}{\begin{fmffile}{625lll}
\begin{fmfgraph*}(50,30)
\fmfpen{thin} \fmfsurroundn{e}{10} \fmfvn{}{g}{4} 
\fmf{dashes}{e7,g1} \fmf{dashes}{g1,e5} \fmf{dashes}{g1,g2} \fmf{plain}{e4,g2} \fmf{phantom}{g2,e8} \fmf{phantom}{g3,e9} \fmf{plain}{g2,g3} \fmf{dashes}{e3,g3} \fmf{plain}{g3,g4} \fmf{dashes}{e2,g4} \fmf{plain}{g4,e10}
\end{fmfgraph*}
\end{fmffile}}  & $\mathcal{B}_{\{13|136|45\},{{1}}}={2 \mathcal{B}^y_{4} + -\frac{4}{3}\mathcal{B}^y_{10}  -\frac{2}{3}\mathcal{B}^y_{2}  -\frac{2}{3}\mathcal{B}^y_{7} +  \frac{2}{3}\mathcal{B}^y_{1} +  \frac{2}{3}\mathcal{B}^y_{8}}$ & $\left\{\frac{3}{2},\frac{1}{2},0\right\}$ \\
\cline{2-3} & \multirow{3}{*}{$\mathcal{B}_{\{13|136|45\},{{{2}}, {{3}}, {{4}}}}=\begin{pmatrix}  -\frac{4}{3}\mathcal{B}^y_{11} +  -\frac{4}{3}\mathcal{B}^y_{12}\\
 -\frac{4}{3}\mathcal{B}^y_{9} +  \frac{2}{3}\mathcal{B}^y_{3}\\
 -\frac{4}{3}\mathcal{B}^y_{7}  -\frac{4}{3}\mathcal{B}^y_{8} +  \frac{4}{3}\mathcal{B}^y_{1} + \frac{4}{3}\mathcal{B}^y_{2} \end{pmatrix}$} & \multirow{3}{*}{$\left\{\frac{1}{2},\frac{1}{2},1\right\}$} \\
&&\\
&&\\
\cline{2-3} & \multirow{4}{*}{$\mathcal{B}_{\{13|136|45\},{{{5}}, {{6}}, {{7}}, {{8}}}}=\begin{pmatrix} -\frac{4}{3}\mathcal{B}^y_{12}  +  \frac{4}{3}\mathcal{B}^y_{11}\\
 -\frac{4}{3}\mathcal{B}^y_{10}\\
 -\frac{4}{3}\mathcal{B}^y_{1}  -\frac{4}{3}\mathcal{B}^y_{8} + \frac{4}{3}\mathcal{B}^y_{2}  +  \frac{4}{3}\mathcal{B}^y_{7}\\
-\frac{4}{3}\mathcal{B}^y_{3} \end{pmatrix}$} & \multirow{4}{*}{$\left\{\frac{1}{2},\frac{1}{2},0\right\}$} \\
&&\\
&&\\
&&\\
\hline\multirow{8}{*}{\begin{fmffile}{62lll}
\begin{fmfgraph*}(50,30)
\fmfpen{thin} \fmfsurroundn{e}{10} \fmfvn{}{g}{4} 
\fmf{dashes}{e7,g1} \fmf{dashes}{g1,e5} \fmf{dashes}{g1,g2} \fmf{plain}{e4,g2} \fmf{phantom}{g2,e8} \fmf{phantom}{g3,e9} \fmf{plain}{g2,g3} \fmf{plain}{e3,g3} \fmf{dashes}{g3,g4} \fmf{dashes}{e2,g4} \fmf{dashes}{g4,e10}
\end{fmfgraph*}
\end{fmffile}} \ \  & $\mathcal{B}_{\{34|134|56\},{{1}}}={\mathcal -\frac{4}{3} {B}^y_{10} -\frac{4}{3} \mathcal{B}^y_{11}  -\frac{4}{3}\mathcal{B}^y_{8}   -\frac{4}{3}}\mathcal{B}^y_{9}$ & $\left\{1,\frac{1}{2},1\right\}$ \\
\cline{2-3} & \multirow{2}{*}{$\mathcal{B}_{\{34|134|56\},{{{2}}, {{3}}}}=\begin{pmatrix}  -\frac{4}{3}\mathcal{B}^y_{12} +  \frac{2}{3}\mathcal{B}^y_{6}\\
 -\frac{4}{3}\mathcal{B}^y_{11}   -\frac{4}{3}\mathcal{B}^y_{9} +  \frac{4}{3}\mathcal{B}^y_{10} + \frac{4}{3}\mathcal{B}^y_{8} \end{pmatrix}$} & \multirow{2}{*}{$\left\{1,\frac{1}{2},0\right\}$} \\
&&\\
\cline{2-3} & \multirow{2}{*}{$\mathcal{B}_{\{34|134|56\},{{{4}}, {{5}}}}=\begin{pmatrix} -\frac{4}{3}\mathcal{B}^y_{10}  -\frac{4}{3}\mathcal{B}^y_{11}  +  \frac{4}{3}\mathcal{B}^y_{8} +  \frac{4}{3}\mathcal{B}^y_{9}\\
-\frac{4}{3}\mathcal{B}^y_{7} + \frac{2}{3}\mathcal{B}^y_{1} \end{pmatrix}$} & \multirow{2}{*}{$\left\{0,\frac{1}{2},1\right\}$} \\
&&\\
\cline{2-3} & \multirow{3}{*}{$\mathcal{B}_{\{34|134|56\},{{{6}}, {{7}}, {{8}}}}=\begin{pmatrix}  -\frac{4}{3}\mathcal{B}^y_{11} +  -\frac{4}{3}\mathcal{B}^y_{8} +  \frac{4}{3}\mathcal{B}^y_{10} +  \frac{4}{3}\mathcal{B}^y_{9}\\
 -\frac{4}{3}\mathcal{B}^y_{6}\\
-\frac{4}{3}\mathcal{B}^y_{1} \end{pmatrix}$} & \multirow{3}{*}{$\left\{0,\frac{1}{2},0\right\}$} \\
&&\\
&&\\
\hline
\caption{The j-basis of the type $L^2H^3H^\dagger D^2$ for three different kinds of topologies. Note that the last column only labels the spin of the three internal particles. } \label{tab:llhhhh2}
\end{longtable}

\section{SMEFT Operators: Complete Tree-level UV Resonances}
\label{sec:dim-67}

Having built the j-basis/UV correspondence, in this section we apply the above procedure to the SMEFT operator bases. For the dimension-5 Weinberg operator, the complete UV resonances are well-known: the three types of the seesaw models. In this work, we have proved that there are only three-types of seesaws at tree level. For the dimension-6 operators, the complete tree-level dictionary have been listed in Ref.~\cite{deBlas:2017xtg} assuming the UV particle spin information and exhausting all possible BSM-SM interactions with gauge symmetry selection. Alternatively, we utilize the Casimir method and the j-basis/UV correspondence to write down a complete list of all the tree-level UV resonances, and prove the completeness. For the dimension-7 operators, there was no reference on listing possible UV resonances, although a partial list of the tree-level UV particles generating the neutrino mass from the operator $L^2H^3H^{\dagger}$ has been listed in Ref.~\cite{Bonnet:2009ej}. We list all possible UV resonances of the dimension-7 operators for the first time. For the dimension-8 operators, there are too many UV resonances and here in this article, we only list the UV for the bosonic operators, and present the UV for the fermionic operators in a separate work~\cite{dim8UV}. For the dimension-9 operators, typically listing all the UV resonances is not so interesting, except for some special cases, such as the lepton number violating processes, neutrinoless double beta decay, etc~\cite{Anamiati:2018cuq, Bonnet:2012kh,Gargalionis:2020xvt}. We will present these in future works~\cite{dim9UV}.


\subsection{Effects of Equations of Motion}\label{sec:effEOM}

\hl{
In previous sections, we obtain the angular momentum information for the scattering of certain external particles by analyzing the local on-shell amplitude at a given dimension, and then we interpret such angular momenta as spins of some UV resonances that propagate as intermediate state in a given scattering topology. Such an interpretation is further translated into the operator Language and claims that the operator corresponding to such a local amplitude must be able to be generated by integrating out a heavy field of that spin in the UV theory. 
A subtlety occurs, which may result in missing UV resonance for a given operator if we only take into account the UV resonances obtained by the aforementioned Casimir operator method. The reason is rooted in  our algorithm for the reduction of an arbitrary amplitude onto our y-basis amplitudes, where we set EOM of the external particle to that of an massless free field theory and CDC to zero. 
Let us take an example of converting the $(D_\mu\phi_1)\phi_2\phi_3\phi_4(D^\mu\phi_5)$ to our y-basis operators. Using the amplitude-operator correspondence, such a operator corresponds to a non-standard Young tableau and can be converted to our y-basis as shown in the following equation: 
\eq{
	\begin{array}{lllll}
		\quad \young(21,35,4) 									&= \quad \young(14,25,3) 										&+ \quad \young(12,35,4) 					&- \quad \young(13,25,4) \quad &, \\[1em]
		-\langle 15 \rangle[15] &= \langle 45 \rangle[45] 		&+ \langle 25 \rangle[25]	&+ \langle 35 \rangle[35]& {\color{gray}+\langle 55 \rangle[55]}, \\[1em]
		-D_\mu\phi_1\phi_2\phi_3\phi_4D^\mu\phi_5				&= \phi_1\phi_2\phi_3D_\mu\phi_4D^\mu\phi_5						&+ \phi_1D_\mu\phi_2\phi_3\phi_4D^\mu\phi_5	&+ \phi_1\phi_2D_\mu\phi_3\phi_4D^\mu\phi_5	&{\color{gray}+ \phi_1\phi_2\phi_3\phi_4 D^2\phi_5}.
	\end{array}\label{eq:non-standard}
}
In  Eq.~\eqref{eq:non-standard}, one can find that in the conversion of the amplitude $\langle 15 \rangle[15]$, we use the momentum conservation: $p_1 = p_2+p_3+p_4+p_5$, which is equivalent to the IBP relation in the operator language. However, in the on-shell amplitude language, the last term $\langle 55 \rangle[55]$ vanishes due to the on-shell condition, and the corresponding operator contains $D^2\phi_5$, which in general does not vanish but is converted to operators of other types with the EOM of $\phi_5$. The neglect of the last term does not hurt the completeness and independence in the basis construction, as we set up the convention to always replace such a portion with EOM as long as possible and enumerate the operator basis of each type.  
On the contrary, when enumerating the UV resonance of a specific operator, the effect of last term proportional to $D^2\phi_5$ cannot be neglected, and its replacement by EOM indicates that whatever UV resonances that contribute to $(D_\mu\phi_1)\phi_2\phi_3\phi_4(D^\mu\phi_5)$ actually contribute to operators of other types after conversion to our basis.
For example, when investigating the UV origin of the operator type ${e_{\mathbb{C}}}H{H^\dagger}^2L$, one needs to take into account the UV origins of operator types $e_{\mathbb{C}}e_{\mathbb{C}}^\dagger HH^\dagger D$, $LL^\dagger HH^\dagger D$ and $D^2H^2H^{\dagger 2}$ obtained by the j-basis analysis. Because in the actual integrating out process, they are possible to generate the operators in the type ${e_{\mathbb{C}}}H{H^\dagger}^2L$ by using EOM of $e_{\mathbb{C}}^\dagger$, $L^\dagger$ and $H$ respectively.
On the other hand, our Casimir operator method cannot assure whether the UV origins of operator types $e_{\mathbb{C}}e_{\mathbb{C}}^\dagger HH^\dagger D$, $LL^\dagger HH^\dagger D$ and $D^2H^2H^{\dagger 2}$ really contribute to  operators in the types  ${e_{\mathbb{C}}}H{H^\dagger}^2L$. Therefore, in this sense, we claim that all the UV resonances that can definitely contribute to a specific operator in the type  ${e_{\mathbb{C}}}H{H^\dagger}^2L$ are the UV origins obtained by the j-basis analysis of the type  ${e_{\mathbb{C}}}H{H^\dagger}^2L$ plus a subset of the those for operator types $e_{\mathbb{C}}e_{\mathbb{C}}^\dagger HH^\dagger D$, $LL^\dagger HH^\dagger D$ and $D^2H^2H^{\dagger 2}$. However, similar to the basis construction, if one is interested in the union of UV resonances that are capable to generate operators for a given dimension, then one can neglect the effect of EOM and CDC, and simply combine all the possible UV from the j-basis analysis for each operator. Furthermore, for those operators cannot be converted via other operators with replacements of EOM and CDC, then our j-basis analysis is enough to find their complete set of UV origin. In particular, all the operators with three and four field contents are free from the influence of EOM and CDC subtleties.

To help reader to identify the relation between types of operators, we draw the diagram in figure.~\ref{fig:relD} to illustrate the relations of all the Lorentz classes present in dimension 6 and dimension 7, and connect different types related by EOM and CDC with arrows, where the direction of the arrow indicates that UV resonances that generates operators at the starting points may also generate operators at the ending point.
Following the logic described above,  one can assert that for operators that are not pointed by any arrows, e.g. $D^2\phi^4$ and $D\bar{\psi}\phi^2\psi$, their UV origins are completely fixed by the our j-basis analysis using the Casimir operator method\footnote{EOM of Higgs contains the mass term $m^2 H$, which may induce operator of lower dimension, but the Wilson coefficients generated by this way are in addition suppressed by a factor of $m^2/\Lambda^2$.}, since they cannot be generated from operators of other type using EOM and CDC\footnote{If one can developed a reduction algorithm including the information of EOM, then by constructing j-basis with the bridge method discussed in section 2 and reducing them to our y-basis will eliminate the subtlety of EOM at least for the resonance with spin up to 1.}. 

\begin{figure}[h]
\centering
\includegraphics[width=0.4\textwidth]{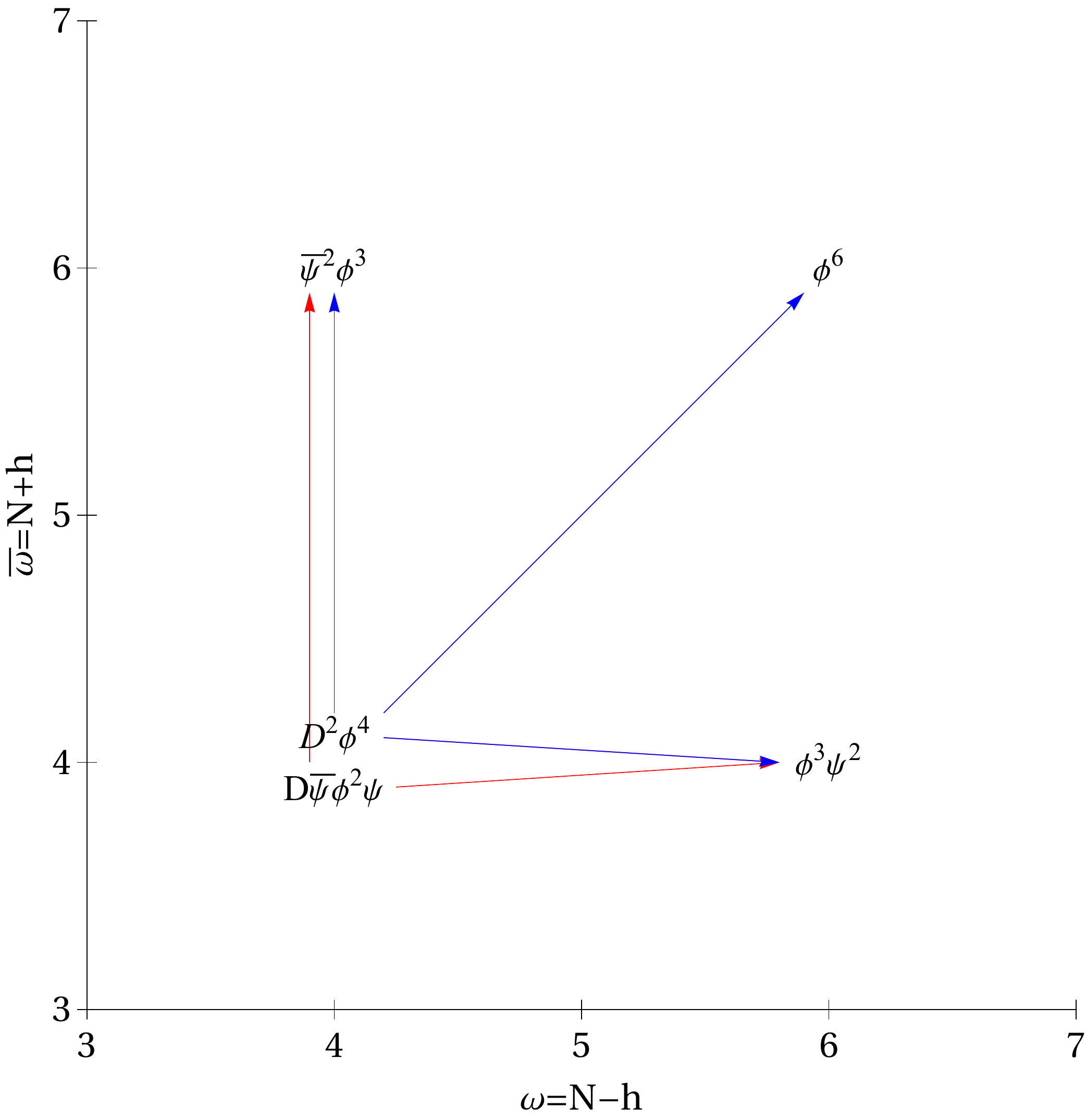}
\includegraphics[width=0.4\textwidth]{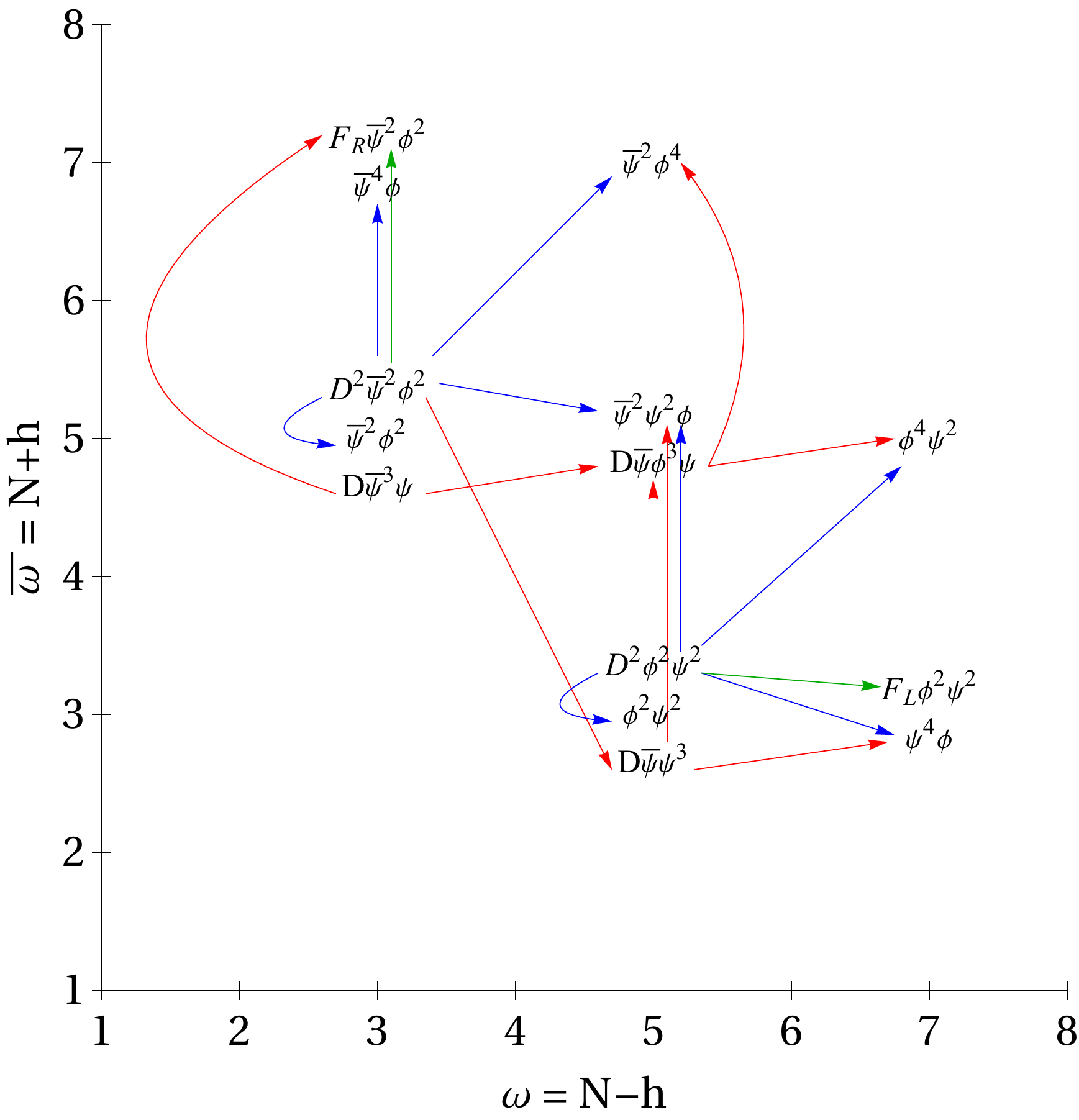}
\caption{The Lorentz classes related by the EOM for dimension 6 (left) and dimension 7 (right). Blue arrows relate different operators by the fermion EOM, red arrows indicate the use of the Higgs EOM, green arrows represent the replacement of $[D,D]$ by field strength tensors. We did not take into account the Lorentz classes that are not involving any conversion relations, e.g. $\psi^2\bar{\psi}^2$ and $\psi^4$ at dimension 6. }\label{fig:relD}
\end{figure}

Finally, we want to emphasis that the UV origins for a specific operator is dependent on the choice of the bases for certain relevant types of operators that are connect with EOM and CDC, thus in some sense discussing the UV origins of a specific operator without referring to the whole basis is not physical. A more appropriate statement would be the UV origins of a physical process or the corresponding physical amplitude at a certain dimension. In the Lagrangian language, after specifying the operator basis, it corresponds to the union of the UV origins of the set of operators that contribute to the given physical process. In this UV-finding procedure, one does not need to take into account the subtlety of EOM and CDC, and the results obtained with our j-basis analysis for each operator is enough, because the subtlety of EOM and CDC is taken care of when combining all the UV origins for the set of operators.
It can be seen more clearly in the language of on-shell amplitudes. The physical amplitude of the given physical process at a particular mass dimension consists of a local and a non-local part, where ``non-local" means the part containing poles of light particles. These non-local part of the amplitudes can be iteratively deconstructed into building blocks of lower-point local sub-amplitudes in our amplitude basis. Therefore the UV origin of the process can be taken as the union of UV origins of the original local amplitude and those of the relevant lower-point local sub-amplitudes, all of which can be directly obtained by our j-basis analysis. 
}

\subsection{Complete UV Resonances of the SMEFT Operators}\label{sec:smeft}
Here we list  all the possible UV resonances that can generate the dimension-6 and dimension-7 SMEFT operators.
We organize all the resonances into two sets of tables --- table.~\ref{tab:scalarreso} to table.~\ref{tab:vectorreso} and table.~\ref{tab:scalarreso7} to table.~\ref{tab:vectorreso7} for dimension-6 and dimension-7 operators respectively, and each set contains three tables for scalar, fermion and vector resonances respectively.
In each table, the left column records the name and the gauge quantum numbers of the corresponding resonance, and the starting letters ``S", ``F", ``V" are abbreviations of scalar, fermion and vector respectively, indicating the spin of the resonance, the right column summarizes the types of SMEFT operators that the corresponding resonance can generate. 
Particularly, for dimension-7, we have two additional spin-3/2 resonances listed in table.~\ref{tab:spin32reso7}, where we denoted as $R1$ and $R2$. 
For the operator types in right columns, additional information for the usage of EOM and the need of extra resonances are recorded in the color and the list of set of fields appended in the square bracket respectively. For example, for the first scalar resonance $S1$, the operator $H^2 H^\dagger Q u_{\mathbb{C}}[(S4)]$ cannot be generated solely with $S1$, but it could be generated if UV includes $S1$ and $S4$ simultaneously as one can in Table~\ref{tab:scalarreso}. On the other hand, the three operator types in the red color --- ${e_{\mathbb{C}} H H^\dagger{}^2 L}$, $
{d_{\mathbb{C}} H H^\dagger{}^2 Q}$, ${H {}^2H^\dagger Q u_{\mathbb{C}}}$  indicate that they are generated by conversions of other types of operators,  here the $D^2H^2H^{\dagger 2}$, with the EOM of the Higgs boson.
In general, the operators in  red, blue and green, represent that the EOM of the Higgs boson, the EOM of fermions and the replacement of covariant derivative commutator are used in the generation of such operators. 
In this sense the same operator type can appear multiple times in the right column for a specific resonance, which indicates different ways to generate such an operator.


From table.~\ref{tab:scalarreso} to table.~\ref{tab:vectorreso}, we have listed all possible UV resonances for the dimension 6 operators from the Poincare and gauge symmetry analysis, which provides a proof and validation on the tree-level dictionary provided in Ref.~\cite{deBlas:2017xtg}.  However, our analysis does not include one of the resonance listed in the Ref.~\cite{deBlas:2017xtg} --- ${\cal W}_1$ with $\left(SU(3)_c, SU(2)_2, U(1)_y\right)$ quantum numbers $(\mbf{1},\mbf{3},3/2)$. The reason is that the renormalizable UV Lagrangian involving ${\cal W}_1$ reads ${\cal W}^\mu_{1\,a} (D_\mu \tilde{H})\sigma^a H+h.c.$, which is equivalent to $(D_\mu {\cal W}^\mu_{1\,a}) \tilde{H}\sigma^a H+h.c.$ by integration by part, and can be converted to higher dimensional operators using the EOM of ${\cal W}_1$. It stems from the fact that higher spin fields can not be fundamental and thus the Lagrangian is intrinsically nonrenormalizable. Actually this operator does not contribute to amplitudes involving the on-shell massive vector particle $\mc{W}_1$. Hence in our prescription, we do not count such operators as an interaction term involving this heavy particle.
In other words, although some effective operators are generated by this Lagrangian term, they are considered as originated by the EOM in the UV rather than produced in the matching procedure related to the spin-1 resonance.
Compared with Ref.~\cite{deBlas:2017xtg}, our method is capable to find the UV origin for a single type without explicitly writing down the BSM-Lagrangian including all the possible UV resonances and performing the matching to select UV resonances that contribute to the operator type of interest.  Therefore, in this sense, our method is a pure bottom-up technique without any top-down matching procedure.

From table.~\ref{tab:scalarreso7} to table.~\ref{tab:spin32reso7} , we enumerate all the possible UV resonances that can generate dimension-7 operators for the first time, and they are all capable to generate dimension-6 operators except for the fermion $F7$ and two spin-3/2 resonances $R1$ and $R2$.
Among all the resonance, only the following resonances are purely real:
$S1$, $S5$, $F1$, $F5$, $V1$, $V4$, $V12$, $V14$, $R1$, $R2$, all other resonances are complex.
Consequently, only fermion resonances $F1$, $F5$ can be Majorana type. 



\label{tab:fermionreso}
\caption{Fermion resonances which could generate the dim-6 operators. The name of the resonances is organized by the gauge quantum number. The resonance appeared in the squared bracket of the operator indicate that both resonances are needed to generate such operator.}
\end{center}
\end{table}


\vspace{2cm}

\begin{longtable}{c|c}
    \hline
    \multicolumn{2}{c}{Vector}\\
    \hline
    \multicolumn{1}{c}{$\left(SU(3)_c, SU(2)_2, U(1)_y\right)$} & \multicolumn{1}{c}{} \\
    \hline
$V1\ (\mbf{1},\mbf{1},0)$ & \makecell{$d_{\mathbb{C}}{}^2 d_{\mathbb{C}}^\dagger{}^2\quad d_{\mathbb{C}} d_{\mathbb{C}}^\dagger e_{\mathbb{C}} e_{\mathbb{C}}^\dagger\quad e_{\mathbb{C}}{}^2 e_{\mathbb{C}}^\dagger{}^2\quad D d_{\mathbb{C}} d_{\mathbb{C}}^\dagger H H^\dagger$\\ $D e_{\mathbb{C}} e_{\mathbb{C}}^\dagger H H^\dagger\quad D{}^2 H{}^2 H^\dagger{}^2\quad d_{\mathbb{C}} d_{\mathbb{C}}^\dagger L L^\dagger\quad e_{\mathbb{C}} e_{\mathbb{C}}^\dagger L L^\dagger$\\ $D H H^\dagger L L^\dagger\quad L{}^2 L^\dagger{}^2\quad d_{\mathbb{C}} d_{\mathbb{C}}^\dagger Q Q^\dagger\quad e_{\mathbb{C}} e_{\mathbb{C}}^\dagger Q Q^\dagger$\\ $D H H^\dagger Q Q^\dagger\quad L L^\dagger Q Q^\dagger\quad Q{}^2 Q^\dagger{}^2\quad d_{\mathbb{C}} d_{\mathbb{C}}^\dagger u_{\mathbb{C}} u_{\mathbb{C}}^\dagger$\\ $e_{\mathbb{C}} e_{\mathbb{C}}^\dagger u_{\mathbb{C}} u_{\mathbb{C}}^\dagger\quad D H H^\dagger u_{\mathbb{C}} u_{\mathbb{C}}^\dagger\quad L L^\dagger u_{\mathbb{C}} u_{\mathbb{C}}^\dagger\quad Q Q^\dagger u_{\mathbb{C}} u_{\mathbb{C}}^\dagger$\\
${\color{blue}d_{\mathbb{C}} H H^\dagger{}^2 Q}\quad  {\color{blue}e_{\mathbb{C}} H H^\dagger{}^2 L}\quad {\color{blue}H {}^2 H^\dagger Q u_{\mathbb{C}}}$\\
$ {\color{red}e_{\mathbb{C}} H H^\dagger{}^2 L}\quad 
{\color{red}d_{\mathbb{C}} H H^\dagger{}^2 Q}\quad {\color{red}H {}^2 
H^\dagger Q u_{\mathbb{C}}}$}\\ 
 \hline
$V2\ (\mbf{1},\mbf{1},1)$ & \makecell{$D{}^2 H{}^2 H^\dagger{}^2\quad D d_{\mathbb{C}} H^\dagger{}^2 u_{\mathbb{C}}^\dagger\quad d_{\mathbb{C}} d_{\mathbb{C}}^\dagger u_{\mathbb{C}} u_{\mathbb{C}}^\dagger$\\
$ {\color{red}e_{\mathbb{C}} H H^\dagger{}^2 L}\quad 
{\color{red}d_{\mathbb{C}} H H^\dagger{}^2 Q}\quad {\color{red}H {}^2 
H^\dagger Q u_{\mathbb{C}}}$\\${\color{blue}d_{\mathbb{C}} H H^\dagger{}^2 Q}$}\\
\end{longtable}


\begin{longtable}{c|c}
    \hline
    \multicolumn{2}{c}{Scalar}\\
    \hline
    \multicolumn{1}{c}{$\left(SU(3)_c, SU(2)_2, U(1)_y\right)$} & \multicolumn{1}{c}{} \\
    \hline
    $S1\ (\mbf{1},\mbf{1},0)$ & \makecell{$H{}^3 H^\dagger L{}^2[(S6),(F5),(F1),(S4,S6),$\\$(S4,F5),(S4,F1),(F3,F5),(F1,F3),(S6,F3)]$}\\ 
 \hline
$S2\ (\mbf{1},\mbf{1},1)$ & \makecell{$ \quad e_{\mathbb{C}} H L{}^3[(S4),(F4),(F1)]\quad d_{\mathbb{C}} H L{}^2 Q[(S4),(F10),(F9)]\quad$\\$ H L{}^2 Q^\dagger u_{\mathbb{C}}^\dagger[(S4),(F8),(F12)]\quad D e_{\mathbb{C}} H^\dagger{}^3 L^\dagger[(F1),(F3),(V3)]$}\\ 
 \hline
$S4\ (\mbf{1},\mbf{2},\frac{1}{2})$ & \makecell{$e_{\mathbb{C}} H L{}^3[(S6),(S2),(F5),(F1)]\quad d_{\mathbb{C}} H L{}^2 Q[(S6),(S2),(F5),(F1)]$\\$ H L{}^2 Q^\dagger u_{\mathbb{C}}^\dagger[(S6),(S2),(F5),(F1)]\quad H{}^3 H^\dagger L{}^2[(S6),(F5),(F1),(S5,S6),(S1,S6),$\\$(S6,F5),(S6,F1),(S5,F5),(S5,F1),(S1,F5),(S1,F1)]$}\\ 
 \hline
$S5\ (\mbf{1},\mbf{3},0)$ & \makecell{$H{}^3 H^\dagger L{}^2[(S6),(F1,F5),(S6,S7),(S4,S6),(S7,F5),$\\$(S4,F5),(S4,F1),(F5,F7),(F3,F5),(F1,F3),(S6,F7),(S6,F3)]$}\\ 
 \hline
$S6\ (\mbf{1},\mbf{3},1)$ & \makecell{$D{}^2 H{}^2 L{}^2\quad e_{\mathbb{C}} H L{}^3[(S4),(F4),(F5)]\quad d_{\mathbb{C}} H L{}^2 Q[(S4),(F10),(F14)]$\\$\quad H L{}^2 Q^\dagger u_{\mathbb{C}}^\dagger[(S4),(F13),(F12)]$\\ $D e_{\mathbb{C}} H^\dagger{}^3 L^\dagger[(F5),(F3),(V3)]\quad H{}^2 L{}^2 W_L[(F7)]\quad $\\
$ H{}^3 H^\dagger L{}^2[(S4),(S8),(S7),(S5),(S1),(F5,F6),(F1,F6),$\\$(S5,S7),(S4,S5),(S1,S4),(S7,F5),(S4,F5),(S4,F1),(F5,F7),(F3,F5),$\\$(F1,F3),(S8,F6),(F6,F7),(F3,F6),(S5,F7),(S5,F3),(S1,F3)]$\\
${\color{green}H{}^2 L{}^2 W_L}\quad {\color{green}B_L H{}^2 L{}^2}\quad {\color{red}e_{\mathbb{C}} H L{}^3}$\\ ${\color{red}H L{}^2 Q^\dagger u_{\mathbb{C}}^\dagger}\quad {\color{red}d_{\mathbb{C}} H L{}^2 Q}\quad {\color{blue}D e_{\mathbb{C}}^\dagger H{}^3 L}$}\\ 
 \hline
$S7\ (\mbf{1},\mbf{4},\frac{1}{2})$ & \makecell{$H{}^3 H^\dagger L{}^2[(S6),(F5),(S5,S6),(S6,F5),(S5,F5)]$}\\
\end{longtable}


\begin{longtable}{c|c}
    \hline
    \multicolumn{2}{c}{Fermion}\\
    \hline
    \multicolumn{1}{c}{$\left(SU(3)_c, SU(2)_2, U(1)_y\right)$} & \multicolumn{1}{c}{} \\
    \hline
    $F1\ (\mbf{1},\mbf{1},0)$ & \makecell{$D{}^2 H{}^2 L{}^2\quad e_{\mathbb{C}} H L{}^3[(S4),(S2)]\quad d_{\mathbb{C}} H L{}^2 Q[(S4),(S10),(S12)]\quad $\\ $H L{}^2 Q^\dagger u_{\mathbb{C}}^\dagger[(S4),(V5),(V8)]\quad D e_{\mathbb{C}} H^\dagger{}^3 L^\dagger[(F3),(V2)]\quad $\\$d_{\mathbb{C}}{}^2 H L u_{\mathbb{C}}[(S11),(S10)]\quad d_{\mathbb{C}} e_{\mathbb{C}}^\dagger H L u_{\mathbb{C}}^\dagger[(S10),(V5)]\quad d_{\mathbb{C}} H L Q^\dagger{}^2[(S10),(V8)]$\\ $H{}^2 L{}^2 W_L[(F5)]$\\$H{}^3 H^\dagger L{}^2[(S4),(S5,F5),(S1),(S6,F6),(F3,F5),(F3),(F3,F6),$\\ $(S4,S6),(S6,F3),(S4,S5),(S1,S4),(S5,F3),(S1,F3)]$\\
${\color{green}H{}^2 L{}^2 W_L}\quad {\color{green}B_L H{}^2 L{}^2}\quad {\color{red}e_{\mathbb{C}} H L{}^3}$\\ ${\color{red}H L{}^2 Q^\dagger u_{\mathbb{C}}^\dagger}\quad {\color{red}d_{\mathbb{C}} H L{}^2 Q}\quad {\color{blue}D e_{\mathbb{C}}^\dagger H{}^3 L}$}\\ 
 \hline
$F2\ (\mbf{1},\mbf{1},1)$ & \makecell{$d_{\mathbb{C}}{}^3 H^\dagger L[(S11)]$}\\ 
 \hline
$F3\ (\mbf{1},\mbf{2},\frac{1}{2})$ & \makecell{$D e_{\mathbb{C}} H^\dagger{}^3 L^\dagger[(F5),(F1),(S6),(V2)]\quad d_{\mathbb{C}} e_{\mathbb{C}}^\dagger H L u_{\mathbb{C}}^\dagger[(S12),(V8)]\quad $\\$d_{\mathbb{C}}{}^2 e_{\mathbb{C}}^\dagger H Q^\dagger[(V8),(S11)]\quad H{}^3 H^\dagger L{}^2[(F5),(F1,F5),(F1),(F5,F6),(F1,F6),(S6,F5),$\\$(S6,F1),(S5,F5),(S5,F1),(S1,F5),(S1,F1),(S6,F6),(S5,S6),(S1,S6)]$}\\ 
 \hline
$F4\ (\mbf{1},\mbf{2},\frac{3}{2})$ & \makecell{$e_{\mathbb{C}} H L{}^3[(S6),(S2)]$}\\
\end{longtable}


\begin{longtable}{c|c}
    \hline
    \multicolumn{2}{c}{Vector}\\
    \hline
    \multicolumn{1}{c}{$\left(SU(3)_c, SU(2)_2, U(1)_y\right)$} & \multicolumn{1}{c}{} \\
    \hline
$V2\ (\mbf{1},\mbf{1},1)$ & \makecell{$D{}^2 H{}^2 L{}^2\quad D e_{\mathbb{C}} H^\dagger{}^3 L^\dagger[(F1),(V3),(F3)]$
\\
${\color{green}H{}^2 L{}^2 W_L}\quad {\color{green}B_L H{}^2 L{}^2}\quad {\color{red}e_{\mathbb{C}} H L{}^3}$\\ ${\color{red}H L{}^2 Q^\dagger u_{\mathbb{C}}^\dagger}\quad {\color{red}d_{\mathbb{C}} H L{}^2 Q}\quad {\color{blue}D e_{\mathbb{C}}^\dagger H{}^3 L}$}\\ 
 \hline
$V3\ (\mbf{1},\mbf{2},\frac{3}{2})$ & \makecell{$D e_{\mathbb{C}} H^\dagger{}^3 L^\dagger\quad d_{\mathbb{C}} e_{\mathbb{C}}^\dagger H L u_{\mathbb{C}}^\dagger[(F10),(F12)]\quad D e_{\mathbb{C}} H^\dagger{}^3 L^\dagger[(V2),(S6)]$}\\ 
 \hline
$V5\ (\mbf{3},\mbf{1},\frac{2}{3})$ & \makecell{$D d_{\mathbb{C}}{}^3 e_{\mathbb{C}}^\dagger\quad H L{}^2 Q^\dagger u_{\mathbb{C}}^\dagger[(F1),(V8),(F12)]\quad d_{\mathbb{C}} e_{\mathbb{C}}^\dagger H L u_{\mathbb{C}}^\dagger[(F1),(V8),(F12)]$\\$ d_{\mathbb{C}} H L Q^\dagger{}^2[(V8),(F10),(F8)]$\\${\color{blue}d_{\mathbb{C}}^2 e_{\mathbb{C}}^\dagger H Q^\dagger}\quad {\color{blue}d_{\mathbb{C}}^3 H^\dagger L^\dagger}$}\\ 
 \hline
$V8\ (\mbf{3},\mbf{2},\frac{1}{6})$ & \makecell{$D d_{\mathbb{C}} L{}^2 u_{\mathbb{C}}^\dagger\quad D d_{\mathbb{C}}{}^2 L Q^\dagger\quad H L{}^2 Q^\dagger u_{\mathbb{C}}^\dagger[(F5),(F1),(V9),(V5),(F13),(F8)]$\\ $ d_{\mathbb{C}} e_{\mathbb{C}}^\dagger H L u_{\mathbb{C}}^\dagger[(V5),(F10),(F3)]\quad d_{\mathbb{C}}{}^2 e_{\mathbb{C}}^\dagger H Q^\dagger[(V5),(F10),(F3)]$\\$d_{\mathbb{C}} H L Q^\dagger{}^2[(F5),(F1),(V9),(V5),(F13),(F8)]$\\$\color{blue} d_{\mathbb{C}} H L {}^2 Q\quad d_{\mathbb{C}}{}^2 e_ {\mathbb{C}}^\dagger H Q^\dagger\quad d_{\mathbb{C}} H L Q^\dagger {}^2 $\\$\color{blue}d_{\mathbb{C}}{}^2 H L u_{\mathbb{C}}\quad d_{\mathbb{C}} e_ {\mathbb{C}}^\dagger H L u_ {\mathbb{C}}^\dagger\quad H L {}^2 Q^\dagger u_ {\mathbb{C}}^\dagger$}\\
\end{longtable}



\begin{longtable}{c|c}
    \hline
    \multicolumn{2}{c}{Spin-3/2}\\
    \hline
    \multicolumn{1}{c}{$\left(SU(3)_c, SU(2)_2, U(1)_y\right)$} & \multicolumn{1}{c}{} \\
    \hline
    $R1\ (\mbf{1},\mbf{1},0)$ & \makecell{$ D{}^2 H{}^2 L{}^2$\\
${\color{green}H{}^2 L{}^2 W_L}\quad {\color{green}B_L H{}^2 L{}^2}\quad {\color{red}e_{\mathbb{C}} H L{}^3}$\\ ${\color{red}H L{}^2 Q^\dagger u_{\mathbb{C}}^\dagger}\quad {\color{red}d_{\mathbb{C}} H L{}^2 Q}\quad {\color{blue}D e_{\mathbb{C}}^\dagger H{}^3 L}$}\\ 
 \hline
 $R2\ (\mbf{1},\mbf{2},0)$ & \makecell{$ D{}^2 H{}^2 L{}^2$\\
${\color{green}H{}^2 L{}^2 W_L}\quad {\color{green}B_L H{}^2 L{}^2}\quad {\color{red}e_{\mathbb{C}} H L{}^3}$\\ ${\color{red}H L{}^2 Q^\dagger u_{\mathbb{C}}^\dagger}\quad {\color{red}d_{\mathbb{C}} H L{}^2 Q}\quad {\color{blue}D e_{\mathbb{C}}^\dagger H{}^3 L}$}\\ 
 \hline
 \caption{Spin-3/2 resonances which could generate the dim-6 operators.}
\label{tab:spin32reso7}
\end{longtable}

\subsection{UV Completion of Dim-6 Operators}\label{sec:uvdim6}

In this subsection, in Table~\ref{tab:6-4}-\ref{tab:6-6} we list the complete tables of j-basis for all the dimension-6 SMEFT operators, exhausting all possible partitions. The table of each type consists of the following structures:
1. A head indicates the form of the operator type;
2. The f-basis operators for the given type denoted as ${\cal O}^f_i$;
3. The detailed j-basis information for each topology and each partition is contained in a sub-table of two columns. 
In such a sub-table, the left column contains the topology diagrams and all the induced partitions recorded with sets of fields grouped in curly brackets. Each set of grouped fields corresponds to a particular UV resonance, which carries the quantum numbers of the j-basis. The resulting quantum numbers for the UV resonances are listed in different rows in the left column, and the corresponding j-basis operators expressed in terms of p-basis are recorded in the right columns.  

Comments are in order for these series of tables (not only for the dimension 6 operator UV list, but also dimension 7 and 8 ones):
\bet
\item As one can see there are UV resonances decorated with an ``$*$" in the front, this means that the UV interactions contain non-renormalizable vertices. 
Hence we can infer that some of the operators cannot be generated at tree level by a renormalizable UV theory if all the j-basis operators are with the ``$*$'' sign. These operators are sometimes called ``loop-generated'', and are supposed to be suppressed compared to other operators at the same dimension.

\item There are UV resonances colored in grey~\footnote{These UV resonances colored in grey are not shown for dimension 8 bosonic sector. }, which means that the minimal dimension by integrating out the corresponding UV resonance is larger than the given type (which must involve non-renormalizable vertices, and hence must be decorated with ``$*$''), and thus they should be eliminated after the \textit{Dim selection} discussed in section.~\ref{sec:corandtopo}.
However, one should keep in mind that the j-basis amplitude for such scattering channel still exist, it is just the interpretation that the existence of the UV resonance is problematic. 

\item Only the ratios of coefficients of the p-basis operators for a given j-basis operators are fixed, multiplying an overall constant does not alter quantum numbers of the partial waves generated by such a modified j-basis operators. In the meantime, we point out that the j-basis operators and the ones obtained by integrating out the corresponding UV resonance are consistent to the extent of modulo of EOM and $[D,D]$. 

\item We have thrown away the j-basis that are forbidden due to the number of flavor constraints. For example, the scalar UV resonance $(0, 1,2,3/2)$ is forbidden to generate the dimension 6 operators. However, if there are two Higgses, the scalar resonance $(0, 1,2,3/2)$ would contribute.  

\item We do not list the operator types with only three fields, such as $ {W_L}^3 $ and $ {G_L}^3 $, as they cannot be generated by the tree-level topology. Thus these belong to the loop-level UV generated effective operators.  
\eet

\begin{longtable}{cc|c}
\caption{UV resonance of topology dim6-4. ``*'' in front of each row indicates topologies with nonrenormalizable vertices. The UV resonances colored in gray are eliminated by UV selection. }\label{tab:6-4}\\
\hline\hline\multicolumn{3}{l}{\multirow{2}{*}{\large Type: ${d_{\mathbb{C}}}{e_{\mathbb{C}}}{u_{\mathbb{C}}}^2$}}\\
 \multicolumn{3}{c}{}\\
\hline\multicolumn{3}{c}{$\begin{array}{ccl}\mathcal{O}^f_{1}&=&\frac{1}{2}\mathcal{Y}[\young(st)](\epsilon_{abc}({{{d_{\mathbb{C}}}_p}^a}{{{u_{\mathbb{C}}}_s}^b})({e_{\mathbb{C}}}_r{{{u_{\mathbb{C}}}_t}^c})),\\
\mathcal{O}^f_{2}&=&\frac{1}{2}\mathcal{Y}[\young(s,t)](\epsilon_{abc}({{{d_{\mathbb{C}}}_p}^a}{e_{\mathbb{C}}}_r)({{{u_{\mathbb{C}}}_s}^b}{{{u_{\mathbb{C}}}_t}^c})).\end{array}$}\\
\hline
\multicolumn{2}{c|}{group: $(\text{Spin},SU(3)_c,SU(2)_w,U(1)_y)$} & \multirow{2}{*}{$\mathcal{O}^{j}$}\\
\begin{fmffile}{4}
\begin{fmfgraph*}(40,20)
\fmfpen{thin} \fmfleft{e1,e2} \fmfv{}{g1} \fmfv{}{g2} \fmfright{e4,e3}
\fmf{plain}{e1,g1} \fmf{plain}{g1,e2} \fmf{plain}{g1,g2} \fmf{plain}{e4,g2} \fmf{plain}{g2,e3}
\end{fmfgraph*}
\end{fmffile} &\multicolumn{1}{c|}{ ${\{ {d_{\mathbb{C}}}_1, {e_{\mathbb{C}}}_2\}, \{ {u_{\mathbb{C}}}_3, {u_{\mathbb{C}}}_4\}}$ } & \\ 
 \hline* & {\color{gray} $(1,\overline{3},1,\frac{4}{3})$} & {\color{gray} $2 \mathcal{O}^f_{{1}}$}\\ 
& $(0,\overline{3},1,\frac{4}{3})$ & $-\mathcal{O}^f_{{2}}$\\
\hline
\end{longtable}

\begin{longtable}{ccc|c}
\caption{UV resonance of topology dim6-5. ``*'' in front of each row indicates topologies with nonrenormalizable vertices. The UV resonances colored in gray are eliminated by UV selection. }\label{tab:6-5}\\
\hline\hline\multicolumn{4}{l}{\multirow{2}{*}{\large Type: ${e_{\mathbb{C}}}H{H^\dagger}^2L$}}\\
 \multicolumn{4}{c}{}\\
\hline\multicolumn{4}{c}{$\begin{array}{ccl}\mathcal{O}^f_{1}&=&\frac{1}{2}\mathcal{Y}[\yng(2)_{{H^\dagger}}]{{H^\dagger}^i}{{H^\dagger}^j}{H_j}({e_{\mathbb{C}}}_p{{L_r}_i}).\end{array}$}\\
\hline
\multicolumn{3}{c|}{group: $(\text{Spin},SU(3)_c,SU(2)_w,U(1)_y)$} & \multirow{2}{*}{$\mathcal{O}^{j}$}\\
\begin{fmffile}{5-1}
\begin{fmfgraph*}(40,20)
\fmfpen{thin} \fmfleft{e1,e2} \fmftop{e3} \fmfbottom{n} \fmfv{}{g1} \fmfv{}{g2} \fmfv{}{g3} \fmfright{e5,e4}
\fmf{plain}{e1,g1} \fmf{plain}{g1,e2} \fmf{plain}{g1,g2} \fmf{plain}{g2,e3} \fmf{phantom}{g2,n} \fmf{plain}{g2,g3} \fmf{plain}{e4,g3} \fmf{plain}{g3,e5}
\end{fmfgraph*}
\end{fmffile} &\multicolumn{2}{c|}{ ${\{ {e_{\mathbb{C}}}_1, L_2\}, \{ {e_{\mathbb{C}}}_1, L_2, {H^\dagger}_5\}, \{ H_3, {H^\dagger}_4\}}$ } & \\ 
 \hline& $(0,1,2,\frac{1}{2})$ & $(0,1,3,0)$ & $\mathcal{O}^f_{{1}}$\\ 
& $(0,1,2,\frac{1}{2})$ & $(0,1,1,0)$ & $\mathcal{O}^f_{{1}}$\\
\hline
\end{longtable}

\begin{longtable}{cccc|c}
\caption{UV resonance of topology dim6-6. ``*'' in front of each row indicates topologies with nonrenormalizable vertices. The UV resonances colored in gray are eliminated by UV selection. }\label{tab:6-6}\\
\hline\hline\multicolumn{5}{l}{\multirow{2}{*}{\large Type: $H^3{H^\dagger}^3$}}\\
 \multicolumn{5}{c}{}\\
\hline\multicolumn{5}{c}{$\begin{array}{ccl}\mathcal{O}^f_{1}&=&\frac{1}{36}\mathcal{Y}[\yng(3)_{H}]\mathcal{Y}[\yng(3)_{{H^\dagger}}]{{H^\dagger}^i}{{H^\dagger}^j}{{H^\dagger}^k}{H_i}{H_j}{H_k}.\end{array}$}\\
\hline
\multicolumn{4}{c|}{group: $(\text{Spin},SU(3)_c,SU(2)_w,U(1)_y)$} & \multirow{2}{*}{$\mathcal{O}^{j}$}\\
\begin{fmffile}{6-1}
\begin{fmfgraph*}(30,30)
\fmfpen{thin} \fmfsurroundn{e}{6} \fmfvn{}{g}{4} 
\fmf{plain}{e1,g1} \fmf{plain}{g1,e2} \fmf{plain}{g1,g4} \fmf{plain}{e3,g2} \fmf{plain}{g2,e4} \fmf{plain}{g2,g4} \fmf{plain}{e5,g3} \fmf{plain}{g3,g4} \fmf{plain}{g3,e6}
\end{fmfgraph*}
\end{fmffile} &\multicolumn{3}{c|}{ ${\{ H_1, H_2\}, \{ H_3, {H^\dagger}_4\}, \{ {H^\dagger}_5, {H^\dagger}_6\}}$ } & \\ 
 \hline& $(0,1,3,1)$ & $(0,1,3,0)$ & $(0,1,3,-1)$ & $\mathcal{O}^f_{{1}}$\\ 
& $(0,1,3,1)$ & $(0,1,1,0)$ & $(0,1,3,-1)$ & $2 \mathcal{O}^f_{{1}}$\\
\hline
\end{longtable}

\subsection{UV Completion of Dim-7 Operators}
In this subsection, in Table~\ref{tab:7-4}-\ref{tab:7-6} we list all the j-basis and the corresponding UV resonances for all the dim-7 SMEFT operators for the first time. The convention and the meaning of the table are described in the beginning of section.~\ref{sec:uvdim6}.

\begin{longtable}{cc|c}
\caption{UV resonance of topology dim7-4. ``*'' in front of each row indicates topologies with nonrenormalizable vertices. The UV resonances colored in gray are eliminated by UV selection. }\label{tab:7-4}\\
\hline\hline\multicolumn{3}{l}{\multirow{2}{*}{\large Type: $D{d_{\mathbb{C}}}L^2{{u_{\mathbb{C}}}^\dagger}$}}\\
 \multicolumn{3}{c}{}\\
\hline\multicolumn{3}{c}{$\begin{array}{ccl}\mathcal{O}^f_{{1}}&=& i\epsilon^{ij}({{L_r}_i}({D^\mu}{{L_s}_j}))({{{d_{\mathbb{C}}}_p}^a}{\sigma_\mu}{{{{u_{\mathbb{C}}}^\dagger}_t}_a}).\end{array}$}\\
\hline
\multicolumn{2}{c|}{group: $(\text{Spin},SU(3)_c,SU(2)_w,U(1)_y)$} & \multirow{2}{*}{$\mathcal{O}^{j}$}\\
\begin{fmffile}{4}
\begin{fmfgraph*}(40,20)
\fmfpen{thin} \fmfleft{e1,e2} \fmfv{}{g1} \fmfv{}{g2} \fmfright{e4,e3}
\fmf{plain}{e1,g1} \fmf{plain}{g1,e2} \fmf{plain}{g1,g2} \fmf{plain}{e4,g2} \fmf{plain}{g2,e3}
\end{fmfgraph*}
\end{fmffile} &\multicolumn{1}{c|}{ ${\{ {d_{\mathbb{C}}}_1, L_2\}, \{ L_3, {{u_{\mathbb{C}}}^\dagger}_4\}}$ } & \\ 
 \hline* & $(1,\overline{3},2,-\frac{1}{6})$ & $\mathcal{O}^f_{1}$\\ 
\hline
& \multicolumn{1}{c|}{ ${\{ {d_{\mathbb{C}}}_1, {{u_{\mathbb{C}}}^\dagger}_4\}, \{ L_2, L_3\}}$ } & $\mathcal{O}^{j}$\\
\hline
\end{longtable}

\begin{longtable}{ccc|c}
\caption{UV resonance of topology dim7-5. ``*'' in front of each row indicates topologies with nonrenormalizable vertices. The UV resonances colored in gray are eliminated by UV selection. }\label{tab:7-5}\\
\hline\hline\multicolumn{4}{l}{\multirow{2}{*}{\large Type: ${e_{\mathbb{C}}}HL^3$}}\\
 \multicolumn{4}{c}{}\\
\hline\multicolumn{4}{c}{$\begin{array}{ccl}\mathcal{O}^f_{1}&=&\frac{1}{6}\mathcal{Y}[\young(rst)]\epsilon^{ik}\epsilon^{jl}{H_l}({e_{\mathbb{C}}}_p{{L_r}_i})({{L_s}_j}{{L_t}_k}),\\
\mathcal{O}^f_{2}&=&\frac{1}{3}\mathcal{Y}[\young(rs,t)]\epsilon^{ik}\epsilon^{jl}{H_l}({e_{\mathbb{C}}}_p{{L_r}_i})({{L_s}_j}{{L_t}_k}),\\
\mathcal{O}^f_{3}&=&\frac{1}{6}\mathcal{Y}[\young(r,s,t)]\epsilon^{ik}\epsilon^{jl}{H_l}({e_{\mathbb{C}}}_p{{L_r}_i})({{L_s}_j}{{L_t}_k}).\end{array}$}\\
\hline
\multicolumn{3}{c|}{group: $(\text{Spin},SU(3)_c,SU(2)_w,U(1)_y)$} & \multirow{2}{*}{$\mathcal{O}^{j}$}\\
\begin{fmffile}{5-1}
\begin{fmfgraph*}(40,20)
\fmfpen{thin} \fmfleft{e1,e2} \fmftop{e3} \fmfbottom{n} \fmfv{}{g1} \fmfv{}{g2} \fmfv{}{g3} \fmfright{e5,e4}
\fmf{plain}{e1,g1} \fmf{plain}{g1,e2} \fmf{plain}{g1,g2} \fmf{plain}{g2,e3} \fmf{phantom}{g2,n} \fmf{plain}{g2,g3} \fmf{plain}{e4,g3} \fmf{plain}{g3,e5}
\end{fmfgraph*}
\end{fmffile} &\multicolumn{2}{c|}{ ${\{ {e_{\mathbb{C}}}_1, L_2\}, \{ {e_{\mathbb{C}}}_1, L_2, H_5\}, \{ L_3, L_4\}}$ } & \\ 
 \hline* & {\color{gray} $(1,1,2,\frac{1}{2})$} & {\color{gray} $(1,1,3,-1)$} & {\color{gray} $6 \mathcal{O}^f_{{3}} + (3 -3 (s  t))\circ \mathcal{O}^f_{{2}}$}\\ 
& $(0,1,2,\frac{1}{2})$ & $(0,1,3,-1)$ & $2 \mathcal{O}^f_{{1}} + (3 + 3 (s  t))\circ \mathcal{O}^f_{{2}}$\\
\hline
\end{longtable}

\begin{longtable}{cccc|c}
\caption{UV resonance of topology dim7-6. ``*'' in front of each row indicates topologies with nonrenormalizable vertices. The UV resonances colored in gray are eliminated by UV selection. }\label{tab:7-6}\\
\hline\hline\multicolumn{5}{l}{\multirow{2}{*}{\large Type: $H^3{H^\dagger}L^2$}}\\
 \multicolumn{5}{c}{}\\
\hline\multicolumn{5}{c}{$\begin{array}{ccl}\mathcal{O}^f_{1}&=&\frac{1}{12}\mathcal{Y}[\yng(3)_{H}]\mathcal{Y}[\young(pr)]\epsilon^{ki}\epsilon^{mj}{{H^\dagger}^l}{H_k}{H_l}{H_m}({{L_p}_i}{{L_r}_j}).\end{array}$}\\
\hline
\multicolumn{4}{c|}{group: $(\text{Spin},SU(3)_c,SU(2)_w,U(1)_y)$} & \multirow{2}{*}{$\mathcal{O}^{j}$}\\
\begin{fmffile}{6-1}
\begin{fmfgraph*}(30,30)
\fmfpen{thin} \fmfsurroundn{e}{6} \fmfvn{}{g}{4} 
\fmf{plain}{e1,g1} \fmf{plain}{g1,e2} \fmf{plain}{g1,g4} \fmf{plain}{e3,g2} \fmf{plain}{g2,e4} \fmf{plain}{g2,g4} \fmf{plain}{e5,g3} \fmf{plain}{g3,g4} \fmf{plain}{g3,e6}
\end{fmfgraph*}
\end{fmffile} &\multicolumn{3}{c|}{ ${\{ H_3, {H^\dagger}_6\}, \{ H_4, H_5\}, \{ L_1, L_2\}}$ } & \\ 
 \hline& $(0,1,3,0)$ & $(0,1,3,1)$ & $(0,1,3,-1)$ & $-\mathcal{O}^f_{{1}}$\\ 
& $(0,1,1,0)$ & $(0,1,3,1)$ & $(0,1,3,-1)$ & $2 \mathcal{O}^f_{{1}}$\\
\hline
\end{longtable}

\subsection{UV Completion of Dim-8 Operators: Bosonic Sector}
In this section, in Table~\ref{tab:8-4}-\ref{tab:8-6} we list the j-basis and the corresponding UV resonances for purely bosonic dim-8 SMEFT operators, except for the type $H^4 H^{\dagger 4}$, of which the number of topology and partition is too huge~\footnote{Another reason that we neglect the UV resonances of the $H^4 H^{\dagger 4}$ is typically the $H^4 H^{\dagger 4}$ has the sub-leading contribution on the Higgs potential compared with  the dimension-6 Higgs operator $H^3 H^{\dagger 3}$}. 
We will present the UV for the fermionic operators in a separate work~\cite{dim8UV}.
The convention and the meaning of the table are described in the beginning of section.~\ref{sec:uvdim6}.     
\begin{longtable}{cc|c}
\caption{UV resonance of topology dim8-bosonic-4. ``*'' in front of each row indicates topologies with nonrenormalizable vertices.}\label{tab:8-4}\\
\hline\hline\multicolumn{3}{l}{\multirow{2}{*}{\large Type: ${B_L}^4$}}\\
 \multicolumn{3}{c}{}\\
\hline\multicolumn{3}{c}{$\begin{array}{ccl}\mathcal{O}^f_{1}&=&\frac{1}{24}\mathcal{Y}[\yng(4)_{{B_L}}]{{B_L}^{\lambda\mu}}{{B_L}^{\lambda\rho}}{{B_L}^{\mu\nu}}{{B_L}^{\nu\rho}}.\end{array}$}\\
\hline
\multicolumn{2}{c|}{group: $(\text{Spin},SU(3)_c,SU(2)_w,U(1)_y)$} & \multirow{2}{*}{$\mathcal{O}^{j}$}\\
\begin{fmffile}{4}
\begin{fmfgraph*}(40,20)
\fmfpen{thin} \fmfleft{e1,e2} \fmfv{}{g1} \fmfv{}{g2} \fmfright{e4,e3}
\fmf{plain}{e1,g1} \fmf{plain}{g1,e2} \fmf{plain}{g1,g2} \fmf{plain}{e4,g2} \fmf{plain}{g2,e3}
\end{fmfgraph*}
\end{fmffile} &\multicolumn{1}{c|}{ ${\{ {B_L}_1, {B_L}_2\}, \{ {B_L}_3, {B_L}_4\}}$ } & \\ 
 \hline* & $(0,1,1,0)$ & $16 \mathcal{O}^f_{{1}}$\\ 
\hline\hline\multicolumn{3}{l}{\multirow{2}{*}{\large Type: ${B_L}^2{W_L}^2$}}\\
\hline
\end{longtable}

\begin{longtable}{ccc|c}
\caption{UV resonance of topology dim8-bosonic-5. ``*'' in front of each row indicates topologies with nonrenormalizable vertices.}\label{tab:8-5}\\
\hline\hline\multicolumn{4}{l}{\multirow{2}{*}{\large Type: ${B_L}D^2H^2{H^\dagger}^2$}}\\
 \multicolumn{4}{c}{}\\
\hline\multicolumn{4}{c}{$\begin{array}{ccl}\mathcal{O}^f_{1}&=&\frac{1}{4}\mathcal{Y}[\yng(2)_{H}]\mathcal{Y}[\yng(2)_{{H^\dagger}}]{{B_L}^{\mu\nu}}{{H^\dagger}^i}{H_i}({D_\mu}{H_j})({D_\nu}{{H^\dagger}^j}).\end{array}$}\\
\hline
\multicolumn{3}{c|}{group: $(\text{Spin},SU(3)_c,SU(2)_w,U(1)_y)$} & \multirow{2}{*}{$\mathcal{O}^{j}$}\\
\begin{fmffile}{5-1}
\begin{fmfgraph*}(40,20)
\fmfpen{thin} \fmfleft{e1,e2} \fmftop{e3} \fmfbottom{n} \fmfv{}{g1} \fmfv{}{g2} \fmfv{}{g3} \fmfright{e5,e4}
\fmf{plain}{e1,g1} \fmf{plain}{g1,e2} \fmf{plain}{g1,g2} \fmf{plain}{g2,e3} \fmf{phantom}{g2,n} \fmf{plain}{g2,g3} \fmf{plain}{e4,g3} \fmf{plain}{g3,e5}
\end{fmfgraph*}
\end{fmffile} &\multicolumn{2}{c|}{ ${\{ {B_L}_1, H_2\}, \{ {B_L}_1, H_2, {H^\dagger}_5\}, \{ H_3, {H^\dagger}_4\}}$ } & \\ 
 \hline 
 \multirow{2}{*}{*} & \multirow{2}{*}{ $(1,1,2,\frac{1}{2})$} & \multirow{2}{*}{$(1,1,3,0)$} & $-12 \mathcal{O}^f_{{1}}$\\ 
 &&& $-12 \mathcal{O}^f_{{1}}$\\
\hline
\end{longtable}

\begin{longtable}{cccc|c}
\caption{UV resonance of topology dim8-bosonic-6. ``*'' in front of each row indicates topologies with nonrenormalizable vertices.}\label{tab:8-6}\\
\hline\hline\multicolumn{5}{l}{\multirow{2}{*}{\large Type: ${B_L}^2H^2{H^\dagger}^2$}}\\
 \multicolumn{5}{c}{}\\
\hline\multicolumn{5}{c}{$\begin{array}{ccl}\mathcal{O}^f_{1}&=&\frac{1}{8}\mathcal{Y}[\yng(2)_{{B_L}}]\mathcal{Y}[\yng(2)_{H}]\mathcal{Y}[\yng(2)_{{H^\dagger}}]{{B_L}^{\mu\nu}}{{B_L}^{\mu\nu}}{{H^\dagger}^i}{{H^\dagger}^j}{H_i}{H_j}.\end{array}$}\\
\hline
\multicolumn{4}{c|}{group: $(\text{Spin},SU(3)_c,SU(2)_w,U(1)_y)$} & \multirow{2}{*}{$\mathcal{O}^{j}$}\\
\begin{fmffile}{6-1}
\begin{fmfgraph*}(30,30)
\fmfpen{thin} \fmfsurroundn{e}{6} \fmfvn{}{g}{4} 
\fmf{plain}{e1,g1} \fmf{plain}{g1,e2} \fmf{plain}{g1,g4} \fmf{plain}{e3,g2} \fmf{plain}{g2,e4} \fmf{plain}{g2,g4} \fmf{plain}{e5,g3} \fmf{plain}{g3,g4} \fmf{plain}{g3,e6}
\end{fmfgraph*}
\end{fmffile} &\multicolumn{3}{c|}{ ${\{ {B_L}_1, {B_L}_2\}, \{ H_3, H_4\}, \{ {H^\dagger}_5, {H^\dagger}_6\}}$ } & \\ 
 \hline* & $(0,1,1,0)$ & $(0,1,3,1)$ & $(0,1,3,-1)$ & $4 \mathcal{O}^f_{{1}}$\\ 
\hline
& \multicolumn{3}{c|}{ ${\{ {B_L}_1, {B_L}_2\}, \{ H_3, {H^\dagger}_5\}, \{ H_4, {H^\dagger}_6\}}$ } & $\mathcal{O}^{j}$\\
\hline
\end{longtable}

\section{Summary and Outlook}
\label{sec:sum}


Although it is known that in the SMEFT framework, the effective operators parametrize all possible new physics contributions above the electroweak scale, it is not an easy task to identify and classify various new physics resonances in a systematic way. In this work, we adopt the bottom-up approach, in which the spin and gauge quantum numbers of the BSM particles and the BSM-SM interactions are neither presumed nor exhausted beforehand. We are able to systematically classify all the possible ultraviolet resonances responsible for the EFT operators only based on the Poincare and gauge symmetries.


First, we construct the generalized partial wave amplitude basis via the spinor bridge method and the Casimir method. The generalization include amplitudes with arbitrary number of particles and we consider general partitions of them. Together with a similar treatment of the gauge factor via the $SU(N)$ Casimir operators, we define the j-basis amplitudes consisting of both Lorentz and gauge partial wave basis, which induce the corresponding j-basis operators. The j-basis amplitudes/operators have definite spin and gauge quantum numbers in the channels specified by a particular tree topology, which treats all the particles/fields as distinguishable, thus called flavor-blind basis. With the Young tensor method, we are able to decompose the j-basis operators in the flavor-specified f-basis, the operator basis in the usual sense, or vice versa.



Second, we build the correspondence between the j-basis operators and the ultraviolet resonances, named the j-basis/UV correspondence. 
Consider a tree diagram with heavy resonances generating effective operators in the infrared, the quantum numbers of the resonances could fix the result to the corresponding j-basis operator, which further converts to a combination of the f-basis operators. Thus whether an operator could be generated from a particular UV resonance lies in whether it appears in the decomposition of the corresponding j-basis operator.
This provides a truly bottom-up approach to investigate the tree-level UV completions of the SMEFT. 
We take the Weinberg operator as a benchmark, and systematically obtain the three types of seesaw models that generate it at tree level. Further examples on the dimension-7 operators for the neutrino mass generation are also investigated.   
Applying the j-basis/UV correspondence to the SMEFT operators, we obtain all the UV resonances that may be responsible for the dimension -5, -6, -7 and -8 operators, without using any information from the UV Lagrangian. 

In the following we summarize the advantages of the bottom-up approach, together with several theoretical developments:
\bit

\item We propose a new and systematic method of decomposing the gauge structure of any scattering amplitude into the gauge eigen-basis using the gauge Casimir action, similar to obtaining the Lorentz eigen-basis using the Poincare Casimir action.  %

\item We propose the j-basis/UV correspondence, by which a j-basis operator is associated with the UV resonances with designated quantum numbers. There is no need to exhaust all kinds of resonances and their UV Lagrangian.



\item For the first time, we list the complete UV resonances for the dimension-7 and higher dimensional SMEFT operators systematically and efficiently. Otherwise, the top-down exhaustive search method would be very hard to list all possible UV resonances beyond the dimension-6 operators.

\item With the complete list of the tree-level origin of the effective operators, it is possible to further distinguish the loop-level generated UVs from the tree-level generated UVs.

\eit

The primary task of the LHC is finding new resonances beyond the standard model, instead of constraining the Wilson coefficients of the effective operators. 
Prior to the LHC run, {\it the LHC inverse problem} was proposed~\cite{Arkani-Hamed:2005qjb} to determine the underlying physical theory giving rise to the signals expected to be seen at the LHC from the LHC data.
Similarly, in the SMEFT program, if there are observables with anomalous behavior, one would identify the relevant SMEFT operators, list all possible underlying UV physics, and then determine which UV physics is preferred from the anomaly. Borrowed from the LHC inverse problem, determining the underlying ultraviolet resonances from the SMEFT operators can be viewed as {\it the SMEFT inverse problem}. For a process with specific final states, we first identify possible effective operators contributing to this process, then list all possible UV resonances contributing to such operators, and finally using the data to discriminate among the UV particles and pin down the unique new physics. This procedure of solving the SMEFT inverse problem is essentially our ultimate goal in the bottom-up EFT framework, and would provide a new and systematic way of searching for new physics at the current and future colliders.

\section*{Acknowledgments}

We would like to thank Qing-Hong Cao for valuable discussion. J.H.Y. is supported by the National Science Foundation of China under Grants No. 12022514, No. 11875003 and No. 12047503, and National Key Research and Development Program of China Grant No. 2020YFC2201501, No. 2021YFA0718304, and CAS Project for Young Scientists in Basic Research YSBR-006, the Key Research Program of the CAS Grant No. XDPB15.
M.-L.X. is supported in part by the U.S. Department of Energy under contracts No. DE-AC02-06CH11357 at Argonne and No.DE-SC0010143 at Northwestern.
H.-L.L. is supported by F.R.S.-FNRS through the IISN convention "Theory of Fundamental Interactions” (N : 4.4517.08).



\appendix

\section{Massless and Massive Spinor-Helicity Notation}\label{app:spinor}

It is well known that single-particle states are defined as irreducible representations of the Poincare group, transformed under the little group as
$$
 U(W)|k,j,\sigma,\mathbf{r}\rangle =\sum_{\sigma'}D_{\sigma'\sigma}(W) |k,j,\sigma',\mathbf{r}\rangle,
$$
in which the little-group transformations of $k$, $\{W(\Lambda,p;k)\text{ }|\text{ }W(\Lambda,p;k)=L^{-1}(\Lambda p,k)\Lambda L(p,k)\}$, form a subgroup of Poincare group named the little group with the group generators
\bea
 w^\mu= \frac12\epsilon^{\mu\nu\rho\sigma} k_\nu M_{\rho\sigma}.
\eea
Such a generator $w^\mu$ can be expanded by polarizations of $k$ labelled as $\epsilon_{0,\pm}$, 
\begin{equation}
\begin{split}
  w_0&\equiv w\cdot\epsilon_0=imM_{\rho\sigma}\epsilon_+^\rho\epsilon_-^\sigma,\\
 w_+&\equiv\sqrt{2}w\cdot\epsilon_+=\sqrt{2}imM_{\rho\sigma}\epsilon_0^\rho\epsilon_+^\sigma,\\
 w_-&\equiv\sqrt{2}w\cdot\epsilon_-=-\sqrt{2}imM_{\rho\sigma}\epsilon_0^\rho\epsilon_-^\sigma.
\end{split}
\end{equation}
The corresponding little-group algebra is
\begin{equation}
  [w_0,w_\pm]=\pm mw_\pm,\quad
 [w_+,w_-]=2mw_0.
\end{equation}
Therefore the little group of a massless particle is locally isomorphic to $U(1)$, and $SU(2)$ for massive particles. 

The homomorphism between $SL(2,\mathbb{C})$ and $SO(3,1)$ is implemented by the spinor maps
\bea
 p_{\alpha\dot{\alpha}}\equiv p_\mu\sigma^\mu_{\alpha\dot{\alpha}},\quad
 p^{\dot{\alpha}\alpha}\equiv p_\mu{\bar{\sigma}}^{\mu \dot{\alpha}a},
\eea
where the Lorentz transformation of $p_{\alpha\dot{\alpha}}$ is
\bea
 p_{\alpha\dot{\alpha}}\rightarrow
 \zeta(\Lambda)_\alpha^\beta p_{\beta\dot{\beta}}
 {\zeta(\Lambda)^\dagger}^{\dot{\beta}}_{\dot{\alpha}}
 =\Lambda^\mu_\nu p^\nu {\sigma_\mu}_{\alpha\dot{\alpha}}.
\eea
The spinor indices $\alpha$ and $\dot{\alpha}$ can be raised and lowered by Levi-Civita tensors $\epsilon_{\alpha\beta},\epsilon^{\alpha\beta},\tilde{\epsilon}_{\dot{\alpha}\dot{\beta}},\tilde{\epsilon}^{\dot{\alpha}\dot{\beta}}$, where $\epsilon^{12}=\epsilon_{21}=\tilde{\epsilon}^{\dot{1}\dot{2}}=\tilde{\epsilon}_{\dot{2}\dot{1}}=1$. 

For massless particles, $\det[p_{\alpha\dot{\alpha}}]=0$, so $p_{\alpha\dot{\alpha}}$ is rank 1 and thus can be written as a direct product of two 2-vectors
\bea
 p_{\alpha\dot{\alpha}}=\lambda_{\alpha}\tilde{\lambda}_{\dot{\alpha}}.
\eea
The little group transformation of a massless spinor is 
\bea
 \zeta(W)_a^b\lambda_{\beta}=e^{-i\theta/2}\lambda_{\alpha},\quad
 \lambda_{\dot{\beta}}{\zeta(W)^\dagger}^{\dot{\beta}}_{\dot{\alpha}} =e^{i\theta/2}\lambda_{\dot{\alpha}}.
\eea

For massive particles, $p_{\alpha\dot{\alpha}}$ is rank 2, so their spinor variables need an extra index $I$,
\bea
 p_{\alpha\dot{\alpha}}=\lambda_{\alpha}^I\tilde{\lambda}_{\dot{\alpha}I}.
\eea
Since the little group of a massive particle is $SU(2)$, we can set $I$ as the little group index,
\bea
 \zeta(W)_{\alpha}^{\beta}\lambda_{\beta}^I=W^I_K\lambda_{\alpha}^K,\quad
 \lambda_{\dot{\beta}I}{\zeta(W)^\dagger}^{\dot{\beta}}_{\dot{\alpha}} =(W^\dagger)^K_I\lambda_{\dot{\alpha}K}.
\eea
The little group index $I$ can also be raised and lowered by $\epsilon^{IJ},\epsilon_{IJ}$. 
It can be set that
\bea
\lambda_\alpha^I\lambda^{\alpha J}=m\epsilon^{JI},\quad \tilde{\lambda}_{\dot{\alpha}I}\tilde{\lambda}^{\dot{\alpha}}_J =m\epsilon_{IJ}
\eea
based on the equation of motion of a massive particle
\bea
m^2=p^2=p_{\alpha\dot{\alpha}}p^{\dot{\alpha}\alpha} =(\lambda_\alpha^I\lambda^{\alpha J}) (\tilde{\lambda}_{\dot{\alpha}I}\tilde{\lambda}^{\dot{\alpha}}_J). 
\eea
The corresponding Dirac equation is
\bea
p_{\alpha\dot{\alpha}}\tilde{\lambda}^{\dot{\alpha}I} =m\lambda_{\alpha}^I,\quad 
p^{\dot{\alpha}\alpha}\lambda_{\alpha}^I =-m\tilde{\lambda}^{\dot{\alpha}I}, 
\eea
which implies that $\lambda^I$ and $\tilde{\lambda}^I$ can convert to each other. 

A Lorentz invariant building block satisfies that there is no free helicity indices $\alpha,\dot{\alpha}$, and for simplicity the bracket notation is applied,
\bea
{\lambda_i}^\alpha {\lambda_j}_\alpha\equiv \langle ij\rangle = \langle i|^\alpha |j\rangle_\alpha,\quad 
\tilde{\lambda}_i{}_{\dot{\alpha}} {\tilde{\lambda}_j}^{\dot{\alpha}}\equiv [ij] = [i|_{\dot{\alpha}} |j]^{\dot{\alpha}};\\
{\lambda_i}^{\alpha I} {\lambda_j}_{\alpha}^J =\langle i^I j^J\rangle \equiv \langle\mathbf{ij}\rangle,\quad
{{\tilde{\lambda}}_i}^I {}_{\dot{\alpha}} {{\tilde{\lambda}}_j}^{\dot{\alpha}J} =[i^I j^J] \equiv [\mathbf{ij}]. 
\eea
The labels of massive particles are in \textbf{bold}, denoting their free little group indices. Also, if a massive particle appears more than once within a Lorentz invariant term of an expression, its free little group indices are symmetrical. For example $\langle \mathbf{i}j\rangle [\mathbf{i}k] \equiv \langle i^{\{I}j\rangle [i^{J\}}k]$. 
It is obvious that $\langle ii\rangle=[ii]=0$ and $\langle\mathbf{ii}\rangle=[\mathbf{ii}]=0$.

\section{Generalized Procedure for j-basis Construction}\label{app:jbasis}


The diagonalization of the Poincar\'e Casimir $W^2$ may be subtle when we deal with scattering of 3 or more particles. We shall begin with the linear space of Lorentz structure of a class of external particles in dimension $d$, labeled as $V^d$, with a basis $\mc{B}_i$, and look at the scattering channel $\{\mc{I}|\bar{\mc{I}}\}$.  We used to assume that Eq.~\eqref{eq:W2action} holds,
but it requires that the image space $W_{\mathcal{I}}^2V^d$ is a subspace of $s_{\mathcal{I}}V^d$, which is itself a subspace of $V^{d+2}$, the space of $(d+2)$-dim local amplitudes among the same particles, so that we can express the images of $W^2$ as superposition of $s_{\mathcal{I}}\mc{B}^{(d)}_i$ and find $\mc{W}$. However, in general it is only guaranteed that $W_{\mathcal{I}}^2V^d \subset V^{d+2}$, which may or may not have intersection with $s_{\mathcal{I}}V^d$. The eigenbasis of $W_{\mathcal{I}}^2$, which we call the generalized partial waves, constitute a subspace of $V^d$, denoted as $\tilde{V}^d$, whose image under both maps form exactly the same subspace of $V^{d+2}$
\eq{\label{eq:iteration_end}
    W_{\mathcal{I}}^2 \tilde{V}^d = s_{\mc{I}} \tilde{V}^d \subset V^{d+2} ,\quad \tilde{V}^d = {\rm span}\{\mc{B}^{ j}_i\} \subset V^d.
}
In this section, we demonstrate the general steps of getting $\tilde{V}^d$.

We demonstrate with an example of the $\psi_1\psi_2\psi^{\dagger}_3\psi^{\dagger}_4\psi^{\dagger}_5\psi^{\dagger}_6$ class with $\{134|256\}$ 2-partite partial wave. Table~\ref{tab:4f{134|256}} provides y-basis for both $V^{d=9}$ and $V^{d=11}$ as $\mc{B}^y_{\psi^2\psi^{\dagger4}}$ and $\mc{B}^y_{\psi^2\psi^{\dagger4}D^2}$. It is straightforward to compute the basis for $W_{\mathcal{I}}^2V^{d=9}$ and $s_{\mathcal{I}}V^{d=9}$, both with dimension\footnote{In this case we don't have zero eigenvalues because the channel is fermionic. In case there may be zero eigenvalues, $W_{\mathcal{I}}^2V^d$ should be understood as including the null space of $W_{\mc{I}}^2$, as is shown in the next example. } $\mathfrak{d}_1=2$, by computing the coordinates of $W_{\{134\}}^2\mathcal{B}^y_{\psi^2\psi^{\dagger4}}$ and $s_{134}\mathcal{B}^y_{\psi^2\psi^{\dagger4}}$ on $V^{d=11}$. 
In this case, we find $W_{\mathcal{I}}^2V^d \not\subset s_{\mathcal{I}}V^d$, and we define $B_1 = W_{\mathcal{I}}^2V^d \cap s_{\mathcal{I}}V^d$, with dimension $\mathfrak{d}'_1=1 < \mathfrak{d}_1$ and a basis $\mc{B}^{(B_1)}$. We can explicitly find the projections from $W_{\mathcal{I}}^2V^d$ and $s_{\mathcal{I}}V^d$ to $B_1$ by solving
\eq{
    & \sum_i^{\mathfrak{d}_1}c_i W^2_{\{134\}}\mc{B}^y_{\psi^2\psi^{\dagger4},i} = \mc{B}^{(B_1)} \quad c_i = (\frac43,0), \\
    & \sum_i^{\mathfrak{d}_1}c'_i s_{134}\mc{B}^y_{\psi^2\psi^{\dagger4},i} = \mc{B}^{(B_1)} \quad c'_i = (-1,0).
}
Fortunately, they are proportional to each other, and that makes it the end of the iteration. We can define the basis of $\tilde{V}^{d=9}\equiv A_2$ as $\mc{B}^{(A_1)} = \mc{B}^y_{\psi^2\psi^{\dagger4},1}$, such that
\eq{
    W^2_{\{134\}} \mc{B}^{(A_2)} = -\frac43 s_{134} \mc{B}^{(A_2)},
}
which means that it is a $J=1/2$ partial wave. 

\begin{figure}
\centering
\includegraphics[width=1\textwidth]{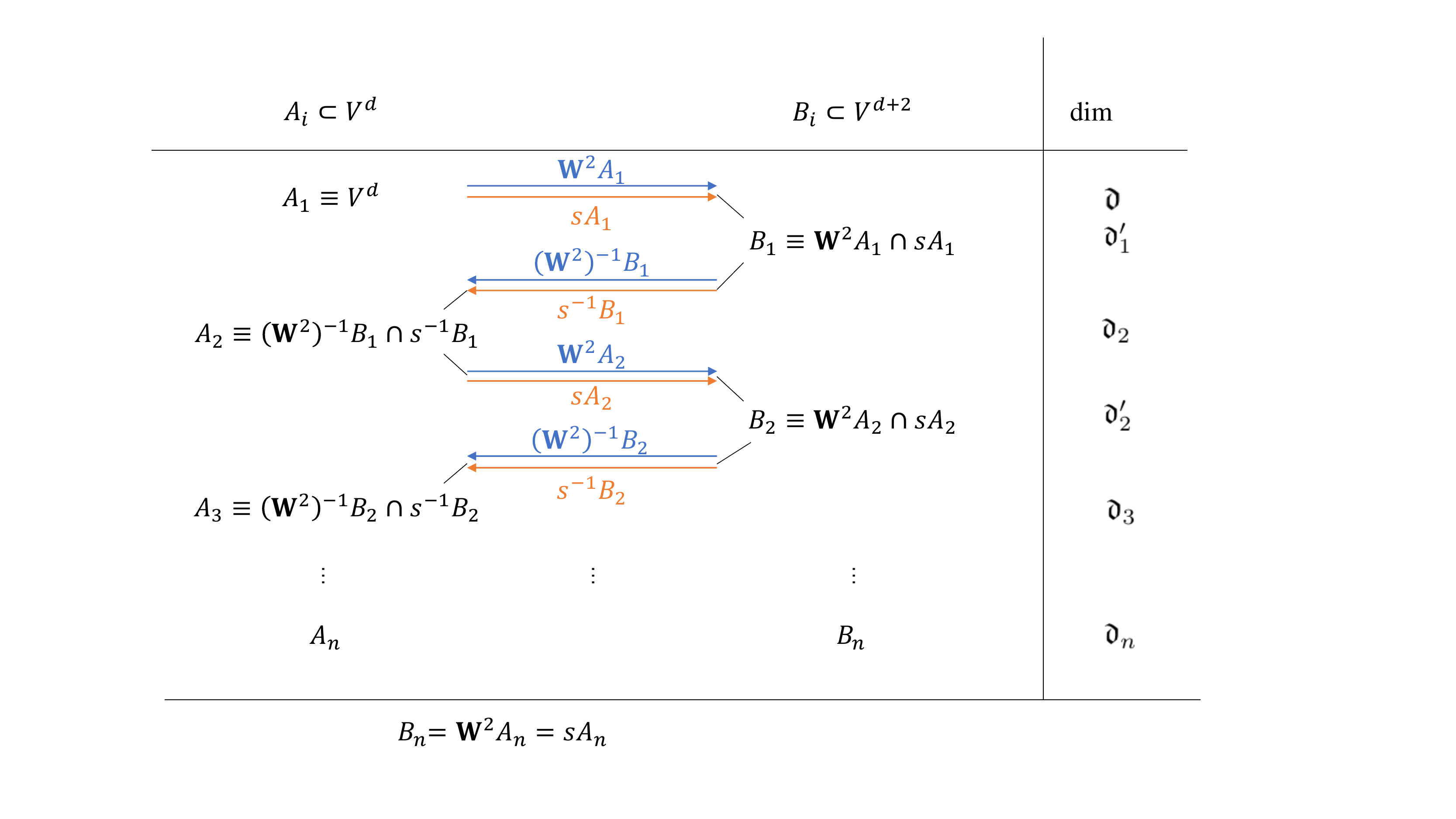}
\caption{The general procedure of Poincar\'e method for constructing Lorentz j-basis at dimension $d$. The spaces of amplitudes at dimension $d$ and $d+2$ are involved, whose subspaces are named $A_i$ and $B_i$. Starting from $A_1=V^d$, we apply two maps -- $W^2$ and the multiplication of $s$ -- to the $A_i$ space towards $V^{d+2}$, where we take the intersection as $B_i$; then we apply the inverse of the two maps backwards to $V^d$, where we also take the intersection as $A_{i+1}$. In each iteration, the dimensions of the subspaces $\mathfrak{d}^{(\prime)}$ get smaller, until we get to a pair of subspaces $A_n$, $B_n$ such that the two maps share the same image. The representation of $W^2$ restricted to $A_n$ can thus be solved. This procudure is implemented by the function \texttt{W2Diagonalize} in the package \href{https://abc4eft.hepforge.org/}{ABC4EFT}.}
\label{fig:caveat}
\end{figure}

There are two subtleties. First, we were lucky to have coinciding inverse image spaces for $B_1$ in the above case. In general we may have distinct inverse image spaces, whose intersection can be defined as $A_2$ with dimension $\mathfrak{d}_2$, while $A_1$ is of course denoting the original $V^d$. This sets up a well defined iteration with decreasing dimensional linear spaces of amplitudes $A_1^{\mathfrak{d}_1} \to B_1^{\mathfrak{d}'_1} \to A_2^{\mathfrak{d}_2} \to B_2^{\mathfrak{d}'_2} \to \dots A_n^{\mathfrak{d}} \to B_n^{\mathfrak{d}}$, with $A_i \subset V^d$ and $B_i \subset V^{d+2}$, which should end at either zero space $\mathfrak{d}=0$, or a finite-dimension space that satisfies Eq.~\eqref{eq:iteration_end} and thus we denote $\tilde{V}^d \equiv A_n$. The iteration is presented in the flow chart figure~\ref{fig:caveat}. Second, the zero eigenfunctions of $W^2$ will not appear in $B_1$ and its inverse images. We should manually include the kernel into $W^2 V^d$ by multiplying $s$, so that we would not miss anything in finding $A_2$ and so on. For example the $V^{d=8}$ for the $\phi^6D^2$ amplitudes have $B_1 = W^2_{\{123\}} V^{d=8} \cap s_{123} V^{d=8} = 0$, but it does have three zero eigenfunctions as $s_{45},s_{56},s_{64}$, obtained by identifying the null space of the operator $W^2$, which should be included in the final partial waves.

To understand why $\tilde{V}^d < V^d$ in some cases, we remind the readers that the partial wave bases are well defined as dimensionless angle distributions among the particles.
Take $\mc{B}^y_{\psi^2\psi^{\dagger4},2}$ as an example, the previous analysis showed that it is not a part of any j-basis at dimension 9. But the nondimensionalization of it must be some superposition of partial wave basis
\eq{
    s_{134}^{-3/2}\mc{B}^y_{\psi^2\psi^{\dagger4},2} = s_{134}^{-3/2}\vev{12}[35][46] = \frac{1}{3}\overline{\mc{B}}^{J=1/2}_1 -\frac{1}{4} \overline{\mc{B}}^{J=1/2}_2 + \overline{\mc{B}}^{J=3/2}\,,
}
where $\overline{\mc{B}}^{J=1/2}_1 = s_{134}^{-3/2}\vev{12}[34][56]$ is the nondimensionalization of $\mc{B}^y_{\psi^2\psi^{\dagger4},1}$. 
When lifted to local amplitudes at dimension $d$, several powers of $s_{\mc{I}}$ have to be multiplied to the partial wave basis, and there is a least number of that for certain partial wave. It turns out that the last two terms in the above equation needs at least $s_{134}^{5/2}$, and thus do not turn to local amplitudes at $d=9$. It forbids the decomposition of $\mc{B}^y_{\psi^2\psi^{\dagger4},2}$ onto some j-basis at $d=9$ level, and implies a smaller $\tilde{V}^{d=9}$ than $V^{d=9}$. It is all because we work in subspaces of local amplitudes at given dimensions, which is not necessarily good subspaces for partial waves. But this feature has interesting phenomenological implication: some operators at given dimension cannot be generated at tree level by resonances in certain channels because of lack of j-basis amplitudes.

\begin{table}[htbp]
    \centering
    \begin{tabular}{c|c|c|c}
 & $\mathcal{B}^y_{\psi^2\psi^{\dagger4}}$ & $s_{134}\mathcal{B}^y_{\psi^2\psi^{\dagger4}}$ & $W_{134}^2\mathcal{B}^y_{\psi^2\psi^{\dagger4}}$\\
 \hline
 \makecell{$d=9$\\ $\psi^2\psi^{\dagger4}$} & \makecell{ $[34] [56] \langle 12\rangle$, $[35] [46] \langle 12\rangle$. } & &\\
 \hline
 \makecell{$d=11$\\ $\psi^2\psi^{\dagger4} D^2$ }& \makecell{ $-[34] [56] s_{56} \langle 12\rangle$, $-[35] [46] s_{56} \langle 12\rangle$,\\$[34] [56] s_{46} \langle 12\rangle$, $-[34] [56] s_{45} \langle 12\rangle$,\\$[35] [46] s_{46} \langle 12\rangle$, $-[35] [46] s_{45} \langle 12\rangle$,\\$-[35] [46] s_{36} \langle 12\rangle$, $[35] [46] s_{35} \langle 12\rangle$,\\$-[34] [56] s_{36} \langle 12\rangle$, $[34] [56] s_{35} \langle 12\rangle$,\\$-[34] [56] s_{34} \langle 12\rangle$, $-[35] [46] s_{34} \langle 12\rangle$,\\$[34] [56]^2 \langle 15\rangle  \langle 26\rangle$, $-[34] [45] [46] \langle 14\rangle  \langle 24\rangle$,\\$-[36] [46] [56] \langle 16\rangle  \langle 26\rangle$, $-[35] [45] [56] \langle 15\rangle  \langle 25\rangle$,\\$[35] [46] [56] \langle 15\rangle  \langle 26\rangle$, $-[34] [46] [56] \langle 14\rangle  \langle 26\rangle$,\\$[34] [45] [56] \langle 14\rangle  \langle 25\rangle$, $-[35] [46]^2 \langle 14\rangle  \langle 26\rangle$,\\$[35] [45] [46] \langle 14\rangle  \langle 25\rangle$, $-[34] [35] [36] \langle 13\rangle  \langle 23\rangle$,\\$[35] [36] [46] \langle 13\rangle  \langle 26\rangle$, $-[35]^2 [46] \langle 13\rangle  \langle 25\rangle$,\\$[34] [36] [56] \langle 13\rangle  \langle 26\rangle$, $-[34] [35] [56] \langle 13\rangle  \langle 25\rangle$,\\$[34]^2 [56] \langle 13\rangle  \langle 24\rangle$, $[34] [35] [46] \langle 13\rangle  \langle 24\rangle$. } & \makecell{ $[34][35][56]\langle 13\rangle \langle 25\rangle$\\ $+[34] [45] [56] \langle 14\rangle  \langle 25\rangle$\\ $+[34] [36] [56] \langle 13\rangle  \langle 26\rangle$\\ $+[34] [46] [56] \langle 14\rangle  \langle 26\rangle$,\\ \\ $[35]^2 [46] \langle 13\rangle  \langle 25\rangle$\\ $+[35] [36] [46] \langle 13\rangle  \langle 26\rangle$\\ $+[35] [45] [46] \langle 14\rangle  \langle 25\rangle$\\ $+[35] [46]^2 \langle 14\rangle  \langle 26\rangle$. } & \makecell{ $-\frac{3}{4} [34] [35] [56] \langle 13\rangle  \langle 25\rangle$\\ $-\frac{3}{4} [34] [36] [56] \langle 13\rangle  \langle 26\rangle$\\ $-\frac{3}{4} [34] [45] [56] \langle 14\rangle  \langle 25\rangle$\\ $-\frac{3}{4} [34] [46] [56] \langle 14\rangle  \langle 26\rangle$,\\ \\ 
 $2 [34] [35] [56] \langle 13\rangle  \langle 25\rangle$\\ $-\frac{15}{4} [35]^2 [46] \langle 13\rangle  \langle 25\rangle$\\ $+[34] [36] [56] \langle 13\rangle  \langle 26\rangle$\\ $-\frac{15}{4} [35] [36] [46] \langle 13\rangle  \langle 26\rangle$\\ $+[34] [45] [56] \langle 14\rangle  \langle 25\rangle$\\ $-\frac{15}{4} [35] [45] [46] \langle 14\rangle  \langle 25\rangle$\\ $+2 [34] [46] [56] \langle 14\rangle  \langle 26\rangle$\\ $-\frac{15}{4} [35] [46]^2 \langle 14\rangle  \langle 26\rangle$. } \\
 \cline{3-4}
 & & \multicolumn{2}{c}{\makecell{$\tilde{V}^9\equiv W_{134}^2[\mathcal{B}^y_{\psi^2\psi^{\dagger4}}] \cap s_{134}[\mathcal{B}^y_{\psi^2\psi^{\dagger4}}] =$\\ $(-[34] [35] [56] \langle 13\rangle  \langle 25\rangle -[34] [36] [56] \langle 13\rangle  \langle 26\rangle$ \\$ -[34] [45] [56] \langle 14\rangle  \langle 25\rangle -[34] [46] [56] \langle 14\rangle  \langle 26\rangle)$. }}
    \end{tabular}
    \caption{Y-basis and j-basis of $\psi^2\bar{\psi}^4$ at partition $\{134|256\}$.}
    \label{tab:4f{134|256}}
\end{table}

\section{Intersection of Two Linear Spaces}\label{app:linearintersection}
We provide here a general algorithm for obtaining the basis of the intersection space of two linear subspaces $U$ and $V$ of dimension $n$ and $m$ embedded in a larger linear space $K$ of dimension $p$. Suppose the basis vectors of the linear space $K$ are denoted as $\{\hat{\boldsymbol e}_i\}$, and the basis vectors of $U$ and $V$ are denoted as $\{\hat{\boldsymbol u}_i\}$ and $\{\hat{\boldsymbol v}_i\}$ respectively. Since $U$ and $V$ are subspaces of $K$, one can express the basis their vectors in terms of $\{\hat{\boldsymbol e}_i\}$:
\begin{equation}
    \hat{\boldsymbol u}_i=\sum_j^p u_{ij}\hat{\boldsymbol e}_j,\quad \hat{\boldsymbol v}_i=\sum_j^p v_{ij}\hat{\boldsymbol e}_j.
\end{equation}
Vectors lie in the intersection space of $U$ and $V$ must be able to spanned by the basis vectors of $U$ and $V$ simultaneously:
\begin{equation}
    \forall\ {\boldsymbol I}\in U\cap V,\ \exists\  c_i,\ c'_i,\ s.t.\ \  {\boldsymbol I}=\sum_{i=1}^{n}c_i\hat{\boldsymbol u}_i=-\sum_{i=1}^{m}c'_i\hat{\boldsymbol v}_i.
\end{equation}
The above equation, can be written in the basis of the original space $K$ as:
\begin{eqnarray}
   \sum_{i=1}^{n}c_i\hat{\boldsymbol u}_i+\sum_{i=1}^{m}c'_i\hat{\boldsymbol v}_i=0\nonumber \to 
   \sum_{j=1}^p\left(\sum_{i=1}^{n}c_iu_{ij}+\sum_{i=1}^{m}c'_iv_{ij}\right)\hat{\boldsymbol e}_j=0,
\end{eqnarray}
which is equivalent to the following matrix equation:
\begin{equation}
 \mathbf{M}\cdot \mathbf{C}=\begin{pmatrix}
           u_{11} & u_{21} & \cdots & u_{p1}\\
            \vdots & \vdots & \ddots & \vdots \\
 u_{1n}&u_{2n}& \cdots & u_{pn}\\
   v_{11} & v_{21} & \cdots & v_{p1}\\
 \vdots & \vdots & \ddots & \vdots \\
   v_{1m}&v_{2m}& \cdots & v_{pm}\\
        \end{pmatrix}^\intercal \cdot
\begin{pmatrix}
           c_{1}\\
            \vdots  \\
 c_n\\
   c'_{1} \\
 \vdots  \\
   c'_{m}\\
        \end{pmatrix}=0,
\end{equation}
Therefore we just need to find the null space of $\mathbf{M}$ composed of $\{\mathbf{C}_i\}$, and then the basis of the intersection space is obtained by
\begin{equation}
    \hat{\boldsymbol I}_i=\sum_{j=1}^{n}(\mathbf{C}_i)_j\hat{\boldsymbol u}_j=\sum_{j=n+1}^{n+m}(\mathbf{C}_i)_j\hat{\boldsymbol v}_j.
\end{equation}
Here we provide a simple example to illustrate the algorithm. Suppose the linear space $K\sim \mathbbm{R}^3$ is spanned by three basis vectors $\{\hat{\boldsymbol e}_1,\hat{\boldsymbol e}_2,\hat{\boldsymbol e}_3\}$, and the subspace $U$ is the $1-2$ plane spanned by $\{\hat{\boldsymbol u}_1=\hat{\boldsymbol e}_1+\hat{\boldsymbol e}_2,\hat{\boldsymbol u}_2=\hat{\boldsymbol e}_1-\hat{\boldsymbol e}_2\}$, while the subspace $V$ is the $1-3$ plane spanned by $\{\hat{\boldsymbol v}_1=\hat{\boldsymbol e}_1+\hat{\boldsymbol e}_3,\hat{\boldsymbol v}_2=\hat{\boldsymbol e}_1-\hat{\boldsymbol e}_3\}$. From the concrete form of basis vectors, one can read off the matrix $\mathbf{M}$:
\begin{eqnarray}
\mathbf{M}=\begin{pmatrix}
1 & 1& 1& 1\\
1& -1 &0 &0\\
0& 0& 1& -1
\end{pmatrix}
\end{eqnarray}
The null space have only one vector $\mathbf{C}_1=(-1,-1,1,1)$, so the basis vector is:
$-\hat{\boldsymbol u}_1-\hat{\boldsymbol u}_1=-2\hat{\boldsymbol e}_1$ or equivalently $\hat{\boldsymbol v}_1+\hat{\boldsymbol v}_1=2\hat{\boldsymbol e}_1$, indeed, the intersection of the two plane is the $\hat{\boldsymbol e}_1$-axis. 

\section{General Casimir Operator for $SU(N)$ Group}\label{app:casimir}
In the main text, we introduce the first two Casimir operators and use their eigenvalue to classify the irreducible representation for $SU(2)$ and $SU(3)$. Here we discuss the $n$-th order Casimir operators $\mathbbm{C}_n$ for the $SU(N)$ group, which can be expressed as a $n$-th order homogeneous polynomial of the generators $\mathbbm{T}^A$'s. As we have already seen in Eq.~\eqref{eq:CasimirC2} and \eqref{eq:CasimirC3} as examples, $\mathbbm{C}_2$ and $\mathbbm{C}_3$ are 2-nd and 3-rd order polynomial of $\mathbbm{T}^A$'s. It can be proven that for $SU(N)$ group, one can find $N-1$ Casimir operators $\mathbbm{C}_1,\dots,\mathbbm{C}_{N-1}$, such that they commute with all the generators in the group. The general expression for  $\mathbbm{C}_n$ can be expressed in the following:
\bea
 \mathbbm{C}_n =f^{A_1D_1D_2} f^{A_2D_2D_3} ... f^{A_nD_nD_1}  \mathbbm{T}^{A_1}...\mathbbm{T}^{A_n}.
\eea
where the invariant tensor $f^{ABC}$ in the above formula is the structure constant. For example, in the $SU(2)$ group it is the 3-rd rank Levi-Civita tensor $\epsilon^{ABC}$, therefore following the above formula we get $\mathbbm{C}_2$ for $SU(2)$ as:
\begin{eqnarray}
\mathbbm{C}_2=\epsilon^{A_1 D_1D_2}\epsilon^{A_2 D_2 D_1}\mathbbm{T}^{A_1}\mathbbm{T}^{A_2}=-2\delta^{A_1A_2}\mathbbm{T}^{A_1}\mathbbm{T}^{A_2},
\end{eqnarray}
Which is equivalent to what we defined in Eq.~\eqref{eq:CasimirC2}, $\delta^{A_1A_2}\mathbbm{T}^{A_1}\mathbbm{T}^{A_2}$, up to an normalization factor.
While for $\mathbbm{C}_3$ for $SU(3)$, we have:
\begin{eqnarray}
&\mathbbm{C}_3&=f^{A_1 D_1 D_2}f^{A_2 D_2 D_3}f^{A_2 D_3 D_1}\mathbbm{T}^{A_1}\mathbbm{T}^{A_2}\mathbbm{T}^{A_3}\nonumber \\
&&=(if^{A_1A_2A_3}+d^{A_1A_2A_3})\mathbbm{T}^{A_1}\mathbbm{T}^{A_2}\mathbbm{T}^{A_3}\nonumber \\
&&=\frac{i}{2}f^{A_1A_2A_3}[\mathbbm{T}^{A_1},\mathbbm{T}^{A_2}]\mathbbm{T}^{A_3}+d^{A_1A_2A_3}\mathbbm{T}^{A_1}\mathbbm{T}^{A_2}\mathbbm{T}^{A_3}\nonumber \\
&&=\frac{-1}{2}f^{A_1A_2A_3}f^{A_1A_2D}\mathbbm{T}^{D}\mathbbm{T}^{A_3}+d^{A_1A_2A_3}\mathbbm{T}^{A_1}\mathbbm{T}^{A_2}\mathbbm{T}^{A_3}\nonumber \\
&&=\frac{1}{2}\mathbbm{C}_2 +d^{A_1A_2A_3}\mathbbm{T}^{A_1}\mathbbm{T}^{A_2}\mathbbm{T}^{A_3}.
\end{eqnarray}
As one can see from the above formula, the non-trivial part of $\mathbbm{C}_3$ is $d^{A_1A_2A_3}\mathbbm{T}^{A_1}\mathbbm{T}^{A_2}\mathbbm{T}^{A_3}$, this is why we define this term in as $\mathbbm{C}_3$ in Eq.~\eqref{eq:CasimirC3}.   

Since Casimir operators commute with all the generators, for irreducible representation $\mathbf{R}$, its matrix representation must be proportional to the identity matrix, and we call the corresponding coefficients as $C_n(\mathbf{R})$, we have the following iterative method to determine the constant $C_n(\mathbf{R})$:
\bea
 &C_2=S_2,&\quad C_3=S_3-(n-\frac{3}{2})S_2,...,\\
 &S_n=\sum_{i=1}^N({l_i}^n-{\rho_i}^n),&\quad (\rho_i=N-i,\quad l_i=\lambda_i-\sum_k\lambda_k+N-i).\nonumber
\eea
Here we list the value of $C_2$ and $C_3$ of common irreducible representation for $SU(2)$ and $SU(3)$ group for reader's convenience in table.~\ref{tab:c2eigen} and \ref{tab:c3eigen}. 
\begin{table}[H]
\centering
\begin{tabular}{c|c|cc|cc|cc|cc|cc|cc}
\multirow{4}{*}{$\mathbb{C}_3,SU(3)$} & $\mathbf{R}$ & 
\multicolumn{2}{c|}{$\mbf{1,8,27,64,125}$} & $\mbf{3}$ & $\mbf{\bar{3}}$ & $\mbf{ 6}$ & $\bar{\mbf{6}}$ & $\mbf{10}$ & $\bar{\mbf{10}}$ & $\mbf{15}$ & $\overline{\mbf{15}}$ & $\mbf{15'}$ & $\overline{\mbf{15'}}$
\\
\cline{2-14}
& $C_3(\mathbf{R})$ & 
\multicolumn{2}{c|}{$0$} & $\frac{10}{9}$ & $-\frac{10}{9}$ & $-\frac{35}{9}$ & $\frac{35}{9}$ & $9$ & $-9$ & $\frac{28}{9}$ & $-\frac{28}{9}$ & $\frac{154}{9}$ & $-\frac{154}{9}$\\
\cline{2-14}
& $\mathbf{R}$ &
$\mbf{21}$ & $\mbf{\overline{21}}$ & $\mbf{24}$ & $\overline{\mbf{24}}$ & $\mbf{35}$ & $\overline{\mbf{35}}$ & $\mbf{42}$ & $\overline{\mbf{42}}$ & $\mbf{60}$ & $\overline{\mbf{60}}$ & $\mbf{90}$ & $\overline{\mbf{90}}$ \\
\cline{2-14}
& $C_3(\mathbf{R})$ & 
$-\frac{260}{9}$ & $\frac{260}{9}$ & $-\frac{80}{9}$ & $\frac{80}{9}$ & $18$ & $-18$ & $\frac{55}{9}$ & $-\frac{55}{9}$ & $-\frac{143}{9}$ & $\frac{143}{9}$ & $\frac{91}{9}$ & $-\frac{91}{9}$ 
\end{tabular}
\caption{The eigenvalues of $\mathbb{C}_3$ for commonly used representations of $SU(3)$. }
\label{tab:c3eigen}
\end{table}

\begin{table}[H]
\centering
\begin{tabular}{c|c|c|c|c|c|c|c|c|c}
\multirow{4}{*}{$\mathbb{C}_2,SU(3)$} & $\mathbf{R}$ &
$\mbf{1}$ & $\mbf{3,\bar{3}}$ & $\mbf{6,\bar{6}}$ & $\mbf{8}$ & $\mbf{10,\overline{10}}$ & $\mbf{15,\overline{15}}$ & $\mbf{15',\overline{15'}}$ & $\mbf{21,\overline{21}}$
\\
\cline{2-10}
& $C_2(\mathbf{R})$ & $0$ & $\frac{4}{3}$ & $\frac{10}{3}$ & $3$ & $6$ & $\frac{16}{3}$ & $\frac{28}{3}$ & $\frac{40}{3}$\\
\cline{2-10}
& $\mathbf{R}$ & $\mbf{24,\overline{24}}$ & $\mbf{27}$ & $\mbf{35,\overline{35}}$ & $\mbf{42,\overline{42}}$ & $\mbf{60,\overline{60}}$ & $\mbf{64}$ & $\mbf{90,\overline{90}}$ & $\mbf{125}$ \\
\cline{2-10}
& $C_2(\mathbf{R})$ & $\frac{25}{3}$ & $8$ & $12$ & $\frac{34}{3}$ & $\frac{46}{3}$ & $15$ & $\frac{58}{3}$ & $24$ \\
\hline\hline
\multirow{2}{*}{$\mathbb{C}_2,SU(2)$} & $\mathbf{R} $ & $\mbf{1}$ & $\mbf{2}$ & $\mbf{3}$ & $\mbf{4}$ & $\mbf{5}$ & $\mbf{6}$ & $\mbf{7}$ & $\mbf{8}$\\
\cline{2-10}
& $C_2(\mathbf{R})$ & $0$ & $\frac{3}{4}$ & $2$ & $\frac{15}{4}$ & $6$ & $\frac{35}{4}$ & $12$ & $\frac{63}{4}$
\end{tabular}
\caption{The eigenvalues of $\mathbb{C}_2$ for commonly used representations of $SU(3)$ and $SU(2)$. }
\label{tab:c2eigen}
\end{table}

\section{Casimir for Direct Product Representations}\label{app:E}
In Eq.~\eqref{eq:Trep} and \eqref{eq:partialT}, we define the action of the group generators on the tensors, and then derive the matrix representation for Casimirs acting on parts of indices following these  definitions, which finally leads to our master formula Eq.~\eqref{eq:master-gauge}. A caveat in this derivation is that the resulting tensors by the action of the generators on our gauge basis --- a complete basis of invariant tensors for a set of representations $\{{\bf r}_i\}$ is not within the space spanned by the original gauge basis, in other words, the resulting tensor may not be an invariant tensor. Therefore one cannot define the matrix representation of the generators on our gauge basis, though that of Casimir operators is still a valid. This is a bit peculiar since we usually construct the matrix representation of Casimir operators directly with the matrix product of generators. In the following we will show more clearly the meaning of partial generators and partial Casimir operators.

\begin{figure}[htb]
\centering
\includegraphics[width=250pt]{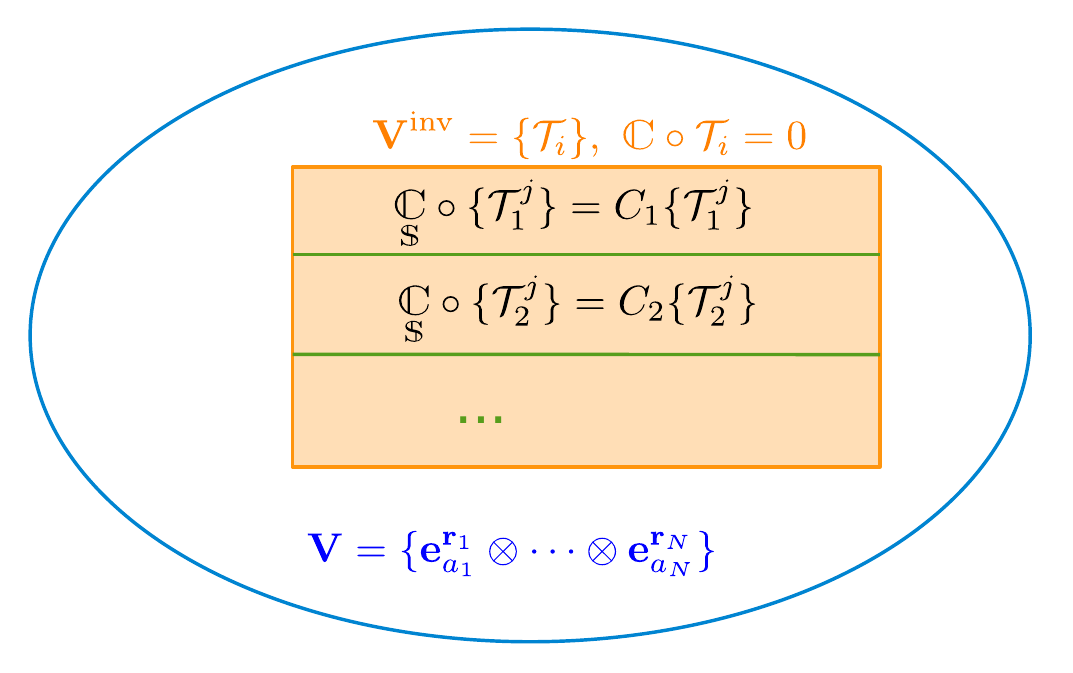}
\caption{A pictorial demonstration of our gauge basis and gauge j-basis in the space of the direct product representations. The blue circle represents the whole space of the direct product representation. The orange box represents the space of gauge invariant tensors spanned by our gauge basis, which is also the null space of the full Casimir operators ${\bf V}^{{\rm inv}}$. Since the partial Casimir and the full Casimir commute, one can further divide the ${\bf V}^{{\rm inv}}$ according to the eigenvalue of $\underset{\mathbbm{S}}{\mathbbm{C}}$, which results in our gauge j-basis}
\label{fig:gaugeapp}
\end{figure}

First we clarify that a general tensor $\Theta_{a_1,\dots,a_N}$ is  the tensor component of a specific vector $\bf \Theta$ in the direct product space $\bf V$ spanned by the abstract basis vectors $\{{\bf e}^{{\bf r}_1}_{a_1}\otimes  {\bf e}^{{\bf r}_2}_{a_2}\dots\otimes  {\bf e}^{{\bf r}_N}_{a_N}\}$, that is:
\begin{eqnarray}
{\bf \Theta} = \Theta_{a_1,\dots,a_N} {\bf e}^{{\bf r}_1}_{a_1}\otimes  \dots\otimes  {\bf e}^{{\bf r}_N}_{a_N}, 
\end{eqnarray}
where we have implicitly used the Einstein summation convention among the indices $\{a_i\}$. The action of (partial) generators on the vector $\bf \Theta$ can either be interpreted as the active transformation of the tensor component $\Theta_{a_1,\dots,a_N}\to \Theta'_{a_1,\dots,a_N}$ or the passive transformation of the basis vectors with components unchanged: 
\begin{eqnarray}
 \underset{\mathbbm{S}}{\mathbbm{T}}^A\circ{\bf e}^{{\bf r}_1}_{a_1}\otimes  \dots\otimes  {\bf e}^{{\bf r}_N}_{a_N}=
\sum_{i\in \mathbbm{S}}^N (T^A_{r_i})_{Za_{i}} {\bf e}^{{\bf r}_1}_{a_1}\otimes \dots {\bf e}^{{\bf r}_i}_{Z} \dots\otimes  {\bf e}^{{\bf r}_N}_{a_N}.
\end{eqnarray}
With the above notation and convention, following statements are easily verified:
\begin{itemize}
    \item In the space $\bf V$, the matrix representations for both the full generators and the partial generators are well defined, i.e. the action of (partial) generators on any vector $\bf \Theta$ is still in the space $\bf V$.
    \item The invariant tensors or the gauge basis ${\cal T}_{a_1,\dots,a_N}$ correspond to vectors  ${\bf {\Theta}}^{{\rm inv}}$ that are invariant under the group transformations or equivalently the null eigenvectors of the full Casimir operators which span the null subspace   ${\bf V}^{\rm inv}$.
    \item The full Casimir operators  ${\mathbbm{C}}$ and the partial Casimir operators  $\underset{\mathbbm{S}}{\mathbbm{C}}$ commute.
\end{itemize}
From the above three statements one can conclude that the null space of the ${\mathbbm{C}}$, which correspondes to the space  spanned by our gauge basis must be invariant under the partial Casimir operators  $\underset{\mathbbm{S}}{\mathbbm{C}}$ and has matrix representations of them:
\begin{equation}
     {\mathbbm{C}}\circ \left( \underset{\mathbbm{S}}{\mathbbm{C}}\circ {\bf {\Theta}}^{{\rm inv}}\right) = \underset{\mathbbm{S}}{\mathbbm{C}}\circ\left({\mathbbm{C}}\circ   {\bf {\Theta}}^{{\rm inv}} \right)= 0\to \left( \underset{\mathbbm{S}}{\mathbbm{C}}\circ {\bf {\Theta}}^{{\rm inv}}\right)\in {\bf V}^{\rm inv}.
\end{equation}
Therefore the null space of ${\bf V}^{{\rm inv}}$ can be further divided into subspace corresponds to the eigenspace of the partial Casimir $\underset{\mathbbm{S}}{\mathbbm{C}}$, the eigenvectors of which is exactly our gauge j-basis. A pictorial demonstration of this relation is shown in figure.~\ref{fig:gaugeapp}.

\bibliographystyle{JHEP}
\bibliography{UVSMEFTref}

\end{document}